\documentclass[12pt,authoryear]{elsarticle}%
\usepackage{epsfig}

\usepackage{nomencl}%

\usepackage{amssymb}
\usepackage{amsthm}

\usepackage{lineno}

\usepackage{etex}
\usepackage{graphicx}
\usepackage{epstopdf, epsfig}
\usepackage[T1]{fontenc}
\usepackage{tikz}
\usepackage{calc}

\usepackage{amssymb}
\usepackage{booktabs}

\usepackage[english]{babel}
\usepackage[T1]{fontenc}
\usepackage[utf8]{inputenc}
\usepackage{tabu}
\usepackage[usestackEOL]{stackengine}
\setstackgap{L}{\normalbaselineskip}
\usepackage{soul} %
\usepackage{notoccite} %
\usepackage{mathtools} %

\usepackage{svg}
\usepackage[hidelinks]{hyperref}
\usepackage{cleveref}
\crefname{figure}{Fig.}{Fig.}
\crefname{equation}{Eq.}{Eq.}

\usepackage{bm}

\usepackage{placeins}

\newcommand{\sci}[1] {\times 10^{#1}} %
 
\newcommand{\gr}{\geq}
\newcommand{\ie}{i.e.\ }%

\newcommand{\ca}{C_A}
\newcommand{\eg}{e.g.\ }

\newcommand{\ppi}{\phi/\pi}

\newcommand{\re}{Re}

\newcommand{\lam}{\lambda^*}

\newcommand{\uinf}{u_{\infty}}

\newcommand{\ctm}{C_{Tm}}
\newcommand{\clrms}{C_{Lrms}}
\newcommand{\etag}{\eta_{group}}

\newcommand{\tdT}{t^*/T}

\newcommand*\diff{\mathop{}\!\mathrm{d}}

\journal{Ocean Engineering}

\begin{document}
	
	\begin{frontmatter}

		\title{How swimming style affects schooling
			\\
			of two fish-like wavy hydrofoils}

		\author[inst1,inst2,inst3]{Zhonglu Lin}

		\author[inst4]{Amneet Pal Singh Bhalla}
		
		\author[inst5]{Boyce Griffith}

		\author[inst1,inst2]{Zi Sheng}
		
		\author[inst1,inst2]{Hongquan Li}

		\author[inst3]{Dongfang Liang}
		
		\author[inst1,inst2]{Yu Zhang\corref{cor1}}
		\ead{yuzhang@xmu.edu.cn}
		\cortext[cor1]{Corresponding author}
		
		\affiliation[inst1]{
			organization={State Key Laboratory of Marine Environmental Science},%
			addressline={College of Ocean and Earth Sciences, Xiamen University}, 
			country={China}
		}

		\affiliation[inst2]{
			organization={Key Laboratory of Underwater Acoustic Communication and Marine Information Technology of the Ministry of Education},%
			addressline={College of Ocean and Earth Sciences, Xiamen University}, 
			country={China}
		}
		
		\affiliation[inst3]{
			organization={Engineering Department, University of Cambridge},%
			country={United Kingdoms}
		}
		
		\affiliation[inst4]{
			organization={Department of Mechanical Engineering, San Diego State University},%
			country={United States}
		}
		
		\affiliation[inst5]{
			organization={Department of Mathematics, The University of North Carolina at Chapel Hill},%
			country={United States}
		}

		\begin{abstract}
			
			Fish swimming style and schooling behaviour are two important aspects of underwater locomotion.
			How swimming style affects fish schooling is investigated by a representative problem setup consisting of two tethered NACA0012 hydrofoils undulating in a free stream flow at various wavelengths $ \lam = 0.8 - 8 $, front-back distance $ D = 0,\ 0.25,\ 0.5,\ 0.75 $, phase difference $ \ppi = 0,\ 0.5,\ 1,\ 1.5 $, and lateral gap distance $ G = 0.25,\ 0.3,\ 0.35 $ with fixed Reynolds number $ Re = 5000 $, Strouhal number $ St = 0.4 $, and maximum amplitude $ A_{max} = 0.1 $. In total, 336 cases were simulated by  open-source software IBAMR based on immersed boundary method.
			The increase in wavelength generally leads to higher thrust and lateral force, consistent with a previous study regarding a single swimmer.
			The highest Froude efficiency is obtained at intermediate wavelength $ \lam = 2 $ for either individual or two foils as a group.
			At side-by-side arrangement $ D = 0 $, the thrust force upon the two foils can be equivalent, indicating a stable formation.
			At staggered arrangement, $ D > 0 $, the follower can take significant advantage of the leader in locomotion performance by tuning phase difference. The follower's benefit decreases with relative distance.
			Various combinations of wavelength and relative distance can lead to distinct flow structures, indicating a tunable stealth capacity of the schooling swimmers.

		\end{abstract}

		\begin{highlights}
			\item It is the first attempt to systematically investigate the combined effect of swimming style and schooling for fish-like swimmers.
			\item By tuning undulation phase difference, the follower can significantly promote its acceleration and energy efficiency in various wavelengths.
			\item Highest Froude efficiency for individual/group is obtained at intermediate wavelength regardless of relative distance.
			\item Distinct flow structure can be formed by various combinations of wavelength and relative distance, indicating tunable stealth capacity.
			\item Thunniform swimmers could be more suitable for schooling compared with Anguilliform swimmers.
		\end{highlights}
		
		
	\end{frontmatter}
	
	%
	%
	\section{Introduction}

	Fish swimming has been an attractive research topic for many years due to its interdisciplinary nature, involving hydrodynamics \citep{Triantafyllou2000,Liao2007}, schooling \citep{Ashraf2017}, swimming styles \citep{Webb1984}, muscle anatomy \citep{Altringham1999}, physiology \citep{HuntvonHerbing2002} and sensory \citep{Liu2016}. The research has been driven by the fish farming \citep{Webb2011} and by biomimetic underwater robotics \citep{Duraisamy2019,Fish2020}.
	\textit{fish schooling} and \textit{fish swimming style} are two interesting sub-categories among the fish swimming problems. Multiple factors can influence the evolved schooling behaviour and swimming style, including hydrodynamic effectiveness, predator defence, feeding, etc.
	Body-caudal fin is one of the most common swimming styles among fish species \citep{Webb1984}, with typical examples of its sub-swimming forms shown in \Cref{fig:4modes}.
	In this study, we focus on the effects of \textit{fish schooling} and \textit{BCF swimming styles} on the hydrodynamics characteristics using simplified physical models. This work is inspired by the biological observation that schooling behaviour of BCF swimmers is, to our best knowledge, only found in \textit{less wavy} swimming forms, \eg thunniform by tunas \citep{Dai2020}, whereas schooling behaviour is undiscovered in \textit{more wavy} swimming forms, \eg anguilliform by eels.

	Fish schooling has been extensively studied in the past decades for its hydrodynamic characteristics \citep{weihs1973hydromechanics,Weihs1975}.
	Cross sections of two schooling fish can be represented by two vibrating cylinders immersed in quasi-static fluid; it has been identified by theoretical \citep{Hlamb1932hydrodynamics,NAIR2007} and numerical \citep{Gazzola2012,Lin2018c,Lin2018b,Lin2019} methods that non-dimensional parameters such as phase difference can have a distinct impact on the flow-mediated interaction between the two cylinders.
	\cite{Shaw1978} estimated that at least 25\% of all fish species demonstrate schooling behaviour.
	Many hydrodynamic studies have focused on the minimal school composed of 2 identical BCF swimmers, using robotic fish \citep{Li2020}, biological fish \citep{Ashraf2016,Li2020}, hydrofoil experiments \citep{Dewey2014,Kurt2020} and numerical simulation \citep{Khalid2016,Li2019}.
	\cite{Ashraf2016} discovered a pair of red nose tetra fish tend to swim either in-phase or anti-phase, with the latter mode being more favourable; lateral and front-back distances are around 0.5 and 0.2 fish body length. For this reason, we place more emphasis on the anti-phase scenarios in a later discussion.
	\cite{Li2020} found that the locomotion efficiency of the followers can be achieved at any relative leader-follower front-back distance by adjusting their tailbeat phase difference.
	As for the schooling size larger than 2, \cite{Ashraf2017} discovered that the phalanx, \ie side-by-side arrangement of multiple fish, formation is most frequently observed in the schooling of red nose tetra fish, \textit{Hemigrammus bleheri}, which is contradictory with the previous idea that a diamond pattern is more efficient \citep{weihs1973hydromechanics}.
	In the present work, we set the front-back and lateral distances in a range similar to the previous works \citep{Ashraf2016,Ashraf2017,Li2020}.

	Fish swimming style is a curious topic that has fascinated many researchers \citep{Sfakiotakis1999,Tytell2010,Cui2018,Thekkethil2017,Thekkethil2018,Thekkethil2020}.
	\cite{Sfakiotakis1999} categorised fish swimming styles into several classes, among which the most common one is the body-caudal fin (BCF) swimming style, featuring the body/tail undulation as the main propulsion generator. BCF styles can be further divided into four types: anguilliform, sub-carangiform, carangiform, and thunniform, as exemplified in \Cref{fig:4modes}.
	\cite{Thekkethil2017,Thekkethil2018,Thekkethil2020} simplified these BCF swimmers into undulating/pitching NACA0012 hydrofoils, representing different swimming forms by non-dimensional wavelength $ \lam = \lambda / C $, where $ \lambda $ is the swimming undulation wavelength and $ C $ is the fish chord length. For example, anguilliform is typically represented by low wavelength $ \lam < 1 $, whereas the characteristics of thunniform swimming can be captured by high wavelength $ \lam \gg 1 $. \cite{Thekkethil2018} discovered that low wavelength $ \lam $ swimmers generate thrust force by the pressure difference between anterior and posterior body parts, whereas high $ \lam $ transfer streamwise momentum by pendulum-like motion; small $ \lam $, \eg anguilliform swimmers, generally causes high locomotion efficiency but low thrust production, and vice versa for large $ \lam $, \eg thunniform swimmers. The numerical results obtained by \cite{Thekkethil2018} are highly coherent with previously reported single fish swimming characteristics. The present study adopts the same simplified geometry and kinematic formula proposed by \cite{Thekkethil2018}, which will be presented later.
	\cite{Nangia2017a} also discovered that optimal wavelength exists for maximum swimming speed and propulsive thrust.

	As \textit{schooling} and \textit{swimming styles} of BCF fish both contain significant hydrodynamic implications, they can have a combined effect on the hydrodynamic characteristics of underwater swimmers. To our best knowledge, the systematic biological discussion does not exist on the behavioural correlation between \textit{schooling} and \textit{swimming styles}. However, while the schooling phenomenon is reported for swimmers of sub-carangiform \citep{Trevorrow1998}, carangiform \citep{Axelsen2001,Guillard2006,Hemelrijk2010} and thunniform \citep{Dai2020,Mitsunaga2013,Uranga2019}, anguilliform species seem never found to exhibit schooling behaviour in the wild.
	Although hydrodynamics may not be the only factor affecting fish's schooling tendency, it is reasonable to hypothesise that sub-carangiform, carangiform, and thunniform styles are more suited for schooling than anguilliform styles from a hydrodynamic perspective.
	To justify our numerical methodology, although robotic fish experiments could be an efficient way \citep{Li2020} to study fish schooling with a fixed swimming style, it will be time-consuming to design and manufacture robotic fish with variable swimming styles. To our best knowledge, existing robotic fish studies have not yet involved such a comparison between swimming styles/wavelengths. For this reason, computational fluid dynamics is utilised to simulate various fish swimming styles using a representative problem setup, thanks to its convenience in varying the swimming style by a unified kinematic formula and the capacity to analyse the flow-mediated interaction mechanism in detail \citep{Thekkethil2018}.

	In summary, for the hydrodynamics of body-caudal fin swimmers, while the mechanism of \textit{schooling} has been extensively studied \citep{weihs1973hydromechanics,Weihs1975,Ashraf2016,Li2020}, research on the \textit{swimming styles} is relatively scarce \citep{Thekkethil2018}. In nature, swimmers of sub-carangiform \citep{Trevorrow1998}, carangiform \citep{Hemelrijk2010} and thunniform \citep{Dai2020} have been reported to exhibit schooling behaviour, whereas the school of travelling anguilliform swimmers seems never been reported in the wild.
	Here, we hypothesise that thunniform swimmers are more adapted for schooling locomotion than anguilliform swimmers.
	In this paper, this hypothesis is tested by a representative problem of two wavy foils interacting in free-stream flow.
	Despite simplification, the present study will be, to our best knowledge, the first study regarding the combined effects of \textit{schooling} and \textit{swimming styles} on the BCF swimmers' locomotion hydrodynamics.
	This work is a continuation of the previous single foil swimmer study conducted by \cite{Thekkethil2018} and has been directly inspired by the previous robotic \citep{Li2020} and biological \citep{Ashraf2017} fish schooling studies.

	\newcommand{\addlabeltrim}[3]{%
		\begin{tikzpicture}
			\node[anchor=south west,inner sep=0] (image) at (0,0) 
			{\includegraphics[width=#1\linewidth, trim={0cm 2.6cm 0cm 2.4cm},clip]{#2}};%
			\begin{scope}[x={(image.south east)},y={(image.north west)}]
				\node[anchor=south west] at (0.00,0.75) {\footnotesize #3};	%
			\end{scope}
		\end{tikzpicture}%
	}
	\newcounter{testdd}
	\setcounter{testdd}{0}
	\newcommand\counterdd{\stepcounter{testdd}\alph{testdd}}
	\begin{figure}
		\centering
		
		\addlabeltrim{0.24}{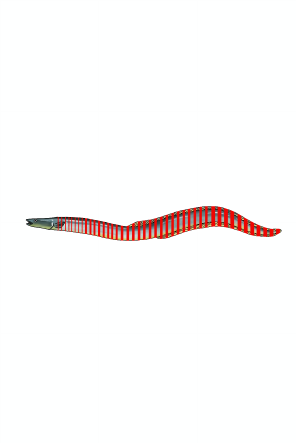}{(\counterdd) Anguilliform}
		\addlabeltrim{0.24}{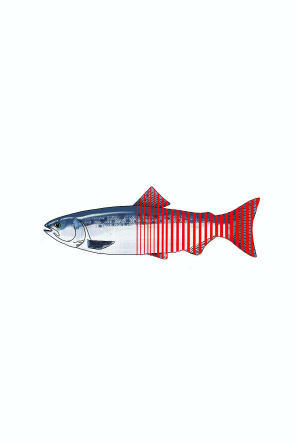}{(\counterdd) Sub-carangiform}
		\addlabeltrim{0.24}{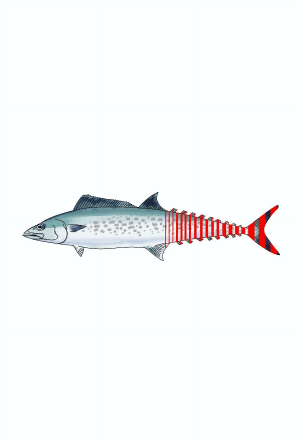}{(\counterdd) Carangiform}
		\addlabeltrim{0.24}{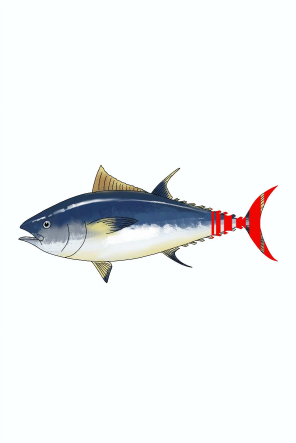}{(\counterdd) Thunniform}
		
		\caption{Four different swimming modes of Body-Caudal-Fin type locomotion
			(a) Anguilliform (body undulation, \eg eel)
			(b) Sub-carangiform (body undulation with caudal fin pitching, \eg salmonid)
			(c) Carangiform (minor body undulation with caudal fin pitching, \eg makrell)
			(d) Thunniform (mainly caudal fin pitching, \eg tuna). The shaded area demonstrates the body parts with the significant lateral motion to generate thrust force (redrawn from figures by \cite{Lindsey1978} and \cite{Sfakiotakis1999}).  }
		\label{fig:4modes}
	\end{figure}

	\section{Methodology}
	
	This section describes the numerical methodology to study the combined effects of fish schooling and swimming styles upon propulsive hydrodynamics.
	This problem is represented by two identical wavy hydrofoils tethered in a free-stream flow, as presented in \Cref{subsec:physical_problem_setup}.
	Numerical simulation is then executed by IBAMR \citep{griffith2013ibamr}, an extensively-validated immersed boundary software, as discussed in \Cref{subsec:immersed_boundary_method}.
	The kinematic model of hydrofoil undulation is formulated by the classic travelling wave equation with additional consideration upon the \textit{wavelengths} as shown in \Cref{subsec:kinematic_model}. Dimensional analysis is conducted in \Cref{subsec:dim_analysis} to formalise the investigated problem. Mesh convergence study and validation can be found in \Cref{subsec:mesh_inde_vali}.

	\subsection{Physical problem setup}
	\label{subsec:physical_problem_setup}
	
\begin{figure}
		\centering
		\includegraphics[width=1\linewidth]{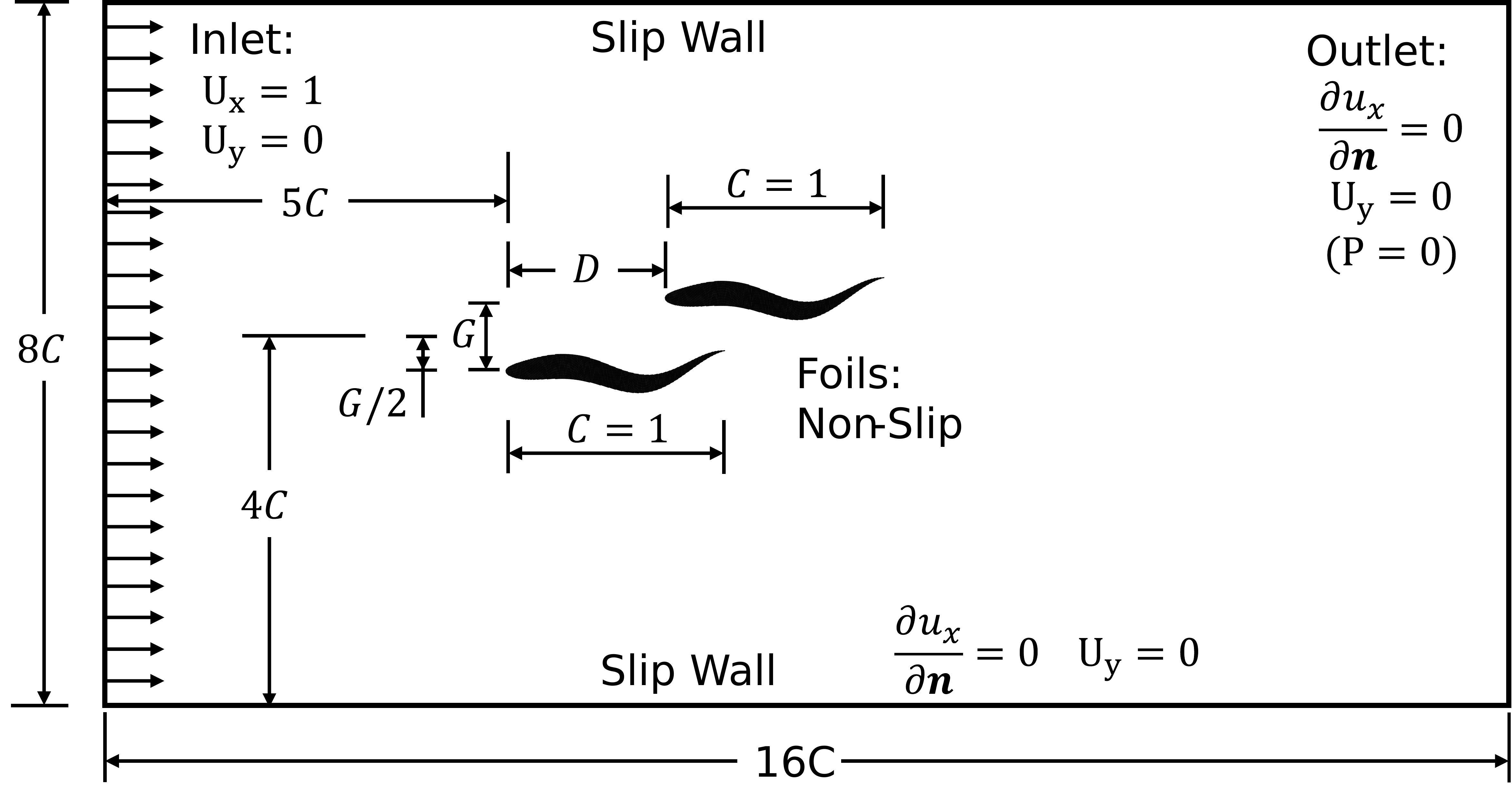}
		\caption{Problem setup of this two fish case}
		\label{fig:problemsetup}
	\end{figure}

	In this paper, the two undulating rigid NACA0012 hydrofoils are fixed at their initial locations, i.e.\ the foils are "tethered".
	The non-dimensional form of the physical problem investigated in the present paper is demonstrated in \Cref{fig:problemsetup}. Here, $ C = 1 $ is the chord length of the two NACA0012 hydrofoils. $ G $ and $ D $ are the lateral and front-back distances between the two hydrofoils, respectively. The \textit{computational domain} is chosen as $ 16C $ in the streamwise direction and $ 8C $ in the transverse direction, which is identical to the domain size chosen by \cite{Thekkethil2018}. The head tip of the leader fish is placed $ 5C $ to the inlet, and the mid-point between the two hydrofoils is placed $ 4C $ to each of the lateral walls.
	As for \textit{boundary conditions}, the two identical hydrofoils are both non-slip on their fluid-solid interface. The inlet free-stream velocity is configured as $ U_{inlet} = (U_x,U_y) = (1,0) $. The outlet boundary is set as $ {\partial u_x}/{\partial \bm{n}} = 0$ and $U_y = 0  $, which is equivalent to the zero pressure outlet boundary condition, i.e.\ $ P_{outlet} = 0 $; here, $ \bm{n} $ is the outward unit vector normal to the boundary. Lateral walls on the left and right sides of the swimming direction are both prescribed as slip wall boundary condition $ {\partial u_x}/{\partial \bm{n}} = 0$ and $U_y = 0 $.
	We note that we are not strictly modelling any specific fish species but are instead seeking underlying principles of how swimming style affects schooling.
	
	\newcommand{\CC}{C\nolinebreak\hspace{-.05em}\raisebox{.4ex}{\tiny\bf +}\nolinebreak\hspace{-.10em}\raisebox{.4ex}{\tiny\bf +}}
	\def\CC{{C\nolinebreak[4]\hspace{-.05em}\raisebox{.4ex}{\tiny\bf ++}}}
	
	\subsection{Immersed boundary method}
	\label{subsec:immersed_boundary_method}
	In the present study, the numerical simulation of the fluid-structure interaction process is achieved by a modified version of the constrained method in the \verb!C++! open-source software IBAMR \citep{griffith2013ibamr}. IBAMR is constructed on the foundation of several libraries, including
	SAMRAI \citep{Hornung2002,Hornung2006},
	PETSc \citep{Balay1997,Balay2010,balay2001petsc},
	hypre \citep{falgout2010hypre,Balay1997},
	and
	libmesh \citep{Kirk2006}.
	The constrained immersed boundary (IB) method of IBAMR has been validated in various scenarios, including fish swimming \citep{Bhalla2013,Griffith2020}, flow past cylinder\citep{Nangia2017}, and free-surface piercing \citep{Nangia2019}. In the present study, the force upon each hydrofoil was obtained by the control volume method \citep{Nangia2017}.

	The IB method uses Eulerian description for the fluid and Lagrangian description for the deforming structure. One of its advantages is the computational efficiency due to circumventing the costly remeshing process encountered in other methods like the finite element method. The implemented formulation is stated as:
	
	\begin{equation}\label{eq:IB1}
		\rho \left( \frac{\partial \bm{u}(\bm{x},t) }{\partial t} + \bm{u}(\bm{x},t) \cdot \nabla \bm{u}(\bm{x},t)\right) = -\nabla p (\bm{x},t) + \mu \nabla^2 \bm{u}(\bm{x},t) + \bm{f}(\bm{x},t)    
	\end{equation}
	\begin{equation}\label{eq:IB2}
		\nabla \cdot \bm{u}(\bm{x},t) = 0
	\end{equation}
	\begin{equation}\label{eq:IB3}
		\bm{f}(\bm{x},t) = \int_{U} \bm{F}(\bm{X},t) \delta(\bm{x} - \bm{\chi}(\bm{X},t)) \diff \bm{X}
	\end{equation}
	\begin{equation}\label{eq:IB4}
		\frac{\partial \bm{\chi} (\bm{X},t)}{\partial t} = \int_{\Omega} \bm{u}(\bm{x},t) \delta(\bm{x} - \bm{\chi}(\bm{X},t)) \diff \bm{x}
	\end{equation}
	
	Here $ \bm{x} = (x,y) \in \Omega $ represents fixed physical Cartesian coordinates, where $ \Omega $ is the physical domain occupied the fluid and the immersed structure. $ \bm{X} = (X,Y) \in U $ means Lagrangian solid structure coordinates, and $ U $ is the Lagrangian coordinate domain. The mapping from Lagrangian structure coordinates to the physical domain position of point $ \bm{X} $ for all time $ t $ can be expressed as $ \bm{\chi}(\bm{X},t) = ( \chi_x(\bm{X},t), \chi_y(\bm{X},t) ) \in \Omega $. In other words, $ \chi(U,t) \subset \Omega $ represents the physical region occupied by the solid structure at time $ t $. $ \bm{u}(\bm{x},t) $ is the Eulerian fluid velocity field and $ p(\bm{b},t) $ is the Eulerian pressure field. $ \rho $ is the fluid density. $ \mu $ is the incompressible fluid dynamic viscosity. $ \bm{f}(\bm{x},t) $ and $ \bm{F}(\bm{X},t) $ is Eulerian and Lagrangian force densities, respectively. $ \delta (\bm{x}) $ is the Dirac delta function. More details regarding the constrained IB formulation and discretisation process can be found in previous literature \citep{Bhalla2013,Griffith2020}.

	\subsection{Kinematic model for fish-like wavy propulsion}
	\label{subsec:kinematic_model}
	The non-dimensional kinematic equations for the centrelines of the two tethered NACA0012 hydrofoils \citep{LangleyResearchCenter2021} are prescribed as \Cref{equ:fish_wave_leader,equ:fish_wave_follower}
	\citep{Thekkethil2018,VIDELER1984}:
	\begin{equation}\label{equ:fish_wave_leader}
		\Delta Y^*_{1} = A_{max} X^*_{1} \sin \left[ 2\pi \left( \frac{X^*_1}{\lam} - \frac{St }{2 A_{max}} t^* \right) \right]
	\end{equation}
	\begin{equation}\label{equ:fish_wave_follower}
		\Delta Y^*_{2} = A_{max} X^*_{2} \sin \left[ 2\pi \left( \frac{X^*_2}{\lam} - \frac{St }{2 A_{max}} t^* \right) + \phi \right]
	\end{equation}
	where $ \Delta Y^*_i = \Delta Y_i/C $ is the lateral displacement of each NACA0012 hydrofoil centreline that varies with streamwise direction $ X^*_i=X_i/C $ and time $ t^* = t\uinf/C $. The wavy undulation period $ T $ is equal to $ St/2A_{max} $, so $ t^*/T = 2t^*A_{max}/St $. $ i = 1 $ and $ i = 2 $ denotes the leader foil and the follower foil, respectively. %

	\subsection{Dimensional analysis}
	\label{subsec:dim_analysis}
	In this paper, we investigate the flow-mediated interaction between two swimming fish with various swimming modes, which can be simplified into two rigid NACA0012 hydrofoils with wavy lateral movement subject to free stream flow. The problem setup can be determined by 7 non-dimensional groups as shown in \Cref{tablecasegroups}, where $ \rho $ is the fluid density, $ \uinf = 1 $ is the free stream velocity, $ C = 1 $ is the hydrofoil chord length, $ \mu $ is the fluid viscosity, $ f $ is the undulation frequency, $ a_{max} $ is the maximum undulation amplitude at tail tip, $ \lambda,\ g,\ d $ are the wavelength, lateral distance and front-back distance to be non-dimensionalised by chord length $ C $.
	Reynolds number $ Re $ and Strouhal number $ St $ are fixed at $ 5000 $ and $ 0.4 $, respectively, since the slowly swimming fish generally swim with moderate $ Re $ and $ St \approx 0.4,\ A_{max} \approx 0.1 $ \citep{Lindsey1978,Thekkethil2018}. The Reynolds number at the order of $ 10^3 $ allows a more economical mesh resolution and computational cost, whereas predominant vortex dynamics still remain understandable \citep{Liu2017}. The chosen Reynolds number is more convenient for comparison with the previous work of single wavy foil by \cite{Thekkethil2018}, which also fixed $ Re $ at $ 5000 $. The lateral and front-back distances vary in the range of $ G = 0.25 - 0.35 $ and $ D = 0 - 0.75 $, respectively, which corresponds to the value range chosen by \cite{Ashraf2017} and \cite{Li2020}.

	\bgroup
	\def\arraystretch{2.2}%
	\begin{table}[thb]
		\caption{Non-dimensional input parameters and the involved range of value}
		\centering
		\label{tablecasegroups}
		\begin{tabular}{l c c c}
			\toprule
			Reynolds number & $ \re $ &{ $ { \rho u_{\infty} C}/{\mu} $} & $ 5000 $ \\
			
			Strouhal number  & $ St $ &{ $ {2 f a_{max}}/{u_{\infty}} $} & $ 0.4 $ \\
			
			Maximum amplitude & $ A_{max} $ &{ $ a_{max}/C $  } & $ 0.1 $ \\
			
			Wavelength  & $ \lam $ &{  $ {\lambda}/{C} $} & $0.8 - 8 $ \\
			
			Lateral gap distance & $ G $ & { $ {g}/{C} $} & $ 0.25,\ 0.3,\ 0.35 $  \\ 
			
			Front-back distance & $ D $ & { $ {d}/{C} $} & $ 0,\ 0.25,\ 0.5,\ 0.75 $  \\ 
			
			Phase difference  & $ \phi $ &{ $ \phi_{2}-\phi_{1} $} & $ 0,\ 0.5\pi,\ \pi,\ 1.5\pi $ \\
			
			\bottomrule
		\end{tabular}
	\end{table}
	\egroup

	To examine the schooling performance of two fish-like hydrofoils, we chose the output parameters as listed in \Cref{tab:output_para}. Here, $ F_{T,i} $ is the net thrust force along streamwise direction and $ F_{L,i} $ is the lateral force on the transverse direction. $ i = 1,\ 2 $ denotes the No.\ $ i $ hydrofoil structure; $ i = 1 $ and $ i = 2 $ represent the leader foil and follower foil respectively. $ V_{body,i} = {\diff \Delta Y^*_i}/{ \diff t^*} $ is the lateral motion velocity of the hydrofoils. $ c_{L,i} $ is the force coefficient density distributed on the surface of the hydrofoil. $ \bm{u}(\bm{x},t) $ is the velocity field of the fluid. %

	\bgroup
	\def\arraystretch{2.2}%
	\begin{table}%
		\caption{Non-dimensional output parameters}
		\centering
		\label{tab:output_para}
		\begin{tabular}{l c c c}
			\toprule
			Cycle-averaged thrust coefficient & $ C_{Tm,i} $ &{\Large $\left( \frac{2F_{T,i}}{\rho \uinf^2 C} \right)_{avg} $}  \\
			
			Root mean square lateral force coefficient  & $ C_{Lrms,i} $ &{\Large $\left( \frac{2F_{L,i}}{\rho \uinf^2 C} \right)_{rms} $}  \\
			
			Froude efficiency \citep{Liu1996} & $ \eta_i $ &{\Large $ \frac{P_{out,i}}{P_{in,i}} = \frac{C_{Tm,i}}{\overline{\int c_{L,i} V_{body,i} dS}} $  }  \\
			
			Group Froude efficiency & $ \eta_{group} $ &{\Large $  \frac{\sum P_{out,i}}{\sum P_{in,i}} $  }  \\
			
			Fluid velocity & $ \bm{u^*} $ &{ $ \bm{u}/\uinf $  }  \\
			
			Fluid vorticity & $ \bm{\omega^*} $ &{ $ \nabla \times \bm{u^*} $  }  \\
			
			\bottomrule
		\end{tabular}
	\end{table}
	\egroup

	\subsection{Mesh independence and validation}
	\label{subsec:mesh_inde_vali}
	IBAMR utilises two sets of "immersed" meshes for numerical simulations, \ie Eulerian mesh for the flow field and Lagrangian mesh for the structure (swimmer), where the Eulerian mesh can be adaptively refined considering the local vorticity strength and adjacency to the Lagrangian described structure.
	In the present study, the Eulerian mesh consists of 3 levels of refinement divided by local vorticity thresholds, with each level being 4 times finer than the coarser level.
	The mesh density of the Lagrangian mesh for the hydrofoil structure is equivalent to that of the finest level Eulerian mesh for the fluid.
	
	Mesh independence study was conducted using 4 meshes with different levels of refinement, as listed in \cref{tab:mesh_val}.
	At $ D = 0.5,\ G = 0.3, \lam = 2.0, \ppi = 1 $, the time history of lateral force coefficient is examined to check mesh convergence, as seen in \Cref{fig:mesh_validation}a. While a large difference can be observed from $ C_L $ yielded by the "Coarsest" mesh and "Coarse" mesh, the time history of $ C_L $ almost overlaps for the output produced by "Normal" and "Refined" meshes. To be conservative, the "Refined" mesh setting is chosen for all the cases in this study. Based on "Refined" mesh, the current result is further validated against results from \cite[pp. 10]{Thekkethil2018} with single fish case of $ Re = 5000,\ St = 0.3 - 0.7,\ A_{max} = 0.1,\ \lam =  1.5$, as seen in \Cref{fig:mesh_validation}b. Excellent coherence can be seen between the results produced by the present IBAMR model and the in-house model by \cite{Thekkethil2018}.

	\newcommand{\addlabel}[3]{%
		\begin{tikzpicture}
			\node[anchor=south west,inner sep=0] (image) at (0,0) 
			{\includegraphics[width=#1\textwidth]{#2}};
			\begin{scope}[x={(image.south east)},y={(image.north west)}]
				\node at (0.2,0.23) {#3};	%
			\end{scope}
		\end{tikzpicture}%
	}
	\begin{figure}
		\centering
		\addlabel{0.49}{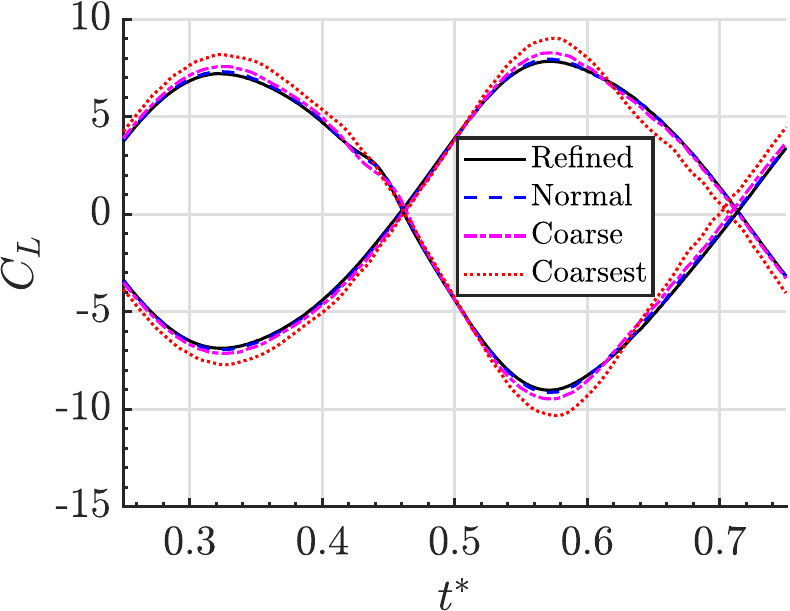}{(a)}
		\addlabel{0.49}{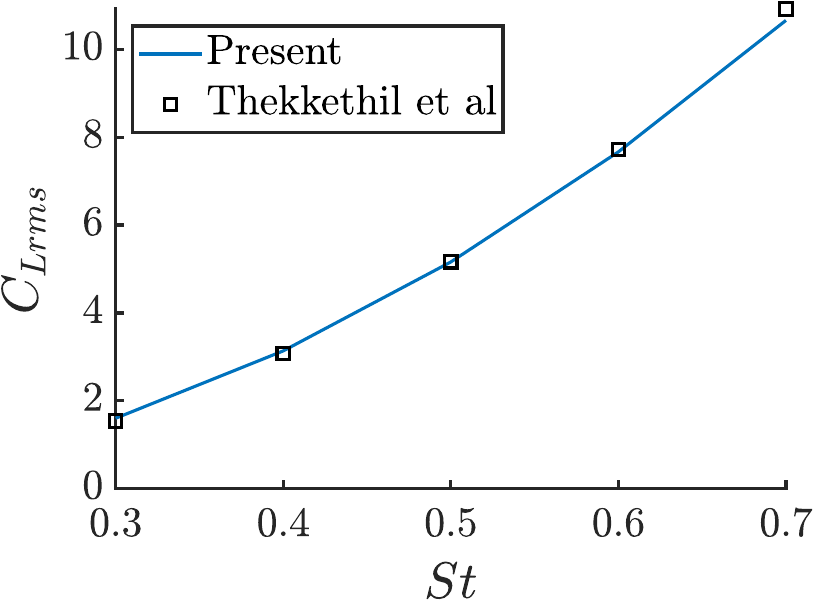}{(b)}
		\caption{
			(a) Mesh independence study for meshes listed in \Cref{tab:mesh_val} with $ Re = 5000 $, $ St = 0.4 $, $ A_{max} = 0.1 $, $ \lam = 2 $, $ \ppi = 1 $, $ G = 0.3 $, $ D = 0.5 $.
			(b) Validation by comparing present results with that from \cite[pp. 10]{Thekkethil2018} with $ Re = 5000,\ St = 0.3 - 0.7,\ A_{max} = 0.1,\ \lam =  1.5$.
		}
		\label{fig:mesh_validation}
	\end{figure}

	\begin{table}
		\caption{Mesh configuration for independence study}
		\centering
		\label{tab:mesh_val}
		\begin{tabular}{ccccc}
			\hline
			{Mesh} 		& Refined & Normal & Coarse & Coarsest \\
			{$ \Delta x^*_{min} $} 	& $2\sci{-3}$ & $4\sci{-3}$ & $8\sci{-3}$ & $16\sci{-3}$ \\
			{$ \Delta t^* $} 	& $2.5\sci{-5}$ & $5\sci{-5}$ & $10\sci{-5}$ & $20\sci{-5}$ \\
			{$ \Delta x^*_{min} / \Delta t^* $} 	&  $80$ & $80$ & $80$ & $80$ \\
			\hline
		\end{tabular}
		\centering
	\end{table}

	\section{Results and Discussion}

	The present paper includes a parametric space of more than 300 combinations with a focus on the variation of wavelengths $ \lam = 0.8 - 8.0 $. The lateral gap distance $ G $ ranges from $ 0.25 $ to $ 0.35 $, the front-back distance $ D = 0 - 0.75 $, leader-follower phase difference $ \ppi = 0, 0.5, 1, 1.5 $. The Reynolds number, Strouhal number and maximum amplitude is fixed at $ Re = 5000 $, $ St = 0.4 $ and $ A_{max} = 0.1 $ throughout this study.
	The following discussion is divided into 2 parts: side-by-side arrangement in \Cref{sec:phalanx_D0} and staggered configuration in \Cref{sec:Staggered_Dg0}.
	The separated discussion of side-by-side arrangement $ D = 0 $ is justified by the rich physics due to its distinguished symmetrical flow structure and potential occurrence of symmetry breaking \citep{Gungor2020} and by the observation of fish's tendency to form a phalanx formation in a free-stream flow \cite{Ashraf2017}. It is also a curious question of how the variation of wavelength affects the flow symmetry and locomotion properties.
	The investigation of staggered arrangement $ D > 0 $ is to understand the effects of swimming style, \ie wavelengths, upon the vortex phase matching mechanism proposed by \cite{Li2020}.

	Several output parameters will be discussed to understand the schooling effect with various swimming styles, as listed in \Cref{tab:output_para}: the thrust force $ \ctm $ is directly relevant to the acceleration of the swimming foils; the lateral force $ \clrms $ is linked to the work done from the undulating foil to the incompressible fluid; Froude efficiency $ \eta_{i} $ and $ \etag $ are the propulsion efficiency converting the input energy to the output locomotion performance as an individual foil and as a grouped system, respectively; vorticity distribution $ \omega^* $ demonstrates the vortex interaction between the two hydrofoils and the vortex shedding pattern in the wake flow, in which the vortex interaction is significant to understand the mechanism resulting the force and efficiency distribution, and vortex wake pattern is important for stealth capacity for fish schooling behaviour.

	\subsection{Overview}

	In this subsection, we offer an overview of the present paper with an example case at $ G = 0.25 $, $ D = 0.75 $, $ \phi = 0 $ and $ \lam = 0.8 - 8.0 $, demonstrating its key results of vorticity distribution $ \bm{\omega^*} $, thrust force $ \ctm $ and the propeller efficiency $ \eta $ at $ t^*/T = 5 $, as seen in \Cref{fig:vorticity_lambda}.
	The irregularity of wake flow generally increases with wavelength $ \lam $, as seen in the vorticity contours in \Cref{fig:vorticity_lambda}a-\ref{fig:vorticity_lambda}f.
	The flow structure near the two foils is relatively regular. In this paper, the output parameters are calculated using the last 3 stable periods.
	The thrust force generally increases with $ \lam $, whereas the propeller efficiency is peaked at $ \lam = 2 $ in most cases, as seen in \Cref{fig:vorticity_lambda}g-\ref{fig:vorticity_lambda}h. The following sections further discuss the inter-relationship between wavelengths $ \lam $ and other parameters.
	Compared with \textit{single} swimmer cases with similar configurations \citep{Thekkethil2018}, the interaction between \textit{two} wavy foils leads to a more complicated flow structure. However, the general trend of thrust force and Froude efficiency is consistent with the single foil cases \citep{Thekkethil2018}.
	For the sake of convenience and conciseness, the vorticity scale in all other figures is identical to the one shown in \Cref{fig:vorticity_lambda}.

	\newcommand{\addlabela}[3]{%
		\begin{tikzpicture}
			\node[anchor=south west,inner sep=0] (image) at (0,0) 
			{\includegraphics[width=#1\linewidth, trim={2cm 0cm 0cm 15.1cm},clip]{#2}};	%
			\begin{scope}[x={(image.south east)},y={(image.north west)}]
				\node at (0.125,0.75) {#3};	%
			\end{scope}
		\end{tikzpicture}%
	}
	\begin{figure}
		\centering
		\includegraphics[width=1.0\linewidth]{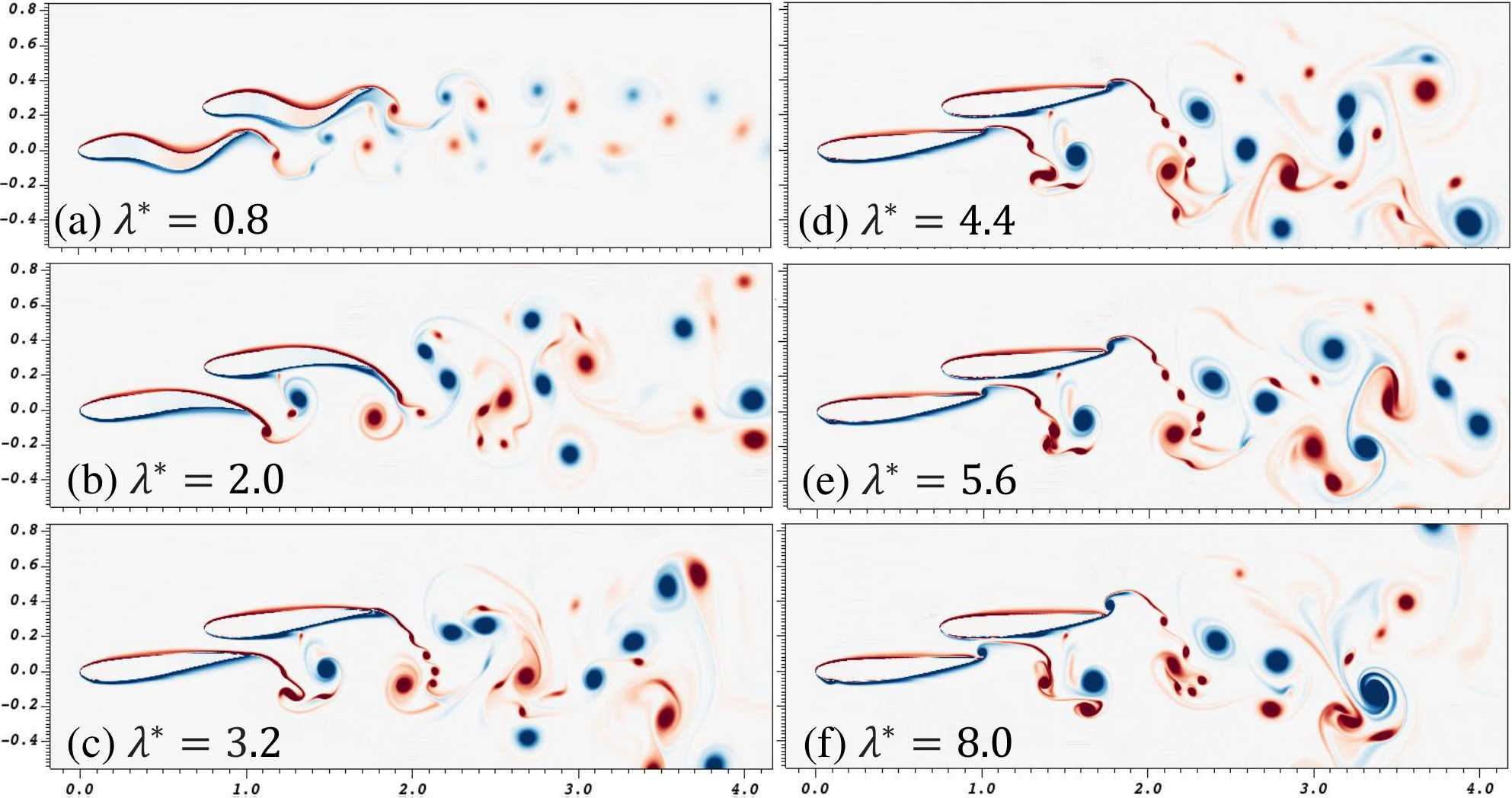}
		\addlabela{0.7}{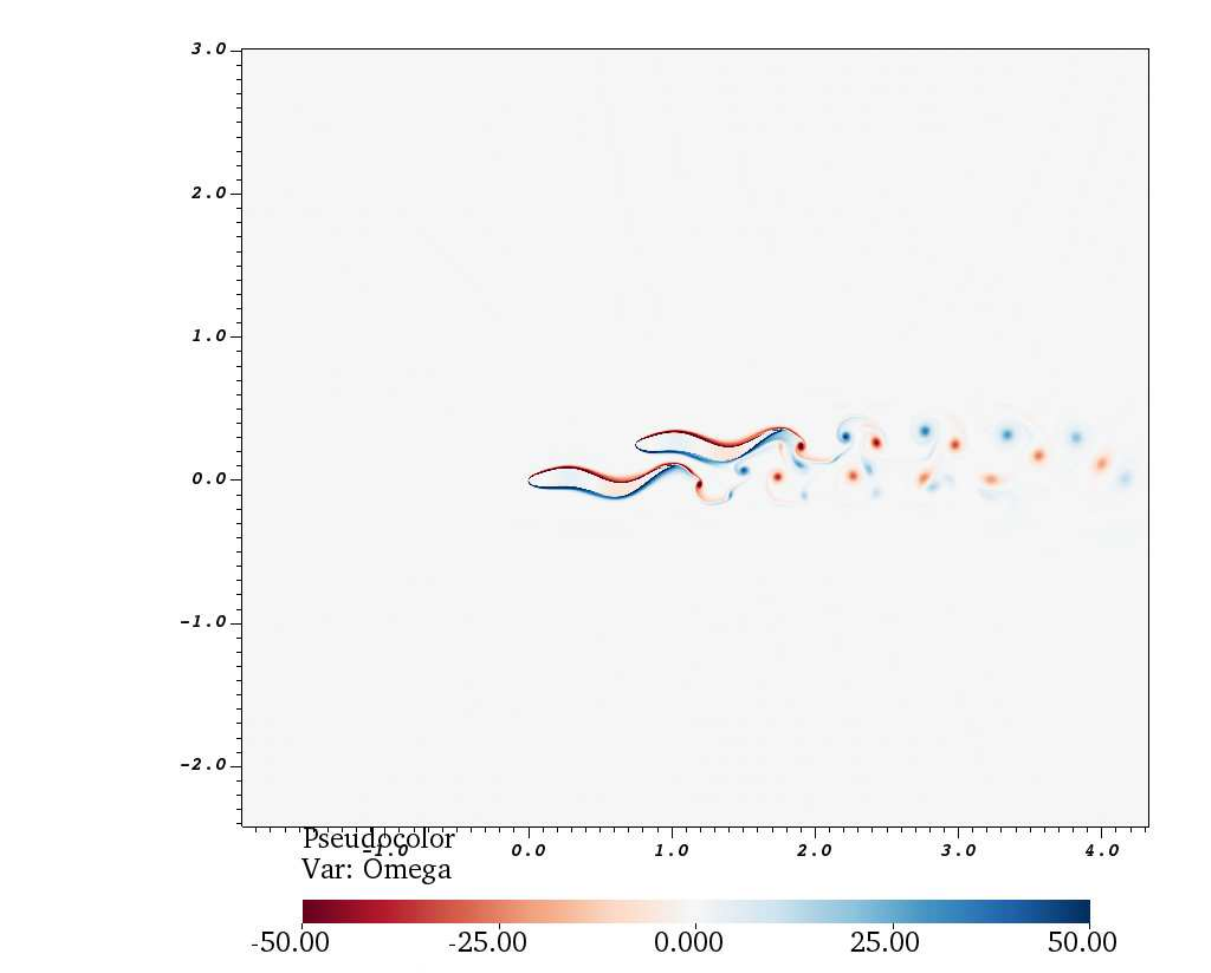}{$\bm{\omega}^*$:}
		\includegraphics[width=0.9\linewidth]{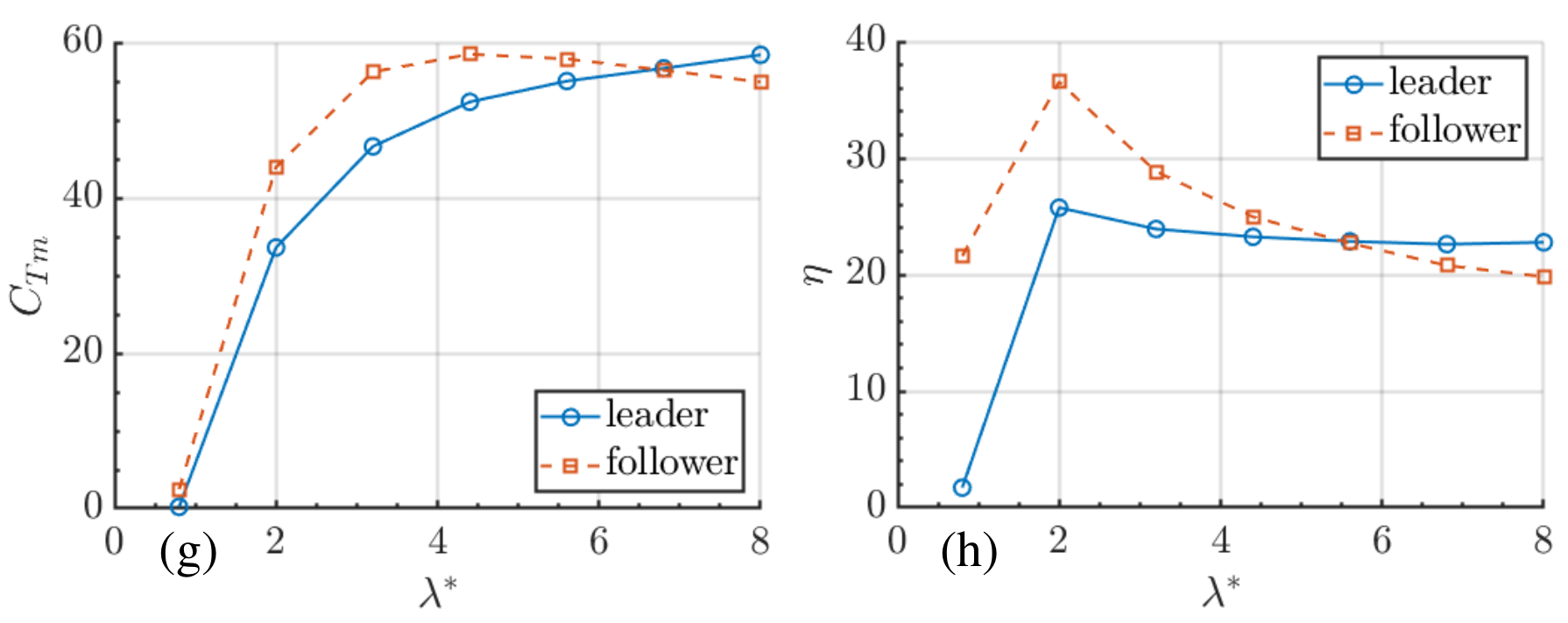}
		\caption{Variation of hydrofoil geometry and vorticity contours at time $ t^*/T = 5 $ with $ G = 0.25 $, $ D = 0.75 $, $ \phi = 0 $, and (a) $ \lam = 0.8 $ (b) $ \lam = 2.0 $ (c) $ \lam = 3.2 $ (d) $ \lam = 4.4 $ (e) $ \lam = 5.6 $ (f) $ \lam = 8.0 $. Variation of (g) thrust force $ C_{Tm} $ and (h) Froude propeller efficiency $ \eta $ corresponding to (a-f).
			The wake structure irregularity increases with $ \lam $. The propeller efficiency $ \eta $ of the \textit{follower} is, in general, higher than the \textit{leader}, corresponding to higher thrust force $ C_{Tm} $. At large $ \lam > 7 $, the \textit{leader} becomes slightly more efficient than the \textit{follower}.
		}
		\label{fig:vorticity_lambda}
	\end{figure}

	\subsection{Side-by-side arrangement $ D = 0 $}
	\label{sec:phalanx_D0}
	
	This section discusses the scenarios with two foils swimming side-by-side $ D = 0 $.
	\Cref{subsec:F_eta_Dis0} examines the variation of swimmers' mean thrust force $ \ctm $, RMS lateral force $ \clrms $, propeller efficiency $ \eta $, and group efficiency $ \eta_{group} $ with a series of wavelength, phase difference and lateral gap distance.
	\Cref{subsec:FSI_Dis0} investigates the variation of flow structure with wavelength $ \lam $, phase difference $ \phi $, front-back distance $ D $ and their inter-relationships with additional discussion regarding the symmetrical anti-phase condition.
	\Cref{subsec:vor_Dis0} discusses the mechanism of flow-mediated interaction between the two swimmers by examining vorticity distribution and hydrodynamic force within one typical cycle of undulation.
	
	\subsubsection{Hydrodynamic force and propulsive efficiency at $ D = 0 $}
	\label{subsec:F_eta_Dis0}
	At side-by-side arrangement with $ D = 0 $, the mean thrust force and RMS lateral force generally increase with $ \lambda^* $ \Cref{fig:lines_D_0_CT_CL}, but the propeller efficiency only slightly changes with wavelength at $ \lambda^* > 2 $, as seen in \Cref{fig:lines_D_0_eta}.
	As the two foils swim in-phase $ \phi = 0 $ and anti-phase $ \phi = \pi $, due to the symmetrical nature of the side-by-side setting, the results of the leader overlap that of the follower. By comparison, results of the leader at $ \phi = 0.5\pi $ tend to coincide with that of the follower at $ \phi = 1.5\pi $.
	These results regarding the undulation phase difference $ \phi $ are in coherence with the observed fish schooling behaviour: when the multiple fish swim side-by-side, they tend to undulate either in-phase $ \phi = 0 $ or in anti-phase $ \phi = \pi $ \citep{Ashraf2017}. While the previous work by \cite{Ashraf2017} is limited to only one species, \ie \textit{Hemigrammus bleheri}, the present results further indicate that the conclusion can be further extended to a wide range of wavelength $ \lambda = 0.8 - 8.0 $.

	The \textit{thrust force} $ \ctm $ generally increases with wavelength $ \lam $ reaching maximum at $ \lam = 4.4 $ while swimming in-phase $ \phi = 0 $ or anti-phase $ \phi = \pi $, as seen in \Cref{fig:lines_D_0_CT_CL}a, \ref{fig:lines_D_0_CT_CL}c and \ref{fig:lines_D_0_CT_CL}e. Thrust force $ \ctm $ in general increases with gap distance $ G $ while swimming in-phase $ \phi = 0 $, but decreases with $ G $ while swimming anti-phase $ \phi = \pi $.
	The flow is less regular at small gap distance $ G < 0.3 $ and anti-phase $ \phi = \pi $ condition, causing the less smooth curves in \Cref{fig:lines_D_0_CT_CL}a and \Cref{fig:lines_D_0_CT_CL}c.
	The \textit{lateral force} $ \clrms $ on the whole increases monotonically with $ \lam $, as demonstrated in \Cref{fig:lines_D_0_CT_CL}b, \ref{fig:lines_D_0_CT_CL}d and \ref{fig:lines_D_0_CT_CL}f.
	As shown in \Cref{fig:lines_D_0_eta}a, \ref{fig:lines_D_0_eta}c and \ref{fig:lines_D_0_eta}e, the \textit{propeller Froude efficiency} $ \eta $ reaches maximum at $ \lam = 2 $ and then only slightly decreases with wavelength $ \lam $ at $ \lam > 2 $ and remains almost constant at $ \lam > 5.6 $.
	One foil can achieve very high propeller efficiency $ \eta $ at the cost of $ \eta $ another foil at phase difference $ \phi = 0.5 \pi $ or $ 1.5 \pi $. This effect is strengthened by stronger flow-mediated interaction through smaller gap distance $ G $. The difference in efficiency can reach $ 50\% $ at $ G = 0.25 $.
	$ \eta $ can be negative at $ \lam = 0.8 $, meaning the leader or the follower foils are not propelling forward.
	
	When the two foils swim side-by-side, their \textit{group efficiency} reaches a maximum of $ \etag = 31.2\% $ at wavelength $ \lam = 2 $ and phase lag $ \phi = 0.5 \pi $, as seen in \Cref{fig:lines_D_0_eta}b, \ref{fig:lines_D_0_eta}d and \ref{fig:lines_D_0_eta}f. Anguilliform swimming with low wavelength $ \lam < 1 $ can lead to negative group efficiency $ \etag < 0 $, indicating that the foils tend to drift along the inlet flow direction; this tendency is strengthened by a narrow gap $ G = 0.25 $ and in-phase undulation $ \phi = 0 $. For Carangiform swimming at high $ \lam $, the group efficiency slightly increases with the gap distance, especially the in-phase $ \phi = 0 $ scenarios. The group efficiency is highly consistent for phase lag $ \phi = 0,\ \pi $ and $ \phi = 0.5\pi,\ 1.5\pi $, at $ G = 0.30,\ 0.35 $ and high wavelength $ \lam > 5 $.

	\newcommand{\addlabelb}[3]{%
		\begin{tikzpicture}
			\node[anchor=south west,inner sep=0] (image) at (0,0) 
			{\includegraphics[width=#1\textwidth]{#2}};
			\begin{scope}[x={(image.south east)},y={(image.north west)}]
				\node [anchor=west] at (0.16,0.9) {#3};	%
			\end{scope}
		\end{tikzpicture}%
	}
	\newcommand{\addlabelbtop}[4]{%
		\begin{tikzpicture}
			\node[anchor=south west,inner sep=0] (image) at (0,0) 
			{\includegraphics[width=#1\textwidth]{#2}};
			\begin{scope}[x={(image.south east)},y={(image.north west)}]
				\node [anchor=west] at (0.16,0.9) {#3};	%
				\node [anchor=west] at (0.16,1.05) {#4};
			\end{scope}
		\end{tikzpicture}%
	}
	\newcounter{testb}
	\setcounter{testb}{0}
	\newcommand\counterb{\stepcounter{testb}\alph{testb}}
	\begin{figure}
		\centering
		\addlabelbtop{0.49}{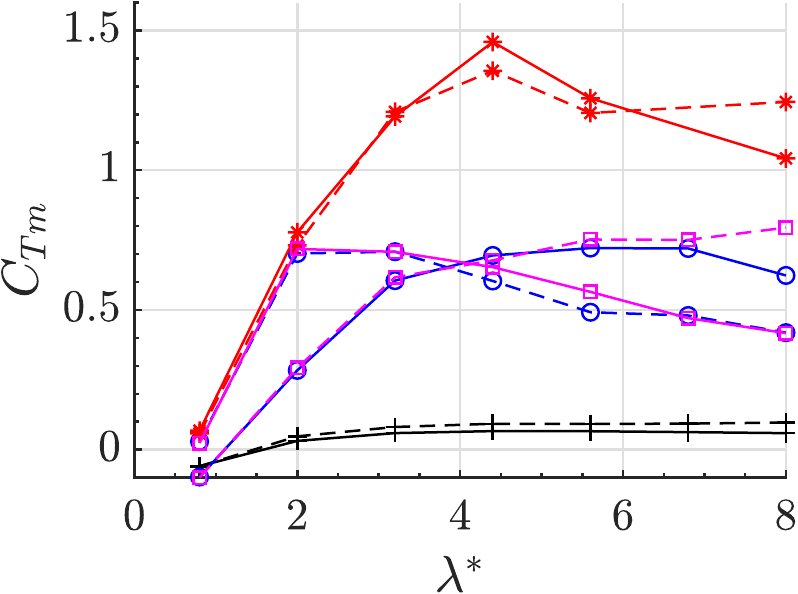}{(\counterb) $ G = 0.25 $}{$ D = 0 $}
		\addlabelb{0.49}{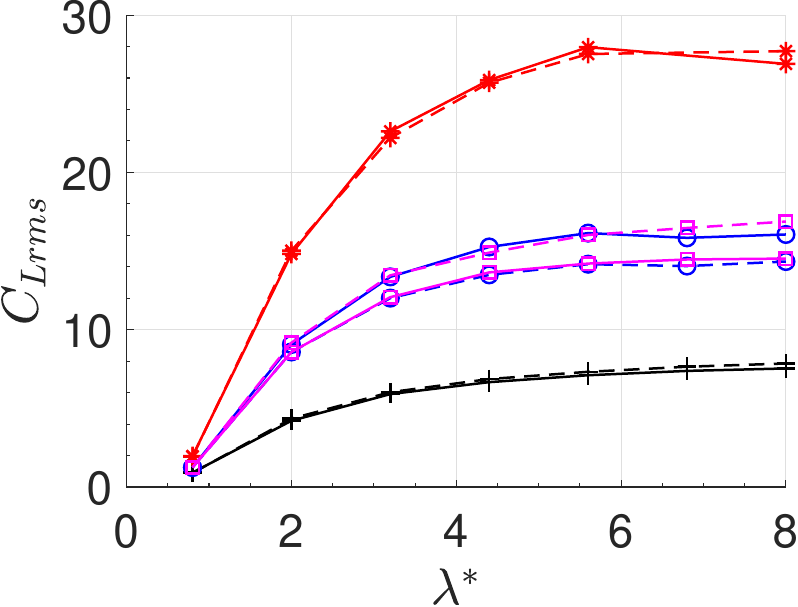}{(\counterb) $ G = 0.25 $}

		\addlabelb{0.49}{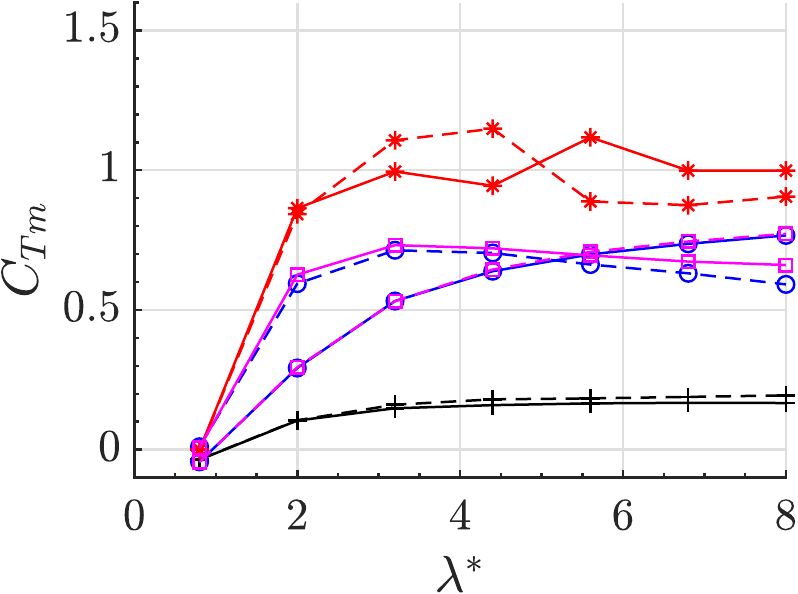}{(\counterb) $ G = 0.30 $}
		\addlabelb{0.49}{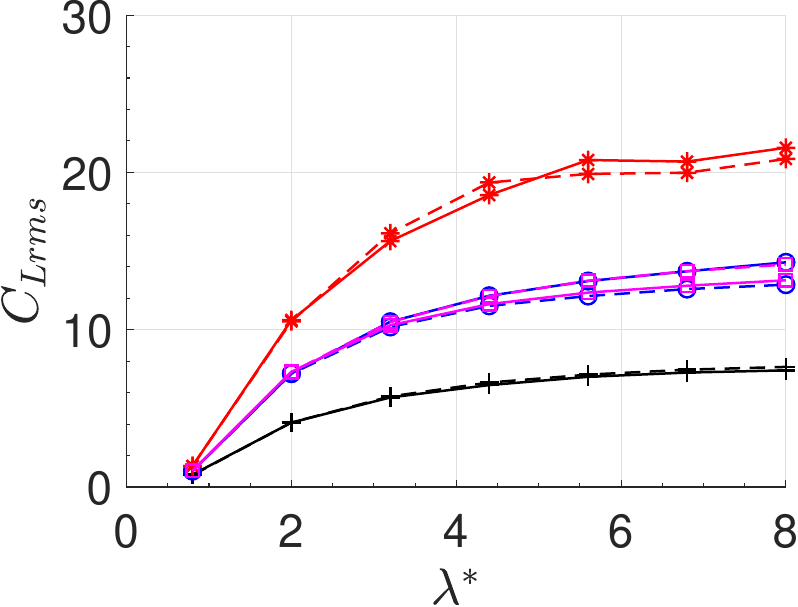}{(\counterb) $ G = 0.30 $}

		\addlabelb{0.49}{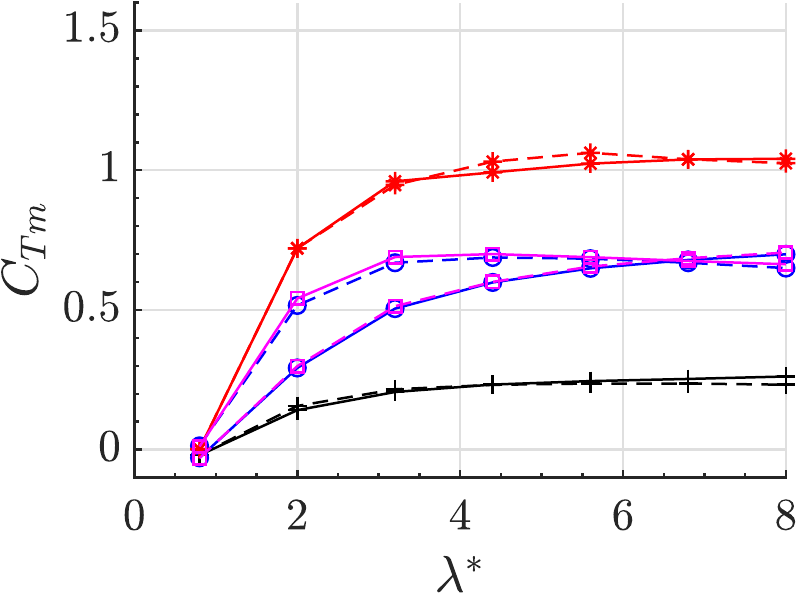}{(\counterb) $ G = 0.35 $}
		\addlabelb{0.49}{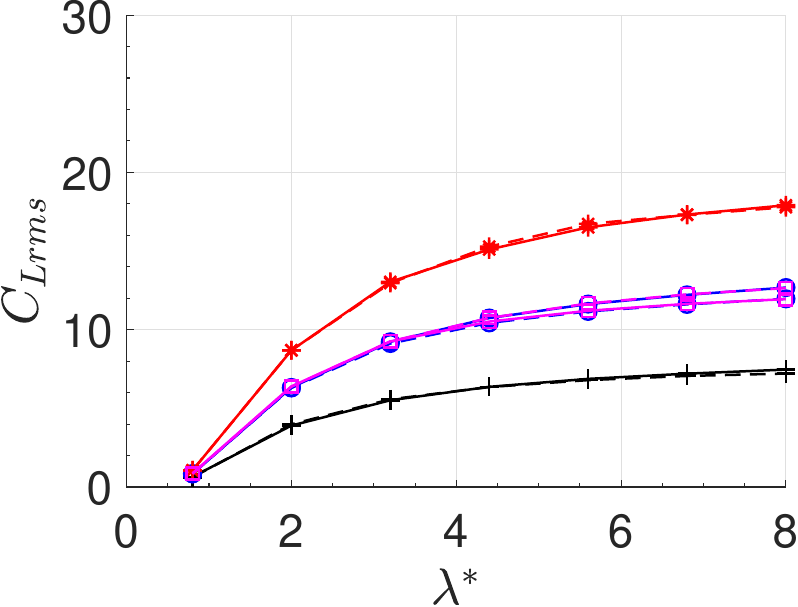}{(\counterb) $ G = 0.35 $}

		\includegraphics[width=0.9\linewidth]{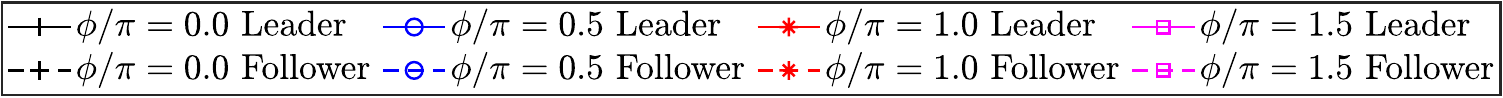}
		
		\caption{Variation of (a \& c \& e) mean thrust force $ C_{Tm} $, (b \& d \& f) RMS lateral force $ C_{Lrms} $ with a series of wavelength $ \lambda^* = 0.8 - 8.0 $, while side-by-side distance $ D = 0 $, phase difference $ \phi = 0 - 1.5 \pi $, and gap distance (a-b) $ G = 0.25 $, (c-e) $ G = 0.30 $, and (e-f) $ G = 0.35 $. Each row of sub-figures demonstrates the results from the same gap distance $ G $ between the two foils. }
		\label{fig:lines_D_0_CT_CL}
	\end{figure}
	
	\setcounter{testb}{0}
	\newcommand{\addlabelbb}[3]{%
		\begin{tikzpicture}
			\node[anchor=south west,inner sep=0] (image) at (0,0) 
			{\includegraphics[width=#1\textwidth]{#2}};
			\begin{scope}[x={(image.south east)},y={(image.north west)}]
				\node at (0.60,0.25) {#3};	%
			\end{scope}
		\end{tikzpicture}%
	}
	\newcommand{\addlabelbbtop}[4]{%
		\begin{tikzpicture}
			\node[anchor=south west,inner sep=0] (image) at (0,0) 
			{\includegraphics[width=#1\textwidth]{#2}};
			\begin{scope}[x={(image.south east)},y={(image.north west)}]
				\node at (0.60,0.25) {#3};	%
				\node [anchor=west] at (0.16,1.05) {#4};
			\end{scope}
		\end{tikzpicture}%
	}
	\begin{figure}
		\centering
		\addlabelbbtop{0.49}{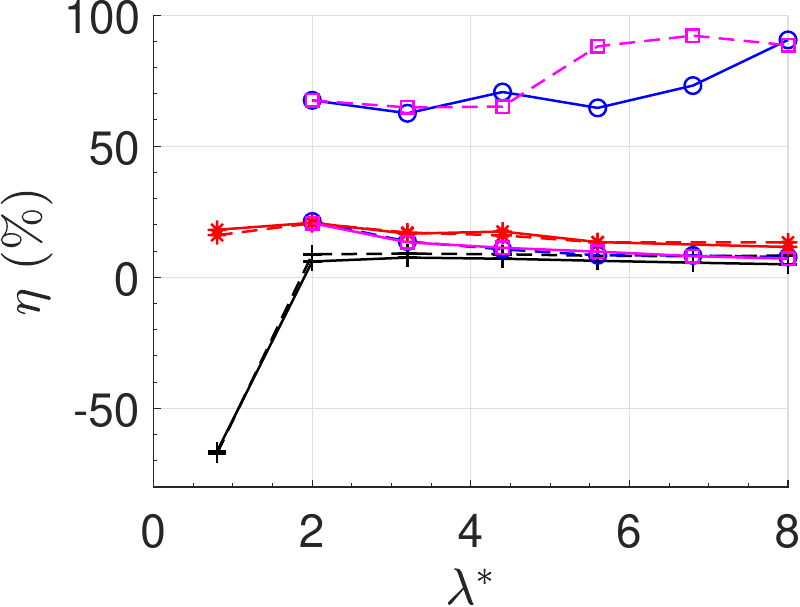}{(\counterb) $ G = 0.25 $}{$ D = 0 $}
		\addlabelbb{0.49}{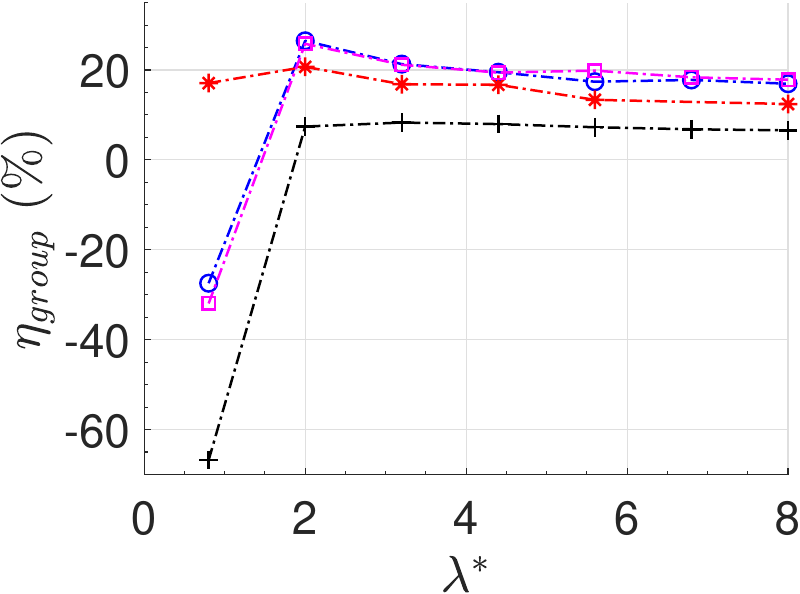}{(\counterb) $ G = 0.25 $}
		
		\addlabelbb{0.49}{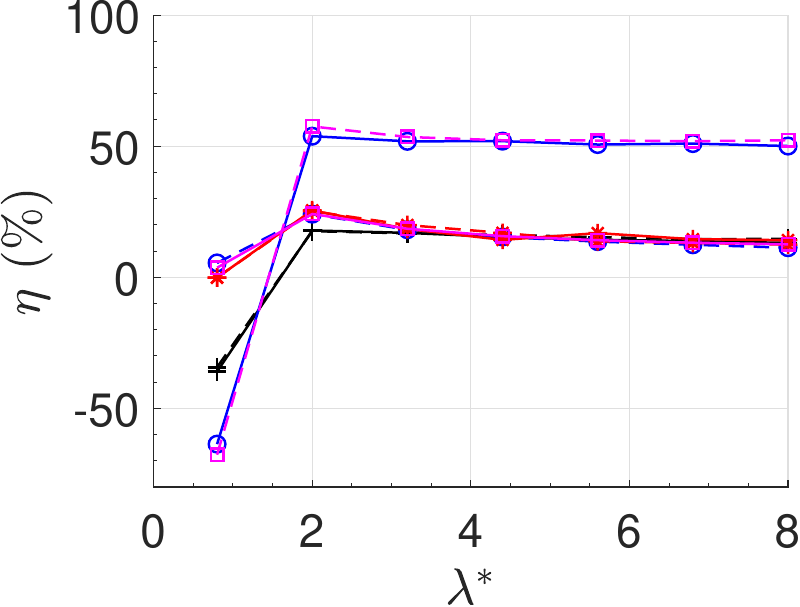}{(\counterb) $ G = 0.30 $}
		\addlabelbb{0.49}{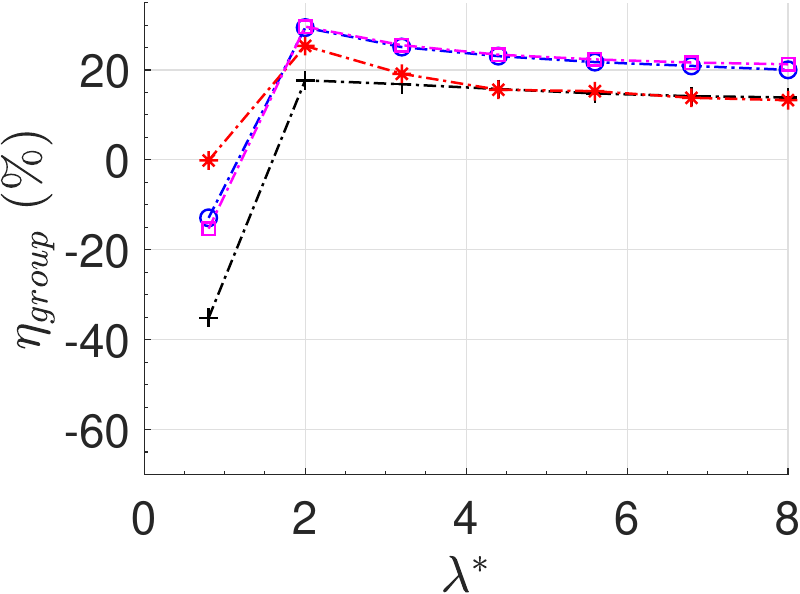}{(\counterb) $ G = 0.30 $}
		
		\addlabelbb{0.49}{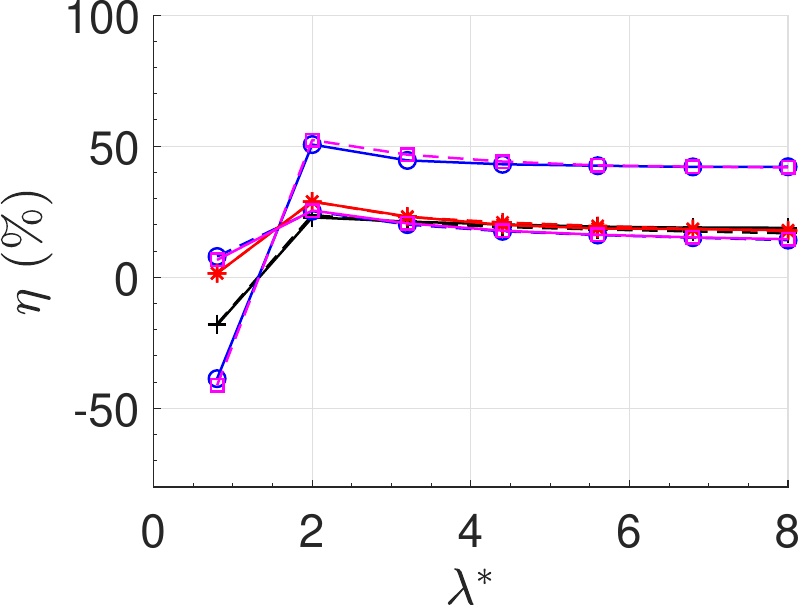}{(\counterb) $ G = 0.35 $}
		\addlabelbb{0.49}{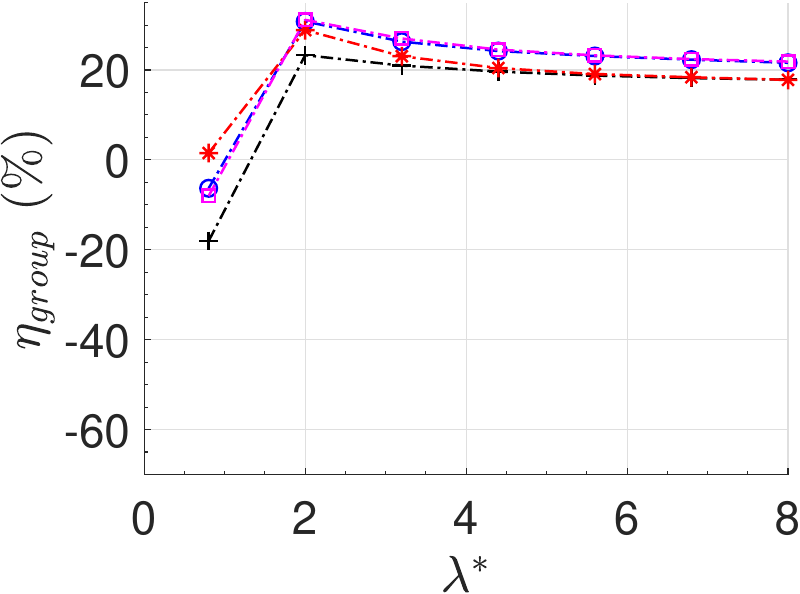}{(\counterb) $ G = 0.35 $}

		\includegraphics[width=1.0\linewidth]{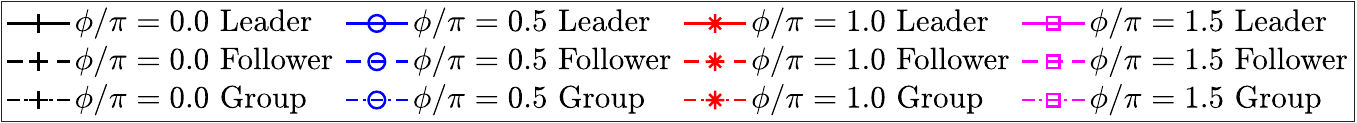}

		\caption{Variation of (a \& c \& e) propeller efficiency $ \eta $ and (b \& d \& f) group efficiency $ \eta_{group} $ with a series of wavelength $ \lambda^* = 0.8 - 8.0 $, while side-by-side distance $ D = 0 $, phase difference $ \phi = 0 - 1.5 \pi $, and gap distance (a-b) $ G = 0.25 $, (c-d) $ G = 0.30 $, and (e-f) $ G = 0.35 $.}
		\label{fig:lines_D_0_eta}
	\end{figure}

	\subsubsection{Vorticity distribution at $ D = 0 $}
	\label{subsec:vor_Dis0}
	In this section, the distribution of vorticity $ \omega^* $ is observed to reveal the fluid-structure interaction mechanism underlying the aforementioned patterns regarding the thrust force and swimming economy of the schooling hydrofoils.
	Wavelength $ \lam $ is positively correlated with shed vortices' intensity and scattering area, as seen in \Cref{fig:vort_overview_lam_var}.
	At low wavelength $ \lam = 0.8 $ in \Cref{fig:vort_overview_lam_var}a, the vortex shedding is distributed in a narrow band at the downstream side of the schooling foils, indicating better stealth performance for low-wave-length swimmers.
	With a longer wavelength $ \lam $, the wake vortices can disturb a larger area, with the flow structure becoming increasingly complicated.
	One vortex dipole, \ie a pair of opposite-sign vortices, is generated in each undulation cycle. The interaction between the shed dipoles becomes more unsteady with a larger wavelength $ \lam $ (see \Cref{fig:vort_overview_lam_var}) and a smaller gap distance $ G $ (see \Cref{fig:vor_wake_G_phi}).

	At phalanx arrangement $ D = 0 $ and anti-phase $ \ppi = 1 $, vorticity distribution is examined in \Cref{fig:vor_inst_phalanx_anti_phase} to seek the underlying mechanism of the irregular force output, aforementioned in \Cref{sec:phalanx_D0}. Here, the two swimmers form a mirror symmetry geometry at any moment during the undulation process.
	However, the consequent flow pattern does not always remain symmetrical through the development of time; this symmetry breaking phenomenon tends to occur with a higher wavelength.
	At low undulation wavelength $ \lam = 0.8 $, the flow pattern is symmetrical in the initial periods, and the symmetry breaking only gradually becomes observed after the 6th period of undulation $ t^*/T = 6 $.
	At high wavelength $ \lam \geq 2.0 $, the wake flow becomes highly irregular within merely 1 or 2 initial periods of undulation; the chaotic flow structure occurs very close to the tails of the undulating fish, thus causing the fluctuation in the output thrust and lateral force (further discussed in \Cref{subsec:FSI_Dis0}). Across various $ \lam $, the propulsive performance measurements such as thrust force are highly consistent with the symmetric features of the near field wake structure.
	\cite{Gungor2020} also discovered similar symmetry breaking phenomenon of two hydrofoils pitching in anti-phase, though only with infinite wavelength $ \lam = + \infty $ at $ Re = 4000 $ and $ St = 0.25 - 0.5 $.
	In summary, the irregularity of the flow structure tends to increase with undulation wavelength $ \lam $.

	At low wavelength $ \lam $, the gap distance $ G $ and phase difference $ \phi $ can affect the skewness, symmetry and regularity of the wake pattern, as demonstrated in \cref{fig:vor_wake_G_phi}.
	It is interesting to observe that at $ \ppi = 0.5 $ and $ \ppi = 1.5 $, the vortex shedding direction is slightly skewed towards the right and left sides of the swimming direction, respectively.
	At $ \ppi = 1.0 $, the flow structure is symmetrical due to the anti-phase undulation of the hydrofoils. The intensity of the vortices decreases with a larger gap distance $ G $. With a small gap distance $ G = 0.25 $, the vortices are distributed in a narrower band of wake flow, i.e. the dynamic energy is more concentrated at $ G = 0.25 $; in contrast, the decrease in gap distance $ G $ is relatively less effective in cases with other phase differences $ \phi $.
	At $ \ppi = 0 $, mixture of vortices is observed at $ G = 0.25 $, whereas at $ G \geq 0.30 $, the mixture does not occur.
	In summary, at low wavelength $ \lam = 0.8 $, the distribution of vortices is subtly affected by the gap distance and phase difference; concentration of dynamic energy is discovered at low gap distance and when the two hydrofoils swim in anti-phase.

	\newcommand{\addlabelvor}[3]{%
		\begin{tikzpicture}
			\node[anchor=south west,inner sep=0] (image) at (0,0) 
			{\includegraphics[width=#1\linewidth, trim={4cm 1cm 0.5cm 1.25cm},clip]{#2}};%
			\begin{scope}[x={(image.south east)},y={(image.north west)}]
				\node[anchor=south west] at (0.00,0.8) {#3};	%
			\end{scope}
		\end{tikzpicture}%
	}
	\newcounter{testaa}
	\setcounter{testaa}{0}
	\newcommand\counteraa{\stepcounter{testaa}\alph{testaa}}
	\begin{figure}
		\centering
		\addlabelvor{0.49}{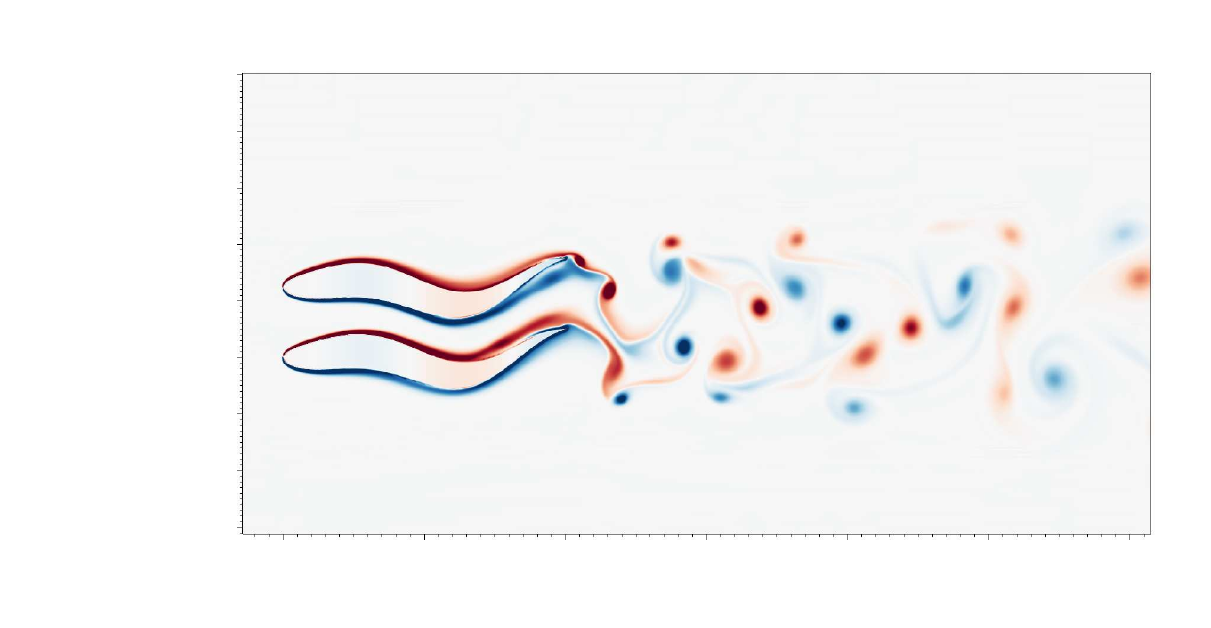}{(\counteraa) $ \lam = 0.8 $}
		\addlabelvor{0.49}{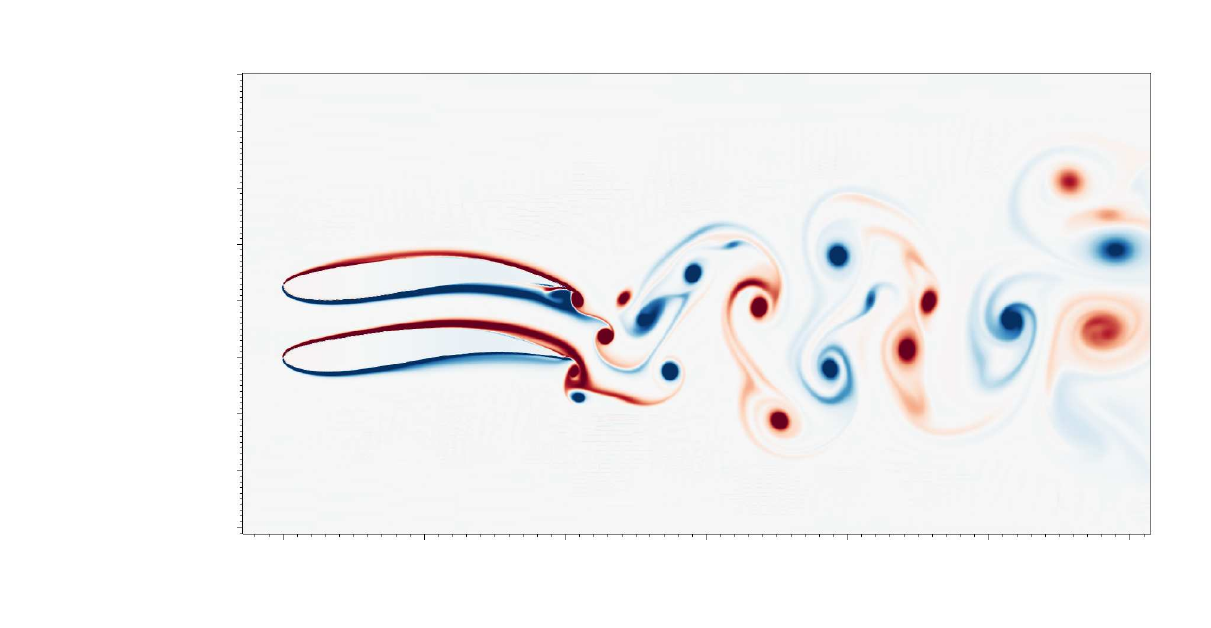}{(\counteraa) $ \lam = 2.0 $}
		\addlabelvor{0.49}{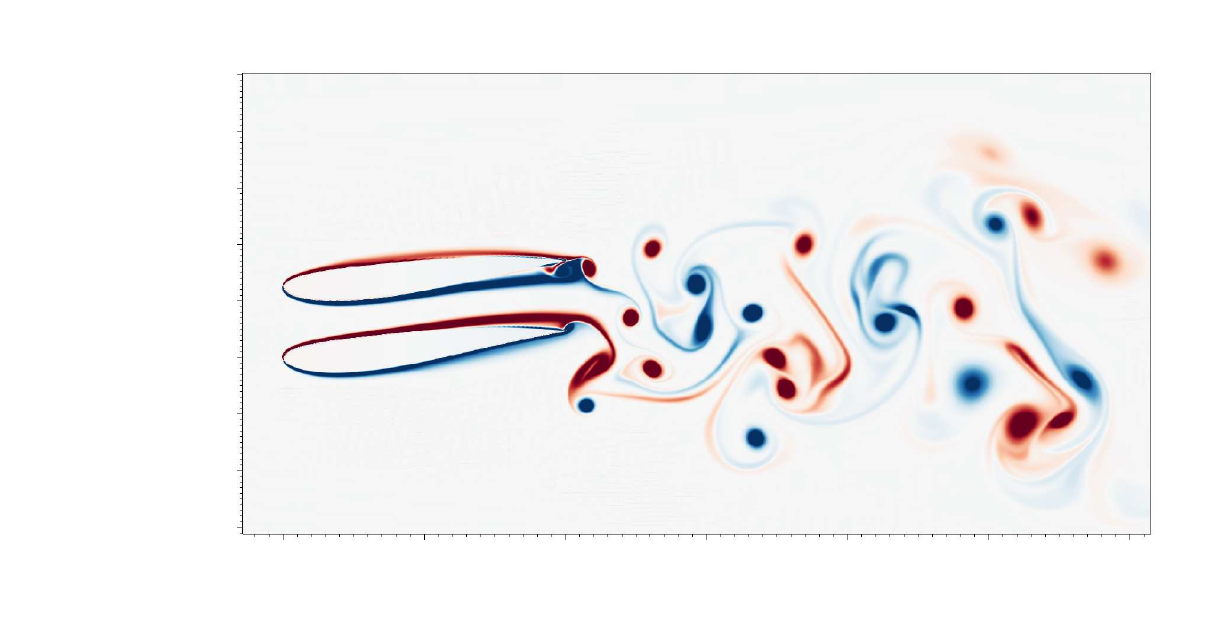}{(\counteraa) $ \lam = 3.2 $}
		\addlabelvor{0.49}{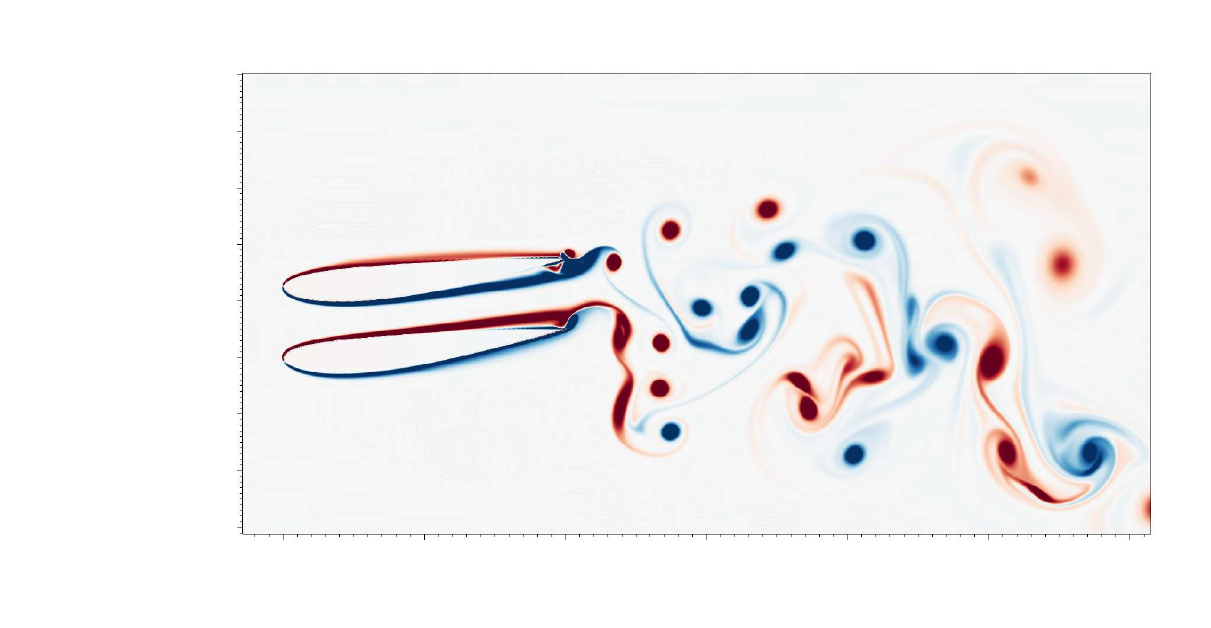}{(\counteraa) $ \lam = 4.4 $}
		\addlabelvor{0.49}{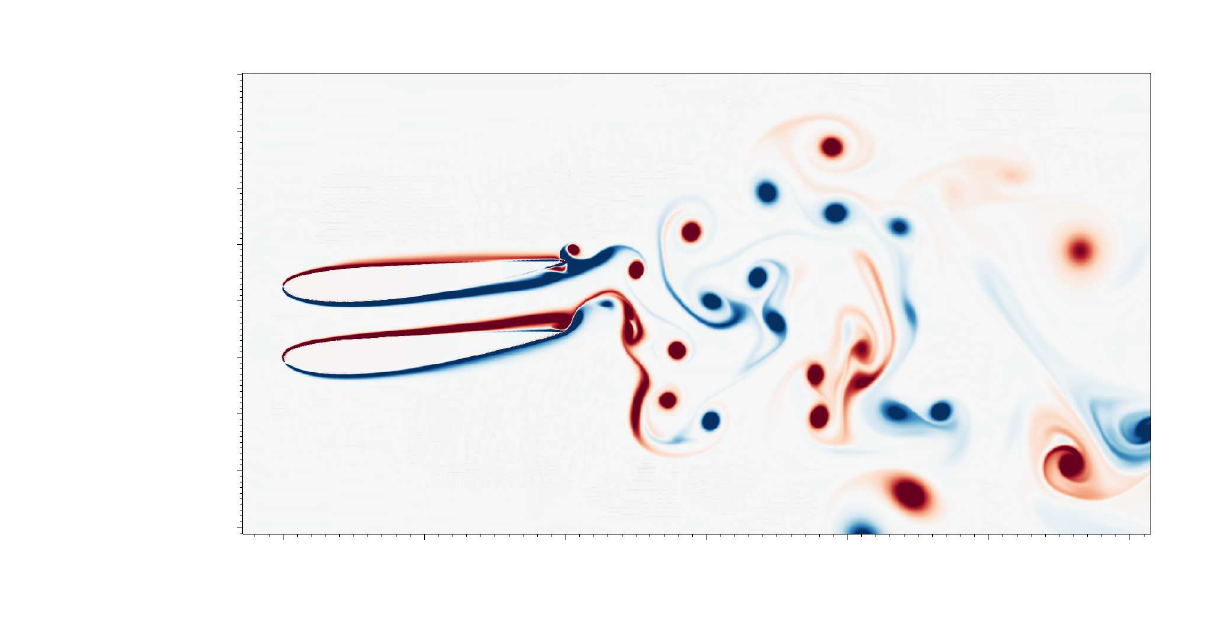}{(\counteraa) $ \lam = 5.6 $}
		\addlabelvor{0.49}{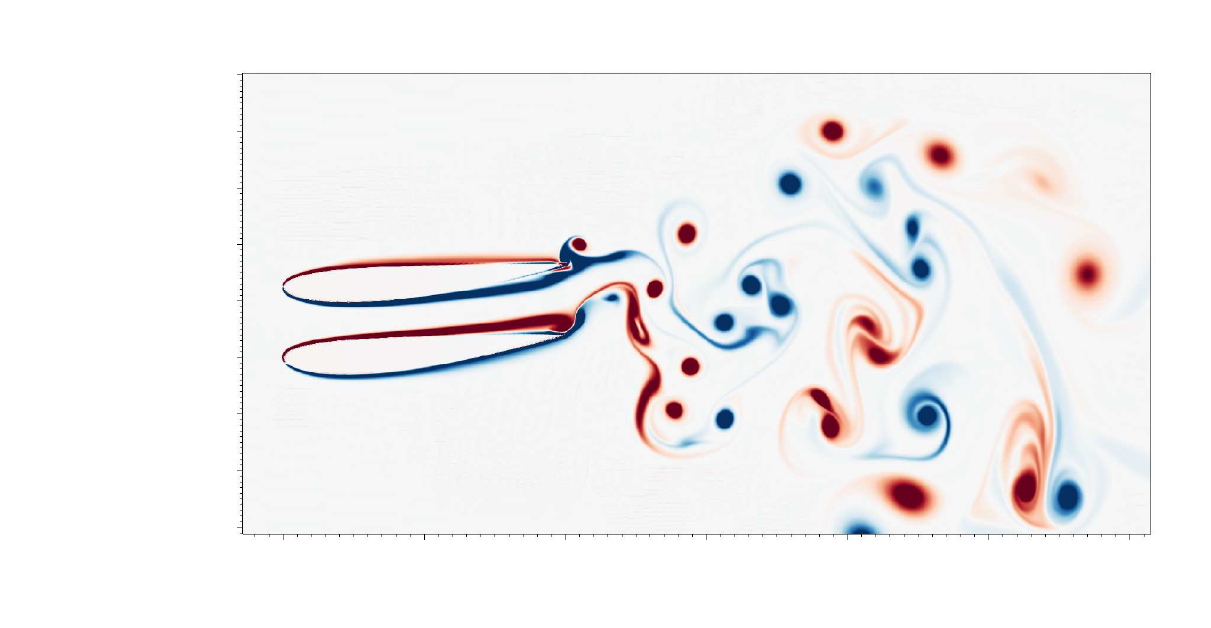}{(\counteraa) $ \lam = 6.8 $}
		\caption{
			Variation of hydrofoil geometry and vorticity contours at time $ t^*/T = 5 $ with $ G = 0.25 $, $ D = 0 $, $ \phi = 0 $, and (a) $ \lam = 0.8 $ (b) $ \lam = 2.0 $ (c) $ \lam = 3.2 $ (d) $ \lam = 4.4 $ (e) $ \lam = 5.6 $ (f) $ \lam = 6.8 $. In general, intensity and scattering angle of wake vorticity distribution increase with wavelength $ \lam $, indicating better stealth performance for low-wave-length swimmers.
		}
		\label{fig:vort_overview_lam_var}
	\end{figure}

	\renewcommand{\addlabelvor}[3]{%
		\begin{tikzpicture}
			\node[anchor=south west,inner sep=0] (image) at (0,0) 
			{\includegraphics[width=#1\linewidth, trim={4.05cm 1cm 1cm 1.25cm},clip]{#2}};%
			\begin{scope}[x={(image.south east)},y={(image.north west)}]
				\node[anchor=south west] at (0.00,0.75) {\footnotesize #3};	%
			\end{scope}
		\end{tikzpicture}%
	}
	\begin{figure}
		\centering
		\setcounter{testaa}{0}
		\addlabelvor{0.32}{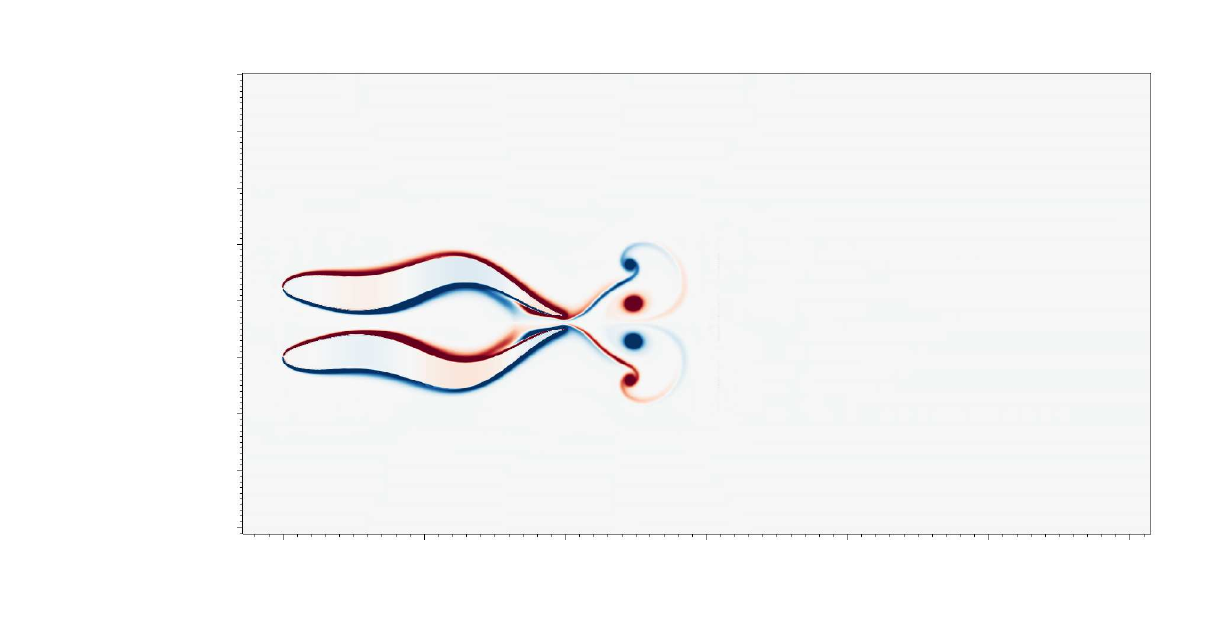}{(\counteraa) $ \lam = 0.8,\ t^*/T = 1 $}
		\addlabelvor{0.32}{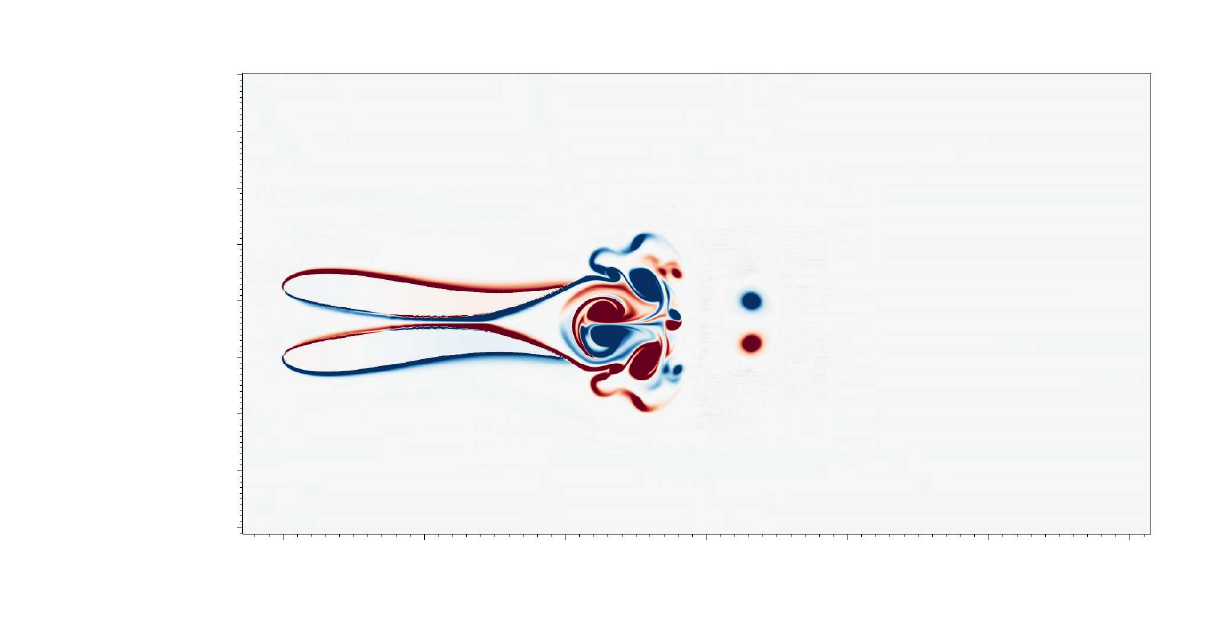}{(\counteraa) $ \lam = 2.0,\ t^*/T = 1  $ }
		\addlabelvor{0.32}{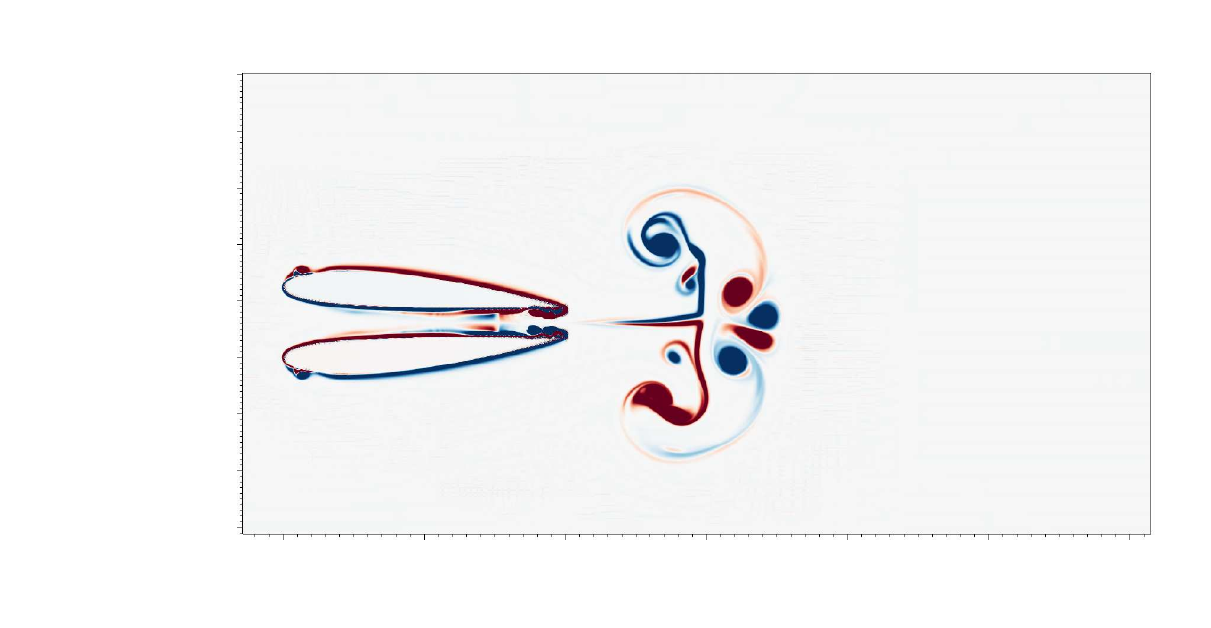}{(\counteraa) $ \lam = 8.0,\ t^*/T = 1  $}
		
		\addlabelvor{0.32}{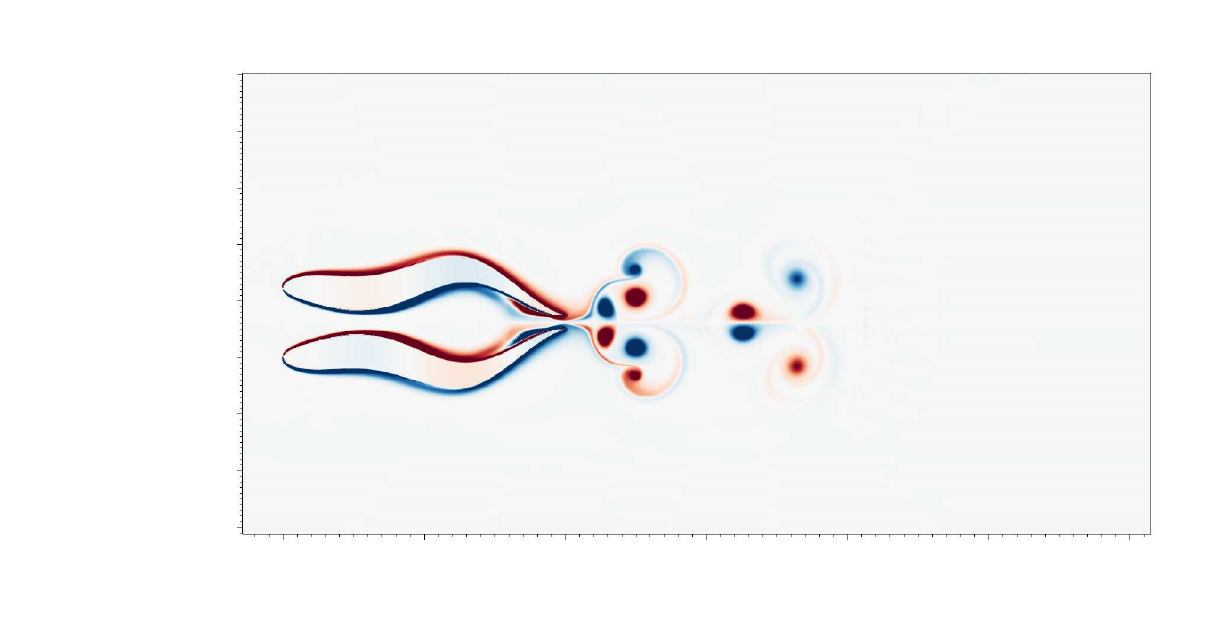}{(\counteraa) $ \lam = 0.8,\ t^*/T = 2  $}
		\addlabelvor{0.32}{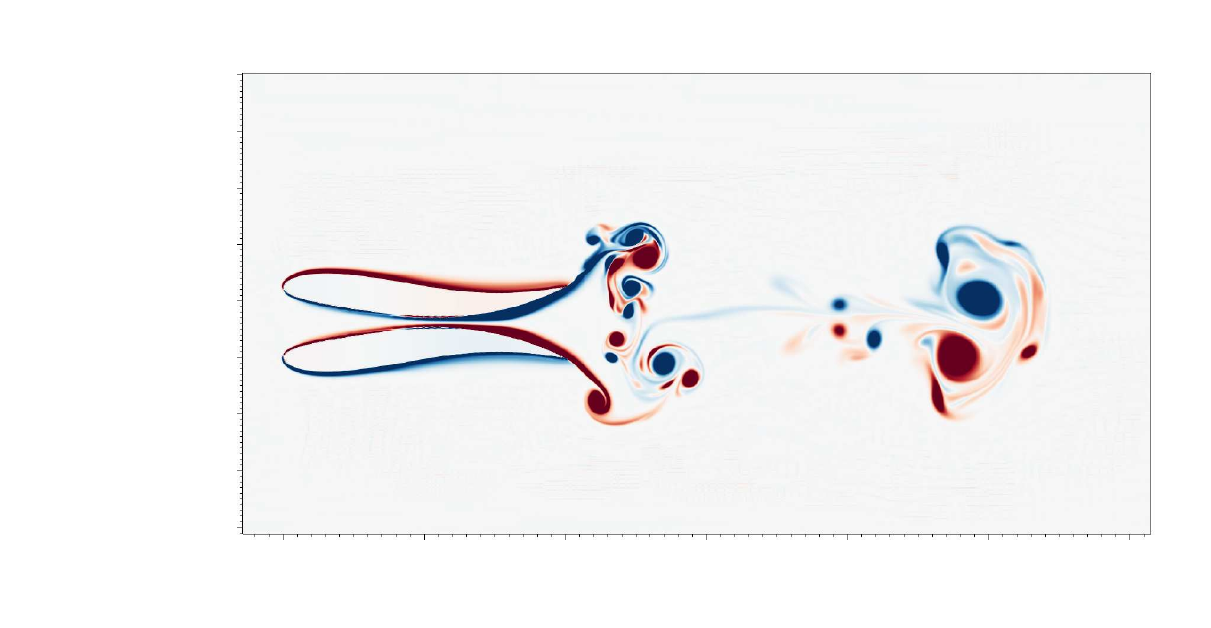}{(\counteraa) $ \lam = 2.0,\ t^*/T = 2  $ }
		\addlabelvor{0.32}{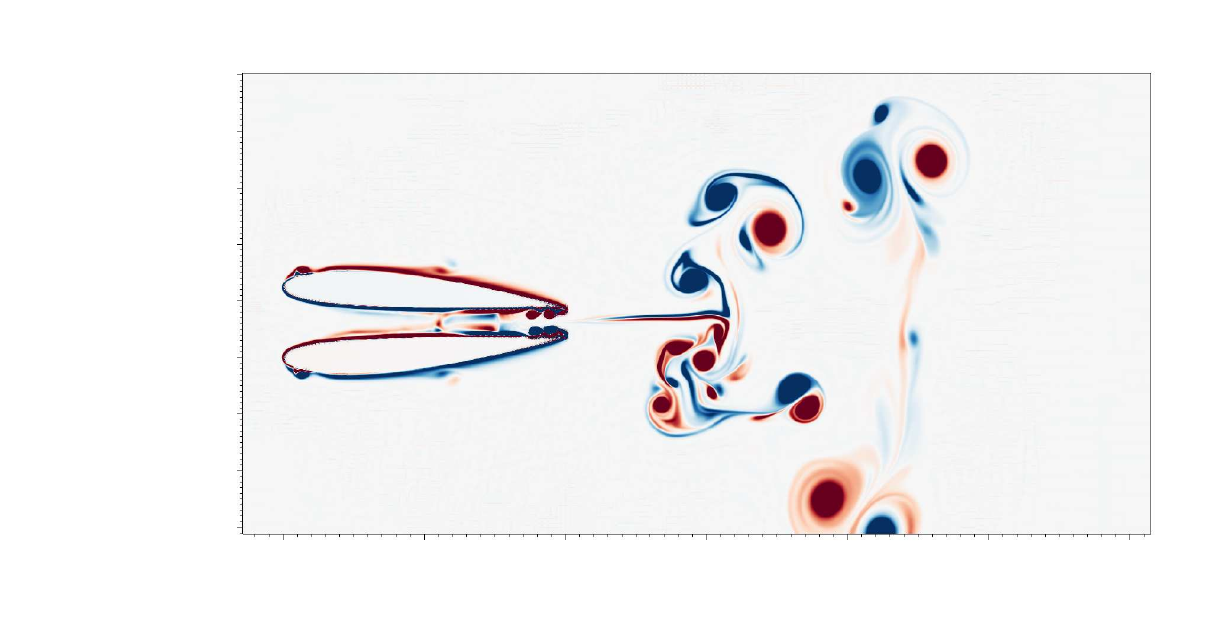}{(\counteraa) $ \lam = 8.0,\ t^*/T = 2  $}
		
		\addlabelvor{0.32}{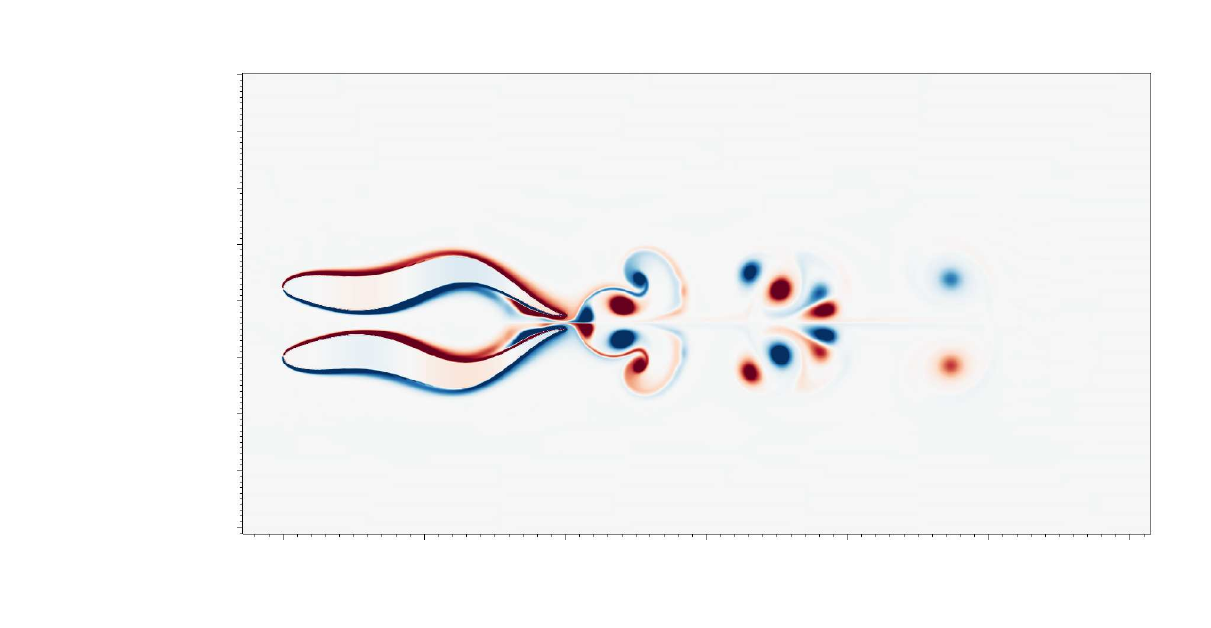}{(\counteraa) $ \lam = 0.8,\ t^*/T = 3  $}
		\addlabelvor{0.32}{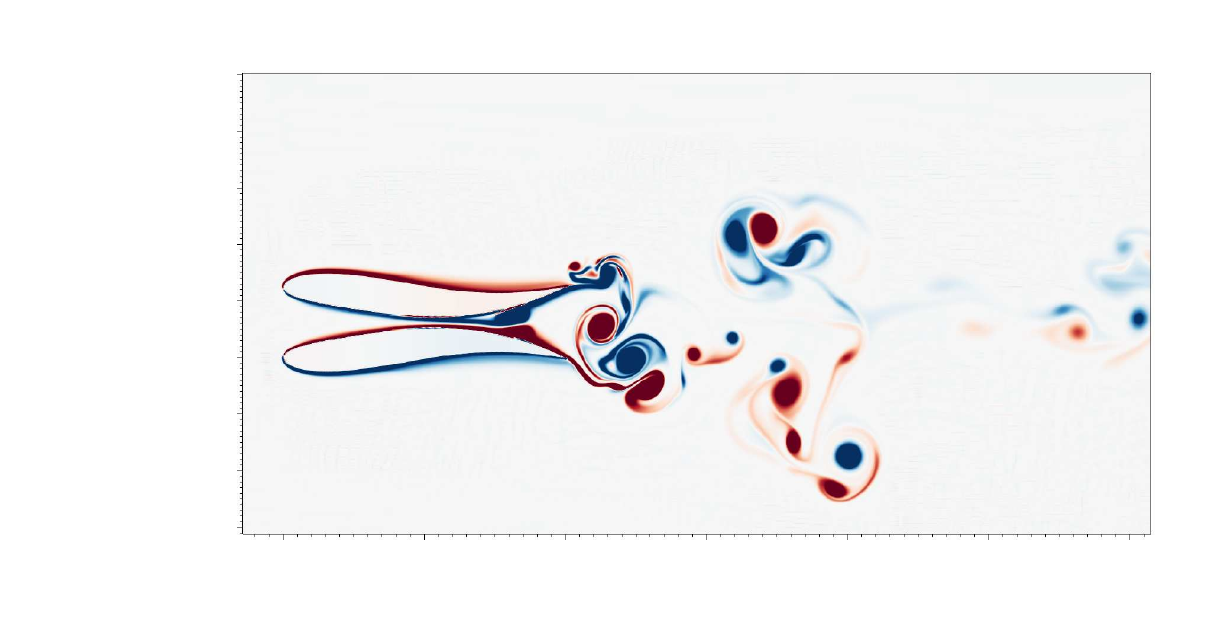}{(\counteraa) $ \lam = 2.0,\ t^*/T = 3  $ }
		\addlabelvor{0.32}{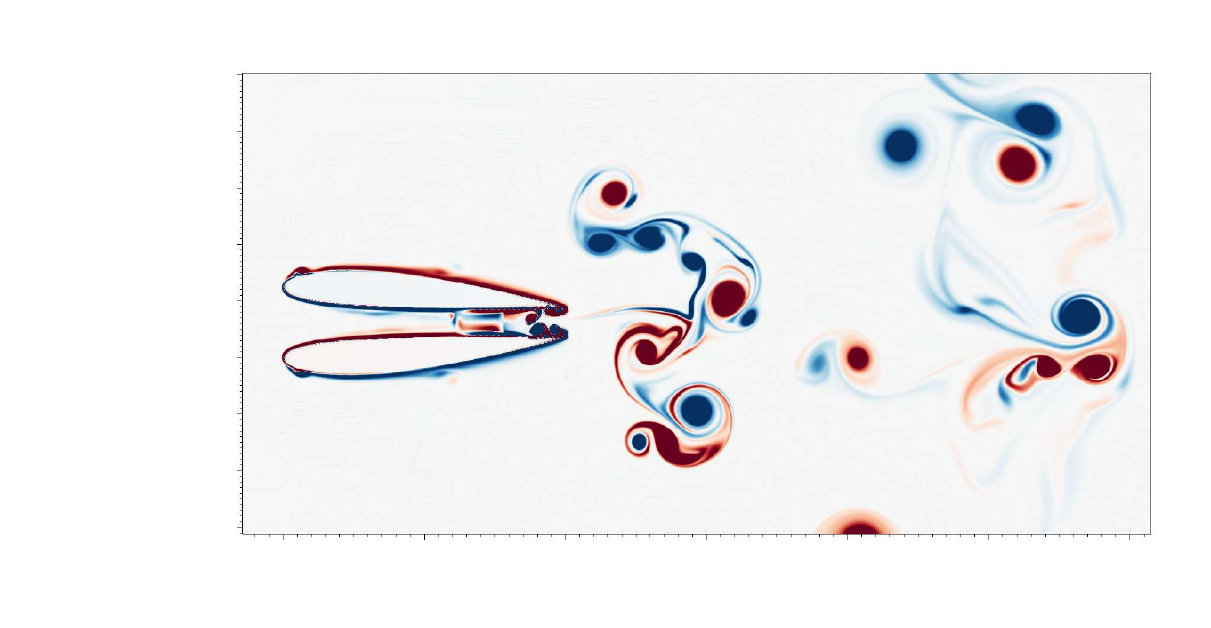}{(\counteraa) $ \lam = 8.0,\ t^*/T = 3  $}
		
		\addlabelvor{0.32}{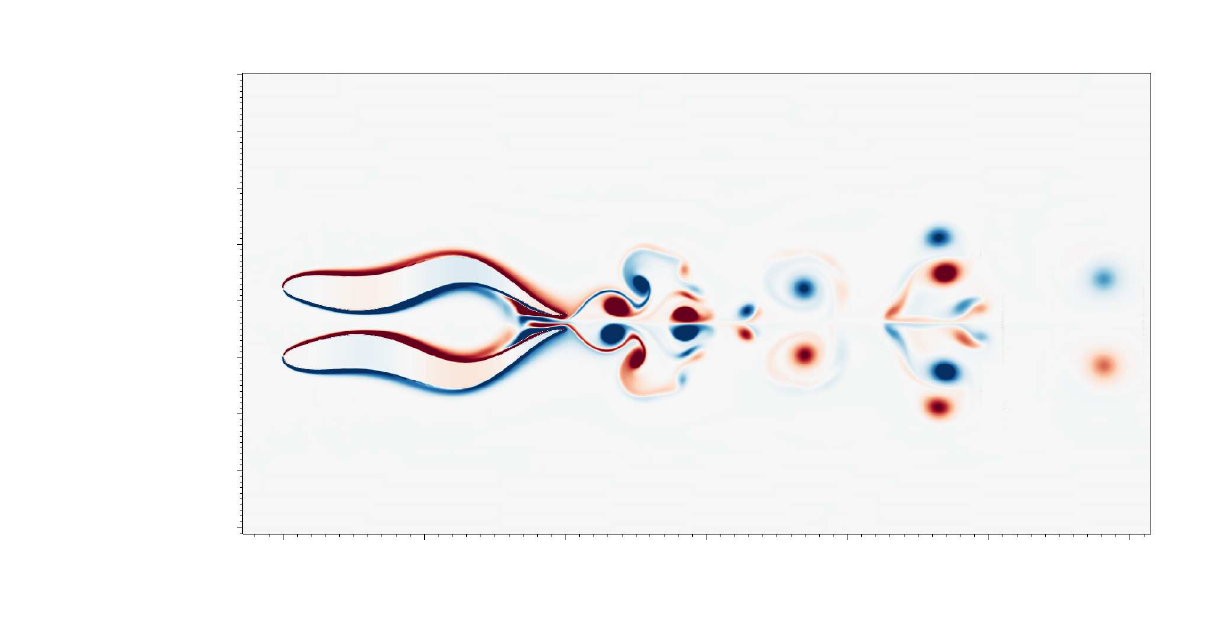}{(\counteraa) $ \lam = 0.8,\ t^*/T = 4  $}
		\addlabelvor{0.32}{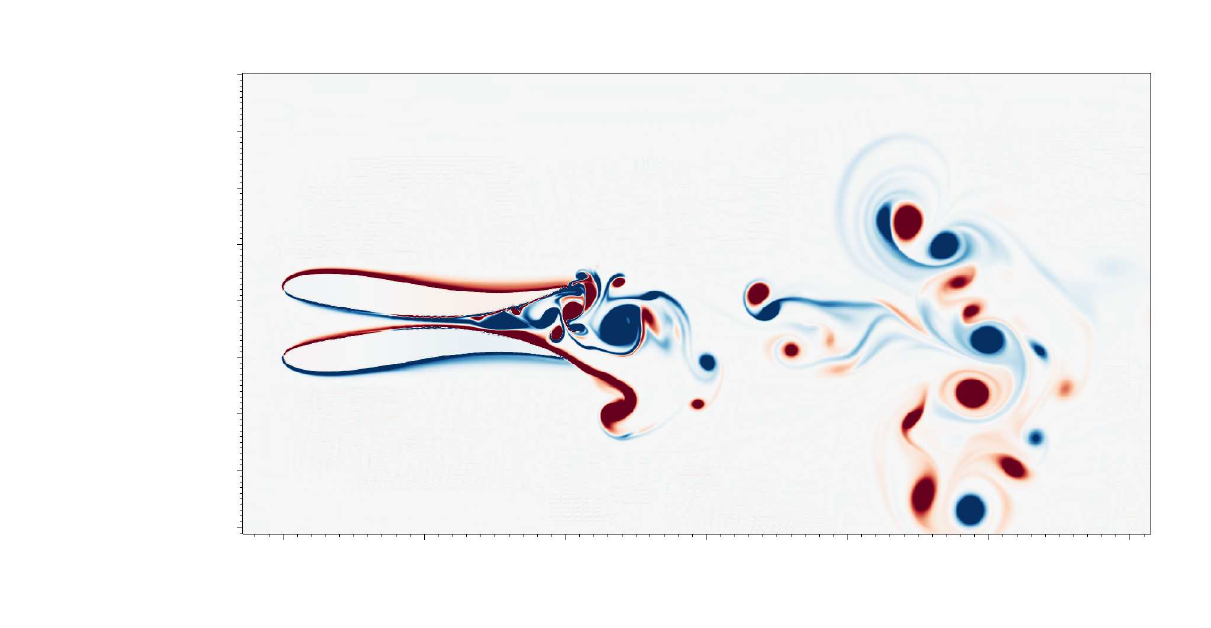}{(\counteraa) $ \lam = 2.0,\ t^*/T = 4  $ }
		\addlabelvor{0.32}{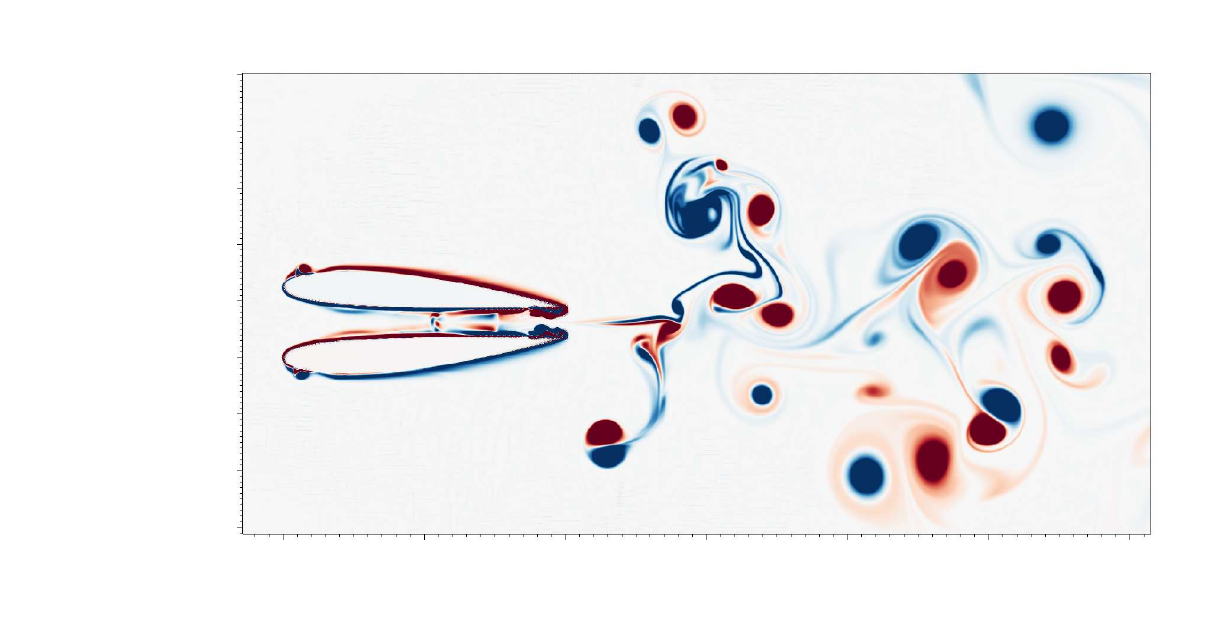}{(\counteraa) $ \lam = 8.0,\ t^*/T = 4  $}
		
		\addlabelvor{0.32}{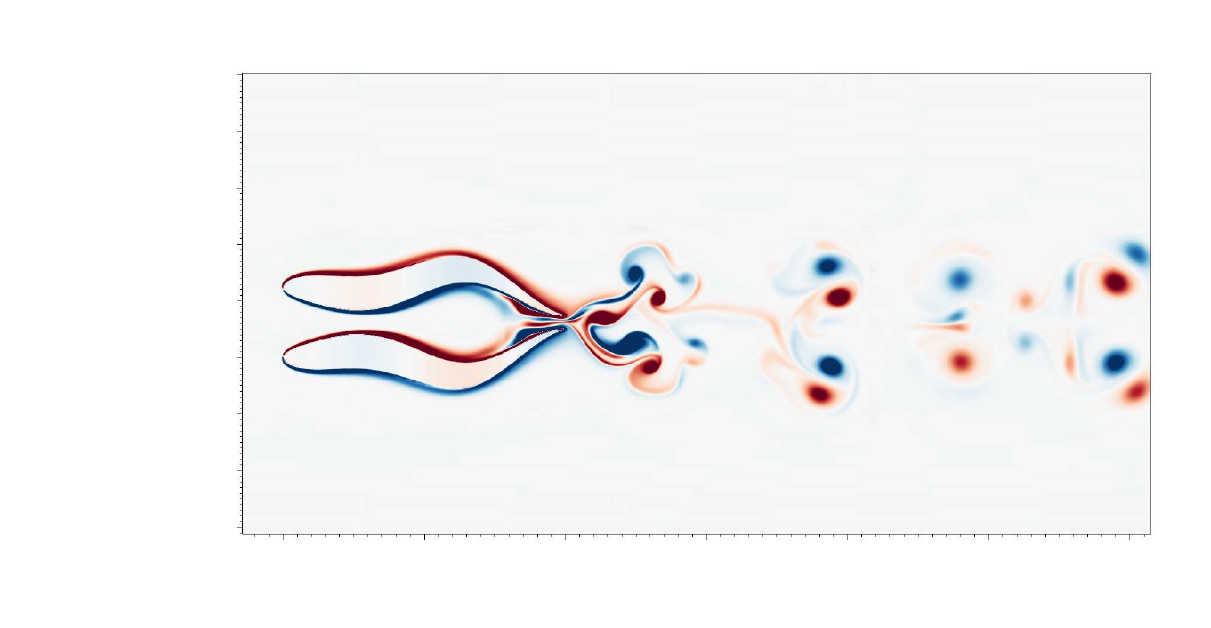}{(\counteraa) $ \lam = 0.8,\ t^*/T = 5  $}
		\addlabelvor{0.32}{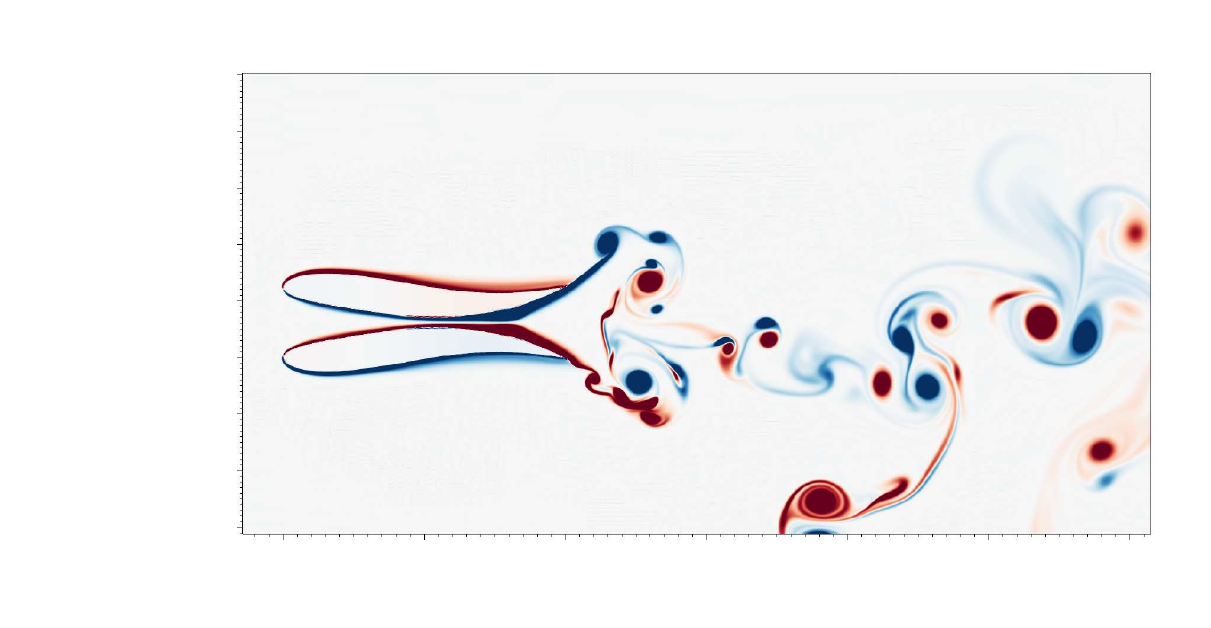}{(\counteraa) $ \lam = 2.0,\ t^*/T = 5  $ }
		\addlabelvor{0.32}{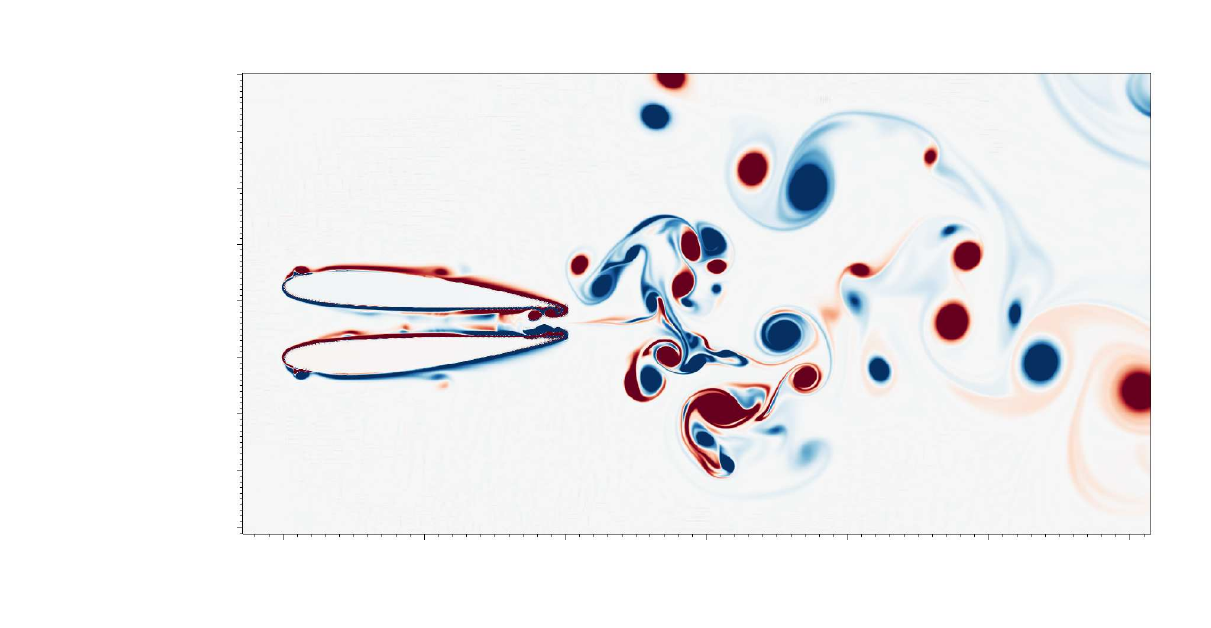}{(\counteraa) $ \lam = 8.0,\ t^*/T = 5  $}
		
		\setcounter{testaa}{0}
		\caption{
			Vorticity contours and hydrofoil deformation at phalanx arrangement $ D = 0 $, lateral gap $ G = 0.25 $, anti-phase $ \ppi = 1.0 $ at instants (a-c) $ t^*/T = 1 $ (d-f) $ t^*/T = 2 $ (g-i) $ t^*/T = 3 $ (j-l) $ t^*/T = 4 $ (m-o) $ t^*/T = 5 $ with various wavelengths (a \& d \& g \& j \& m) $ \lam = 0.8 $ (b \& e \& h \& k \& n) $ \lam = 2.0 $ (c \& f \& i \& l \& o) $ \lam = 8.0 $.
		}
		\label{fig:vor_inst_phalanx_anti_phase}
	\end{figure}
	
	\renewcommand{\addlabelvor}[3]{%
		\begin{tikzpicture}
			\node[anchor=south west,inner sep=0] (image) at (0,0) 
			{\includegraphics[width=#1\linewidth, trim={4.05cm 1cm 1cm 1.25cm},clip]{#2}};%
			\begin{scope}[x={(image.south east)},y={(image.north west)}]
				\node[anchor=south west] at (0.00,0.75) {\footnotesize #3};	%
			\end{scope}
		\end{tikzpicture}%
	}
	\begin{figure}
		\centering
		\setcounter{testaa}{0}
		\addlabelvor{0.32}{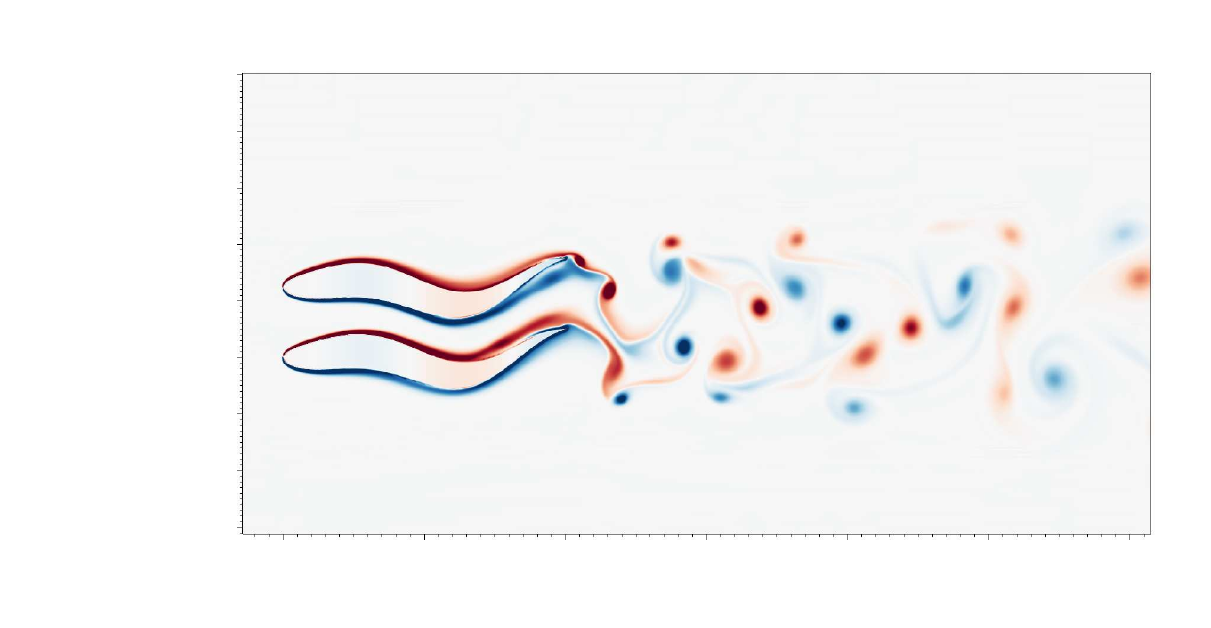}{(\counteraa) $ G = 0.25 $, $ \phi/\pi = 0 $}
		\addlabelvor{0.32}{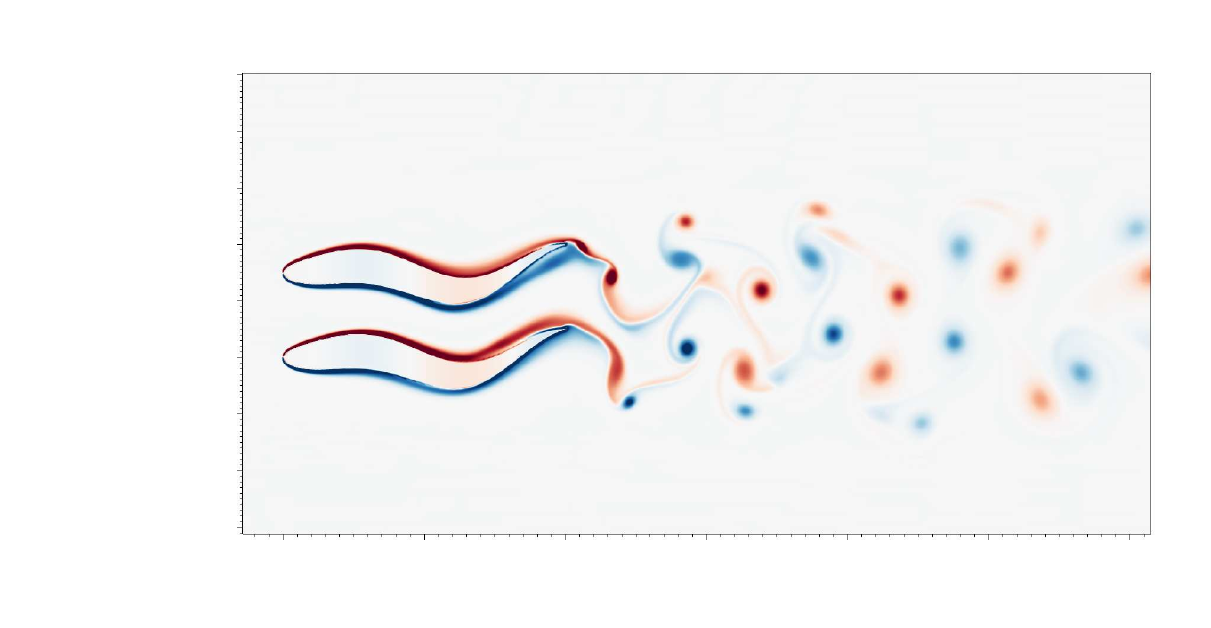}{(\counteraa) $ G = 0.30 $, $ \phi/\pi = 0 $}
		\addlabelvor{0.32}{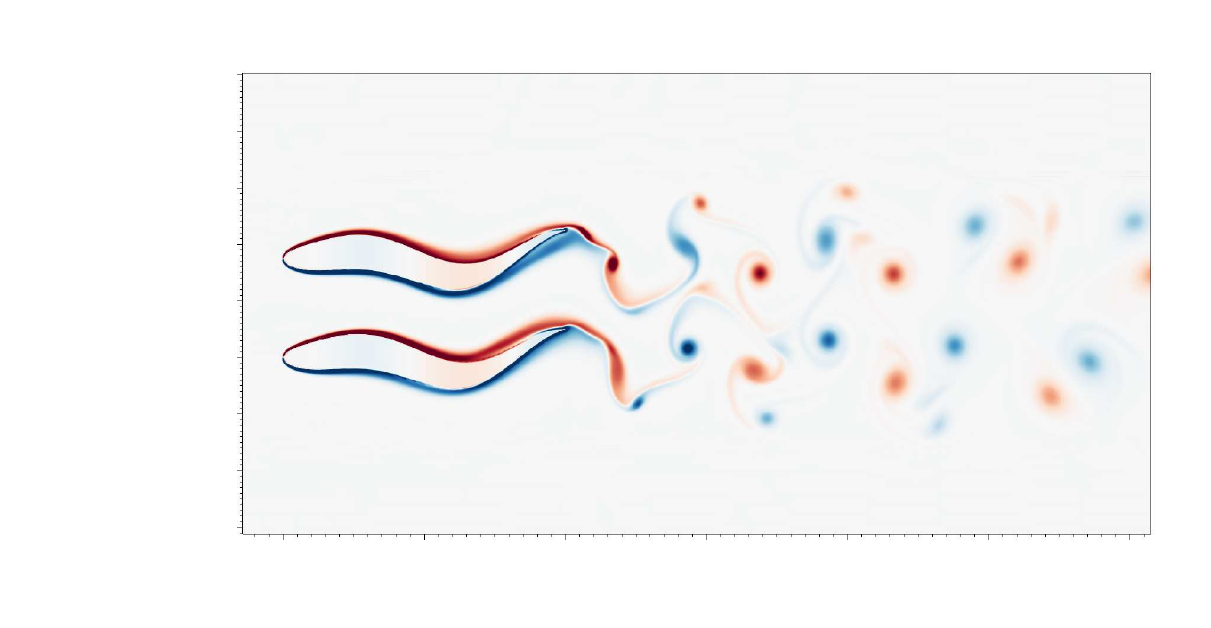}{(\counteraa) $ G = 0.35 $, $ \phi/\pi = 0 $}

		\addlabelvor{0.32}{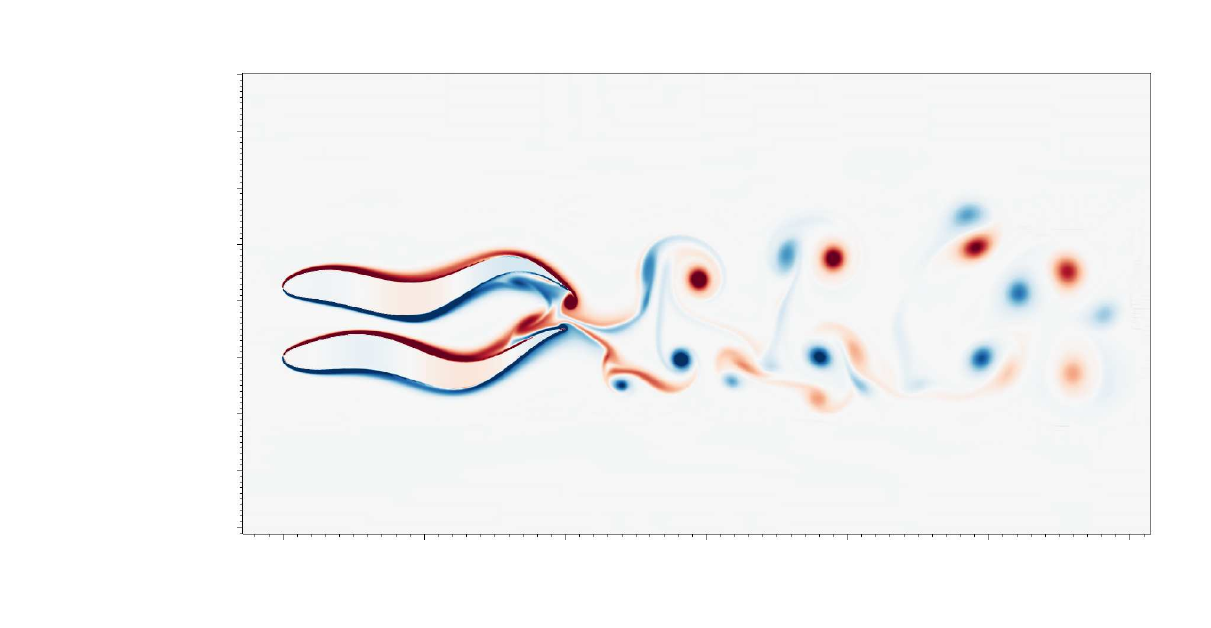}{(\counteraa) $ G = 0.25 $, $ \phi/\pi = 0.5 $}
		\addlabelvor{0.32}{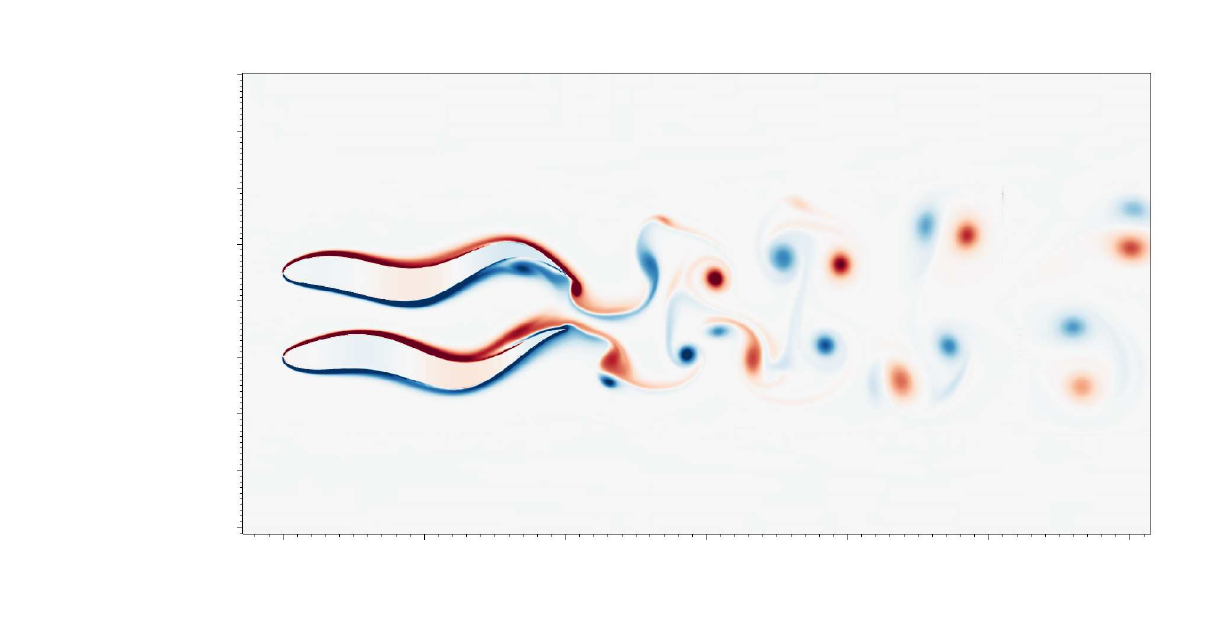}{(\counteraa) $ G = 0.30 $, $ \phi/\pi = 0.5 $}
		\addlabelvor{0.32}{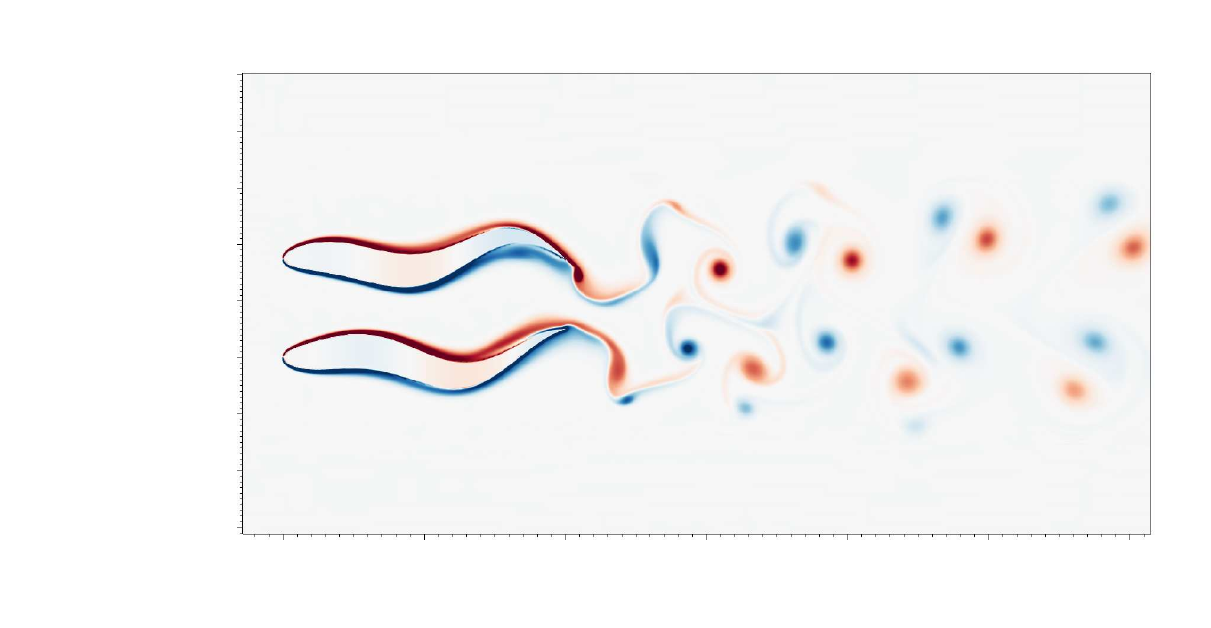}{(\counteraa) $ G = 0.35 $, $ \phi/\pi = 0.5 $}

		\addlabelvor{0.32}{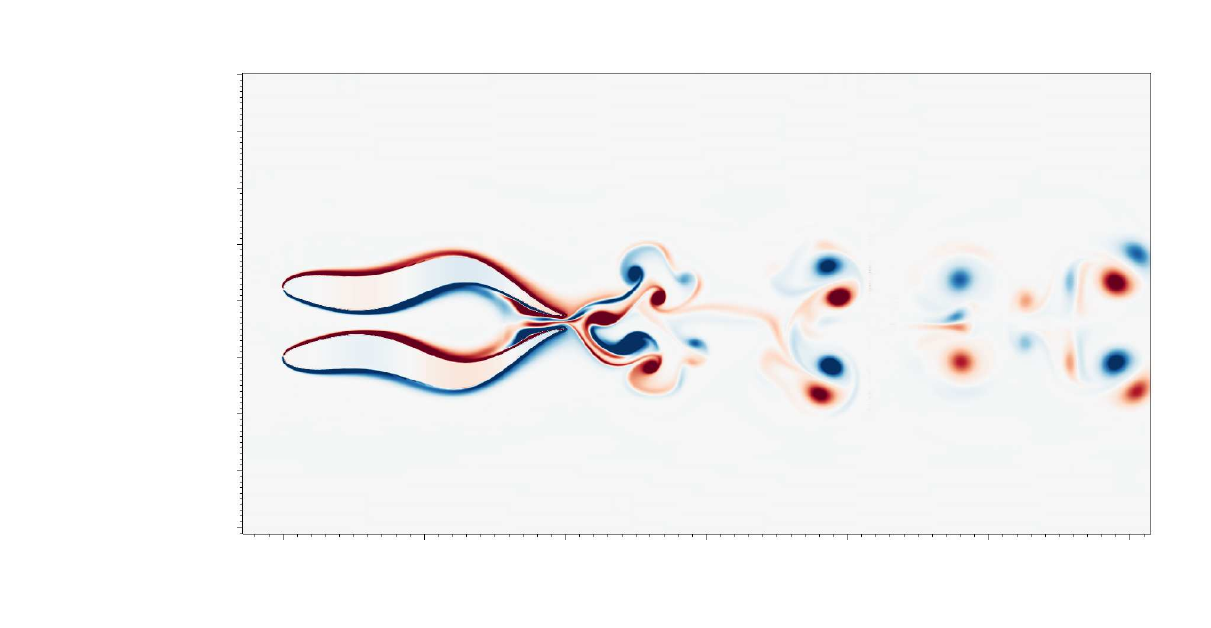}{(\counteraa) $ G = 0.25 $, $ \phi/\pi = 1.0 $}
		\addlabelvor{0.32}{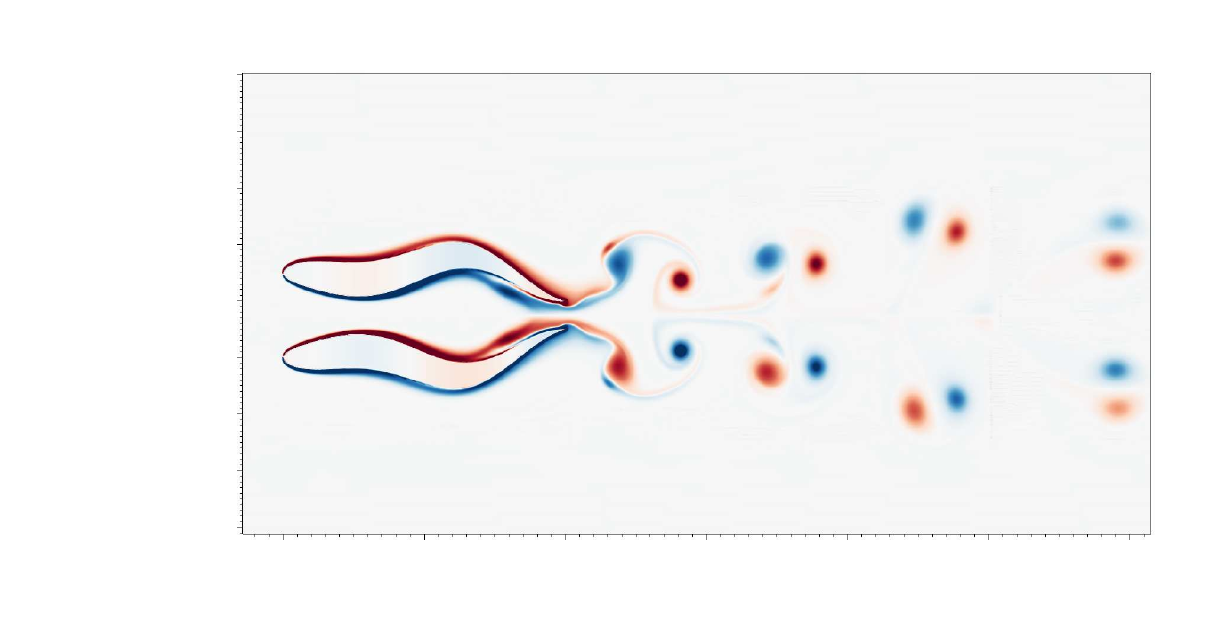}{(\counteraa) $ G = 0.30 $, $ \phi/\pi = 1.0 $}
		\addlabelvor{0.32}{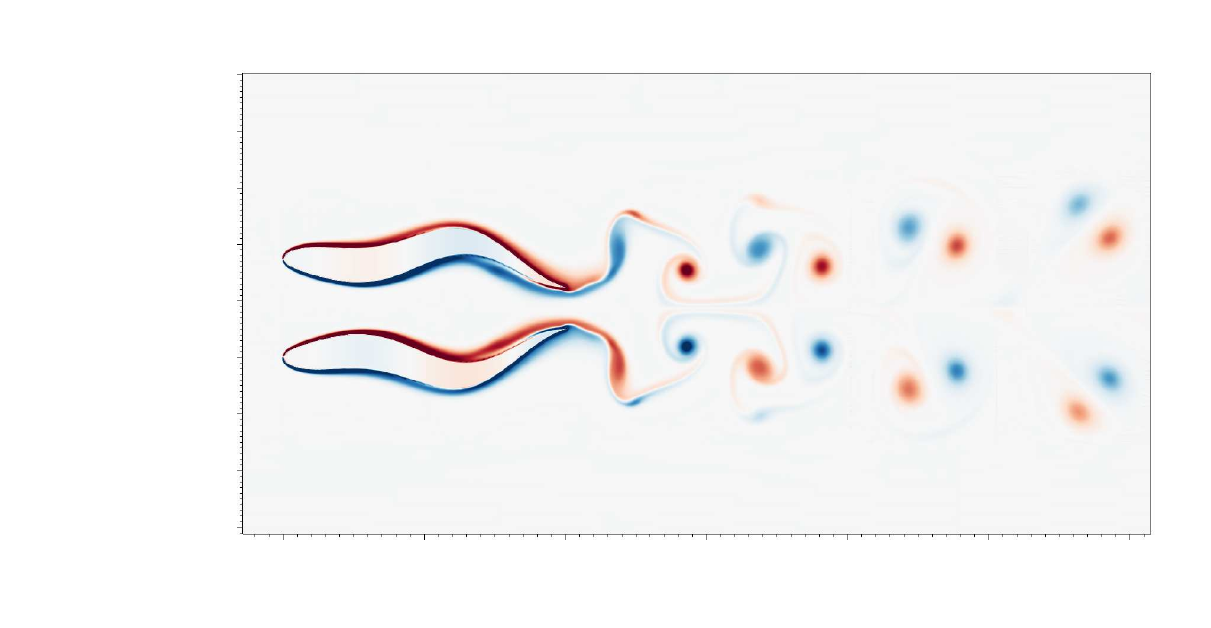}{(\counteraa) $ G = 0.35 $, $ \phi/\pi = 1.0 $}

		\addlabelvor{0.32}{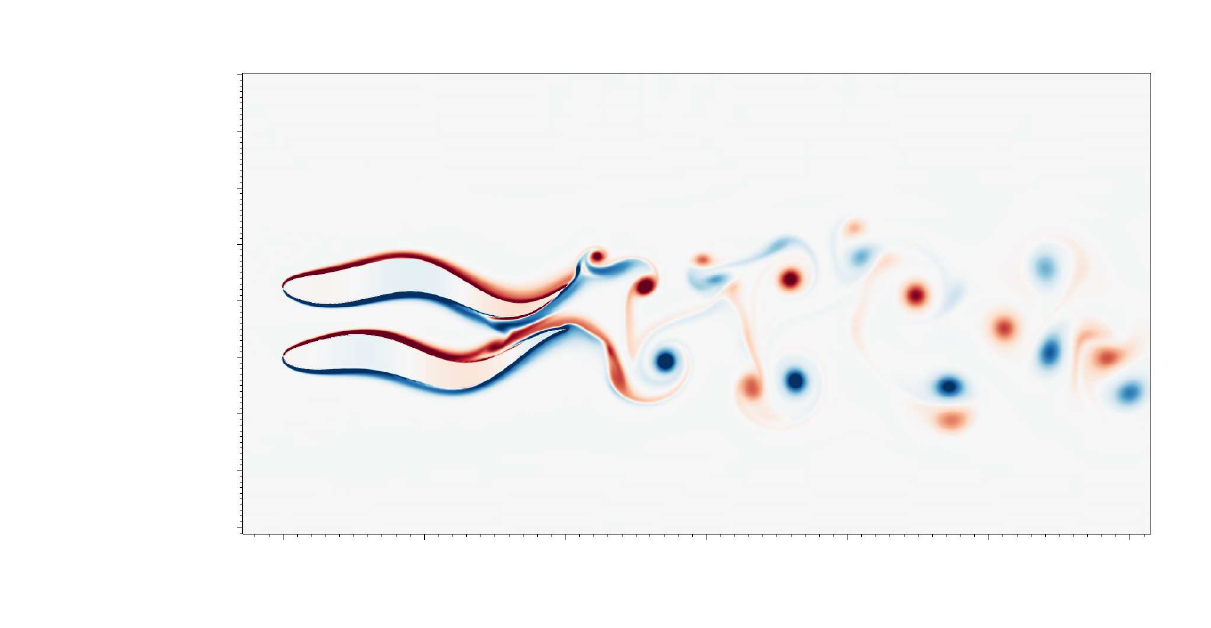}{(\counteraa) $ G = 0.25 $, $ \phi/\pi = 1.5 $}
		\addlabelvor{0.32}{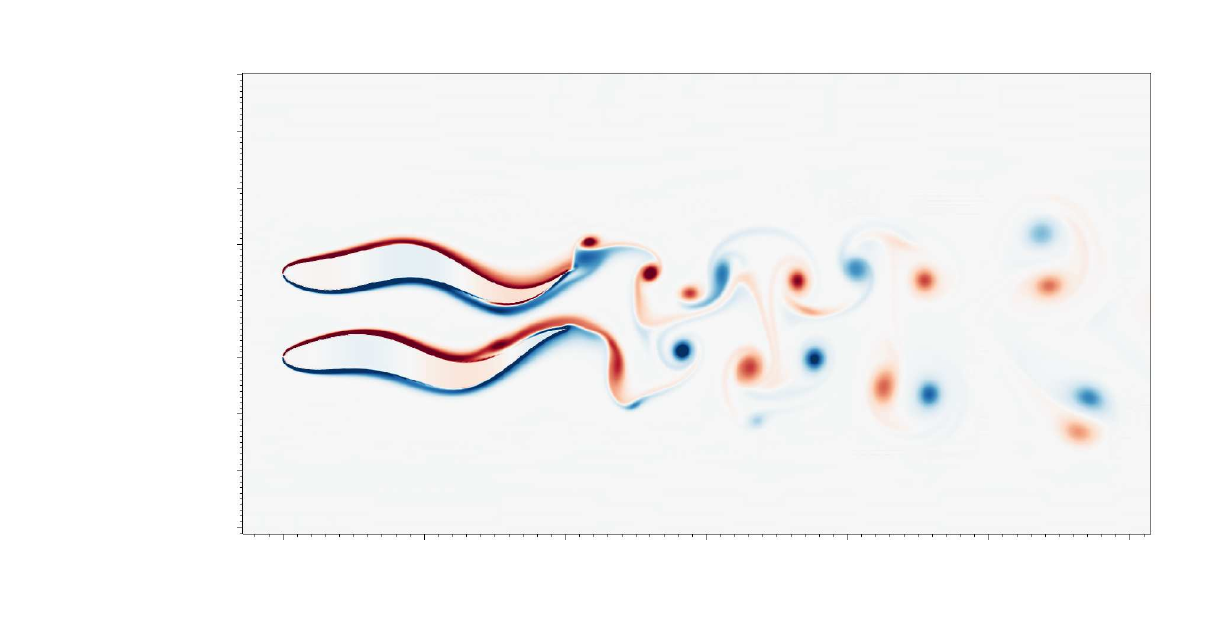}{(\counteraa) $ G = 0.30 $, $ \phi/\pi = 1.5 $}
		\addlabelvor{0.32}{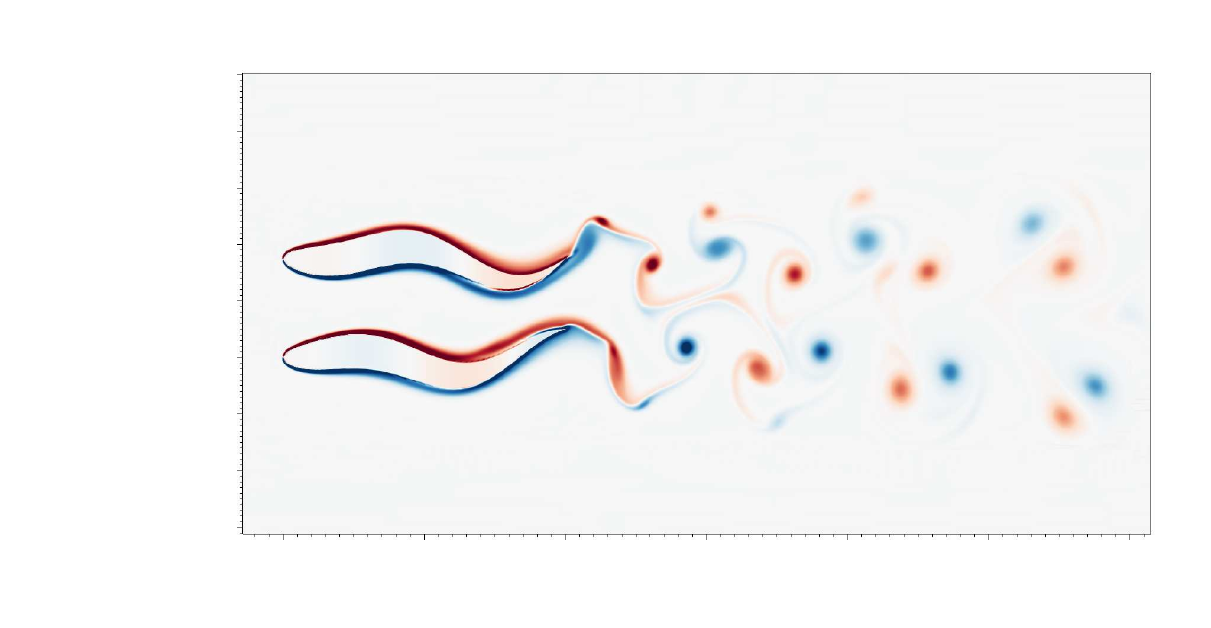}{(\counteraa) $ G = 0.35 $, $ \phi/\pi = 1.5 $}
		
		\setcounter{testaa}{0}
		\caption{
			Vorticity contours and hydrofoil deformation at instant $ t^*/T = 5 $ with $ G = 0.35 $, $ D = 0 $, (a-c) $ \phi/\pi = 0 $ (d-f) $ \phi/\pi = 0.5 $ (g-i) $ \phi/\pi = 1.0 $ (j-l) $ \phi/\pi = 1.5 $. (a \& d \& g \& j) $ G = 0.25 $ (b \& e \& h \& k) $ G = 0.30 $ (c \& f \& i \& l) $ G = 0.35 $.
			At low wavelength $ \lam = 0.8 $, the distribution of vortices is subtly affected by the gap distance and phase difference; concentration of dynamic energy is discovered at low gap distance for anti-phase cases.
		}
		\label{fig:vor_wake_G_phi}
	\end{figure}

	\subsubsection{Flow-mediated interaction between two swimmers at $ D = 0 $}
	\label{subsec:FSI_Dis0}
	For the anti-phase $ \ppi = 1 $ scenarios, the results are analysed in detail for wavelengths $ \lam = 0.8,\ 2.0,\ 8.0 $ with the help of \Cref{fig:vor_1T_D0_lam0d8,fig:vor_1T_D0_lam2d0,fig:vor_1T_D0_lam8d0}, where the time history of thrust $ C_T $ and the lateral force $ C_L $ upon the foils are examined together with vorticity distribution at corresponding instants within 1 period of deformation. This configuration of $ D = 0 $ and $ \ppi = 1 $ is justified by the observation from \cite{Ashraf2017} that schooling fish tend to form a simple side-by-side pattern with the characteristics of synchronised tail-beating with either in-phase or anti-phase swimming modes. \cite{Ashraf2017} previously observed the fish schooling of a single wavelength of red nose tetra \textit{Hemigrammus Rhodostomus}. Strong vortex interaction has been identified at low gap distance $ G = 0.25 $ and anti-phase cases $ \ppi = 1 $ in \Cref{subsec:vor_Dis0}.
	We thus further investigate the effects of wavelengths across 3 typical values $ \lam = 0.8,\ 2,\ 8 $.

	\renewcommand{\addlabelvor}[3]{%
		\begin{tikzpicture}
			\node[anchor=south west,inner sep=0] (image) at (0,0) 
			{\includegraphics[width=#1\linewidth, trim={4.05cm 1cm 1cm 1.25cm},clip]{#2}};%
			\begin{scope}[x={(image.south east)},y={(image.north west)}]
				\node[anchor=south west] at (0.00,0.75) {\footnotesize #3};	%
			\end{scope}
		\end{tikzpicture}%
	}
	\newcommand{\addlabelnotrim}[3]{%
		\begin{tikzpicture}
			\node[anchor=south west,inner sep=0] (image) at (0,0) 
			{\includegraphics[width=#1\linewidth, trim={0cm 0cm 0cm 0cm},clip]{#2}};%
			\begin{scope}[x={(image.south east)},y={(image.north west)}]
				\node[anchor=south west] at (0.1,-0.03) {\footnotesize #3};	%
			\end{scope}
		\end{tikzpicture}%
	}
	\begin{figure}
		\centering
		\setcounter{testaa}{0}
		\addlabelvor{0.49}{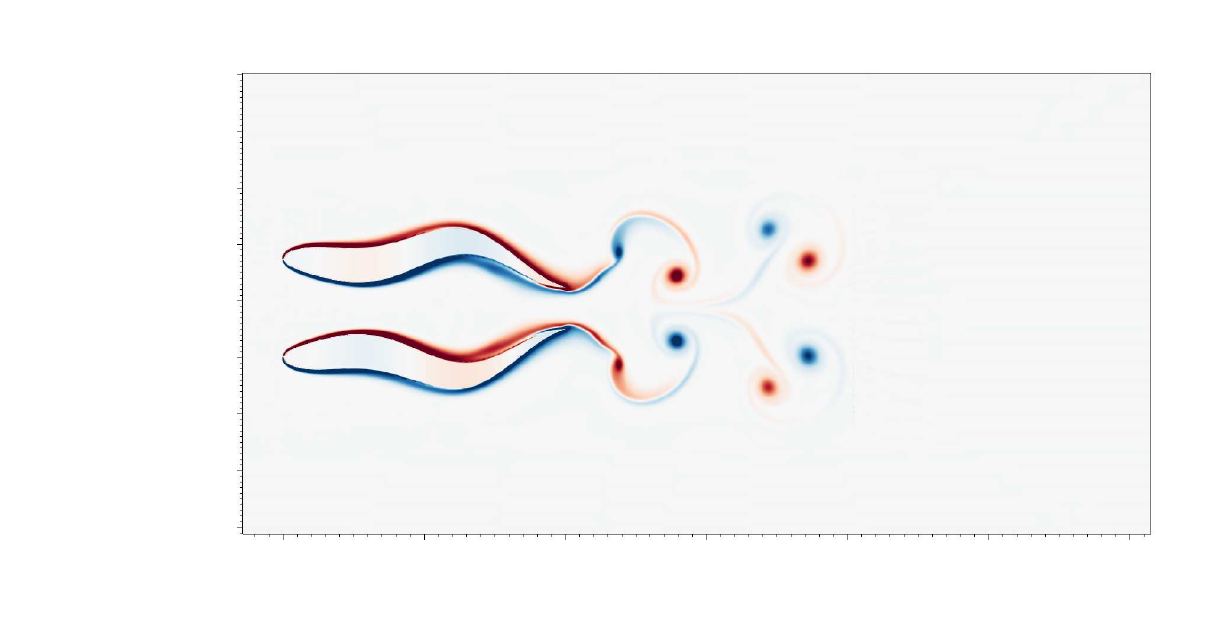}{(a) $ \lam = 0.8 $, $ t^*/T = 2.00 $}
		\addlabelvor{0.49}{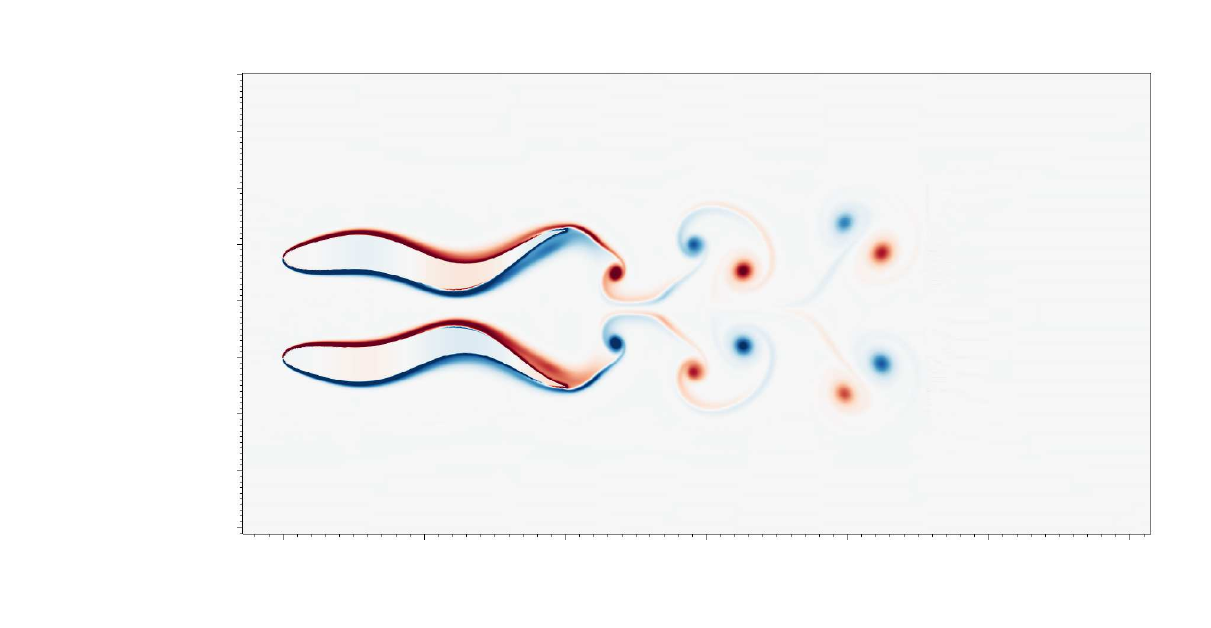}{(e) $ \lam = 0.8 $, $ t^*/T = 2.50 $}
		
		\addlabelvor{0.49}{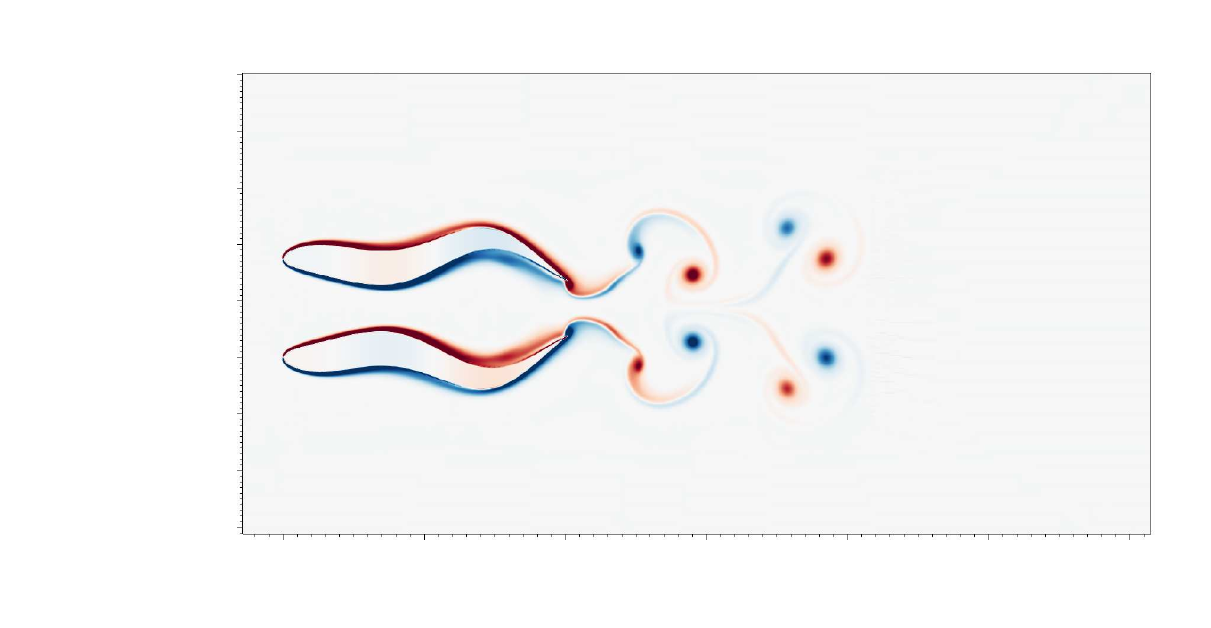}{(b) $ \lam = 0.8 $, $ t^*/T = 2.125 $}
		\addlabelvor{0.49}{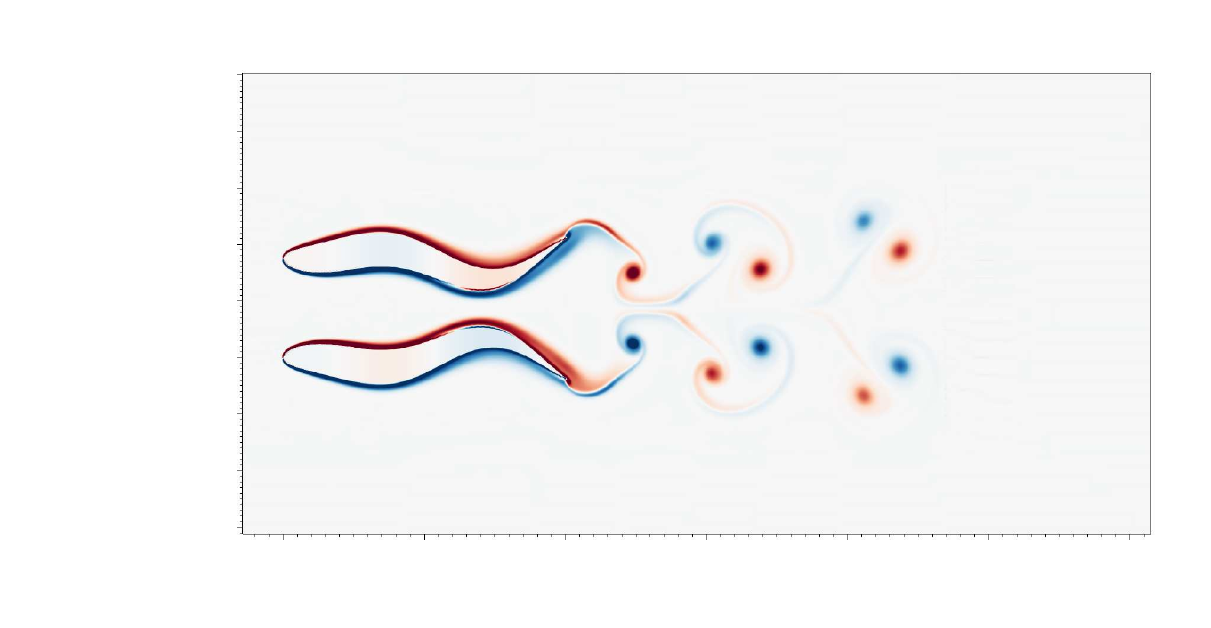}{(f) $ \lam = 0.8 $, $ t^*/T = 2.625 $}
		
		\addlabelvor{0.49}{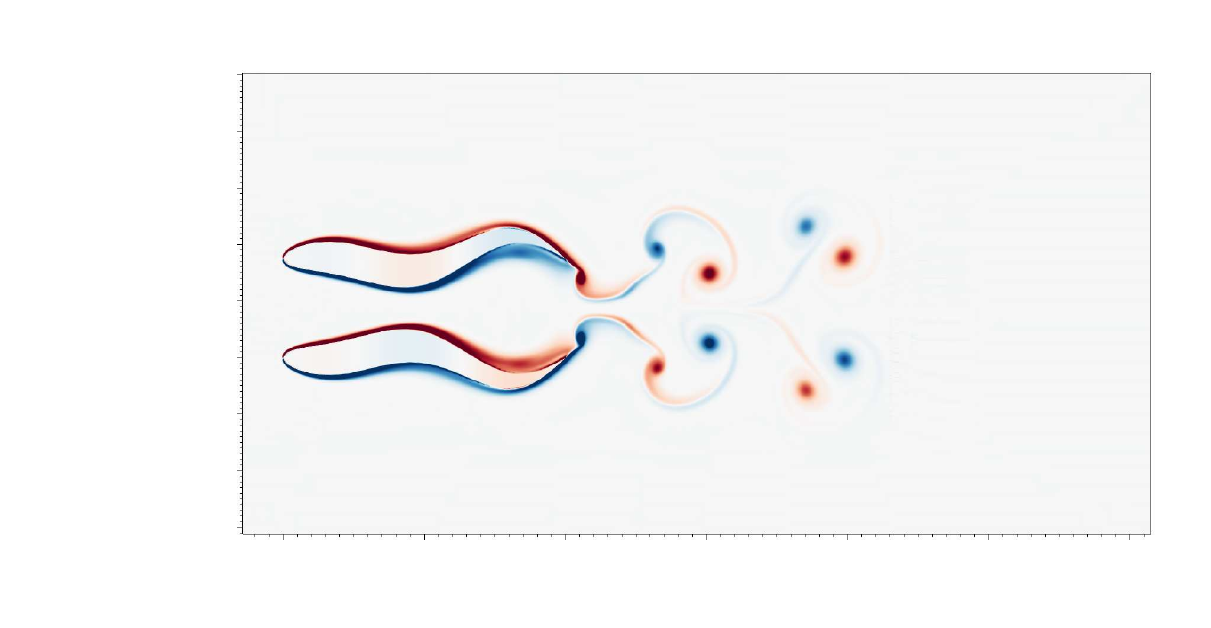}{(c) $ \lam = 0.8 $, $ t^*/T = 2.25 $}
		\addlabelvor{0.49}{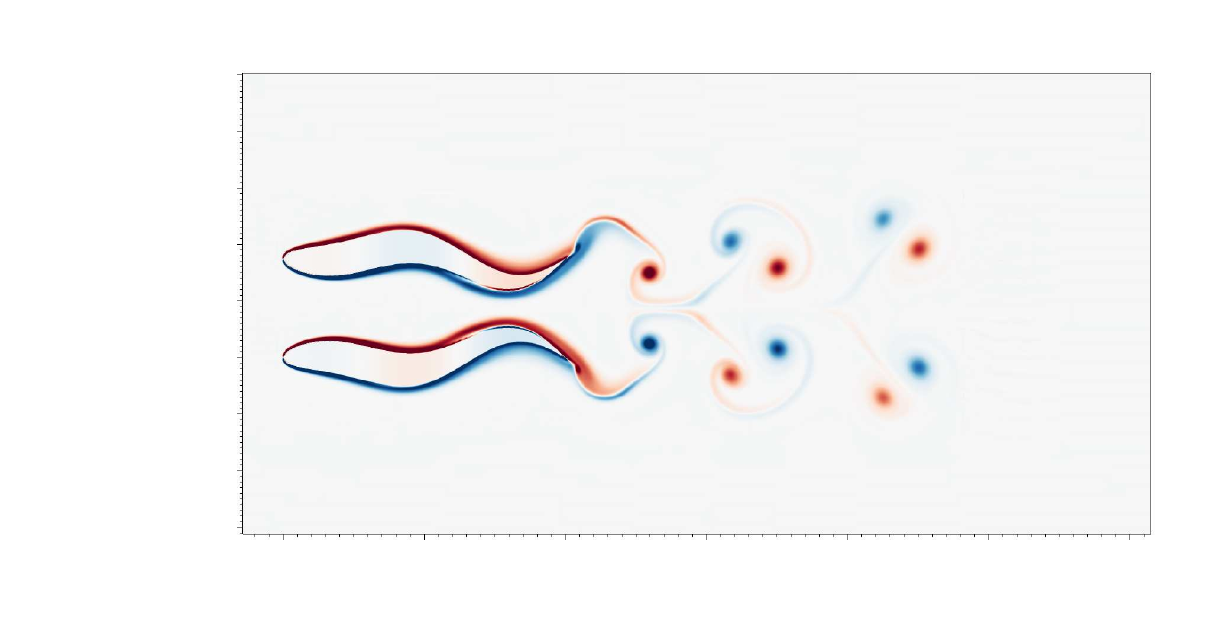}{(g) $ \lam = 0.8 $, $ t^*/T = 2.75 $}
		
		\addlabelvor{0.49}{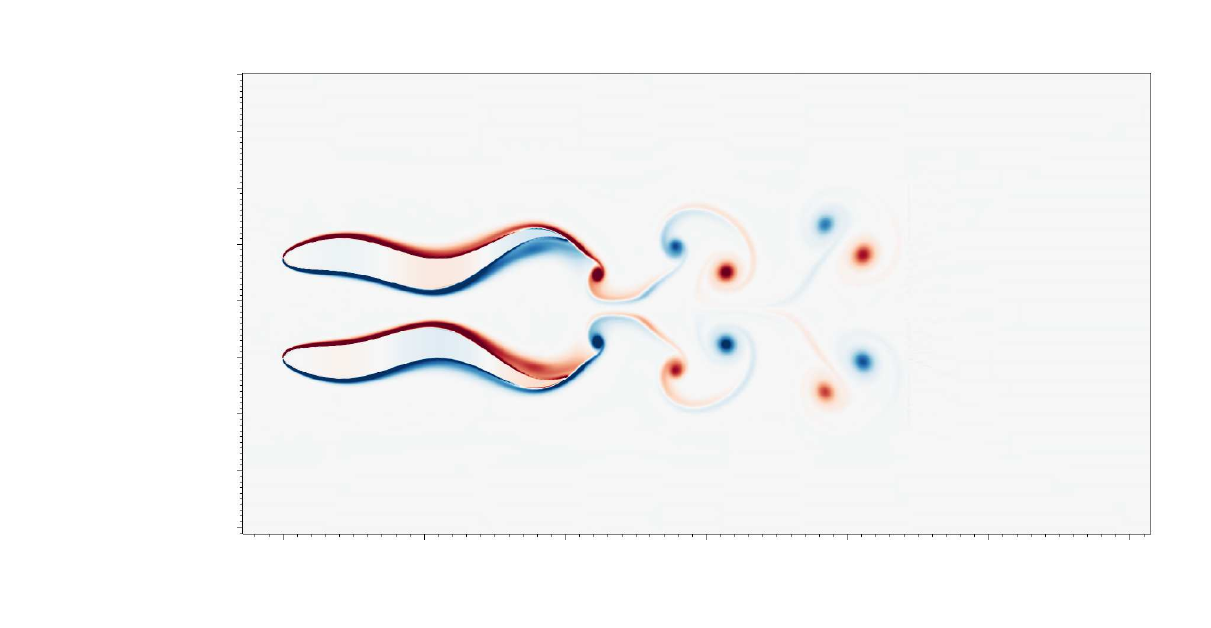}{(d) $ \lam = 0.8 $, $ t^*/T = 2.375 $}
		\addlabelvor{0.49}{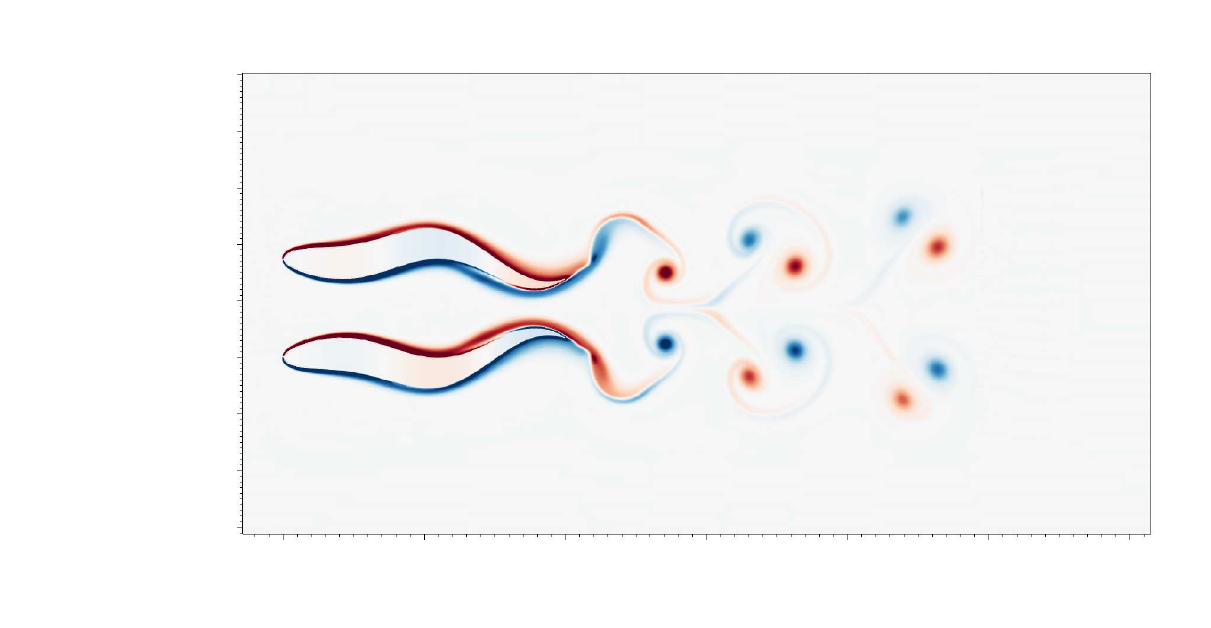}{(h) $ \lam = 0.8 $, $ t^*/T = 2.875 $}
		
		\addlabelnotrim{0.49}{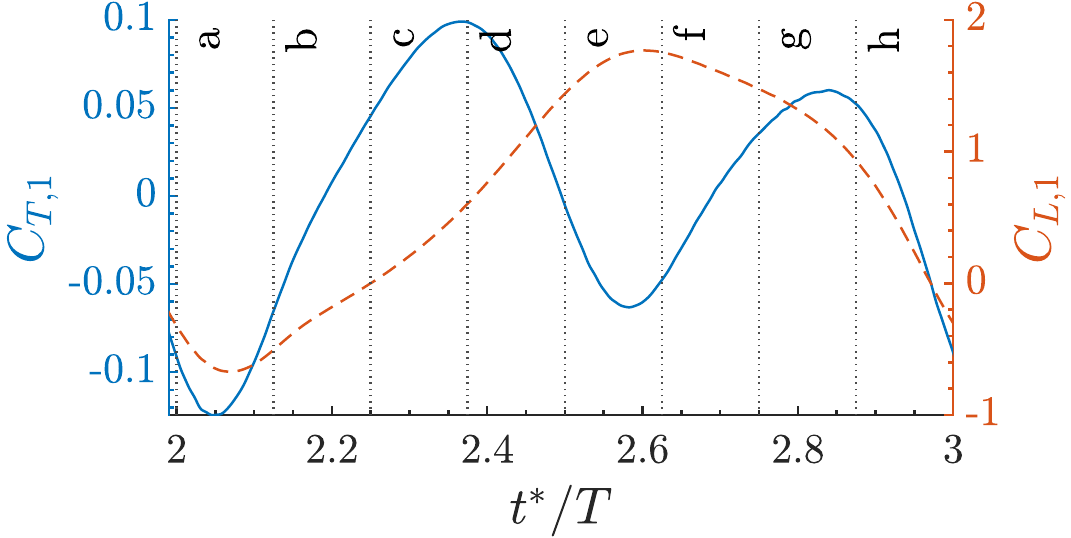}{(i) Bottom}
		\addlabelnotrim{0.49}{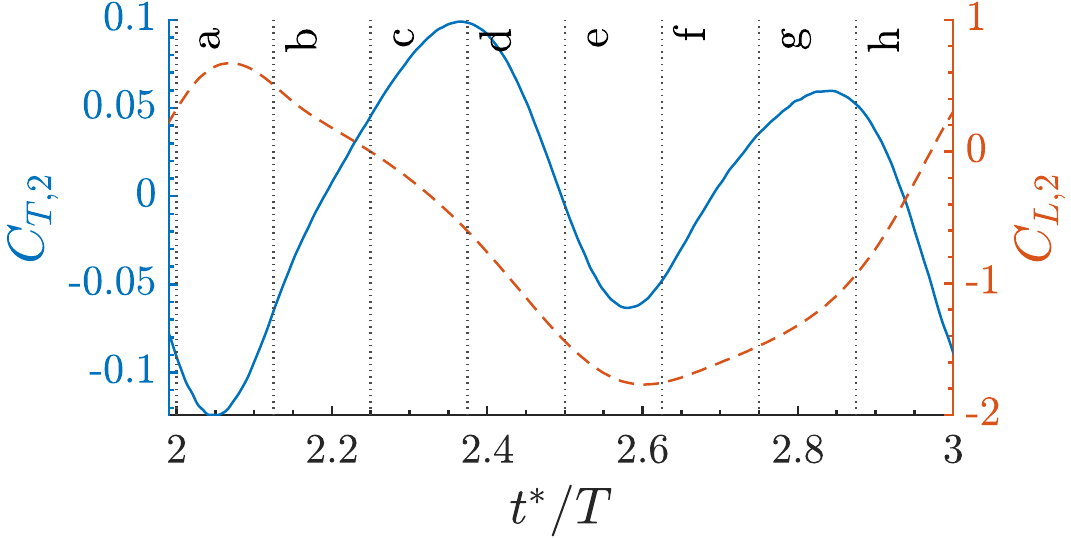}{(j) Top}
		
		\caption{Vorticity contours and hydrofoil deformation with wavelength $ \lam = 0.8 $, side-by-side arrangement $ D = 0 $, lateral gap $ G = 0.35 $, anti-phase $ \ppi = 1.0 $ at instants of a typical period
			(a-h) $ t^*/T = 2.00-2.875 $. Time histories of thrust and lift coefficient for the (i) Bottom and (j) Top swimmers.}
		\label{fig:vor_1T_D0_lam0d8}
	\end{figure}

	The swimming deformation of the foils is an overlap of both lateral pitching motion and travelling sinusoidal wave. At low wavelength $ \lam = 0.8 $, the component of the travelling wave becomes more significant with a wavy appearance shown in \Cref{fig:vor_1T_D0_lam0d8}.
	For the single foil swimming cases \citep{Thekkethil2018}, two vortices of opposite signs are shed in each cycle of undulation. In the present case of two side-by-side foils undulating in anti-phase, strong but symmetrical interference is discovered between the vortices generated by each foil, as seen in \Cref{fig:vor_1T_D0_lam0d8}.
	In the \textit{outward} movement of the swimmers' tail tips, as demonstrated in \Cref{fig:vor_1T_D0_lam0d8}a-\ref{fig:vor_1T_D0_lam0d8}d, each of the two foils produces a vortex of opposite signs, temporarily forming a vortex pair. Meanwhile, the travelling wave deforming the foils propels the fluid between the two foils, pushing these two vortices downstream.
	During the \textit{inward} phase of the tail tip movement, \ie \Cref{fig:vor_1T_D0_lam0d8}e-\ref{fig:vor_1T_D0_lam0d8}h, each foil sheds one more vortex, which, in the next cycle, gradually forms a vortex dipole by pairing with the previous vortex from the same foil, \ie \Cref{fig:vor_1T_D0_lam0d8}a-\ref{fig:vor_1T_D0_lam0d8}d. Eventually, the vortex dipoles from each of the two foils repel their counterparts and travel laterally away from each other, forming a highly symmetrical pattern of vortex dipoles in the downstream area.
	Corresponding to the high level of symmetry in flow structure, the thrust of two foils are identical through the variation of time $ C_{T,1} = C_{T,2} $, indicating the two foils reach a stable formation and propel in a synchronised manner, as seen in \Cref{fig:vor_1T_D0_lam0d8}i and \ref{fig:vor_1T_D0_lam0d8}j. The lift force of the two foils are opposite to each other as $ C_{L,1} = -C_{L,2} $, which also periodically switches direction; so the two foils repel each other at instants \Cref{fig:vor_1T_D0_lam0d8}a-\ref{fig:vor_1T_D0_lam0d8}d while attracting each other at instants \Cref{fig:vor_1T_D0_lam0d8}e-\ref{fig:vor_1T_D0_lam0d8}h.

	\renewcommand{\addlabelvor}[3]{%
		\begin{tikzpicture}
			\node[anchor=south west,inner sep=0] (image) at (0,0) 
			{\includegraphics[width=#1\linewidth, trim={4.05cm 1cm 1cm 1.25cm},clip]{#2}};%
			\begin{scope}[x={(image.south east)},y={(image.north west)}]
				\node[anchor=south west] at (0.00,0.75) {\footnotesize #3};	%
			\end{scope}
		\end{tikzpicture}%
	}
	\begin{figure}
		\centering
		\setcounter{testaa}{0}
		\addlabelvor{0.49}{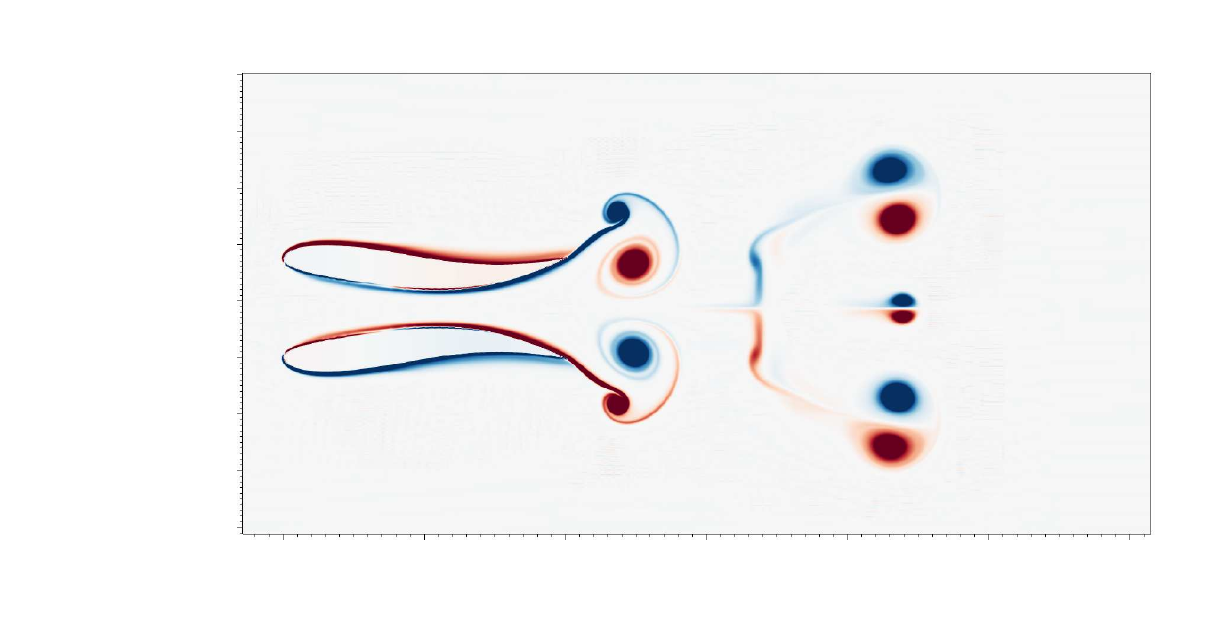}{(a) $ \lam = 2.0 $, $ t^*/T = 2.00 $}
		\addlabelvor{0.49}{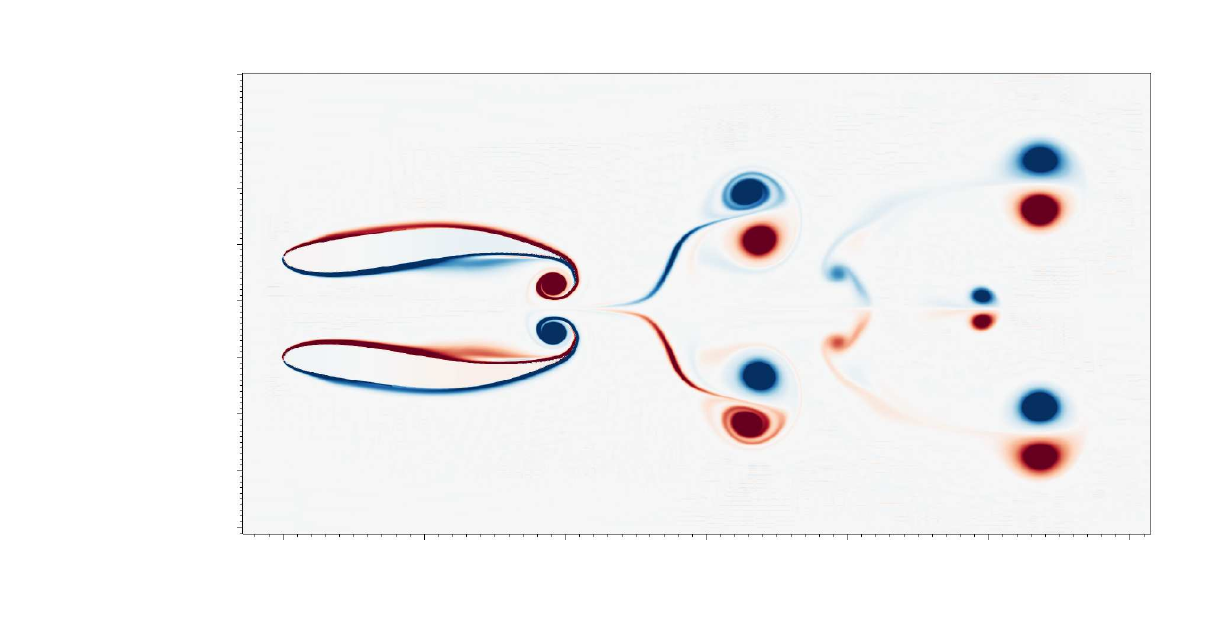}{(e) $ \lam = 2.0 $, $ t^*/T = 2.50 $}
		
		\addlabelvor{0.49}{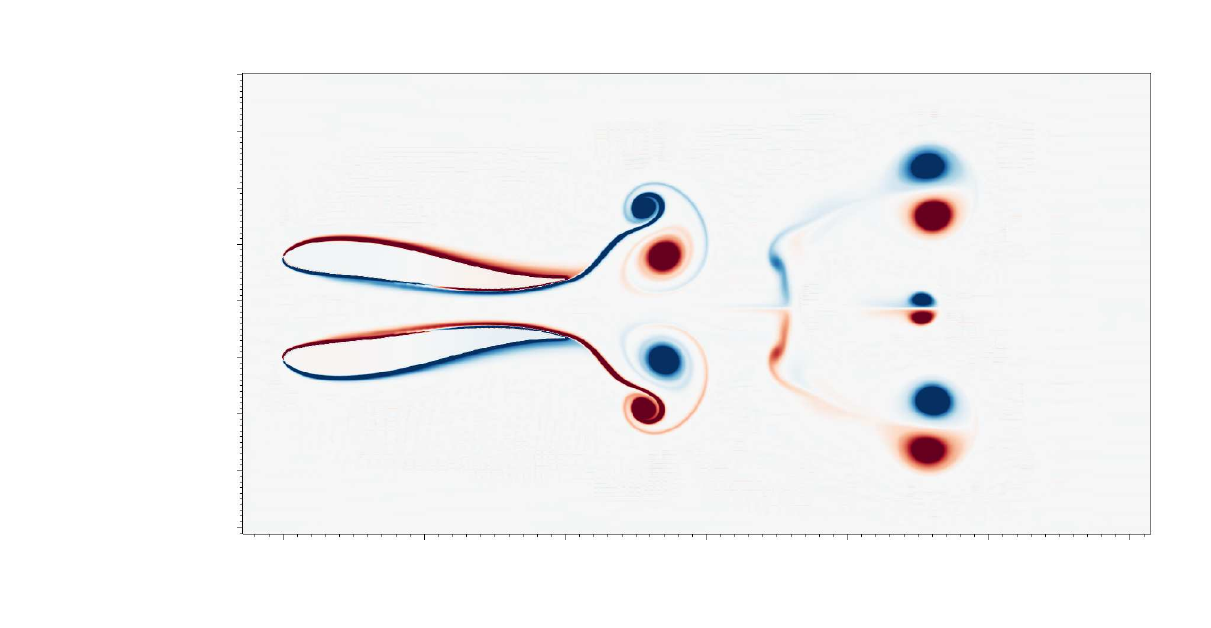}{(b) $ \lam = 2.0 $, $ t^*/T = 2.125 $}
		\addlabelvor{0.49}{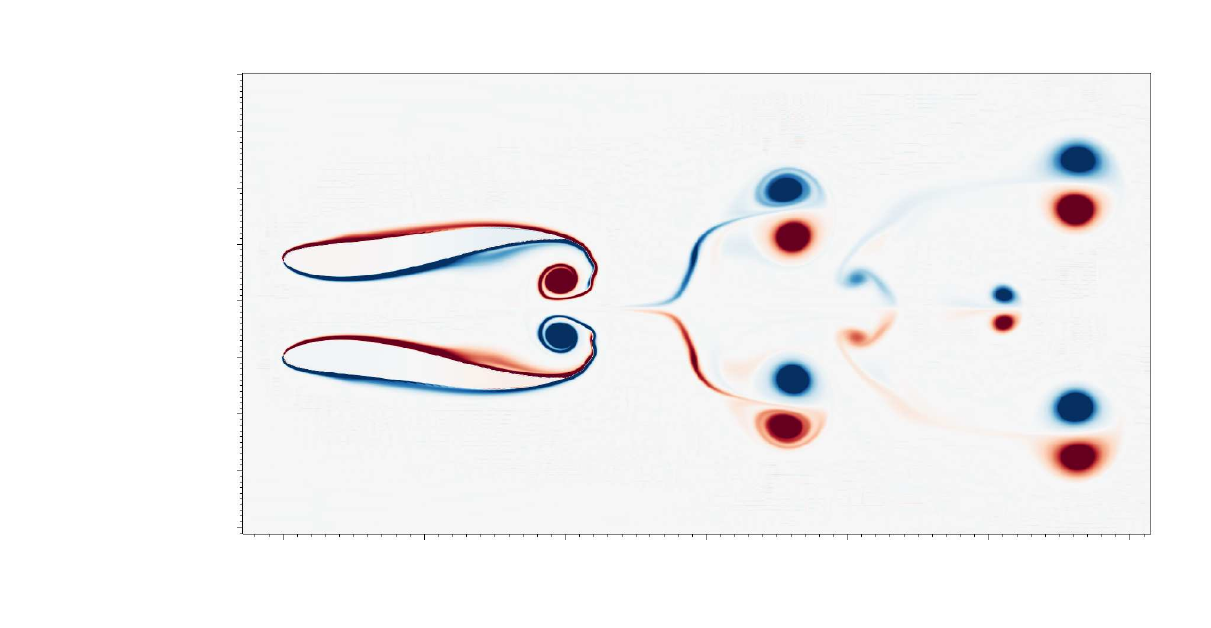}{(f) $ \lam = 2.0 $, $ t^*/T = 2.625 $}
		
		\addlabelvor{0.49}{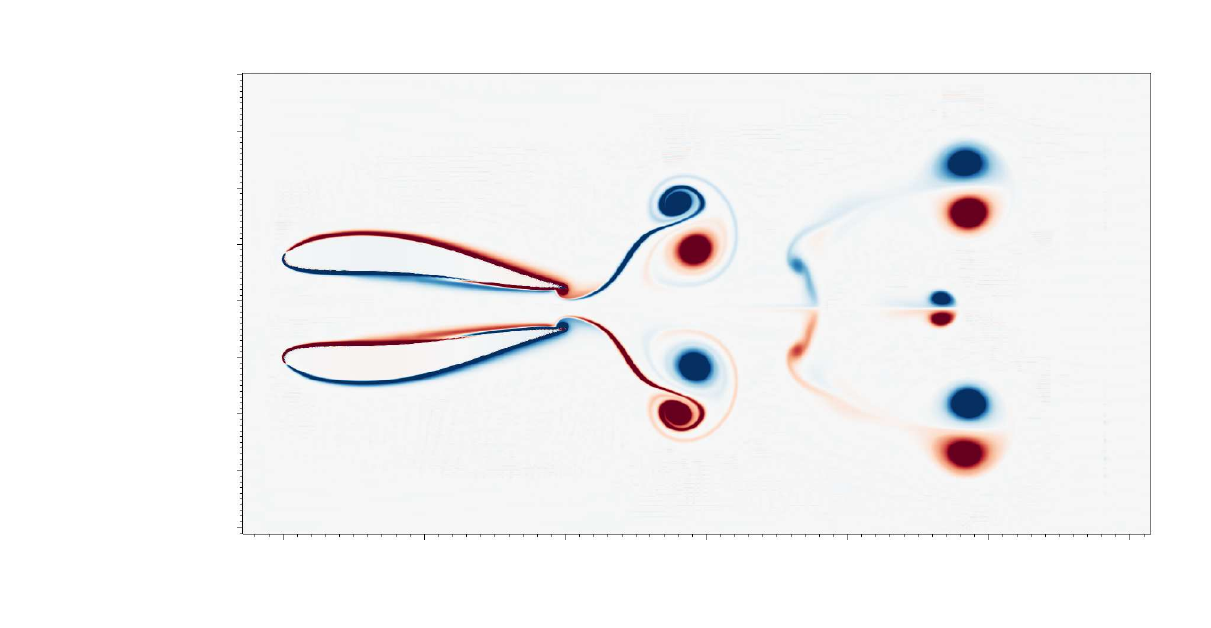}{(c) $ \lam = 2.0 $, $ t^*/T = 2.25 $}
		\addlabelvor{0.49}{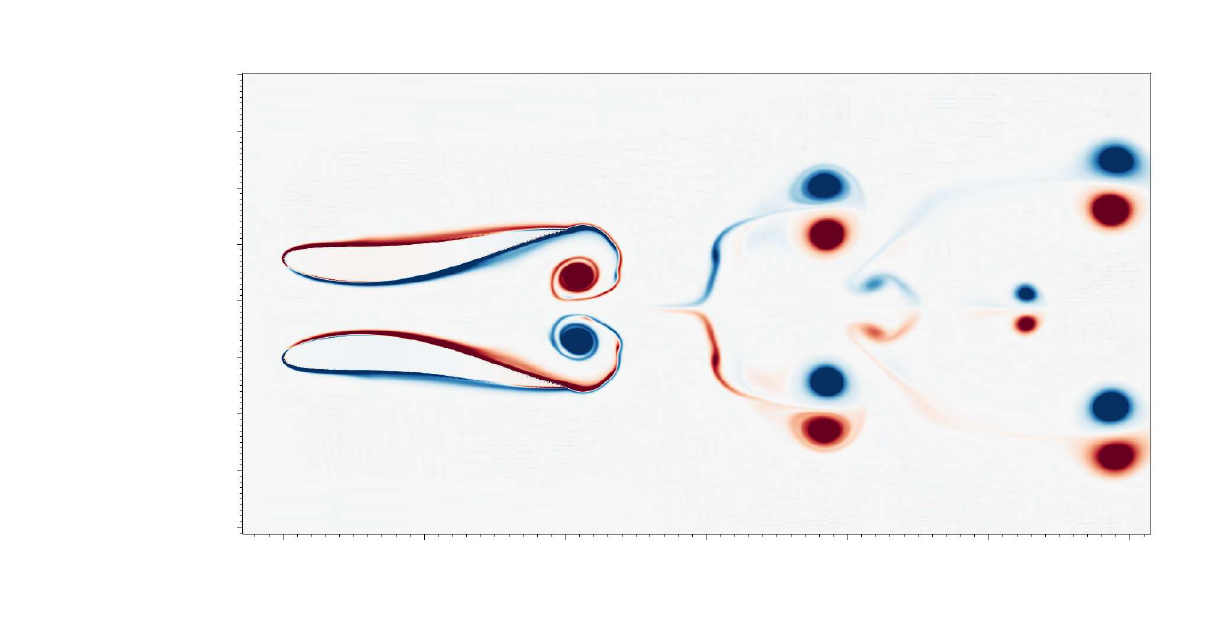}{(g) $ \lam = 2.0 $, $ t^*/T = 2.75 $}
		
		\addlabelvor{0.49}{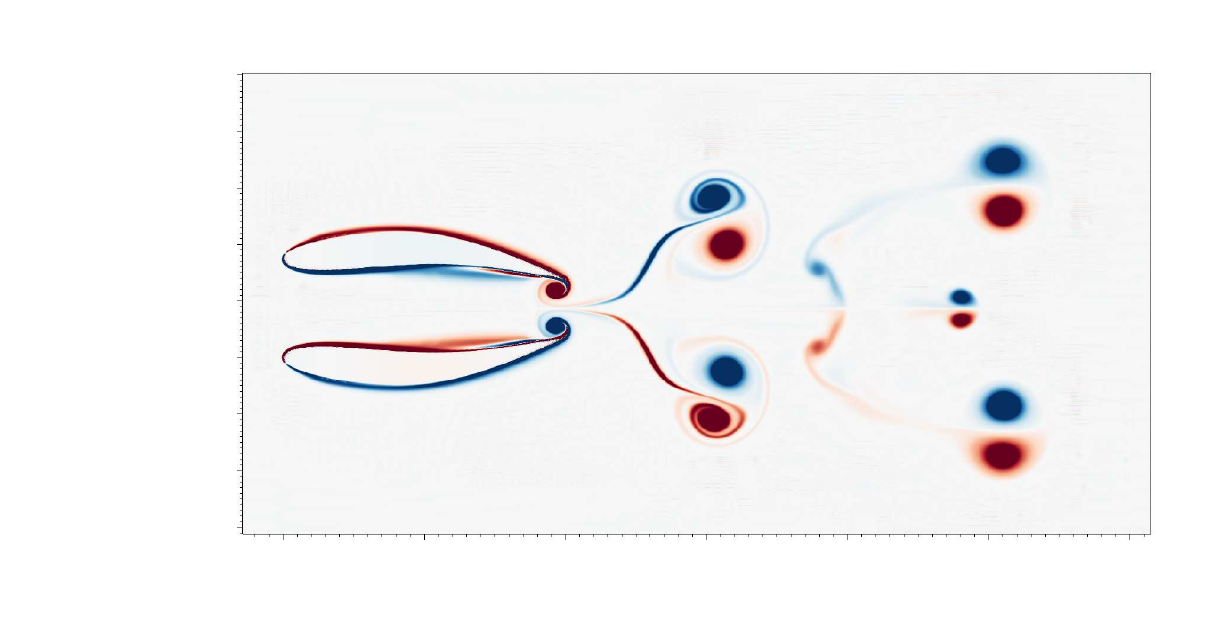}{(d) $ \lam = 2.0 $, $ t^*/T = 2.375 $}
		\addlabelvor{0.49}{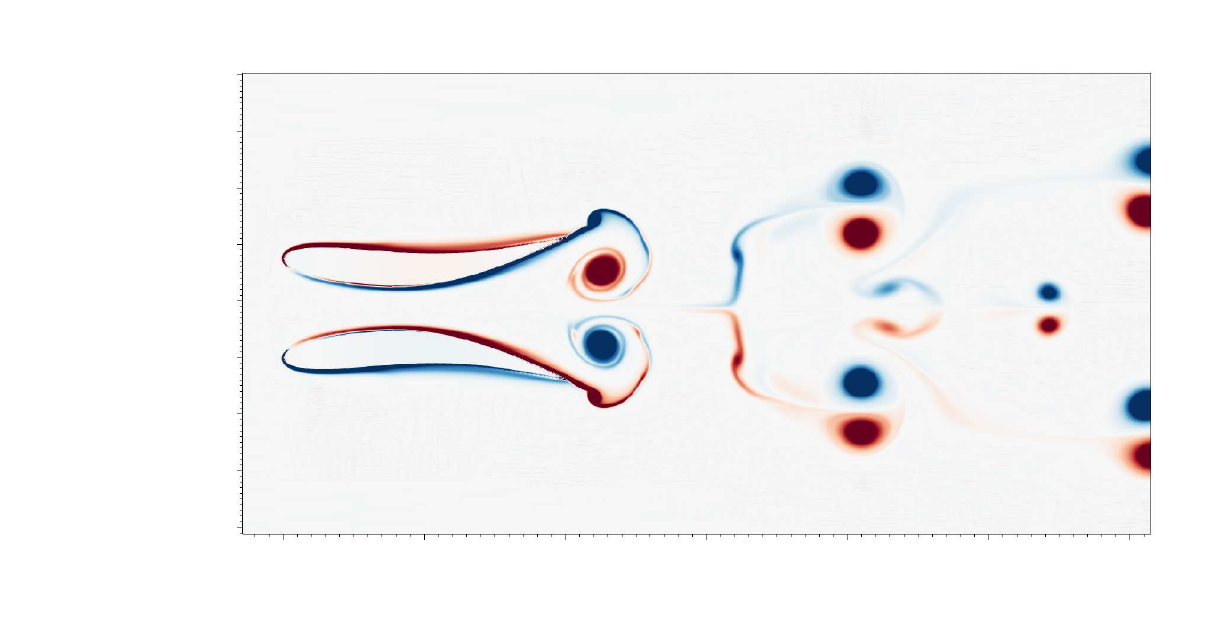}{(h) $ \lam = 2.0 $, $ t^*/T = 2.875 $}
		
		\addlabelnotrim{0.49}{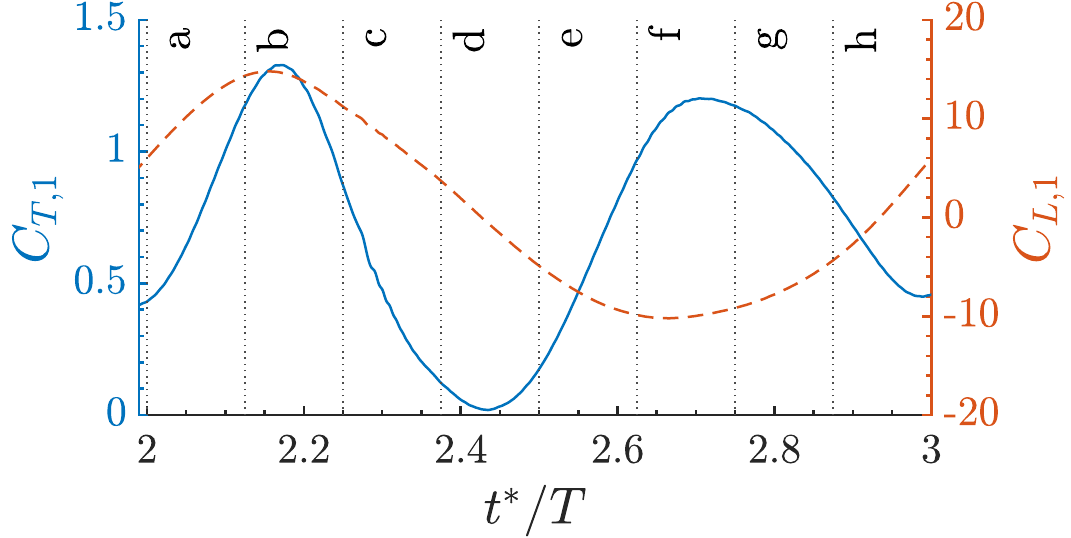}{(i) Bottom}
		\addlabelnotrim{0.49}{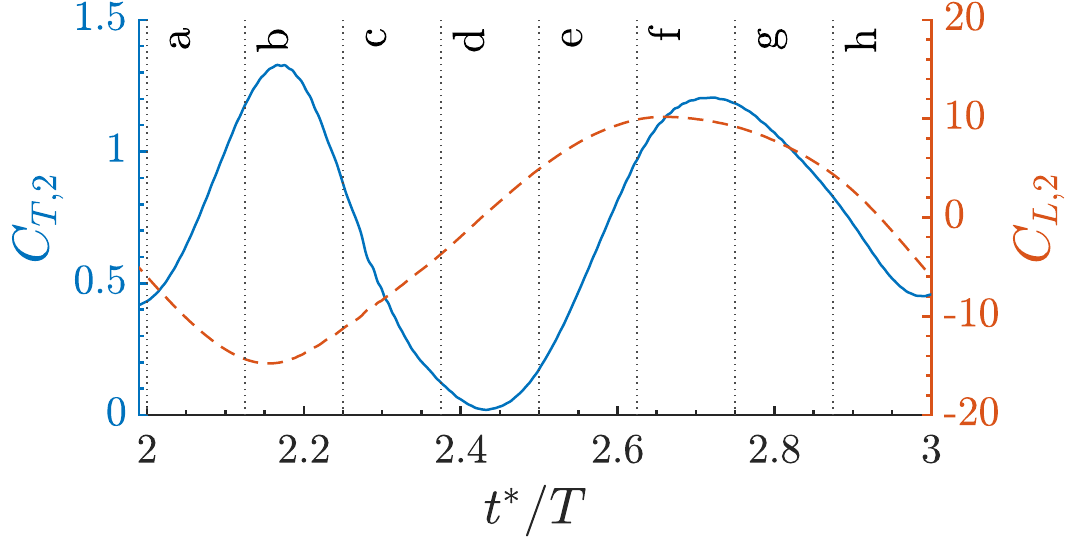}{(j) Top}
		
		\caption{Vorticity contours and hydrofoil deformation with wavelength $ \lam = 2.0 $, side-by-side arrangement $ D = 0 $, lateral gap $ G = 0.35 $, anti-phase $ \ppi = 1.0 $ at instants of a typical period
			(a-h) $ t^*/T = 2.00-2.875 $. Time histories of thrust and lift coefficient for the (i) Bottom and (j) Top swimmers.}
		
		\label{fig:vor_1T_D0_lam2d0}
	\end{figure}
	
	At intermediate wavelength $ \lam = 2 $, the highest energy efficiency can be obtained as previously discussed. Here, we further analyse its vorticity contour in order to study its flow structure and fluid mediated interaction.
	Similar to low wavelength cases, generation of vortices is governed by the tail tip movement. When the tail tip moves \textit{outward} in \Cref{fig:vor_1T_D0_lam2d0}c-\ref{fig:vor_1T_D0_lam2d0}f, each foil generates a vortex in the near-tail region. During the \textit{inward} phase shown in \Cref{fig:vor_1T_D0_lam2d0}g-\ref{fig:vor_1T_D0_lam2d0}h and \Cref{fig:vor_1T_D0_lam2d0}a-\ref{fig:vor_1T_D0_lam2d0}b, the vortices on the outer side of each foil are generated as well. The vortex dipoles eventually form a streaming direction that points downstream, enhancing the propulsion of the swimmers.
	It is also interesting to notice that at $ \lam = 2 $, the velocity of shed vortex dipoles is almost 2 times of that at $ \lam = 0.8 $. 
	The thrust and lateral force generally follows the same pattern as the low wavelength scenario at $ \lam = 0.8 $.

	\renewcommand{\addlabelvor}[3]{%
		\begin{tikzpicture}
			\node[anchor=south west,inner sep=0] (image) at (0,0) 
			{\includegraphics[width=#1\linewidth, trim={4.05cm 1cm 1cm 1.25cm},clip]{#2}};%
			\begin{scope}[x={(image.south east)},y={(image.north west)}]
				\node[anchor=south west] at (0.00,0.75) {\footnotesize #3};	%
			\end{scope}
		\end{tikzpicture}%
	}
	\begin{figure}
		\centering
		\setcounter{testaa}{0}
		\addlabelvor{0.49}{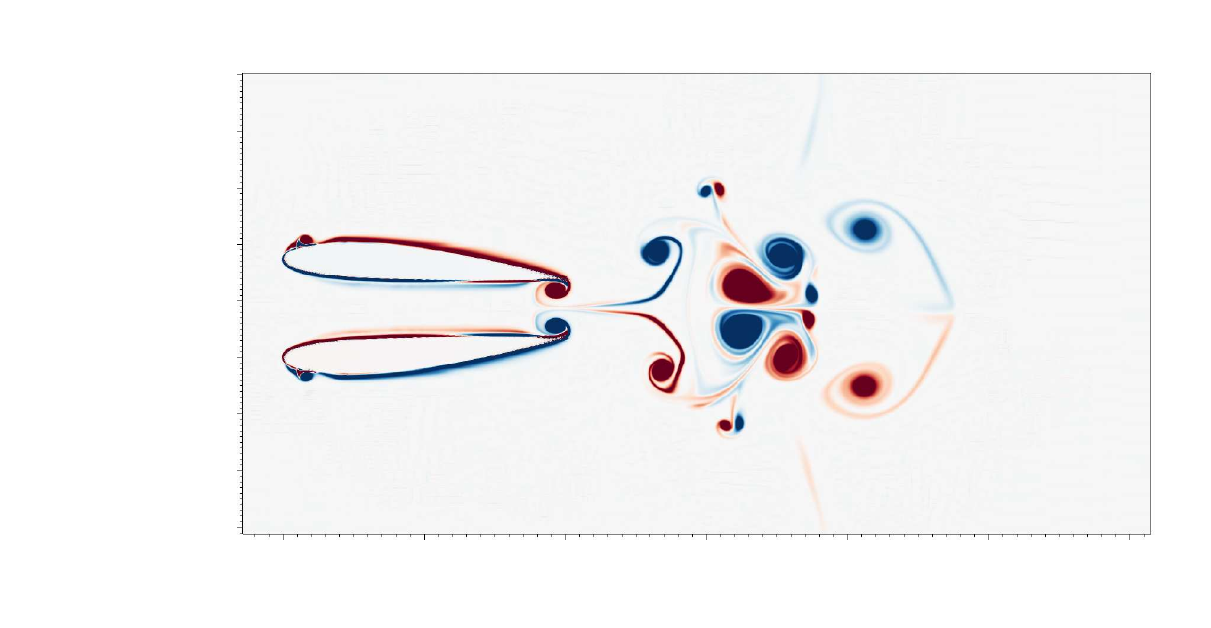}{(a) $ \lam = 8.0 $, $ t^*/T = 2.00 $}
		\addlabelvor{0.49}{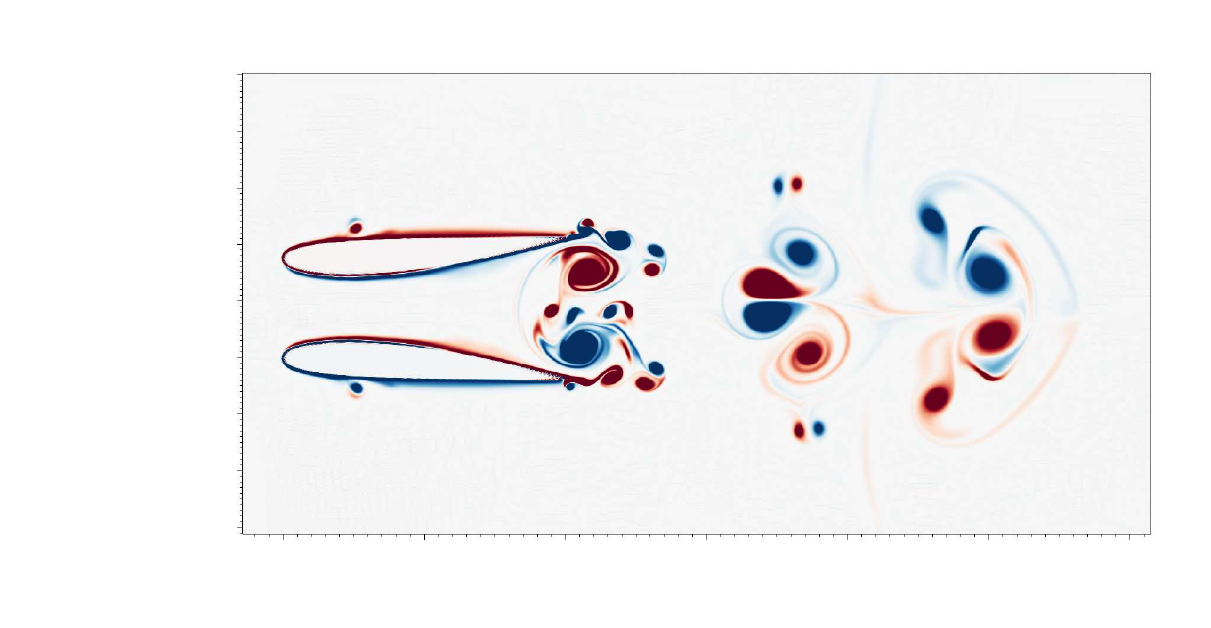}{(e) $ \lam = 8.0 $, $ t^*/T = 2.50 $}
		
		\addlabelvor{0.49}{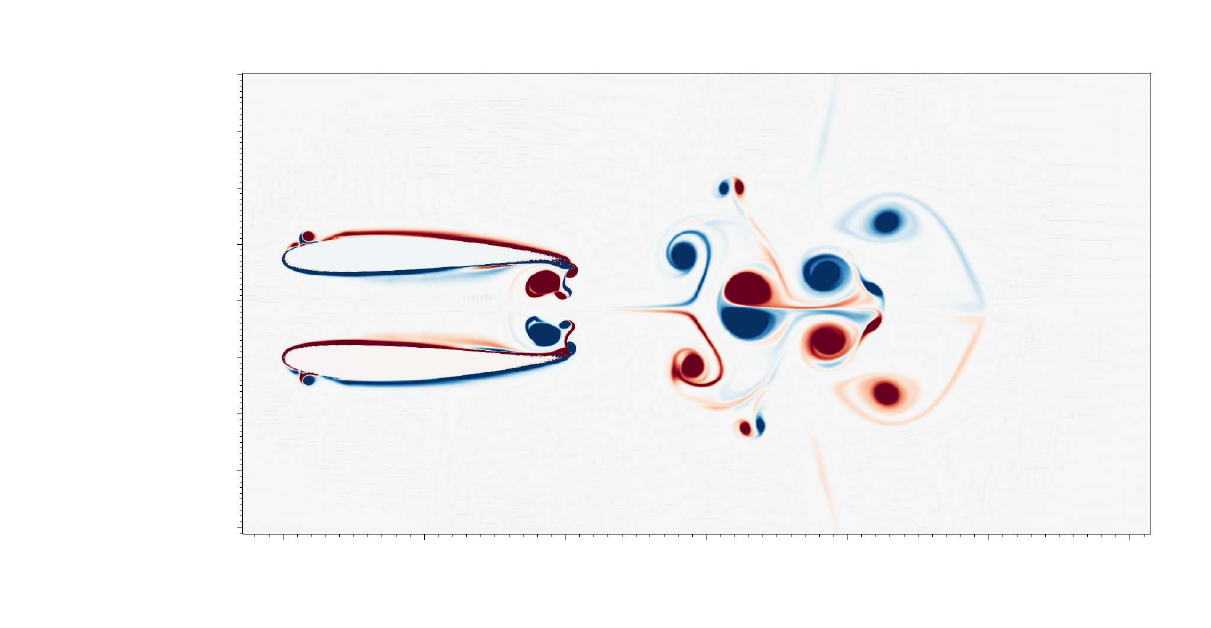}{(b) $ \lam = 8.0 $, $ t^*/T = 2.125 $}
		\addlabelvor{0.49}{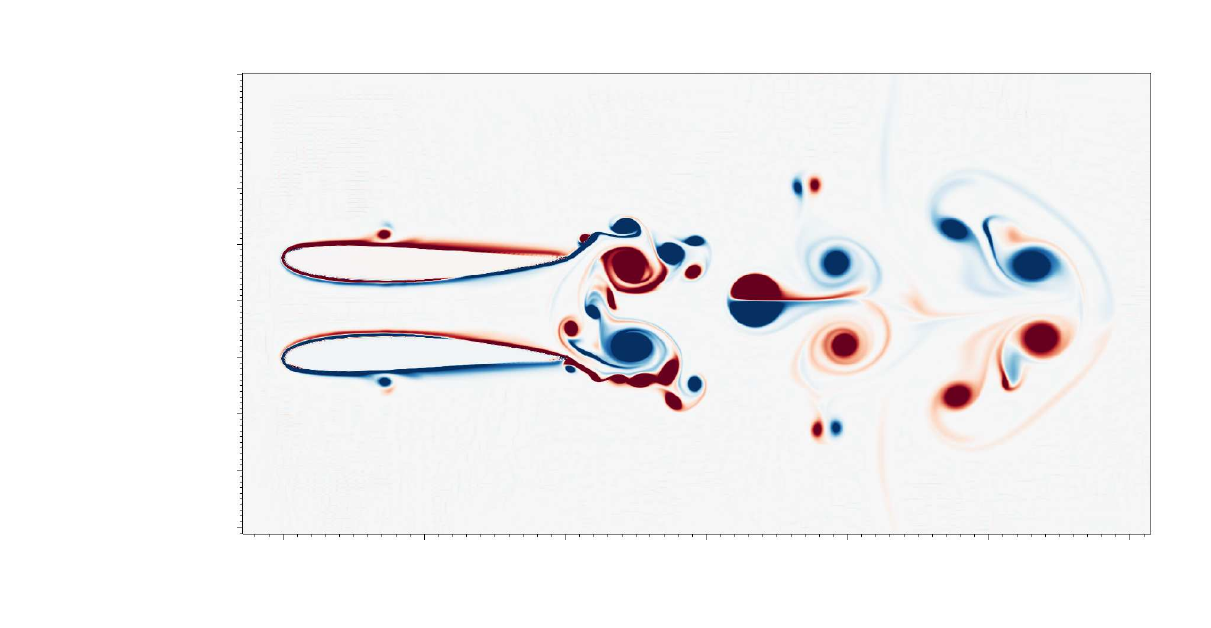}{(f) $ \lam = 8.0 $, $ t^*/T = 2.625 $}
		
		\addlabelvor{0.49}{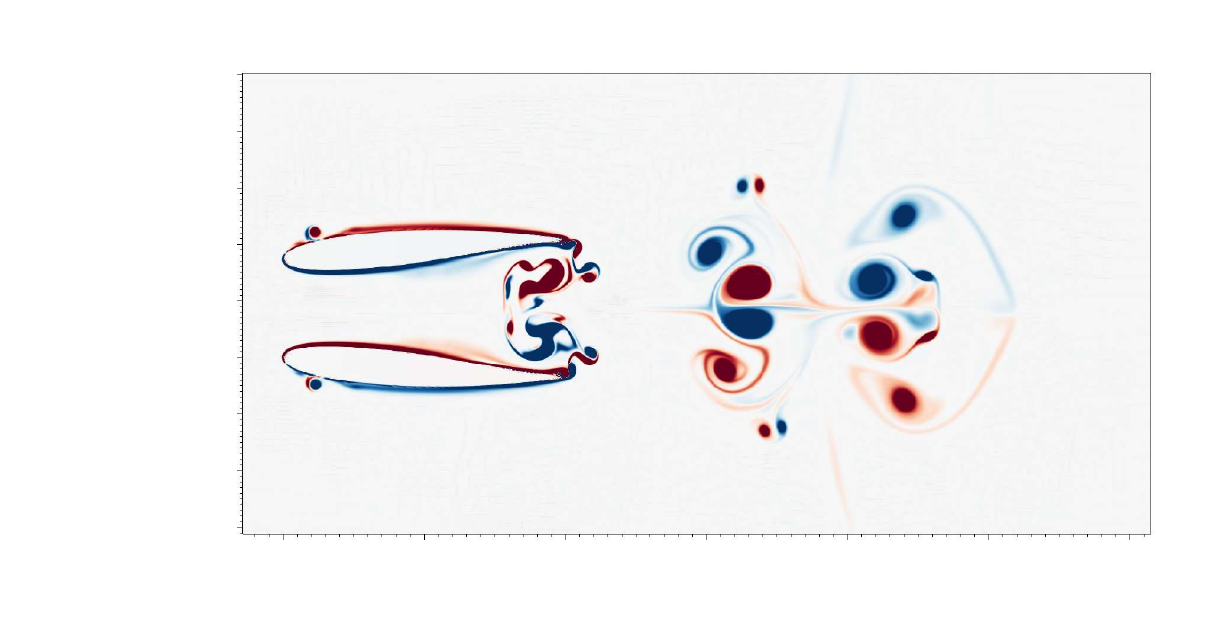}{(c) $ \lam = 8.0 $, $ t^*/T = 2.25 $}
		\addlabelvor{0.49}{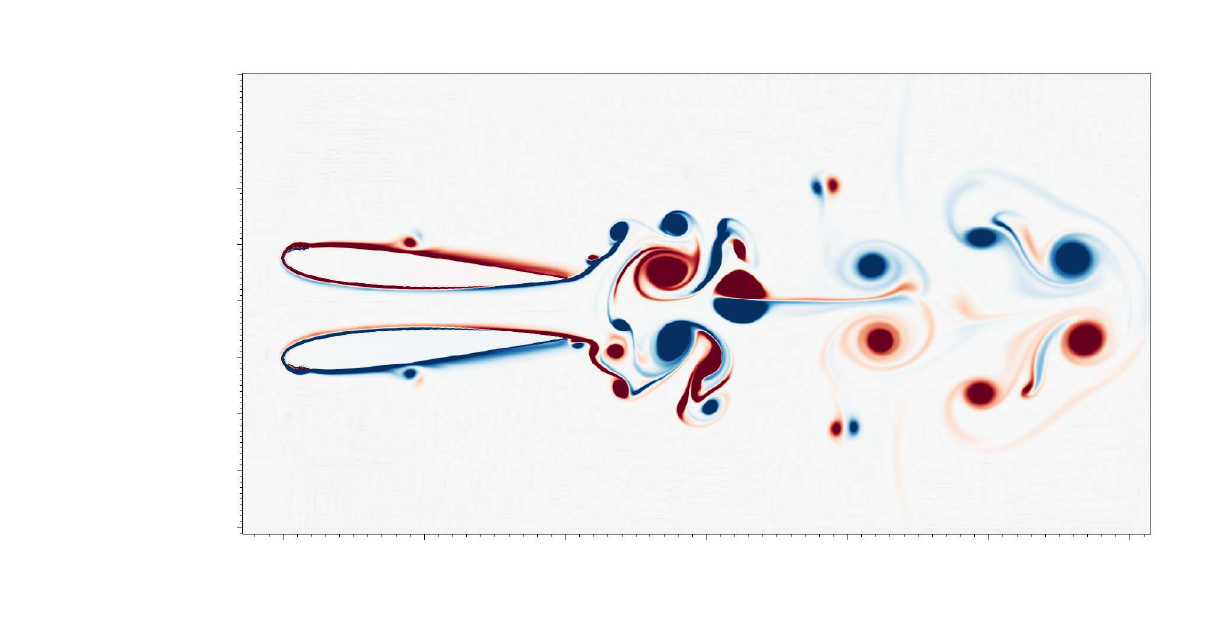}{(g) $ \lam = 8.0 $, $ t^*/T = 2.75 $}
		
		\addlabelvor{0.49}{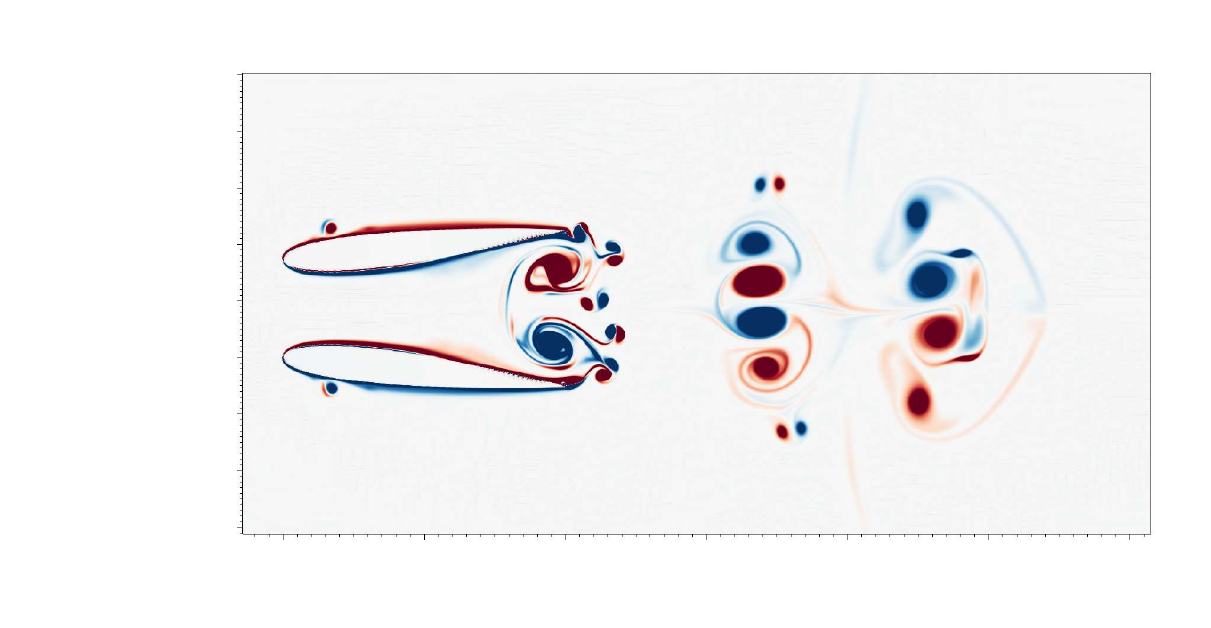}{(d) $ \lam = 8.0 $, $ t^*/T = 2.375 $}
		\addlabelvor{0.49}{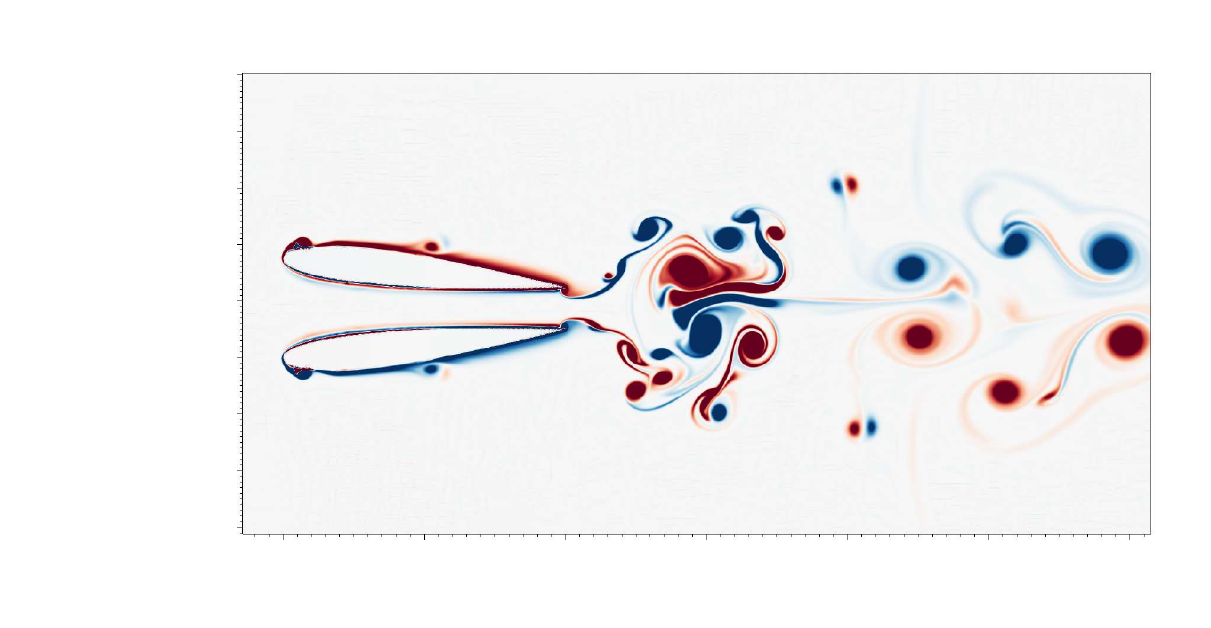}{(h) $ \lam = 8.0 $, $ t^*/T = 2.875 $}
		
		\addlabelnotrim{0.49}{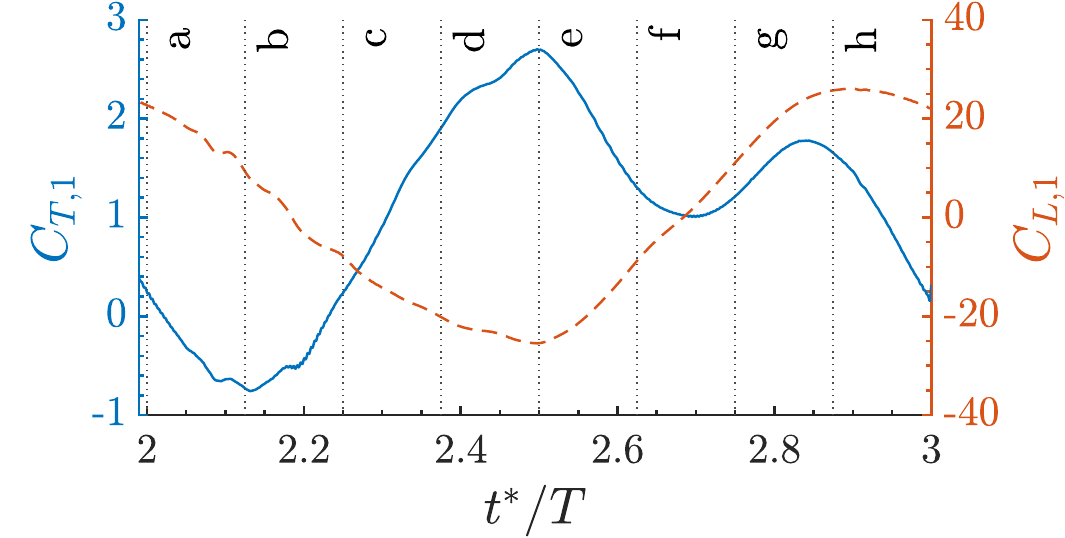}{(i) Bottom}
		\addlabelnotrim{0.49}{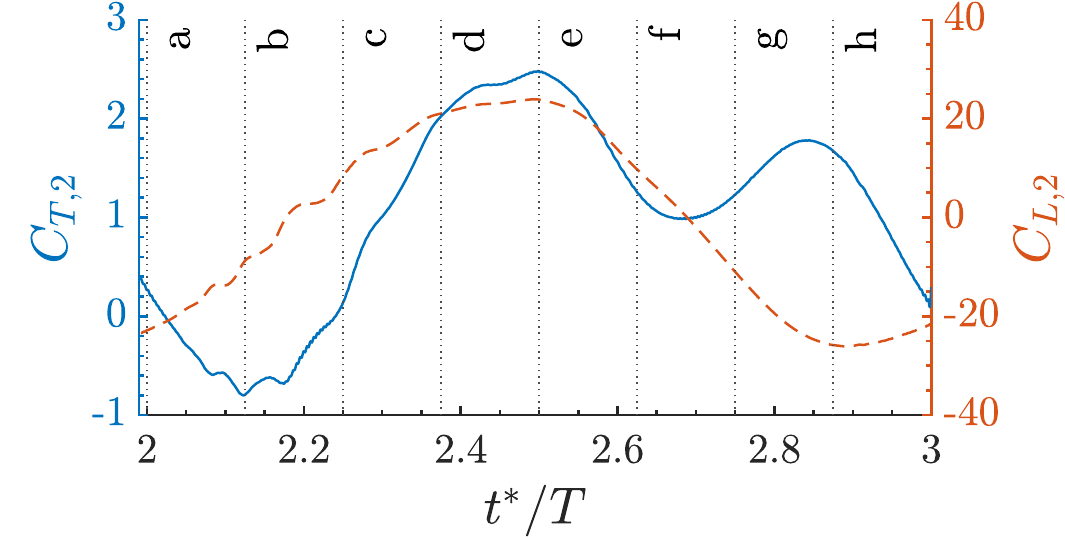}{(j) Top}
		
		\caption{Vorticity contours and hydrofoil deformation with wavelength $ \lam = 8.0 $, side-by-side arrangement $ D = 0 $, lateral gap $ G = 0.35 $, anti-phase $ \ppi = 1.0 $ at instants of a typical period
			(a-h) $ t^*/T = 2.00-2.875 $. Time histories of thrust and lift coefficient for the (i) Bottom and (j) Top swimmers.}
		\label{fig:vor_1T_D0_lam8d0}
	\end{figure}

	At large wavelength $ \lam = 8.0 $, the swimming motion of the foils consists mainly of pitching rather than undulation, as seen in \Cref{fig:vor_1T_D0_lam8d0}.
	Due to the anti-phase setting $ \ppi = 1 $, the pitching of two swimmers periodically switches between \textit{outward} and \textit{inward} movement.
	The \textit{outward} motion creates a vortex dipole between the tails of the two swimmers, as shown in \Cref{fig:vor_1T_D0_lam8d0}a-\ref{fig:vor_1T_D0_lam8d0}d. During the outward movement, the dipole stays near the tail region despite the streaming flow. 
	The \textit{inward} motion, \ie \Cref{fig:vor_1T_D0_lam8d0}e-\ref{fig:vor_1T_D0_lam8d0}h, then pushes out the vortex dipole while creating two vortices at each outer side of the two foils. A strong jet flow is also produced.
	In this period of motion, the flow symmetry gradually breaks, causing an increasingly complicated flow structure, which corresponds to the unsteady thrust $ \ctm $ and lateral $ \clrms $ force as seen in \Cref{fig:vor_1T_D0_lam8d0}i and \ref{fig:vor_1T_D0_lam8d0}j. The symmetry breaking typically occurs at the instant $ \tdT = 2.25 $ in \Cref{fig:vor_1T_D0_lam8d0}c, where the large vortex dipole is broken into multiple small vortices.
	In addition to the near-tail vortex dipole, two relatively small vortices emerge from the outer sides of the each foil, and then travels along the surface of the foils. The generation of these outer minor vortices starts to generate at the later phase of the inward pitching movement at instant $ \tdT = 2.75 $ \Cref{fig:vor_1T_D0_lam8d0}g and instant $ \tdT = 2.875 $ \ref{fig:vor_1T_D0_lam8d0}h, and then remains almost static in the outward pitching motion at \Cref{fig:vor_1T_D0_lam8d0}a-\ref{fig:vor_1T_D0_lam8d0}d; the displacement of these minor vortices takes place during the inward motion at \Cref{fig:vor_1T_D0_lam8d0}e-\ref{fig:vor_1T_D0_lam8d0}h.
	This phenomenon is only observed in the anti-phase cases with high wavelength in the tested parametric space.
	In addition, the broken symmetry corresponds to the irregular thrust force at anti-phase condition, as depicted in \Cref{fig:lines_D_0_CT_CL}.

	\newcommand{\addlabelc}[3]{%
		\begin{tikzpicture}
			\node[anchor=south west,inner sep=0] (image) at (0,0) 
			{\includegraphics[width=#1\textwidth]{#2}};
			\begin{scope}[x={(image.south east)},y={(image.north west)}]
				\node[anchor=west] at (0.16,0.9) {#3};	%
			\end{scope}
		\end{tikzpicture}%
	}
	
	\newcommand{\addlabelctop}[4]{%
		\begin{tikzpicture}
			\node[anchor=south west,inner sep=0] (image) at (0,0) 
			{\includegraphics[width=#1\textwidth]{#2}};
			\begin{scope}[x={(image.south east)},y={(image.north west)}]
				\node[anchor=west] at (0.16,0.9) {#3};	%
				\node[anchor=west] at (0.14,1.05) {#4};
			\end{scope}
		\end{tikzpicture}%
	}
	
	\newcounter{testa}
	\setcounter{testa}{0}
	\newcommand\countera{\stepcounter{testa}\alph{testa}}

	\newcommand{\addlabeld}[3]{%
		\begin{tikzpicture}
			\node[anchor=south west,inner sep=0] (image) at (0,0) 
			{\includegraphics[width=#1\textwidth]{#2}};
			\begin{scope}[x={(image.south east)},y={(image.north west)}]
				\node at (0.60,0.25) {#3};	%
			\end{scope}
		\end{tikzpicture}%
	}
	
	\setcounter{testa}{0}

	\subsubsection{Summary for side-by-side $ D = 0 $ cases}

	In this section, we summarise the hydrodynamic characteristics of side-by-side cases, \ie $ D = 0 $. The discussed content includes the thrust and lateral force, propeller and group efficiency, vorticity distribution and corresponding force time histories.
	\textit{Thrust and lateral force} is identical for the two foils, \ie $ C_{Tm,1} = C_{Tm,2} $ and $ C_{Lrms,1} = C_{Lrms,2} $, at in-phase or anti-phase conditions $ \ppi = 0,\ 1 $. However, such consistency is disrupted at high wavelengths $ \lam > 4 $, due to the high irregularity in flow structure.
	Thrust and lateral force of both foils generally increase with wavelength $ \lam $.
	At $ \lam > 1 $, the hydrodynamic force $ C_{Tm} $ and $ C_{Lrms} $ is highly sensitive to phase difference $ \ppi $. Hydrodynamic force at anti-phase $ \ppi = 1 $ can reach 6 times of that at in-phase $ \ppi = 0 $. When the two swimmers school at a relatively large lateral distance $ G = 0.35 $, the impact of phase difference $ \phi $ is reduced, whereas the flow stability is increased.
	\textit{Propeller efficiency} for each individual foil $ \eta_{1,2} $ or two foils as a group $ \eta_{group} $ generally reaches maximum at $ \lam = 2 $.
	Highest group efficiency is obtained at intermediate wavelength $ \lam = 2 $, phase difference $ \ppi = 0.5,\ 1.5 $ and relatively high lateral distance $ G = 0.35 $.
	The influence of phase difference $ \ppi $ can be more significant than wavelength $ \lam $.

	\textit{Vorticity distribution} $ \omega^* $ is reviewed to understand the flow structure around the two wavy foils.
	Wavelength $ \lam $ is positively correlated with shed vortices' intensity and the disturbed area. Vortex dipoles are generated during foil undulation/pitching.
	At anti-phase $ \ppi = 1 $, vorticity distribution can be symmetrical, since the two swimmers form a mirror symmetry geometry at any moment during the undulation process.
	However, symmetry breaking can occur after a number of initial periods; higher wavelength leads to broken symmetry in fewer starting periods.
	The irregularity of the flow structure tends to increase with undulation wavelength $ \lam $.
	At low wavelength $ \lam $, the gap distance $ G $ and phase difference $ \phi $ can affect the skewness, symmetry and regularity of the wake pattern.
	At $ \ppi = 0.5 $ and $ \ppi = 1.5 $, the vortex shedding direction is slightly skewed towards the right and left sides of the swimming direction, respectively.
	The intensity of the vortices decreases with a larger gap distance $ G $.
	Concentration of dynamic energy is discovered at low gap distance and when the two hydrofoils swim in anti-phase.

	At side-by-side arrangement $ D = 0 $, anti-phase $ \ppi = 1 $ and wavelengths $ \lam = 0.8,\ 2.0,\ 8.0 $, within one cycle of undulation, flow structure at 8 instants is examined with corresponding time histories of thrust $ C_T $ and lateral $ C_L $ force upon the foils.
	At \textit{low wavelength} $ \lam = 0.8 $, strong symmetrical interference exists between the vortices shed by each foil.
	The vortex dipoles from each foil repel their counterparts from another foil and travel laterally away from each other, forming a highly symmetrical pattern of vortex dipoles in the downstream area.
	At \textit{intermediate wavelength} $ \lam = 2 $, vortex dipoles causes a streaming direction that directly points downstream, enhancing the propulsion of the swimmers.
	The velocity of shed vortex dipoles is about 2 times of that at $ \lam = 0.8 $.
	At \textit{large wavelength} $ \lam = 8.0 $, the flow symmetry breaks at $ \tdT = 2.25 $, causing a complicated flow structure.
	Two additional small vortices emerge from the outer sides of each foil and then travels along the foil surface.
	The \textit{thrust} force of two foils is generally identical through the time histories as $ C_{T,1} = C_{T,2} $ with the \textit{lift} force being opposite as $ C_{L,1} = -C_{L,2} $, which also periodically switches direction. The force time histories are smooth at low and intermediate wavelengths $ \lam = 0.8 - 2 $, but fluctuates at high wavelength $ \lam = 8 $ due to irregularity in flow field.

	\subsection{Staggered arrangement $ D > 0 $}
	\label{sec:Staggered_Dg0}

	This section discusses the situation when the two swimmers are placed in a staggered arrangement $ D > 0 $.
	\Cref{subsec:F_eta_Dg0} studies how the non-dimensional parameters affect the leader/follower's mean thrust force $ \ctm $, RMS lateral force $ \clrms $, propeller efficiency $ \eta $, and group efficiency $ \eta_{group} $.
	\Cref{subsec:FSI_Dg0} examines the variation of flow structure with wavelength $ \lam $, phase difference $ \phi $, front-back distance $ D $ and their inter-relationships.
	\Cref{subsec:vor_Dis0} investigates the flow-mediated interaction between the two swimmers with the help of vorticity contours and hydrodynamic force time histories.
	
	\subsubsection{Hydrodynamic force and propulsive efficiency at $ D > 0 $}
	\label{subsec:F_eta_Dg0}

	For both the leader and the follower, mean \textit{thrust force} $ \ctm $ generally increases with the wavelength $ \lam $ in the tested parametric space of $ G = 0.25 - 0.35 $ and $ D = 0.25 - 0.75 $, as demonstrated in \Cref{fig:matrix_CTm}.
	The follower can take great advantages of the schooling interaction through various wavelengths, especially at $ \lam \gr 5.6 $.
	This enhancement of the follower's thrust force $ \ctm $ with $ \lam $ is most significant when the two foils undulate/pitch in anti-phase $ \ppi = 1.0 $. For example, as seen in \Cref{fig:matrix_CTm}a, great difference is observed between the thrust force of the follower and that of the leader at the anti-phase $ \ppi = 1.0 $ cases with close distance $ G = 0.25,\ D = 0.25 $. At high wavelength $ \lam $, follower's thrust force reach 4.5 times as large as the leader's. Generally speaking, this enhancement effect is activated by phase difference of $ \ppi = 1.0,\ 1.5 $ across various lateral $ G $ and front-back $ D $ distances, as shown in \Cref{fig:matrix_CTm}. The leader-follower thrust force difference decreases when the two swimmers are arranged at a further lateral $ G $ and front-back $ D $ distance.
	Although the follower's thrust force is generally larger than the leader's, exceptions are observed in a few cases.
	At $ D = 0.25 $ with relatively high wavelength $ \lam > 3 $, the thrust force upon the leader can become greater than the follower when the two swimmers undulates in-phase $ \ppi = 0 $.
	At $ D = 0.50 $ with $ \lam > 5 $, leader's thrust is higher than the follower at $ \ppi = 0 $ and $ 0.5 $.
	At $ D = 0.75 $ with $ \lam > 5 $, leader's thrust is higher than the follower only at $ \ppi = 0.5 $. 
	Lateral distance $ G $ does not significantly affects this trend. 
	On the contrary, at low wavelength $ \lam \leq 2.0 $, the follower always take a greater advantage upon the thrust force compared with the leader.
	Here, we offer a more general overview regarding the effects of phase difference with the help of \Cref{fig:matrix_CTm}. 
	Phase difference $ \phi $ can significantly affect the follower's net thrust force while its effects on the leader is less prominent. The effect of the phase difference becomes less significant with the enlargement of the lateral gap $ G $ and the front-back distance $ D $, as demonstrated by the converging values of $ \ctm $ from \Cref{fig:matrix_CTm}a to \ref{fig:matrix_CTm}i. The variation of $ \phi $ is more influential at high wavelength $ \lam > 3 $ while being less effective at low wavelength $ \lam \leq 2 $.

	\newcommand{\addlabele}[3]{%
		\begin{tikzpicture}
			\node[anchor=south west,inner sep=0] (image) at (0,0) 
			{\includegraphics[width=#1\textwidth]{#2}};
			\begin{scope}[x={(image.south east)},y={(image.north west)}]
				\node[anchor=south west] at (0.15,0.80) {\footnotesize #3};
			\end{scope}
		\end{tikzpicture}%
	}
	\begin{figure}
		\setcounter{testa}{0}
		\centering
		\addlabele{0.32}{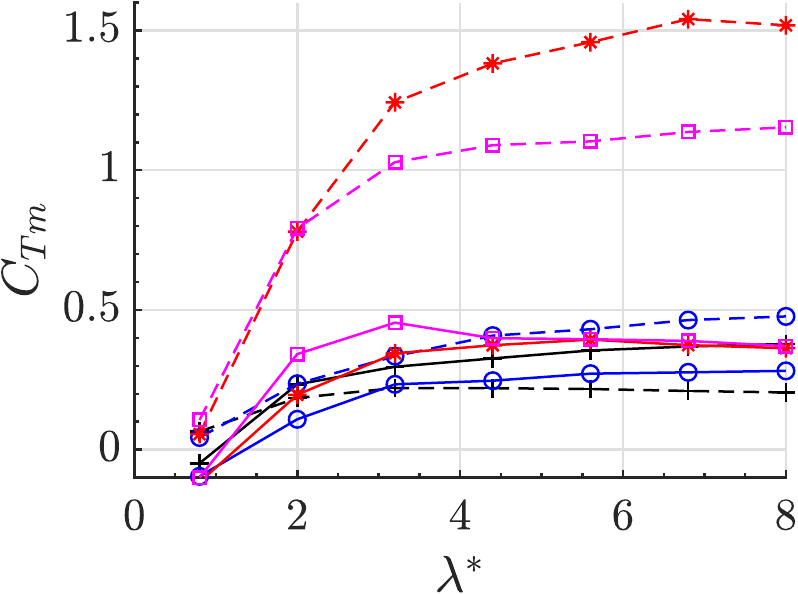}{(\countera) $ G = 0.25, D = 0.25$}
		\addlabele{0.32}{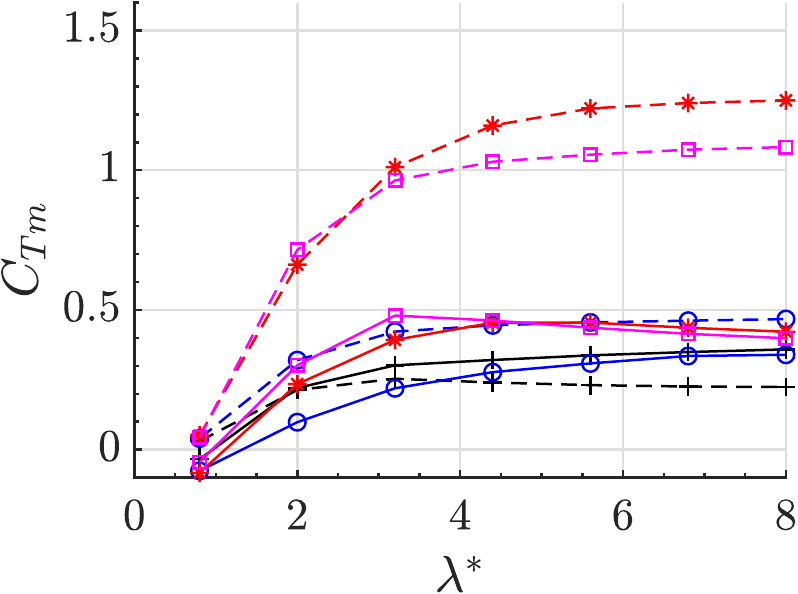}{(\countera) $ G = 0.30, D = 0.25$}
		\addlabele{0.32}{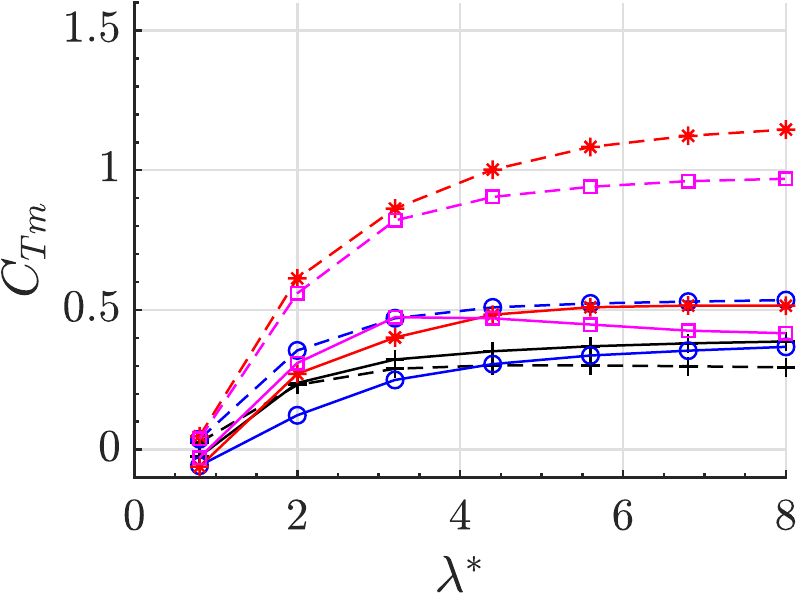}{(\countera) $ G = 0.35, D = 0.25$}
		
		\addlabele{0.32}{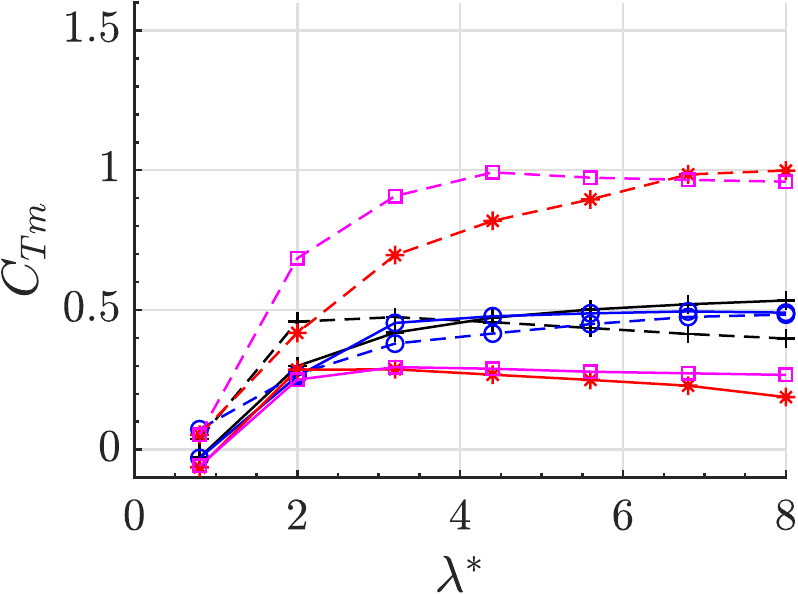}{(\countera) $ G = 0.25, D = 0.50$}
		\addlabele{0.32}{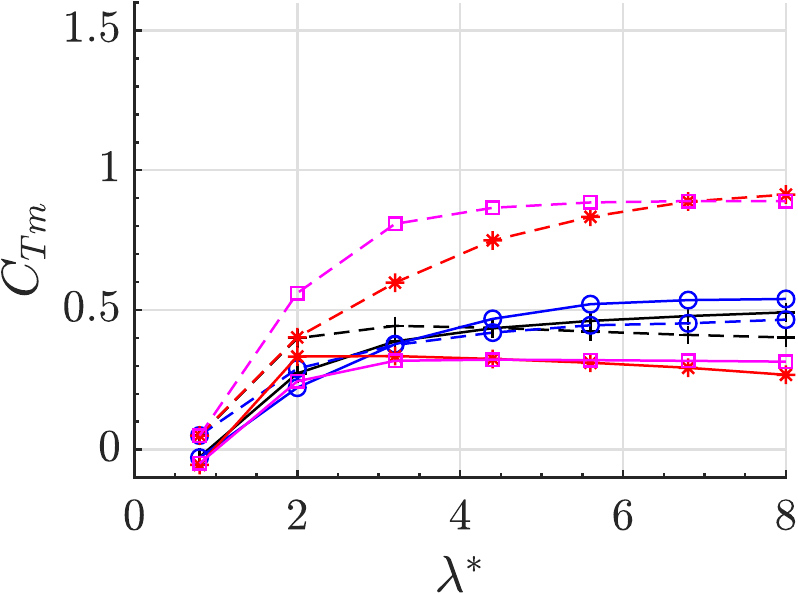}{(\countera) $ G = 0.30, D = 0.50$}
		\addlabele{0.32}{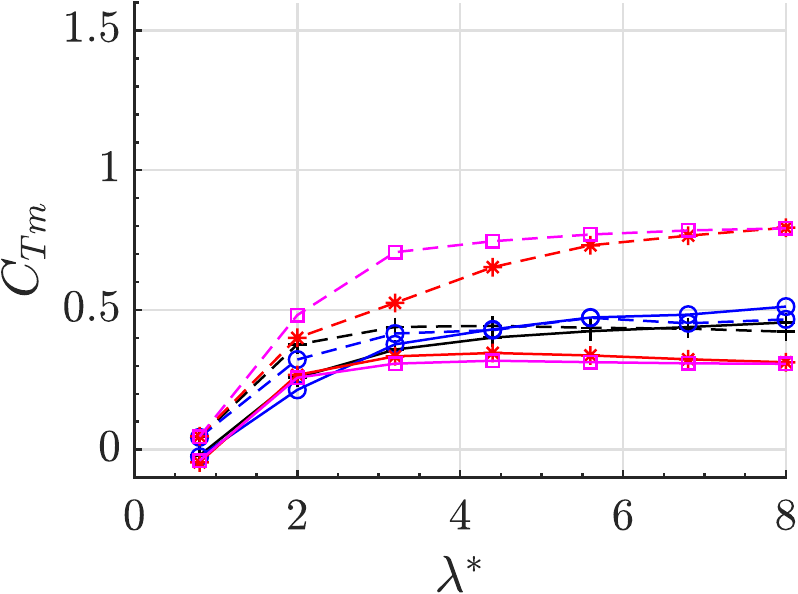}{(\countera) $ G = 0.35, D = 0.50$}
		
		\addlabele{0.32}{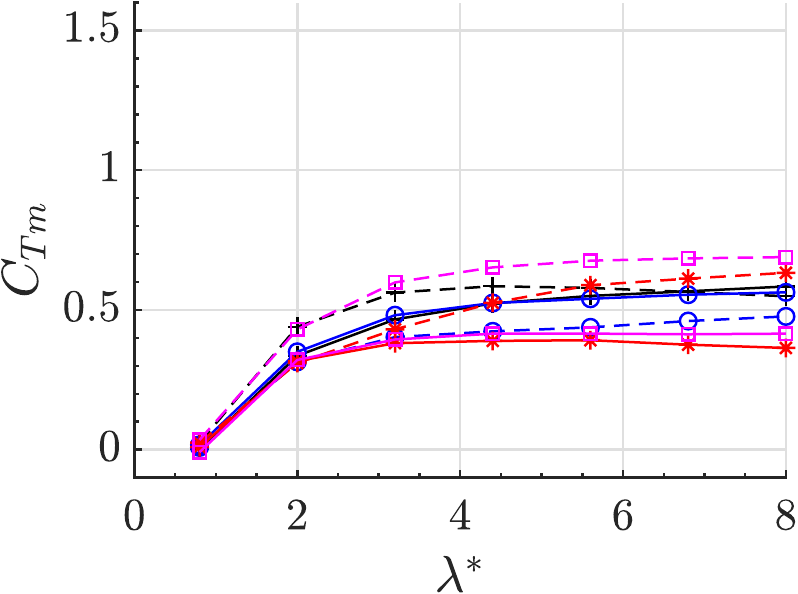}{(\countera) $ G = 0.25, D = 0.75$}
		\addlabele{0.32}{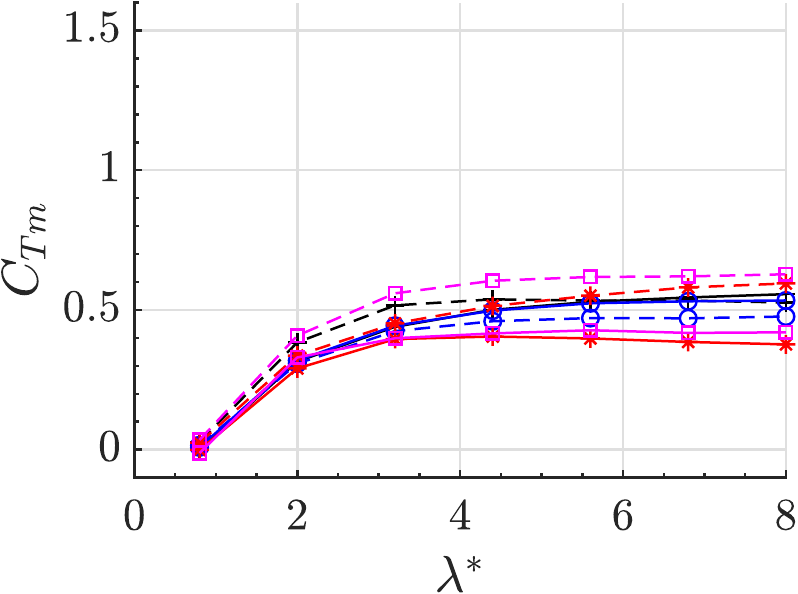}{(\countera) $ G = 0.30, D = 0.75$}
		\addlabele{0.32}{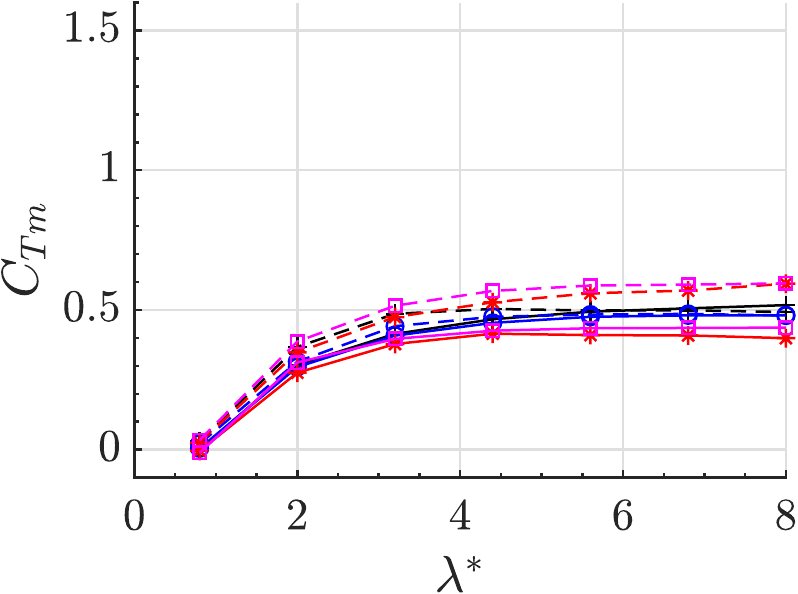}{(\countera) $ G = 0.35, D = 0.75$}
		
		\includegraphics[width=1.0\linewidth]{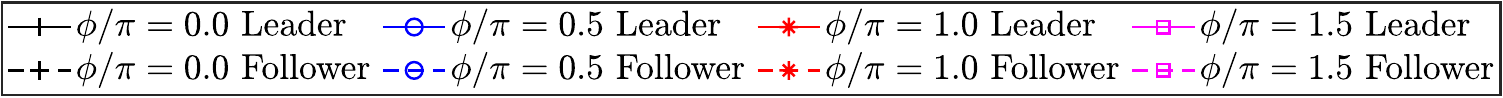}
		
		\caption{
			Variation of \textit{mean thrust force} $ C_{Tm} $ for both leading swimmer (solid lines) and following swimmer (dashed lines) with a series of wavelength $ \lambda^* = 0.8 - 8.0 $, leader-follower phase difference $ \phi / \pi = 0, 0.5, 1.0, 1.5  $ (denoted by marker types), front-back distance (a-c) $ D = 0.25 $, (d-f) $ D = 0.50 $, and (g-i) $ D = 0.75 $; lateral gap distance at (a \& d \& g) $ G = 0.25 $, (b \& e \& h) $ G = 0.30 $, (c \& f \& i) $ G = 0.35 $.
		}
		\label{fig:matrix_CTm}
		
	\end{figure}

\begin{figure}
		\centering
		\setcounter{testa}{0}
		\addlabele{0.32}{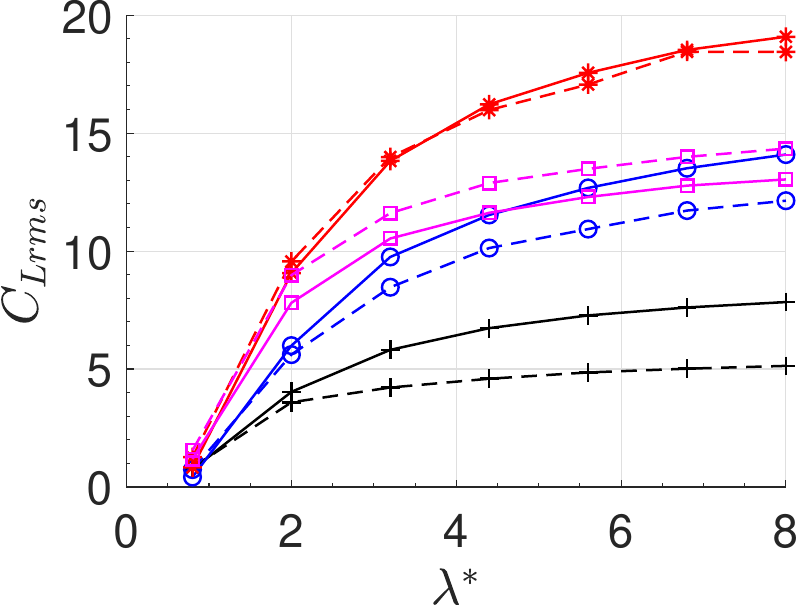}{(\countera) $ G = 0.25, D = 0.25$}
		\addlabele{0.32}{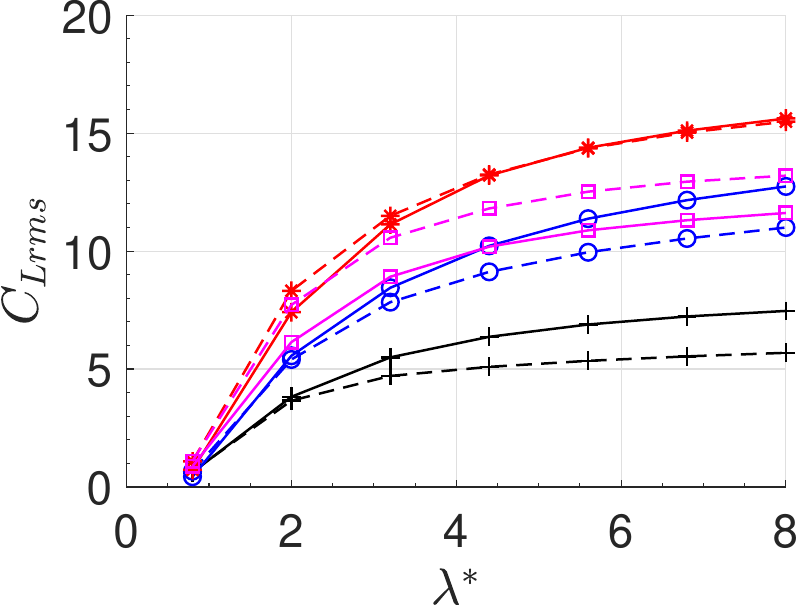}{(\countera) $ G = 0.30, D = 0.25$}
		\addlabele{0.32}{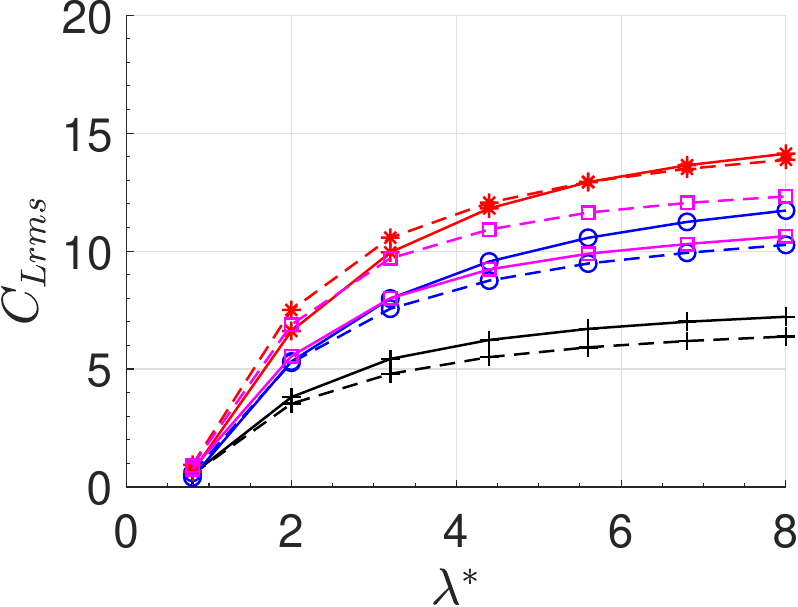}{(\countera) $ G = 0.35, D = 0.25$}
		
		\addlabele{0.32}{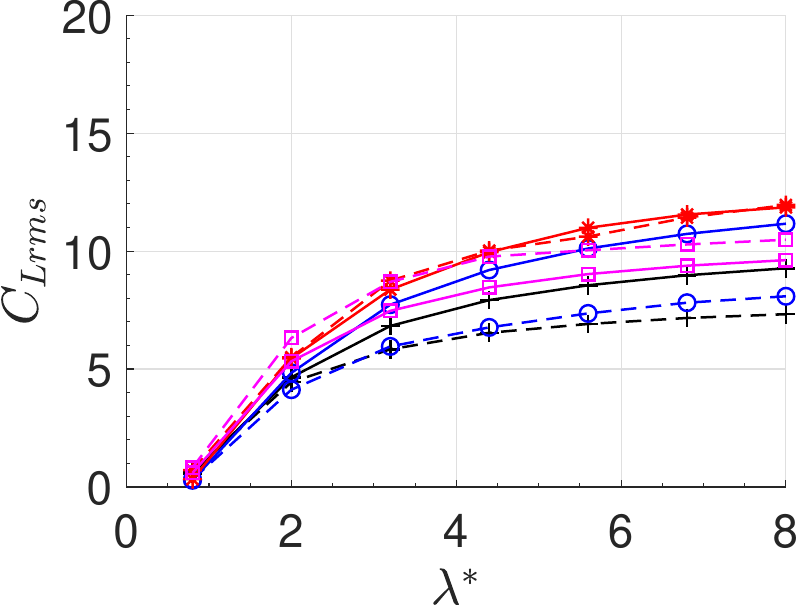}{(\countera) $ G = 0.25, D = 0.50$}
		\addlabele{0.32}{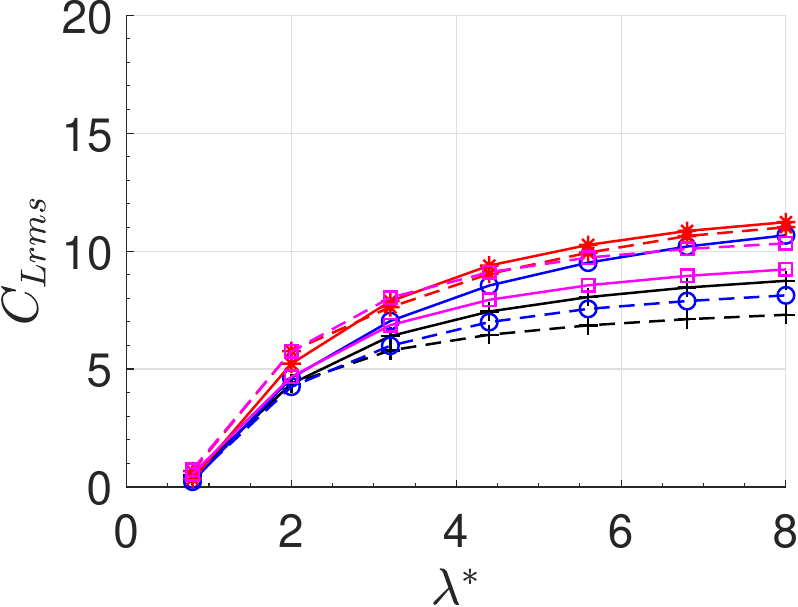}{(\countera) $ G = 0.30, D = 0.50$}
		\addlabele{0.32}{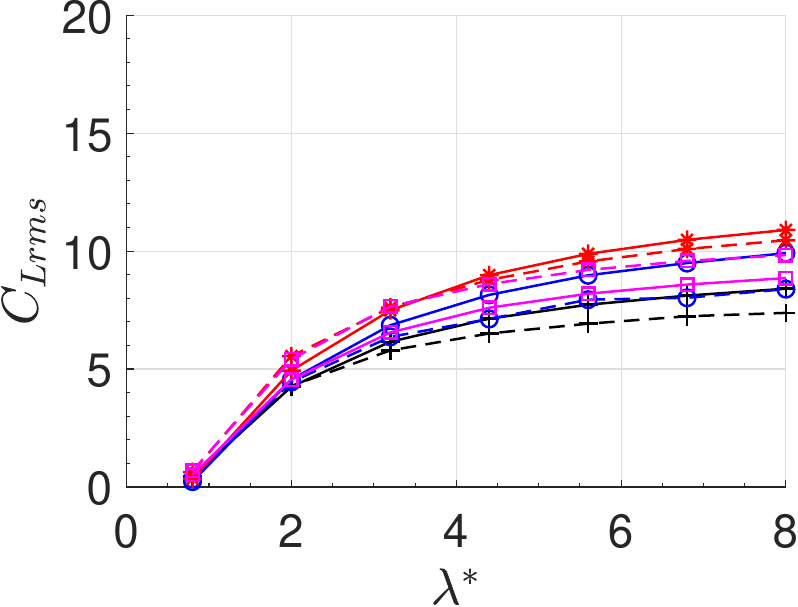}{(\countera) $ G = 0.35, D = 0.50$}
		
		\addlabele{0.32}{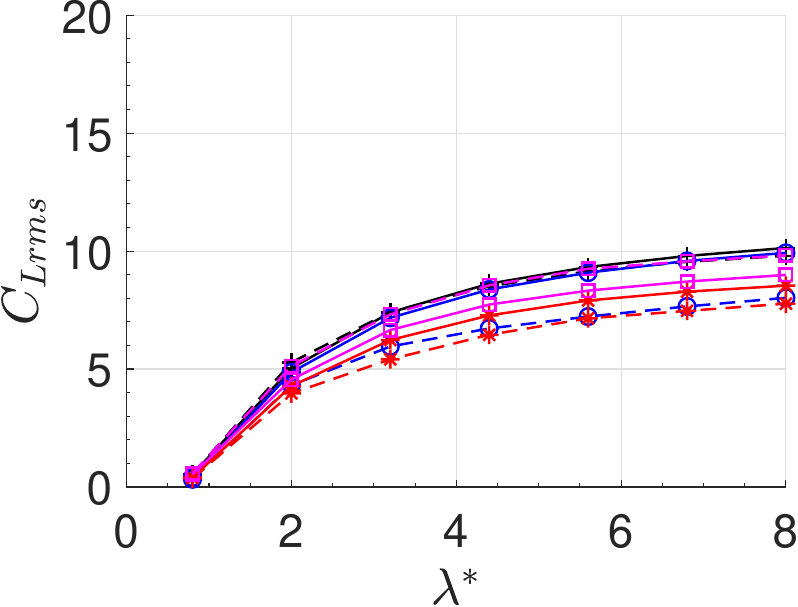}{(\countera) $ G = 0.25, D = 0.75$}
		\addlabele{0.32}{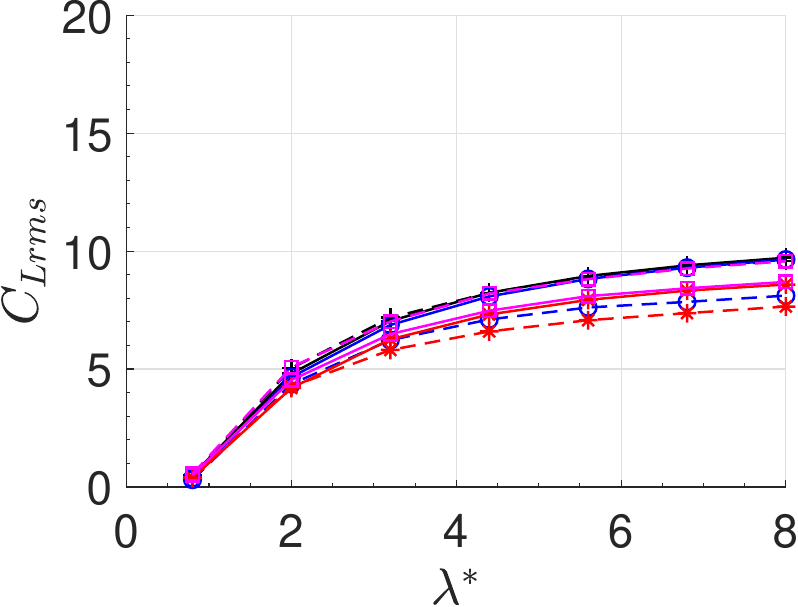}{(\countera) $ G = 0.30, D = 0.75$}
		\addlabele{0.32}{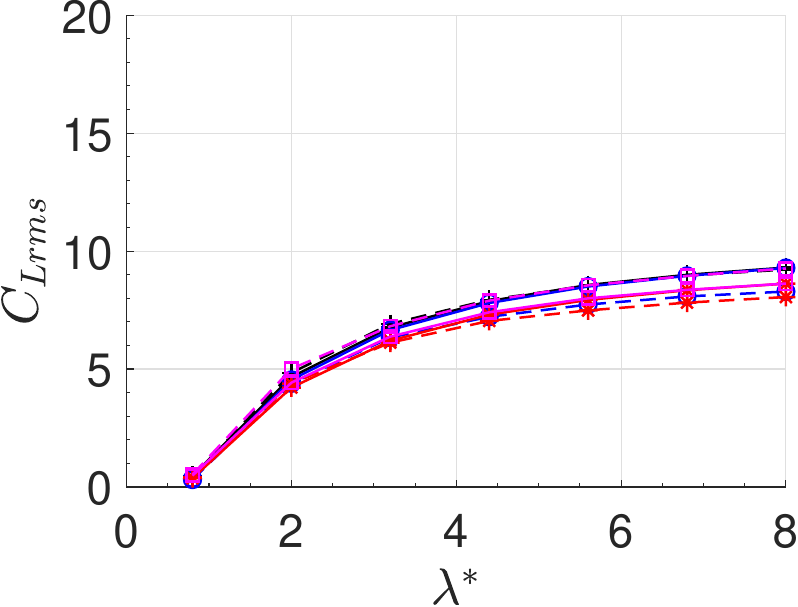}{(\countera) $ G = 0.35, D = 0.75$}

		\includegraphics[width=1.0\linewidth]{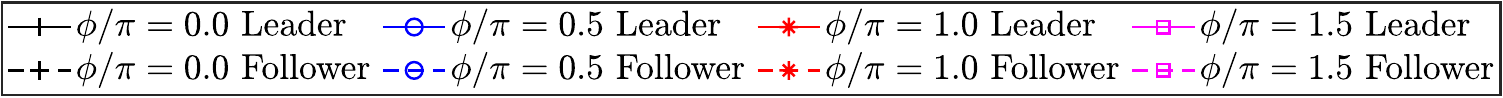}
		
		\caption{Variation of \textit{root mean square of lateral force} $ C_{Lrms} $ for both leader (solid lines) and follower (dashed lines) hydrofoils with a series of wavelength $ \lambda^* = 0.8 - 8.0 $, leader-follower phase difference $ \phi / \pi = 0, 0.5, 1.0, 1.5  $ (denoted by marker types), front-back distance (a-c) $ D = 0.25 $, (d-f) $ D = 0.50 $, and (g-i) $ D = 0.75 $; lateral gap distance at (a \& d \& g) $ G = 0.25 $, (b \& e \& h) $ G = 0.30 $, (c \& f \& i) $ G = 0.35 $.}
		\label{fig:matrix_CLrms}
	\end{figure}

	The RMS of \textit{lateral force} $ \clrms $ increases monotonically with wavelength $ \lam $ despite the variation of other parameters, as seen in \Cref{fig:matrix_CLrms}.
	The lateral force greatly increases with the wavelength $ \lam $, reaching as high as $ \clrms = 18 $ for both swimmers, at short distance $ G = 0.25 $, $ D = 0.25 $ and anti-phase $ \ppi = 1 $, as shown in \Cref{fig:matrix_CLrms}a.
	Regarding the effects of the phase difference $ \phi $, the lateral force $ \clrms $ reaches {minimum} when the two hydrofoils swim {in-phase} $ \ppi = 0 $ while reaching {maximum} at {anti-phase} $ \ppi = 1 $ condition with front-back distance $ D \leq 0.50 $, as seen in \Cref{fig:lines_D_0_CT_CL}b and \ref{fig:matrix_CLrms}. However, at $ D = 0.75 $, this relationship is reversed that the minimal $ \clrms $ is found at anti-phase condition whereas the maximal $ \clrms $ is discovered at in-phase scenarios. This observation can be postulated to be related with vortex shedding and its impingement upon the follower.
	As the distances $ G $ and $ D $ increase, the phase difference $ \phi $ becomes less influential upon the lateral force $ \clrms $ for both the leader and the follower, as shown in \Cref{fig:matrix_CLrms}.
	With the increase of lateral gap $ G $ and front-back distance $ D $, the difference of lateral force $ \clrms $ between the leader and the follower becomes smaller; as the lateral gap increases from $ G = 0.25 $ to $ 0.35 $ and the front-back distance rises from $ D = 0.25 $ to $ 0.75 $, the leader-follower lateral force difference decreases from $ 2.5 $ to $ 0.5 $, as seen in \Cref{fig:matrix_CLrms}a to \ref{fig:matrix_CLrms}i.
	Compared with the large leader-follower difference for the thrust force $ \ctm $ previously discussed, the leader-follower discrepancy in lateral force $ \clrms $ is relatively small, especially at short distances $ G = 0.25,\ D = 0.25 $ indicating the thrust force is more sensitive to the schooling effect than the lateral force, across the tested wavelengths.

	The individual \textit{propeller efficiency} $ \eta $ for each swimmer generally reaches minimum at wavelength $ \lam = 0.8 $ while peaking at $ \lam = 2.0 $, as seen in \Cref{fig:matrix_LF_effi}. With a relatively high wavelength $ \lam > 2.0 $, phase lag $ \phi $ affects how $ \lam $ influences the follower's propeller efficiency $ \eta_{follower} $: at in-phase $ \ppi = 0 $ condition, $ \eta_{follower} $ greatly decreases with wavelength $ \lam $; at $ \ppi = 0.5 $, the negative relationship between $ \lam $ and $ \eta_{follower} $ is less significant than that at $ \ppi = 0 $; at $ \ppi = 1.0 $ and $ 1.5 $, the result generally remains constant regardless of the variation in $ \lam $.
	The increase in front-back distance $ D $ significantly reduces the propeller efficiency of the follower while slightly increasing the leader's efficiency. In the tested range of values, lateral gap $ G $ barely influences the propeller efficiency, being consistent with the conclusion by \cite{Li2020}.
	The follower's efficiency $ \eta_{follower} $ is generally higher than the leader's $ \eta_{leader} $, most significantly at front-back distance $ D = 0.25 $ and phase lag $ \ppi = 1.5 $, as demonstrated in \Cref{fig:matrix_LF_effi}a to \ref{fig:matrix_LF_effi}c; the leader's propeller efficiency can only be slightly higher than the follower's at in-phase condition $ \ppi = 0 $ and high wavelength $ \lam > 7 $.
	In the present combinations of input parameters, the leader efficiency can be negative at $ \lam = 0.8 $ when the front-back distance is small $ D \leq 0.50 $, meaning the leader is moving backwards along the flow direction. At $ D =0.75 $, the swimmers' efficiency is all positive except at $ \ppi = 1.0,\ 1.5 $ with low wavelength $ \lam = 0.8 $.
	
	\renewcommand{\addlabele}[3]{%
		\begin{tikzpicture}
			\node[anchor=south west,inner sep=0] (image) at (0,0) 
			{\includegraphics[width=#1\textwidth]{#2}};
			\begin{scope}[x={(image.south east)},y={(image.north west)}]
				\node[anchor=south west] at (0.16,0.15) {\footnotesize #3};
			\end{scope}
		\end{tikzpicture}%
	}
	\begin{figure}
		\setcounter{testa}{0}
		\centering
		\addlabele{0.32}{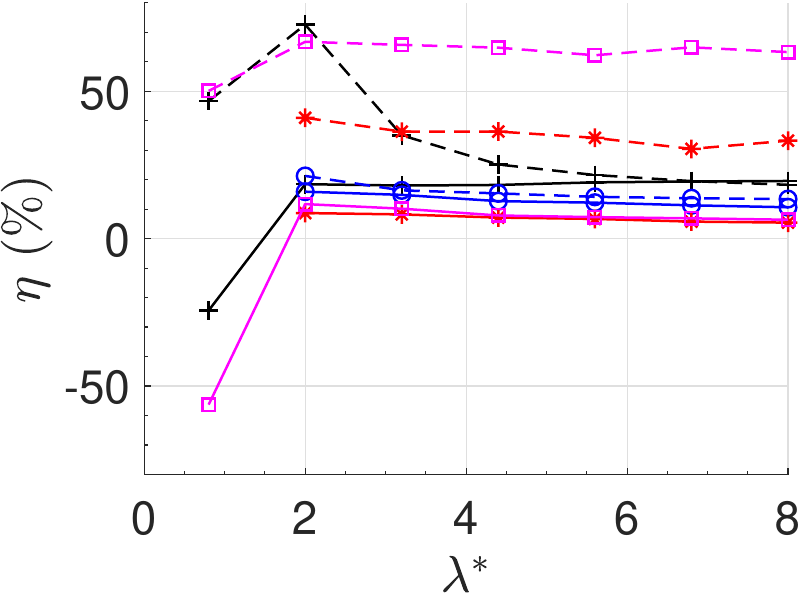}{(\countera) $ G = 0.25, D = 0.25$}
		\addlabele{0.32}{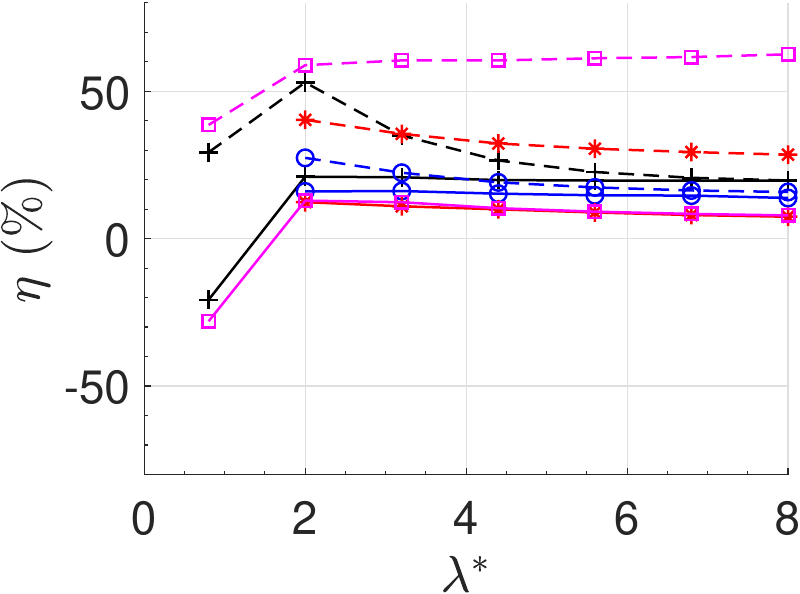}{(\countera) $ G = 0.30, D = 0.25$}
		\addlabele{0.32}{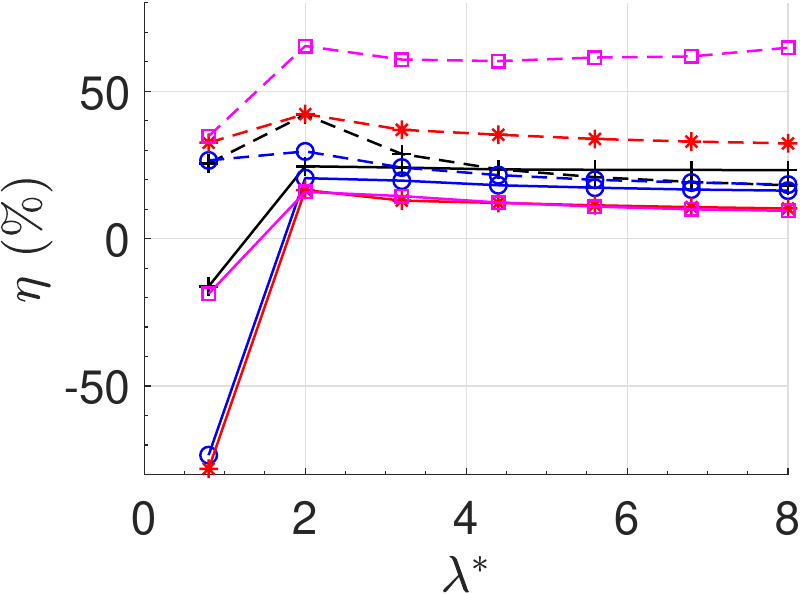}{(\countera) $ G = 0.35, D = 0.25$}
		
		\addlabele{0.32}{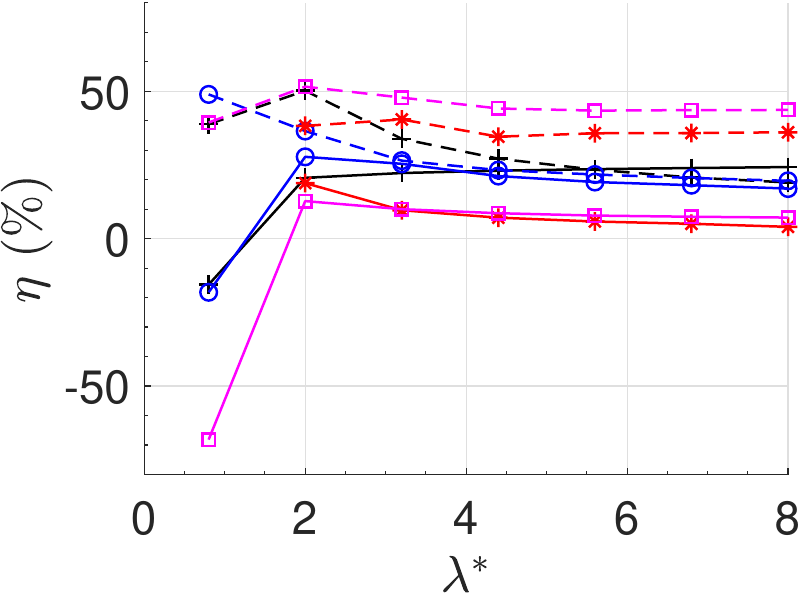}{(\countera) $ G = 0.25, D = 0.50$}
		\addlabele{0.32}{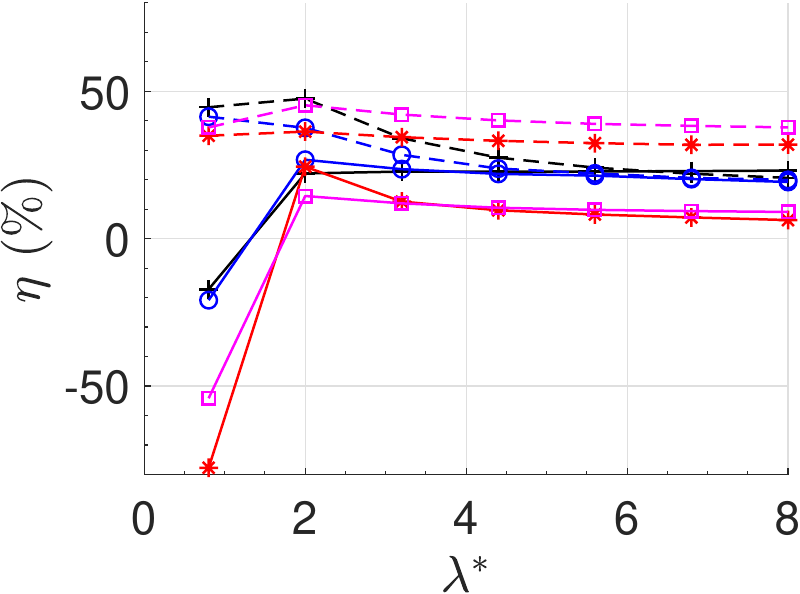}{(\countera) $ G = 0.30, D = 0.50$}
		\addlabele{0.32}{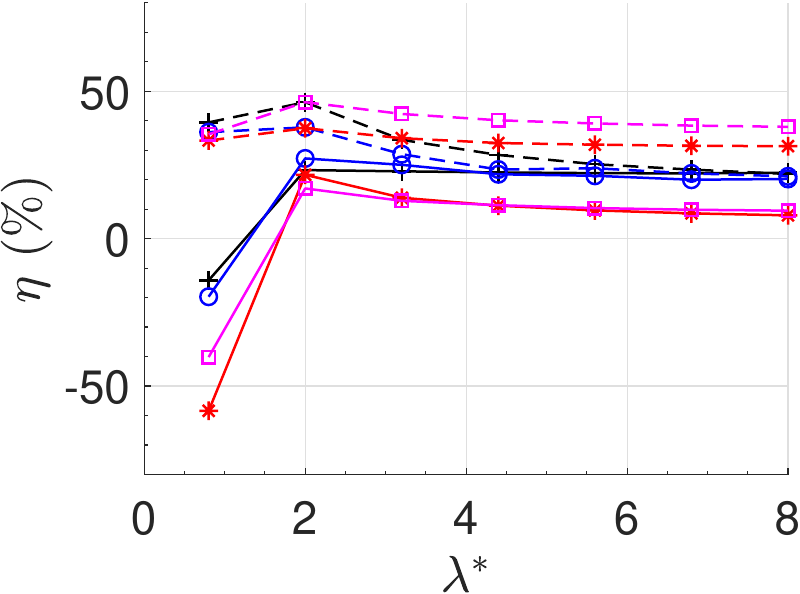}{(\countera) $ G = 0.35, D = 0.50$}
		
		\addlabele{0.32}{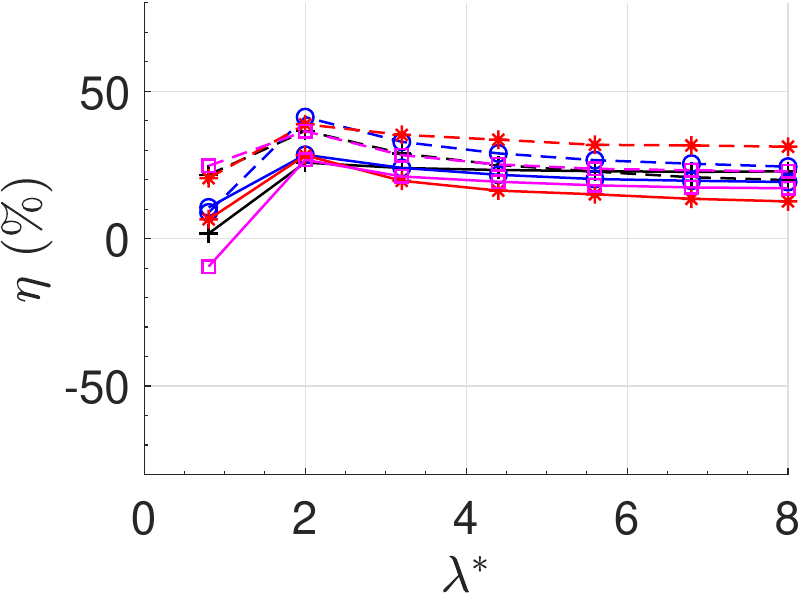}{(\countera) $ G = 0.25, D = 0.75$}
		\addlabele{0.32}{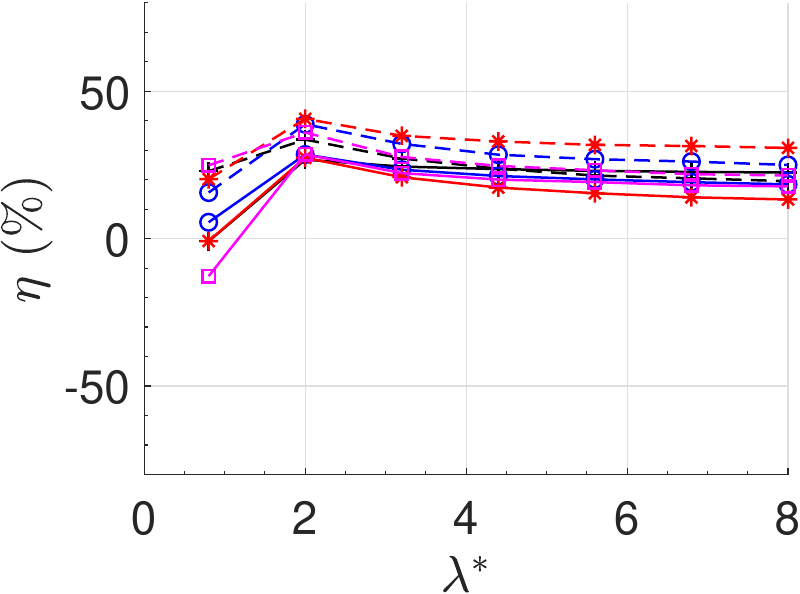}{(\countera) $ G = 0.30, D = 0.75$}
		\addlabele{0.32}{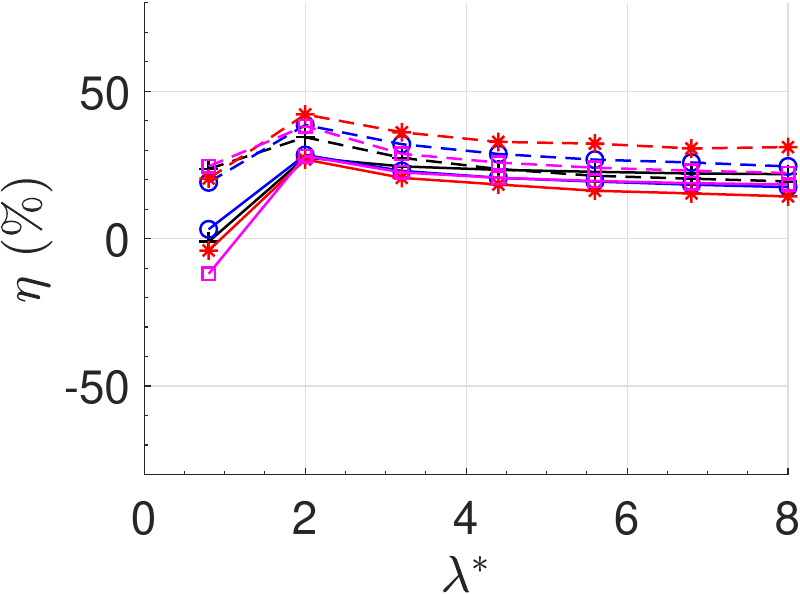}{(\countera) $ G = 0.35, D = 0.75$}

		\includegraphics[width=1.0\linewidth]{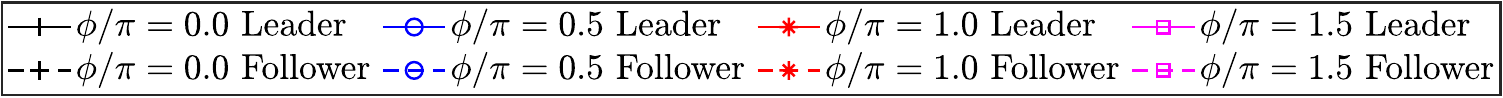}
		\caption{Variation of {propeller efficiency} $ \eta $ for leader (solid lines) and follower (dashed lines) hydrofoils with a series of wavelength $ \lambda^* = 0.8 - 8.0 $, leader-follower phase difference $ \phi / \pi = 0, 0.5, 1.0, 1.5  $ (denoted by marker types), front-back distance (a-c) $ D = 0.25 $, (d-f) $ D = 0.50 $, and (g-i) $ D = 0.75 $; lateral gap distance at (a \& d \& g) $ G = 0.25 $, (b \& e \& h) $ G = 0.30 $, (c \& f \& i) $ G = 0.35 $.}
		\label{fig:matrix_LF_effi}
	\end{figure}

\begin{figure}
		\setcounter{testa}{0}
		\centering
		
		\addlabele{0.32}{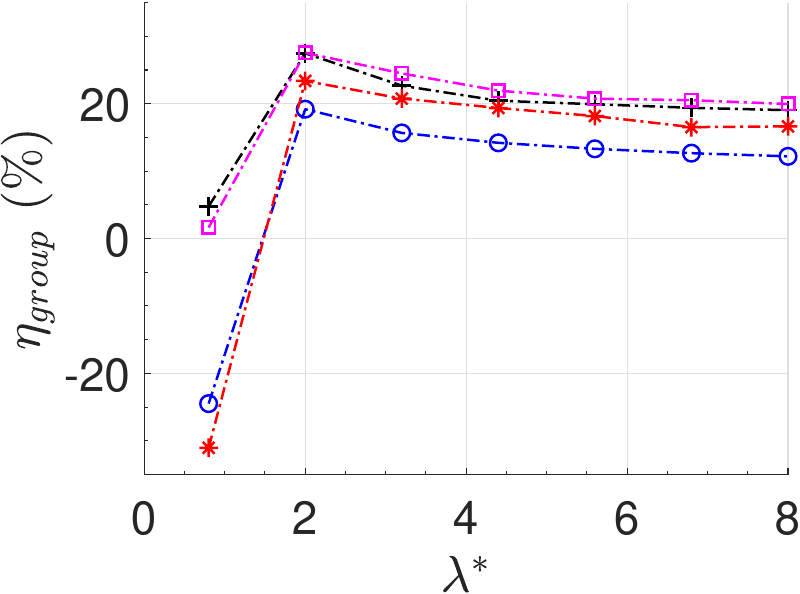}{(\countera) $ G = 0.25, D = 0.25$}
		\addlabele{0.32}{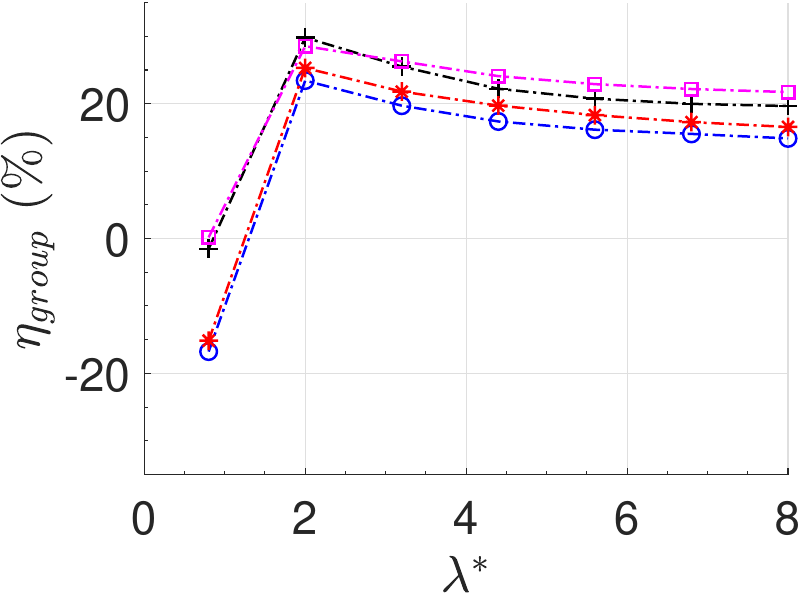}{(\countera) $ G = 0.30, D = 0.25$}
		\addlabele{0.32}{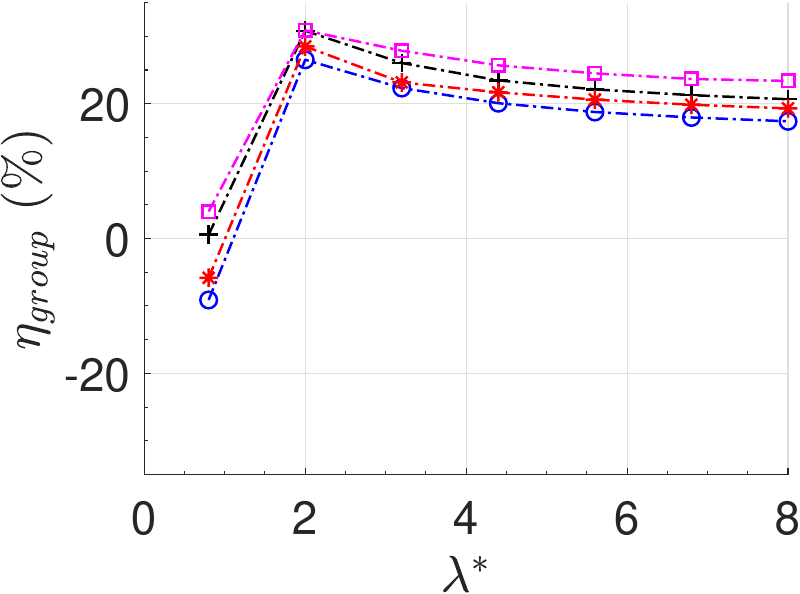}{(\countera) $ G = 0.35, D = 0.25$}
		
		\addlabele{0.32}{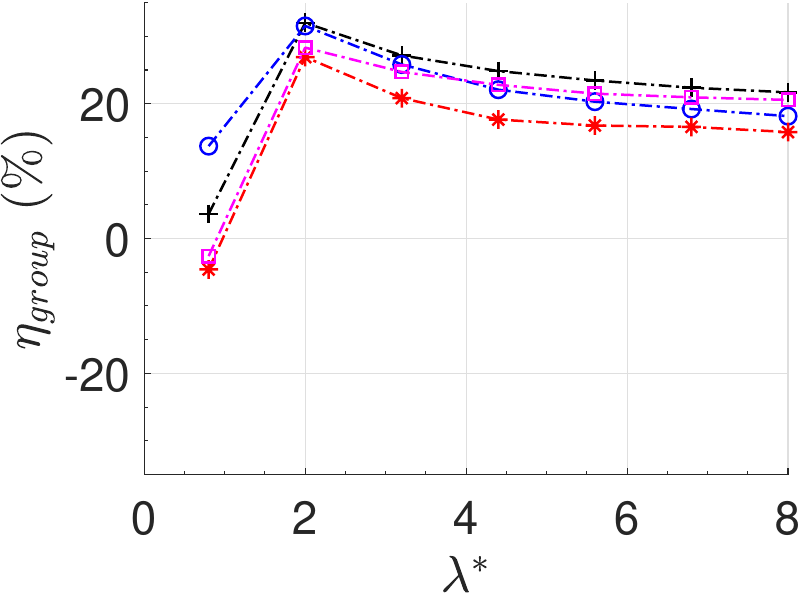}{(\countera) $ G = 0.25, D = 0.50$}
		\addlabele{0.32}{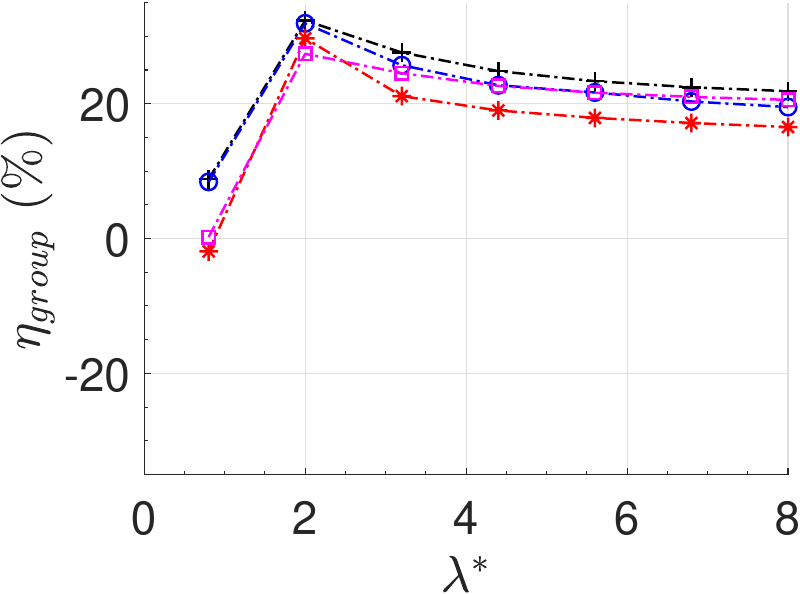}{(\countera) $ G = 0.30, D = 0.50$}
		\addlabele{0.32}{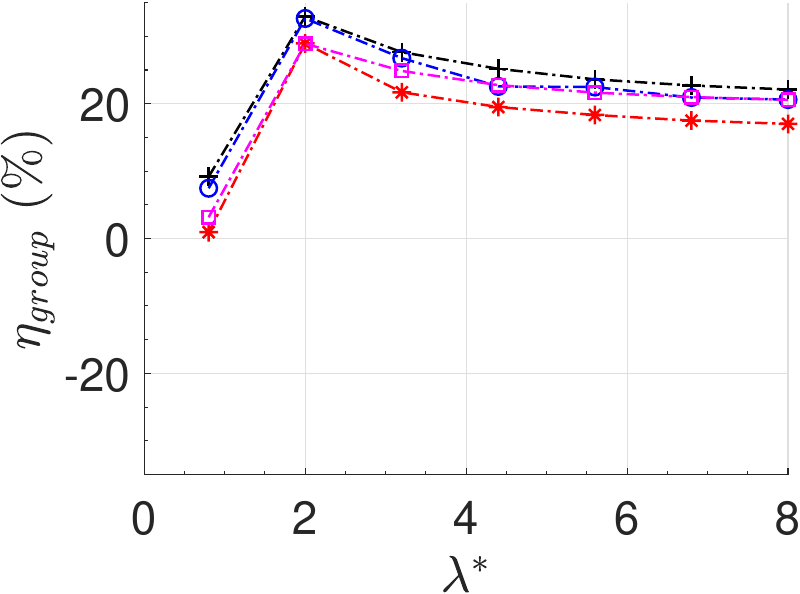}{(\countera) $ G = 0.35, D = 0.50$}
		
		\addlabele{0.32}{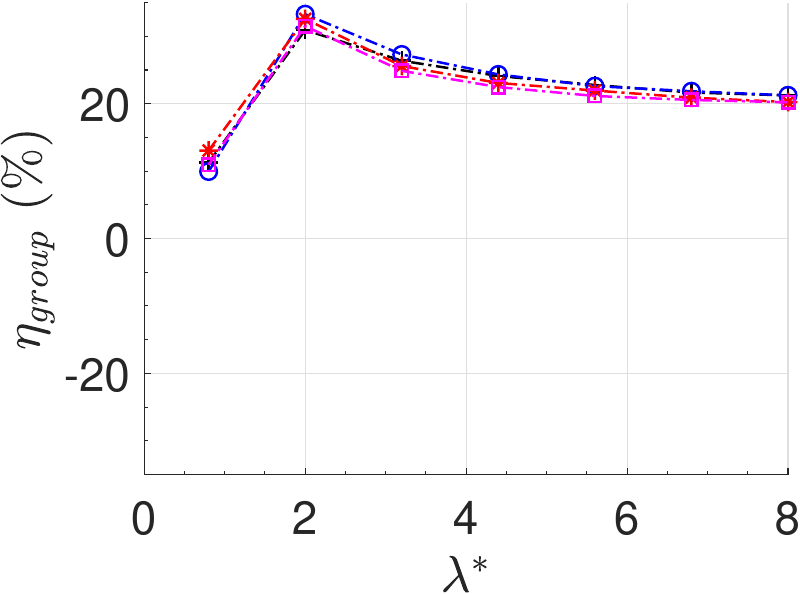}{(\countera) $ G = 0.25, D = 0.75$}
		\addlabele{0.32}{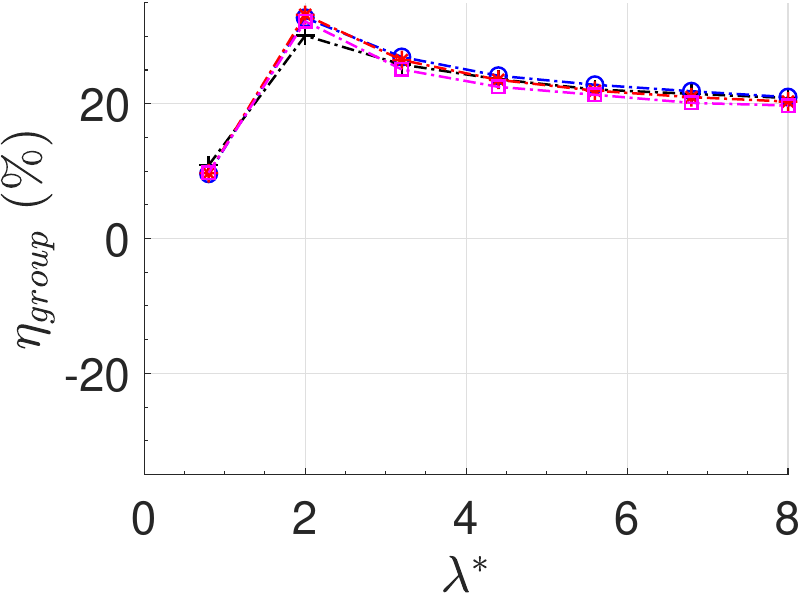}{(\countera) $ G = 0.30, D = 0.75$}
		\addlabele{0.32}{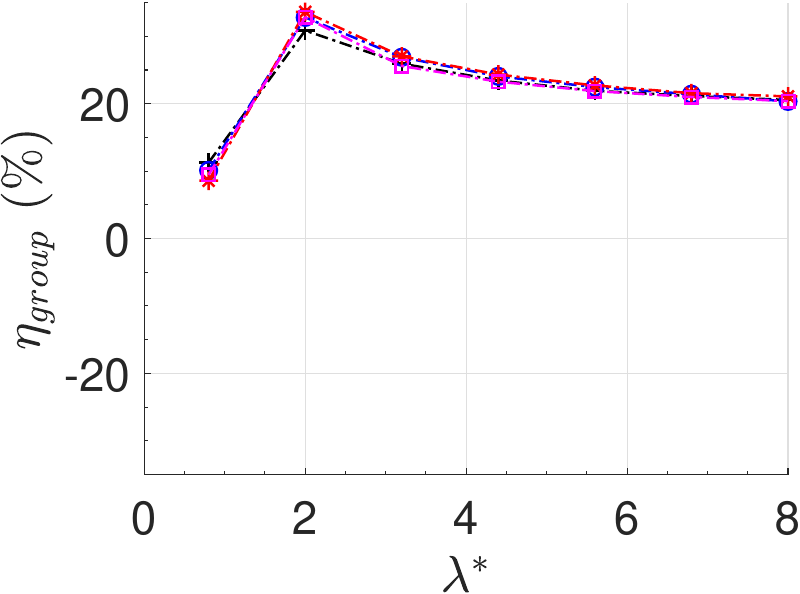}{(\countera) $ G = 0.35, D = 0.75$}
		
		\includegraphics[width=0.9\linewidth]{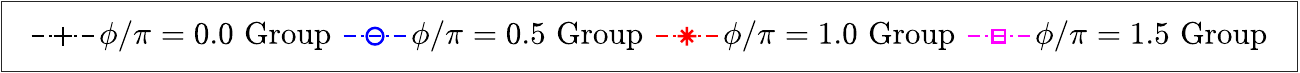}
		\caption{Variation of {group efficiency} $ \eta_{group} $ for the swimming group with a series of wavelength $ \lambda^* = 0.8 - 8.0 $, leader-follower phase difference $ \phi / \pi = 0, 0.5, 1.0, 1.5 $ (denoted by marker types), front-back distance (a-c) $ D = 0.25 $, (d-f) $ D = 0.50 $, and (g-i) $ D = 0.75 $; lateral gap distance at (a \& d \& g) $ G = 0.25 $, (b \& e \& h) $ G = 0.30 $, (c \& f \& i) $ G = 0.35 $.}
		\label{fig:matrix_group_eta}
	\end{figure}

	The \textit{group efficiency} $ \etag $ represents the effectiveness of energy conversion from lateral undulation to the thrust propulsion for the two interacting swimmers as a minimal school.
	The group efficiency $ \etag $ reaches minimum at $ \lam = 0.8 $, peaks at $ \lam = 2.0 $, and then gradually approach a certain value at large wavelength $ \lam = 8.0 $; this pattern can be observed across all simulated cases.
	The {group efficiency} $ \etag $ reaches maximum at 33.3\% with front-back distance $ D = 0.75 $ and wavelength $ \lam = 2.0 $.
	In the explored parametric space, $ \etag $ generally increases with lateral gap $ G $ and front-back distance $ D $ across various wavelengths $ \lam $, as seen in \Cref{fig:matrix_group_eta}. At low wavelength $ \lam = 0.8 $, the group efficiency is especially sensitive to front-back distance $ D $ but less sensitive to lateral distance $ G $, being consistent with the conclusions by \cite{Li2020}.
	The increase of front-back distance $ D $ and lateral gap $ G $ leads to reduced sensitivity regarding phase lag $ \phi $ across various wavelengths $ \lam = 0.8 - 8 $, as seen in \Cref{fig:matrix_group_eta}i for the almost overlapped lines; this trend corresponds to the reduced difference in propeller efficiency $ \eta $ between the leader and the follower, as seen in \Cref{fig:matrix_group_eta}, implying reduced flow mediated interference between the two swimming foils.
	At low wavelength $ \lam = 0.8 $, the negative group efficiency is observed at front-back distance $ D \leq 0.50 $ but not found in cases with $ D = 0.75 $.
	It is an indication that, for schooling Anguilliform swimmers, it can be critical to keep an appropriate front-back distance $ D $, which may even reverse the collective propulsive direction of the swimmer group; in contrast, the schooling performance for Carangiform or Thunniform swimmers with high wavelength $ \lam > 6 $ is more stable; its group efficiency does not vary significantly with phase lag $ \phi $ and front-back distance. This is in support of the hypothesis that the Carangiform and Thunniform swimmers are more suitable for schooling in contrast with Anguilliform swimmers.

	\subsubsection{Vorticity distribution at $ D > 0 $}
	\label{subsec:vor_Dg0}

	The wavelength $ \lam $ and the phase lag $ \phi $ influence the vortex strength and shedding pattern in different ways, as seen in \Cref{fig:vor_lam_phi}.
	At $ \lam = 0.8 $, the general vortex shedding pattern is barely affected by the variation in phase difference $ \phi $. The dipoles shed by the two foils hardly interact with each other, especially when the two foils swim in-phase $ \phi = 0 $, as seen in \Cref{fig:vor_lam_phi}a. The dipoles steadily drift downstream, meaning the streaming direction is stable as well.
	At $ \lam = 2.0 $, significant interference between the dipoles leads to irregular flow structure in the downstream area of the two foils. However, in the area between the two foils, the flow structure is regular and predictable, indicating the interaction between the two foils should largely be periodical despite the complex pattern in the downstream. Phase lag $ \phi $ can significantly affect the flow structure in the area immediately downstream the follower foil.
	At $ \lam = 8.0 $, the flow pattern is similar to that at $ \lam = 2.0 $.
	In summary, despite subtle differences observed, the general wake flow pattern is not significantly affected by the phase difference $ \phi $; this also corresponds to the results in \Cref{fig:matrix_group_eta}.
	\renewcommand{\addlabelvor}[3]{%
		\begin{tikzpicture}
			\node[anchor=south west,inner sep=0] (image) at (0,0) 
			{\includegraphics[width=#1\linewidth, trim={4.05cm 1cm 1cm 1.25cm},clip]{#2}};%
			\begin{scope}[x={(image.south east)},y={(image.north west)}]
				\node[anchor=south west] at (0.00,0.69) {\footnotesize #3};	%
			\end{scope}
		\end{tikzpicture}%
	}
	\begin{figure}
		\centering
		\setcounter{testaa}{0}
		\addlabelvor{0.32}{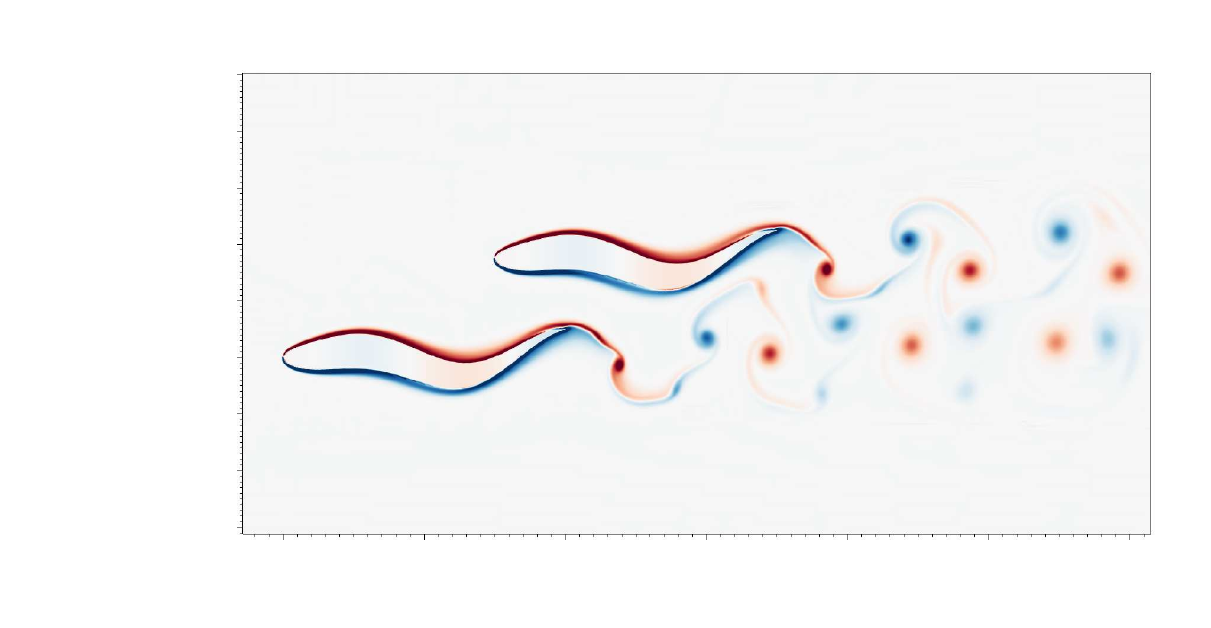}{(\counteraa) $ \lam = 0.8, \ppi = 0 $}
		\addlabelvor{0.32}{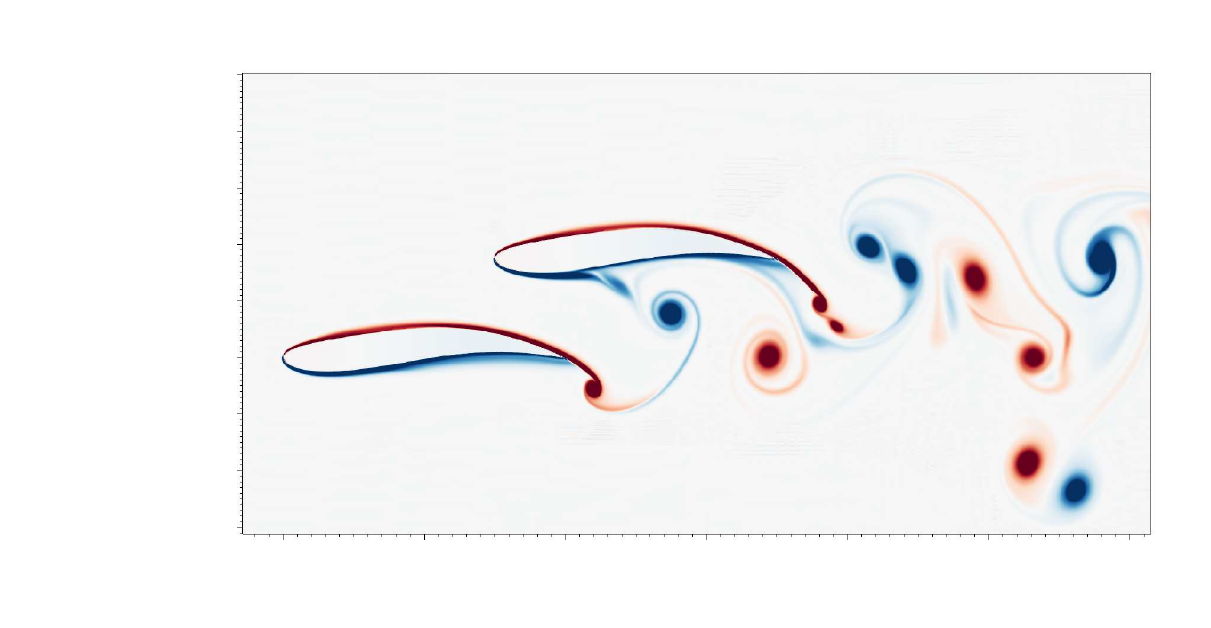}{(\counteraa) $ \lam = 2.0, \ppi = 0 $}
		\addlabelvor{0.32}{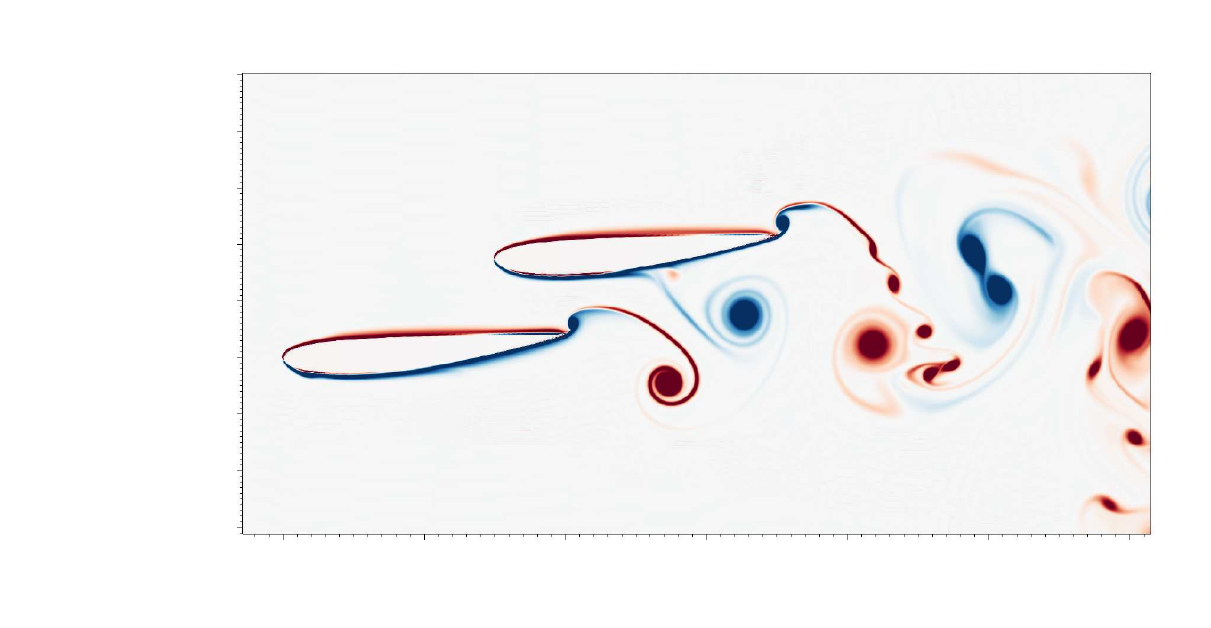}{(\counteraa) $ \lam = 8.0, \ppi = 0 $}
		
		\addlabelvor{0.32}{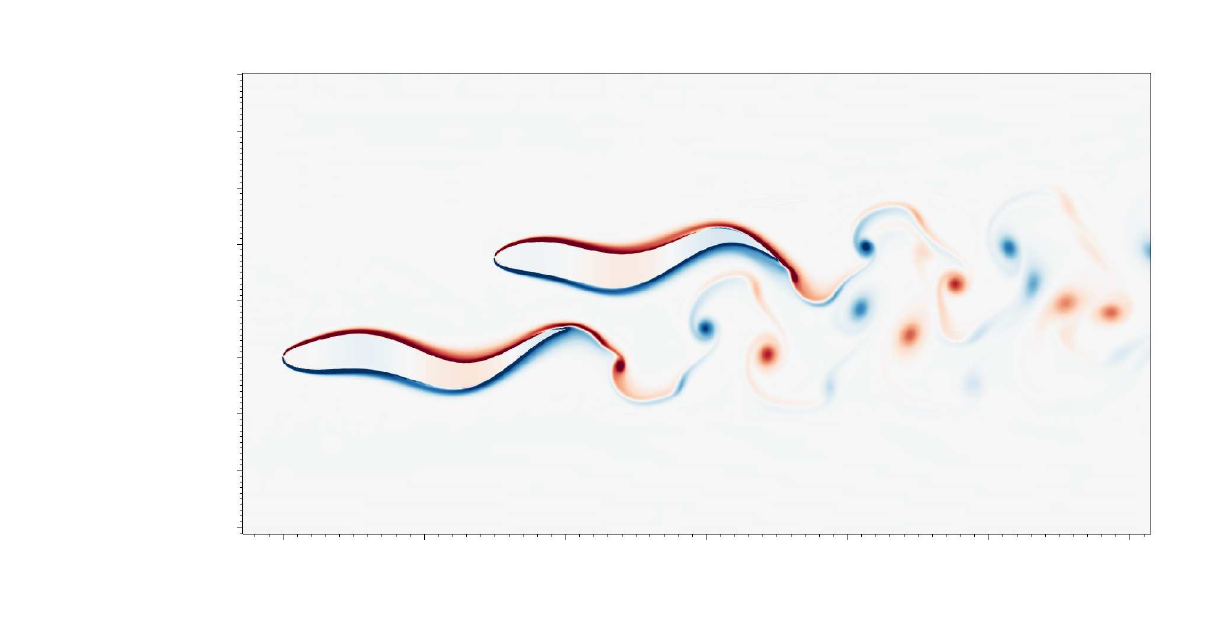}{(\counteraa) $ \lam = 0.8, \ppi = 0.5 $}
		\addlabelvor{0.32}{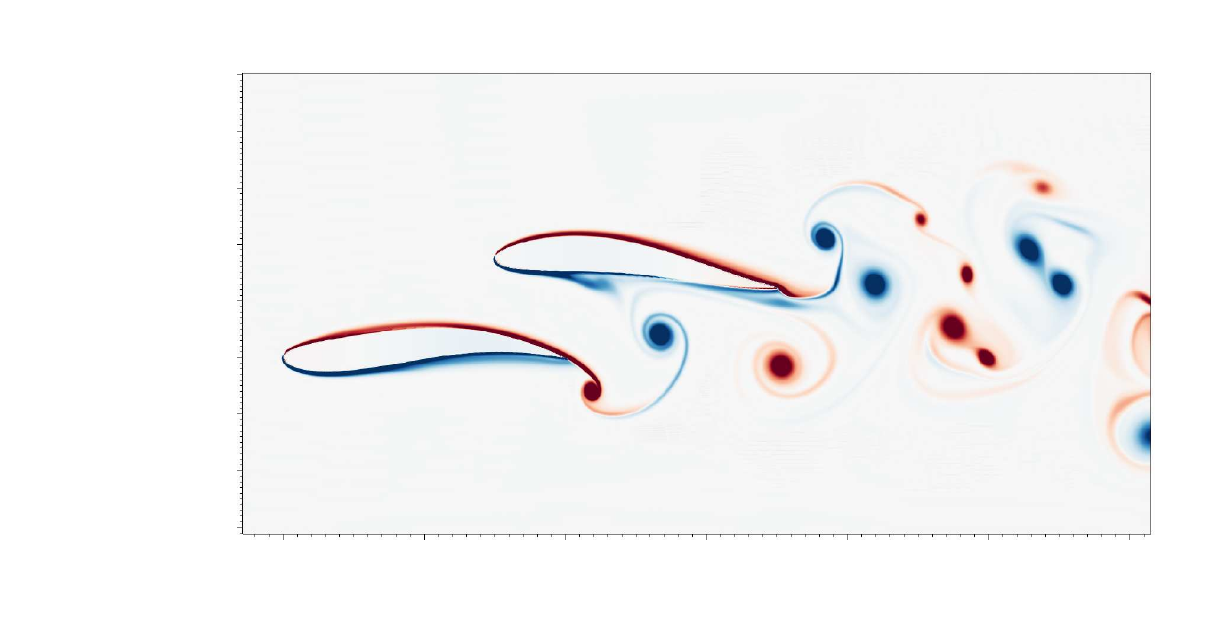}{(\counteraa) $ \lam = 2.0, \ppi = 0.5 $}
		\addlabelvor{0.32}{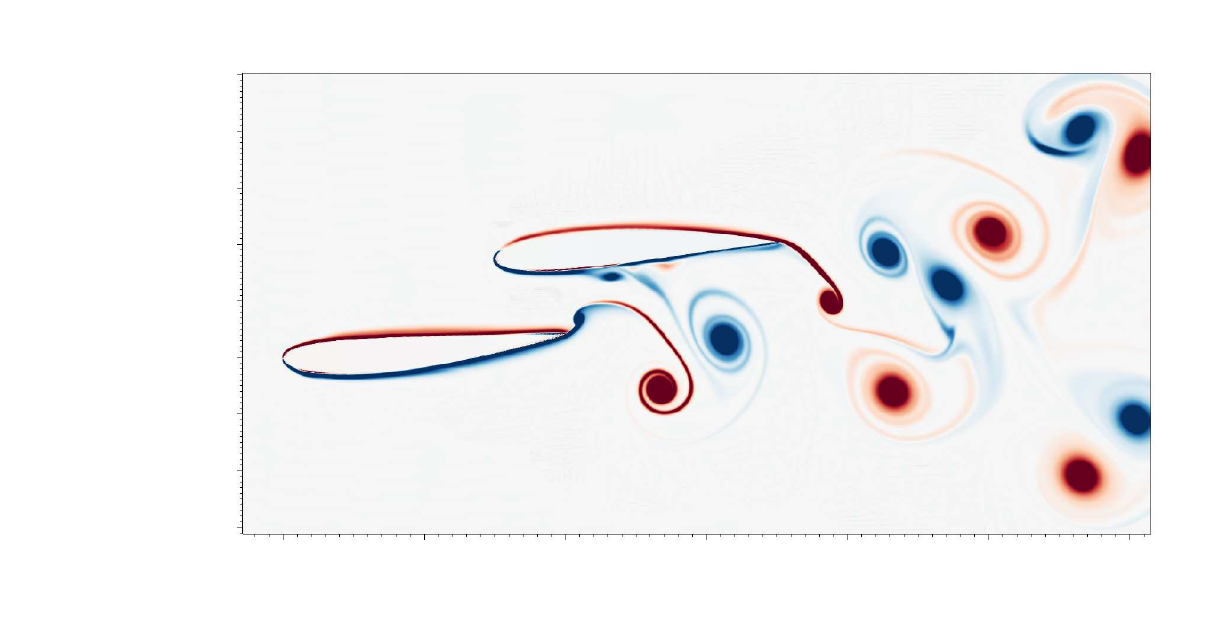}{(\counteraa) $ \lam = 8.0, \ppi = 0.5 $}
		
		\addlabelvor{0.32}{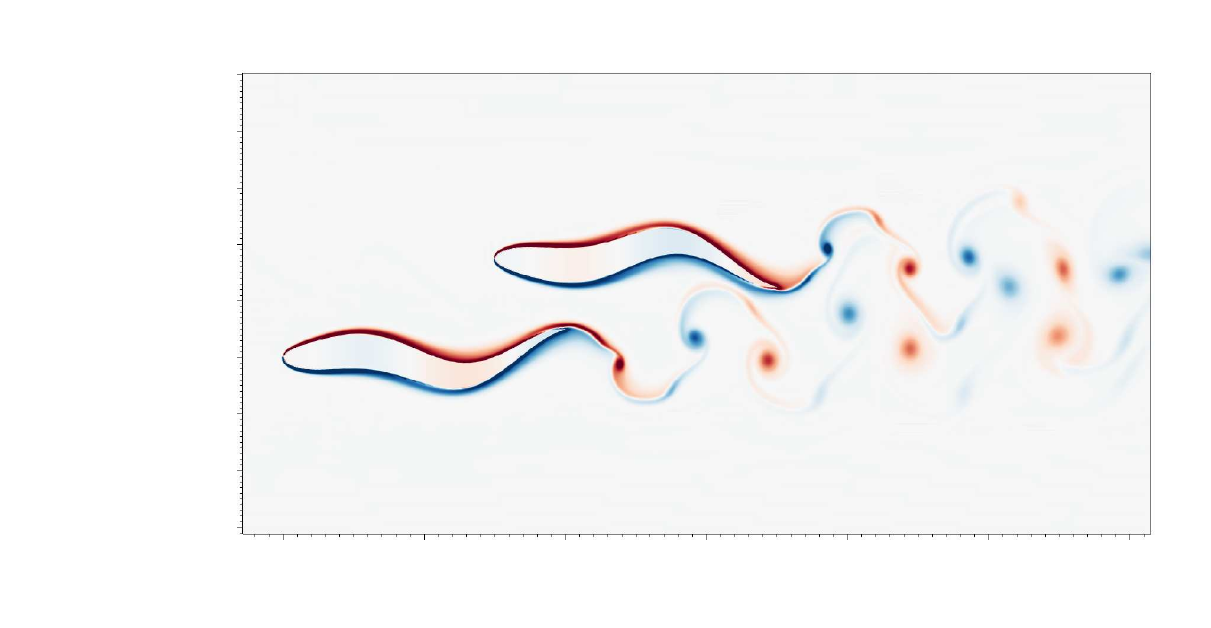}{(\counteraa) $ \lam = 0.8, \ppi = 1 $}
		\addlabelvor{0.32}{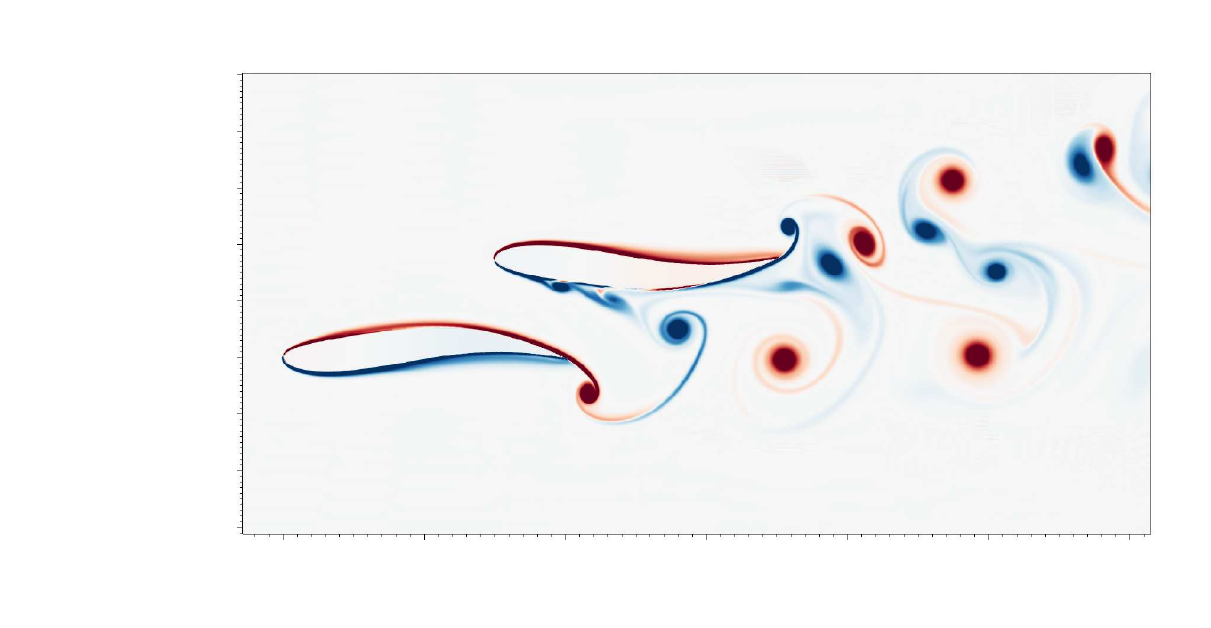}{(\counteraa) $ \lam = 2.0, \ppi = 1 $}
		\addlabelvor{0.32}{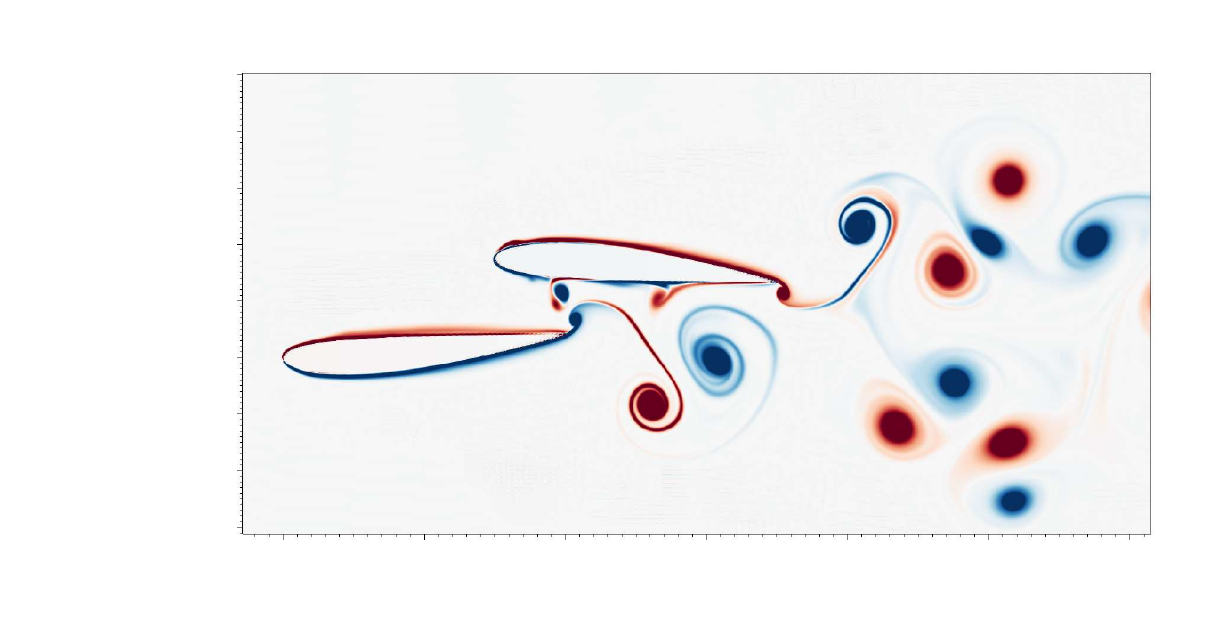}{(\counteraa) $ \lam = 8.0, \ppi = 1 $}
		
		\addlabelvor{0.32}{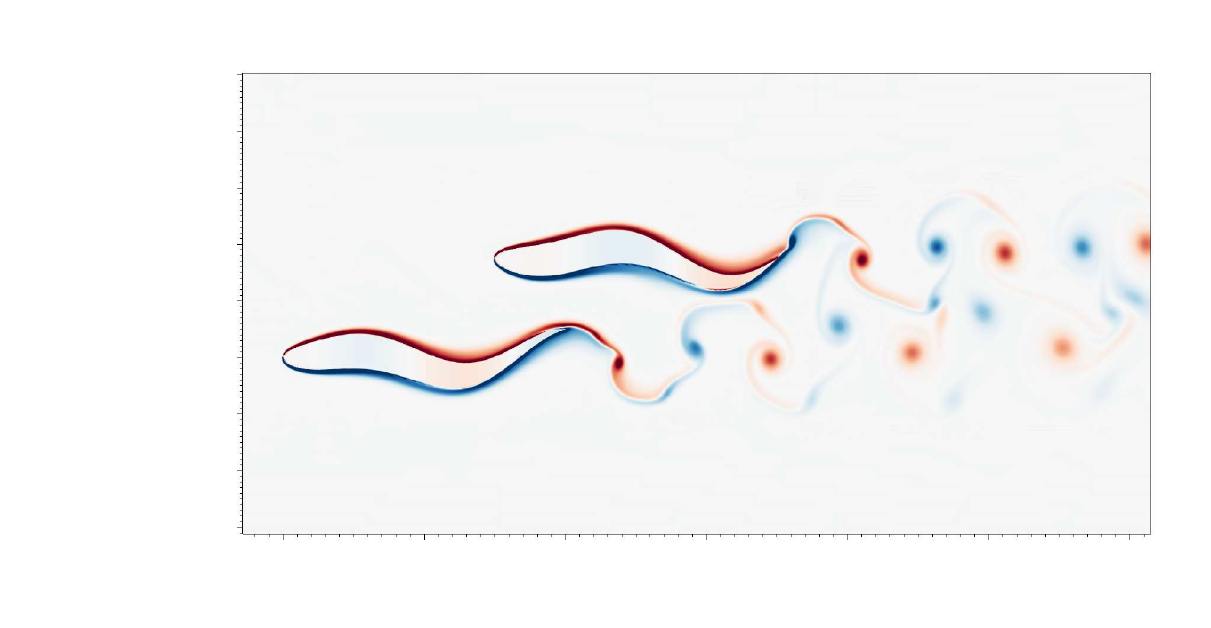}{(\counteraa) $ \lam = 0.8, \ppi = 1.5 $}
		\addlabelvor{0.32}{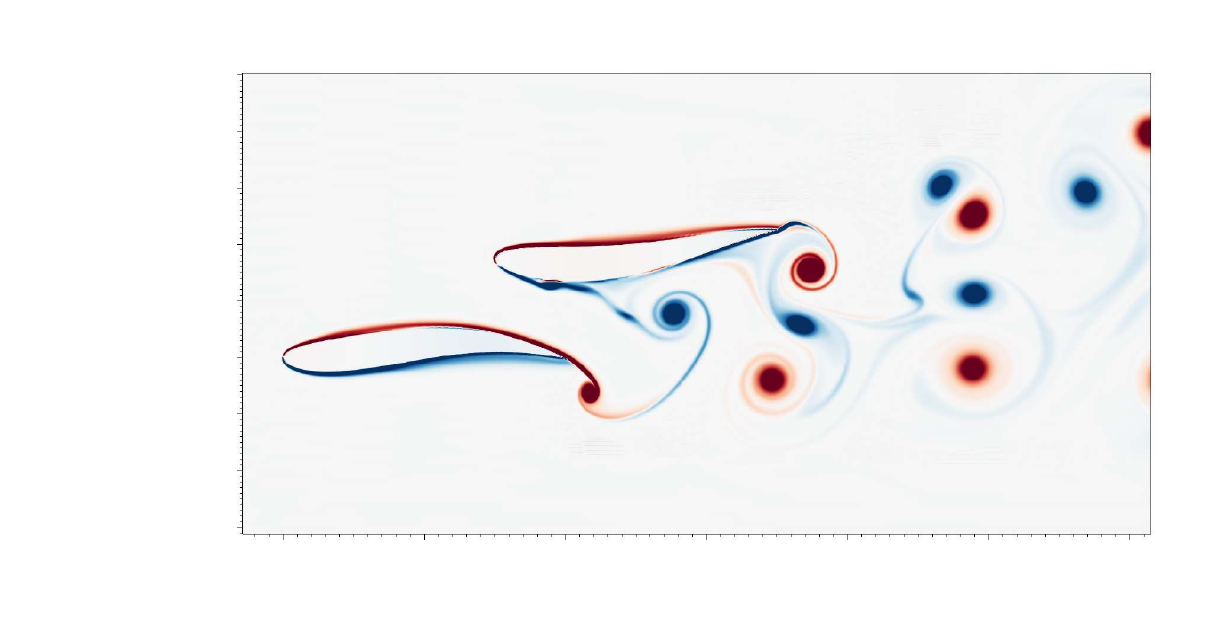}{(\counteraa) $ \lam = 2.0, \ppi = 1.5 $}
		\addlabelvor{0.32}{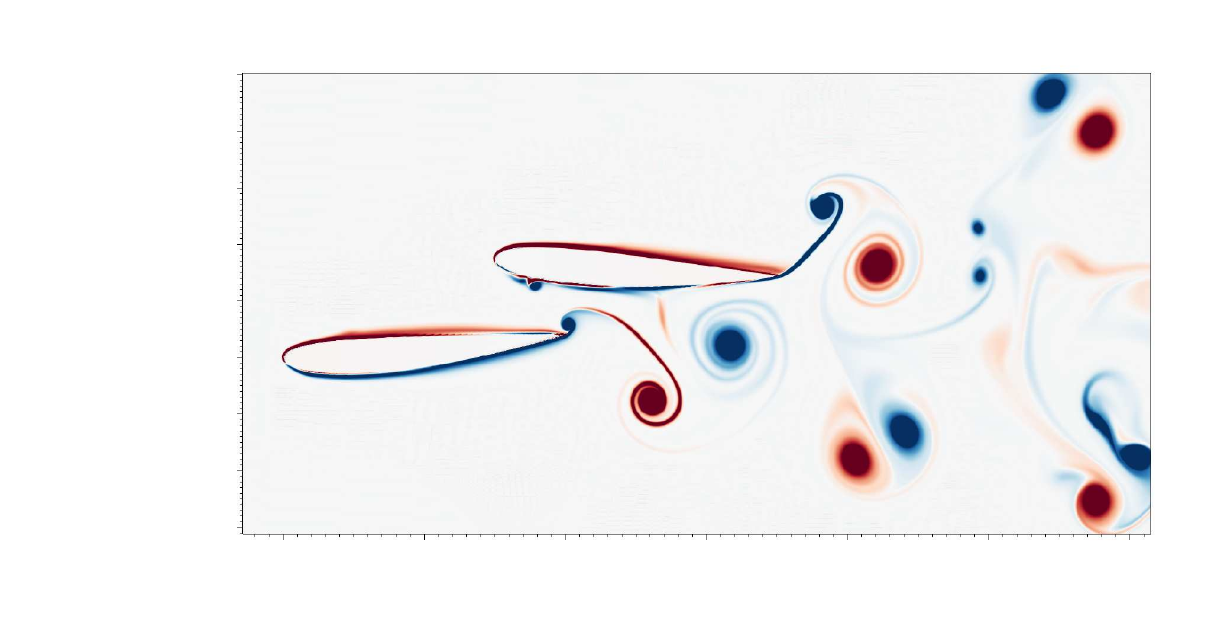}{(\counteraa) $ \lam = 8.0, \ppi = 1.5 $}
		
		\setcounter{testaa}{0}
		\caption{
			Vorticity contours and hydrofoil deformation at instant $ t^*/T = 5 $ with fixed relative distances $ G = 0.35 $, $ D = 0.75 $ and a variety of phase difference (a-c) $ \ppi = 0 $ (d-f) $ \ppi = 0.5 $ (g-i) $ \ppi = 1.0 $ (j-l) $ \ppi = 1.5 $ and various wavelengths (a \& d \& g \& j) $ \lam = 0.8 $ (b \& e \& h \& k) $ \lam = 2.0 $ (c \& f \& i \& l) $ \lam = 6.8 $.
		}
		\label{fig:vor_lam_phi}
	\end{figure}

	In order to examine the inter-relationship between wavelength $ \lam $, front-back distance $ D $, and wake flow structure, we draw the vorticity contours across various wavelengths $ \lam $ and front-back distances $ D $, as seen in \Cref{fig:D0.25_0.75_wake_mixing}.
	At front-back distance $ D = 0.25 $, as seen in \Cref{fig:D0.25_0.75_wake_mixing}a-\ref{fig:D0.25_0.75_wake_mixing}c, the vortex dipoles shed by each hydrofoil do not mix in the wake flow, but bifurcating towards two distinct directions, forming an angle with a near-perfect mirror symmetry about the centreline. This symmetrically stable flow structure persists despite the variation of wavelengths $ \lam $. Such near-symmetrical patterns were only observed in cases with \textit{phalanx} arrangement $ D = 0 $ and \textit{anti-phase} $ \ppi = 1 $ condition, \eg cases discussed in \Cref{subsec:vor_Dis0} and results from another paper focusing on wake symmetry by \cite{Gungor2020}. It is therefore interesting to observe a very similar pattern at a \textit{staggered} placement $ D = 0.25 $ with \textit{in-phase} $ \ppi = 0 $ undulation.
	To explain this phenomenon, we further examine the underlying hydrodynamic mechanism. The front-back distance of $ D = 0.25 $ causes positive vortex from the present half-cycle of the follower to collide with the negative vortex from the previous half-cycle of the leader. These two vortices collide with each other but cannot merge together due to their opposite rotating direction, thus pushing each other away while drifting downstream, eventually leading to a steady flow structure with a certain angle. This periodic flow pattern does not lead to an outstanding thrust force or locomotion efficiency as previously discussed, yet it may contain implications for stealth capacity of the swimmers.
	As for cases at $ D \geq 0.50 $, the wake flow structure is more irregular due to unsteady interaction among dipoles, \eg some vortices merge together to form a larger one. The overall vorticity strength and the degree of irregularity both increase with the wavelengths $ \lam $. Although the wake flow is irregular for the cases at $ D \geq 0.50 $, the flow structure near the two foils is largely predictable, especially in the area between the two foils.
	\renewcommand{\addlabelvor}[3]{%
		\begin{tikzpicture}
			\node[anchor=south west,inner sep=0] (image) at (0,0) 
			{\includegraphics[width=#1\linewidth, trim={4.05cm 1cm 1cm 1.25cm},clip]{#2}};%
			\begin{scope}[x={(image.south east)},y={(image.north west)}]
				\node[anchor=south west] at (0.00,0.75) {\footnotesize #3};	%
			\end{scope}
		\end{tikzpicture}%
	}
	\begin{figure}
		\centering
		\setcounter{testaa}{0}
		\addlabelvor{0.32}{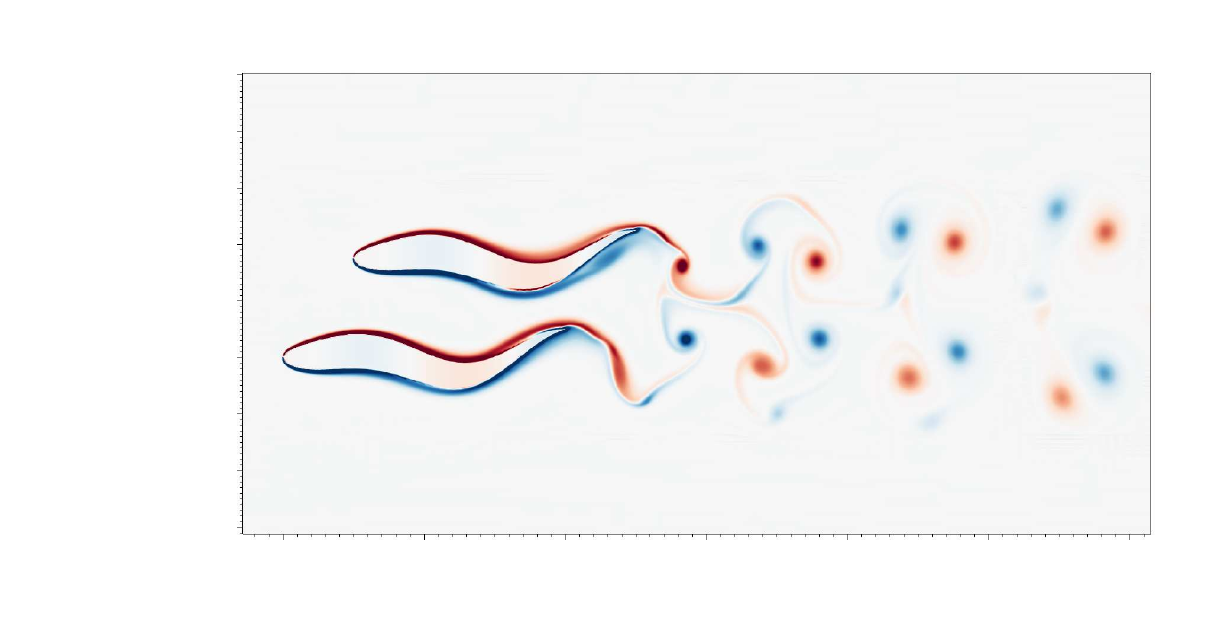}{(\counteraa) $ \lam = 0.8, D = 0.25 $}
		\addlabelvor{0.32}{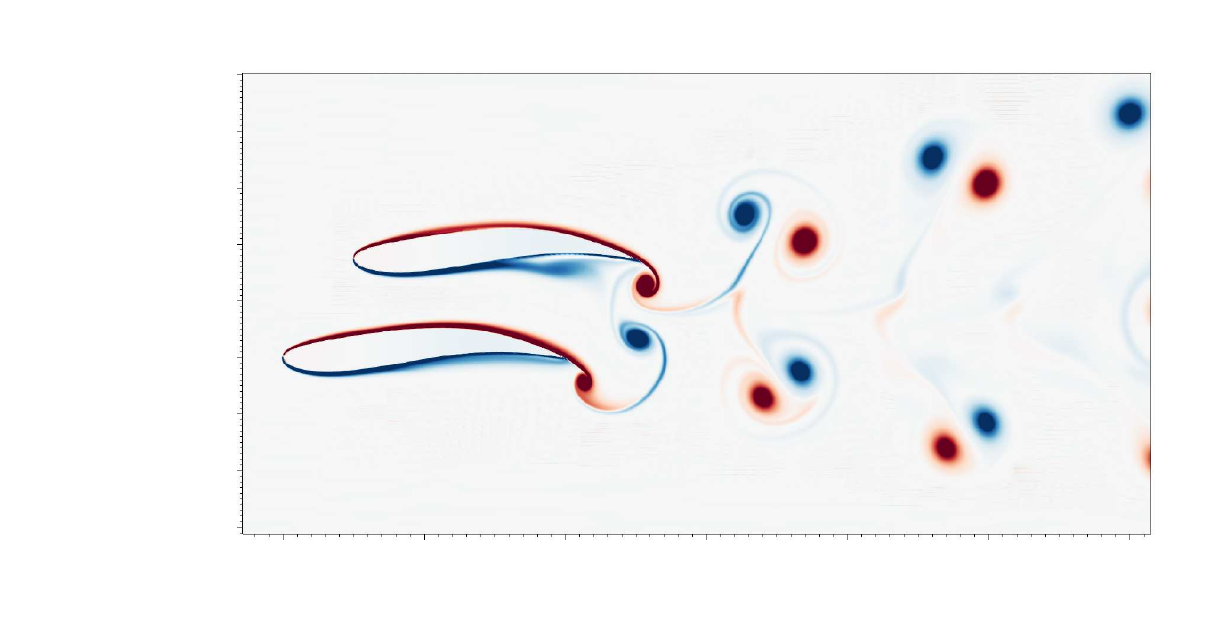}{(\counteraa) $ \lam = 2.0, D = 0.25 $ }
		\addlabelvor{0.32}{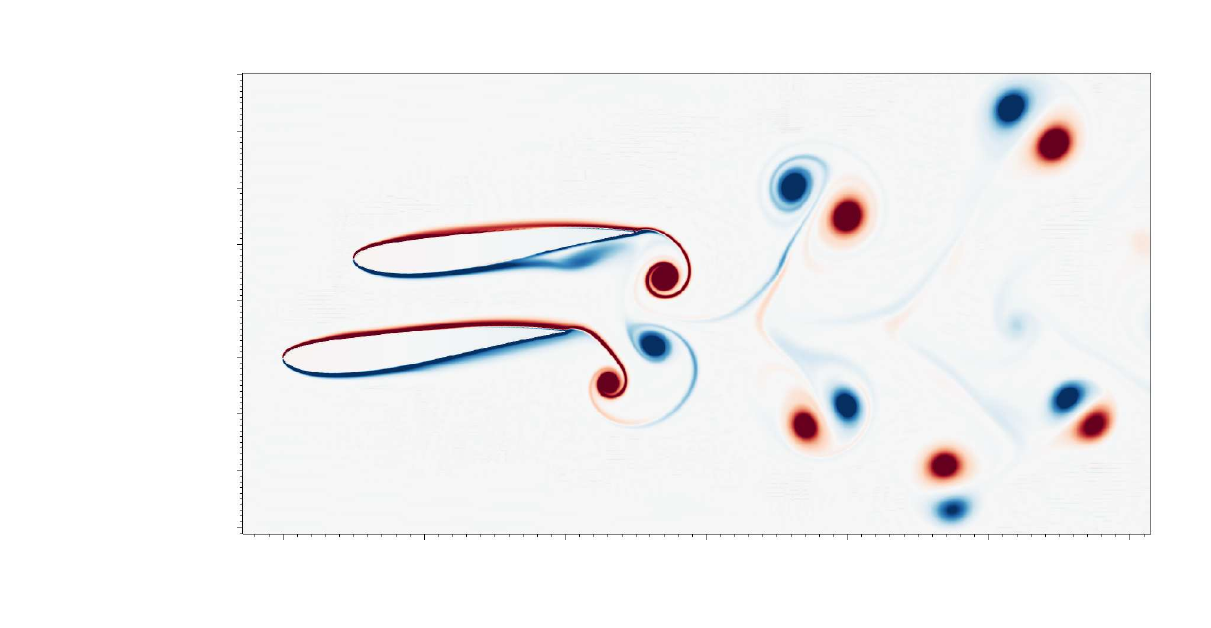}{(\counteraa) $ \lam = 3.2, D = 0.25 $}
		
		\addlabelvor{0.32}{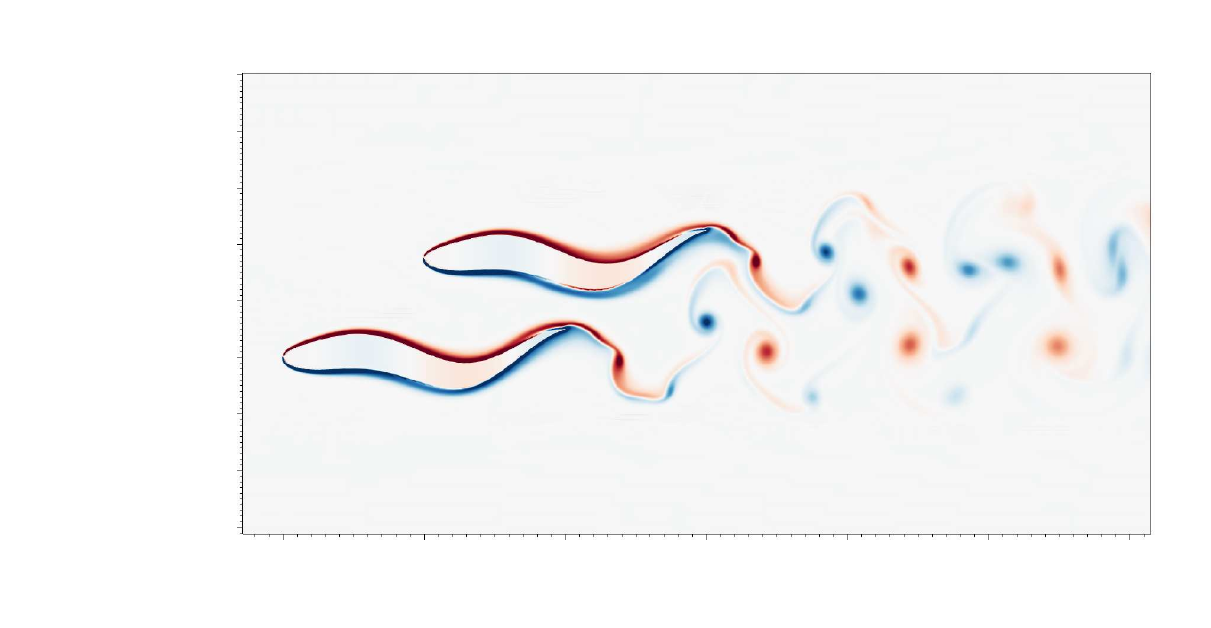}{(\counteraa) $ \lam = 0.8, D = 0.50 $}
		\addlabelvor{0.32}{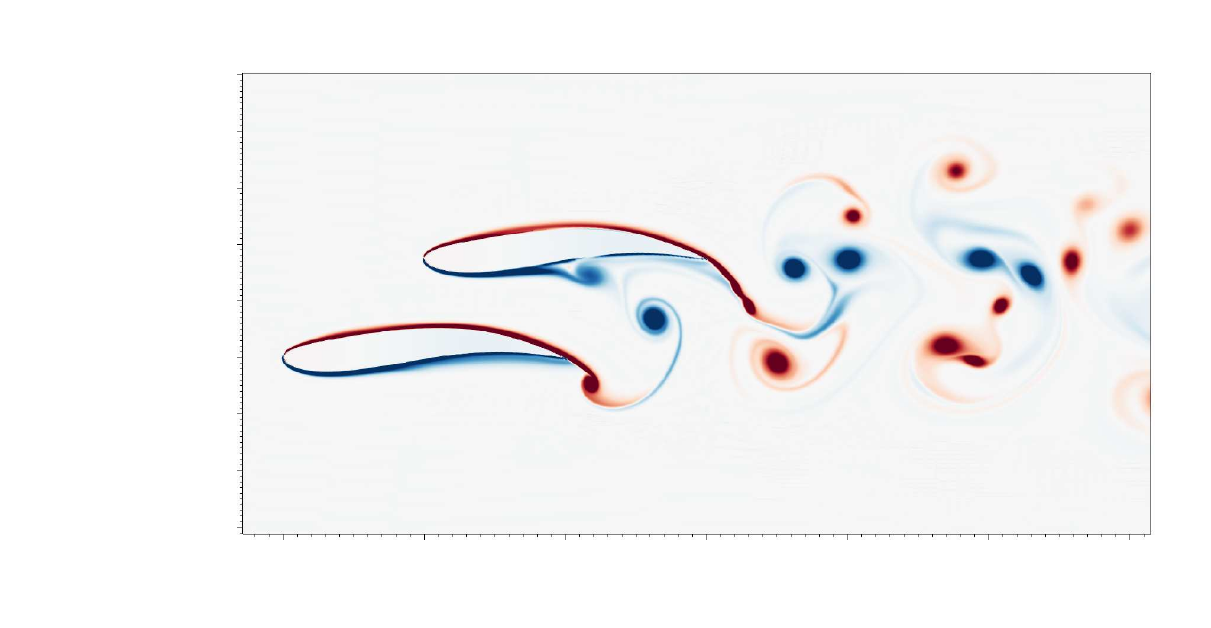}{(\counteraa) $ \lam = 2.0, D = 0.50 $}
		\addlabelvor{0.32}{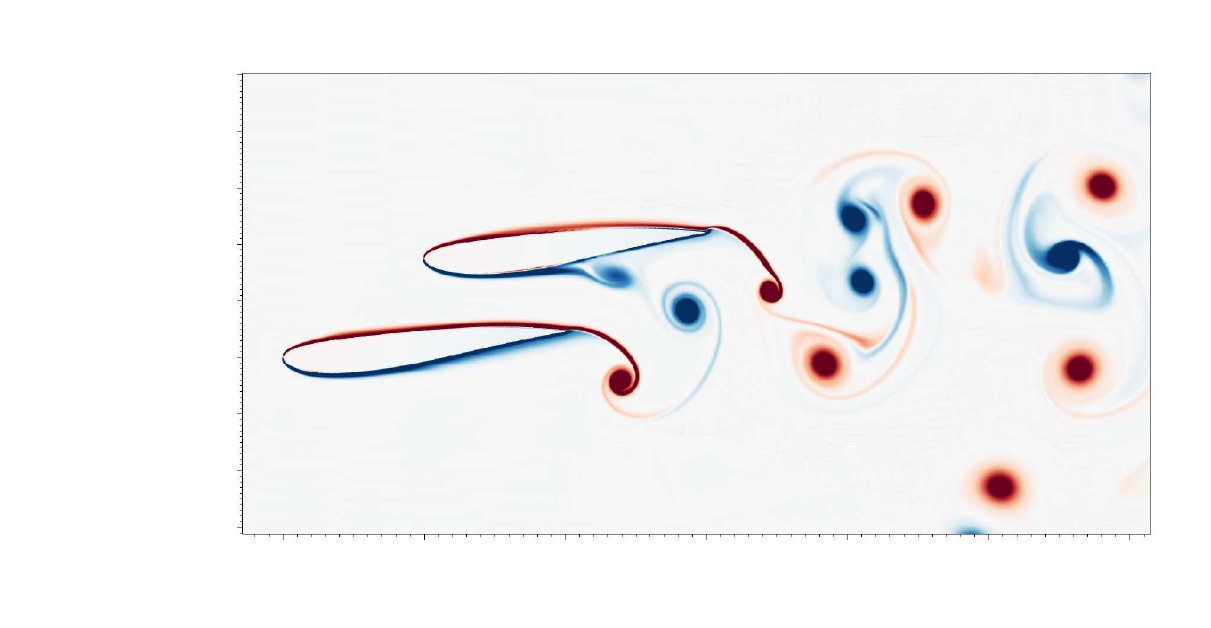}{(\counteraa) $ \lam = 3.2, D = 0.50 $}
		
		\addlabelvor{0.32}{Figure/ps_format/g0.35d0.75p0.00e5000.0s0.4r139_Lam00.800_Num_0000}{(\counteraa) $ \lam = 0.8, D = 0.75 $}
		\addlabelvor{0.32}{Figure/ps_format/g0.35d0.75p0.00e5000.0s0.4r139_Lam02.000_Num_0000}{(\counteraa) $ \lam = 2.0, D = 0.75 $}
		\addlabelvor{0.32}{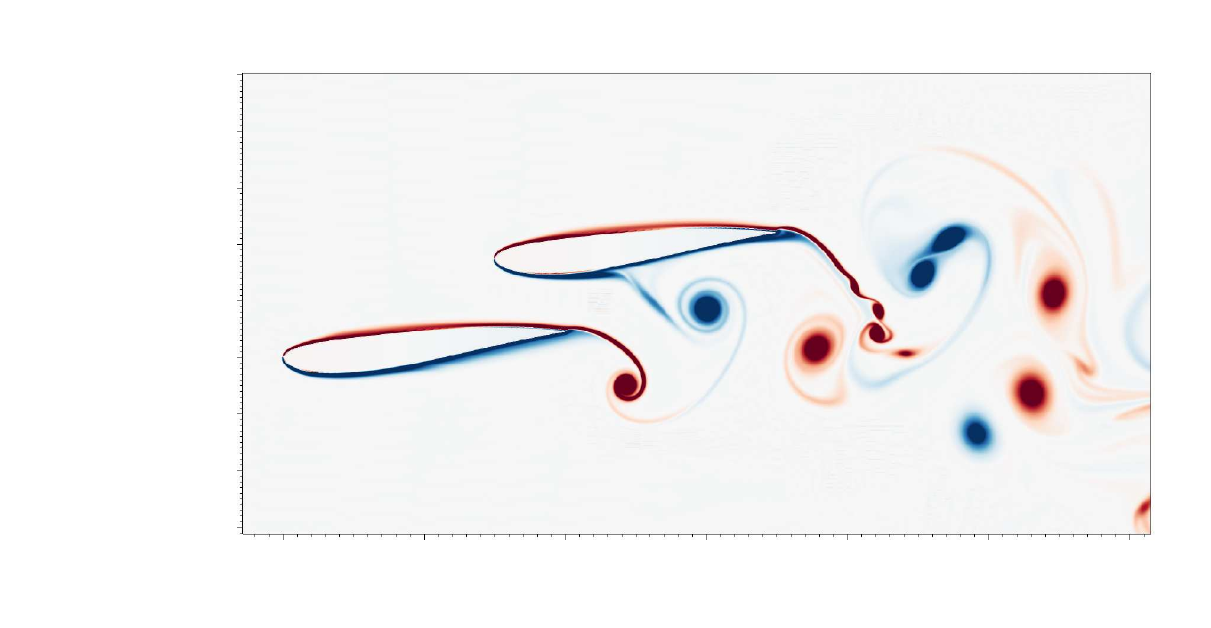}{(\counteraa) $ \lam = 3.2, D = 0.75 $}
		
		\setcounter{testaa}{0}
		\caption{
			Vorticity contours and hydrofoil deformation at instant $ t^*/T = 5 $ with $ G = 0.35 $, $ \phi = 0 $, (a-c) $ D = 0.25 $ (d-f) $ D = 0.50 $ (g-i) $ D = 0.75 $ (a \& d \& g) $ \lam = 0.8 $ (b \& e \& h) $ \lam = 2.0 $ (c \& f \& i) $ \lam = 3.2 $. 
		}
		\label{fig:D0.25_0.75_wake_mixing}
	\end{figure}

	\subsubsection{Flow-mediated interaction between two swimmers at $ D > 0 $}
	\label{subsec:FSI_Dg0}

	\renewcommand{\addlabelvor}[3]{%
		\begin{tikzpicture}
			\node[anchor=south west,inner sep=0] (image) at (0,0) 
			{\includegraphics[width=#1\linewidth, trim={4.05cm 1cm 1cm 1.25cm},clip]{#2}};%
			\begin{scope}[x={(image.south east)},y={(image.north west)}]
				\node[anchor=south west] at (0.00,0.75) {\footnotesize #3};	%
			\end{scope}
		\end{tikzpicture}%
	}
	\renewcommand{\addlabelnotrim}[3]{%
		\begin{tikzpicture}
			\node[anchor=south west,inner sep=0] (image) at (0,0) 
			{\includegraphics[width=#1\linewidth, trim={0cm 0cm 0cm 0cm},clip]{#2}};%
			\begin{scope}[x={(image.south east)},y={(image.north west)}]
				\node[anchor=south west] at (0.1,-0.03) {\footnotesize #3};	%
			\end{scope}
		\end{tikzpicture}%
	}
	\begin{figure}
		\centering
		\setcounter{testaa}{0}
		\addlabelvor{0.49}{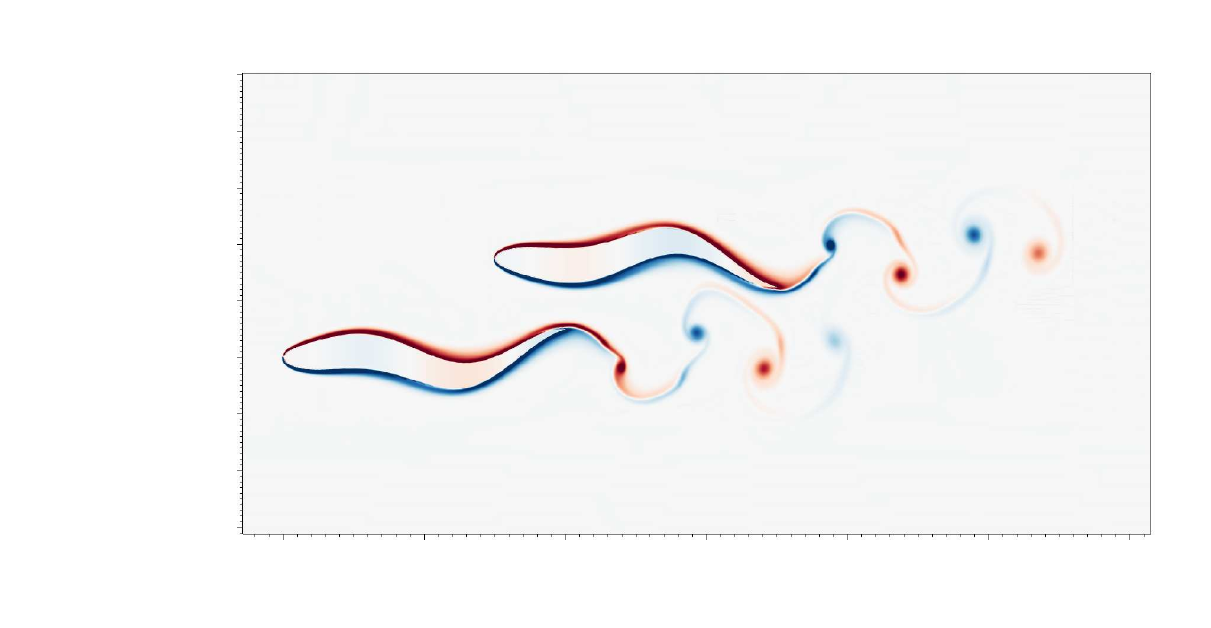}{(a) $ \lam = 0.8 $, $ t^*/T = 2.00 $}
		\addlabelvor{0.49}{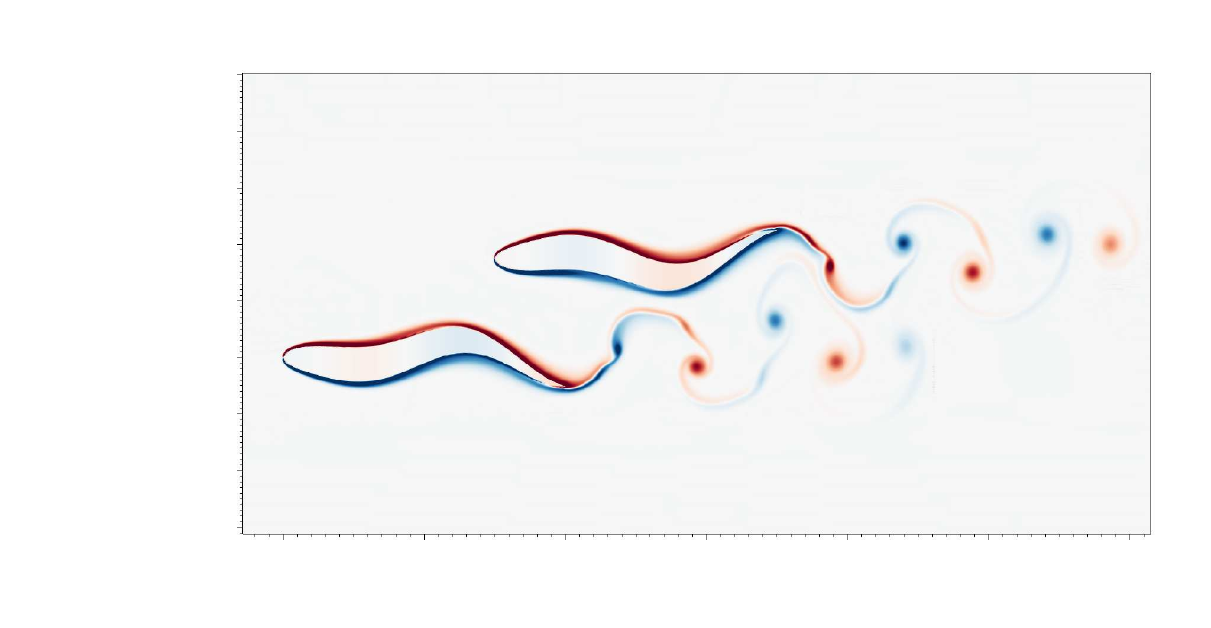}{(e) $ \lam = 0.8 $, $ t^*/T = 2.50 $}
		
		\addlabelvor{0.49}{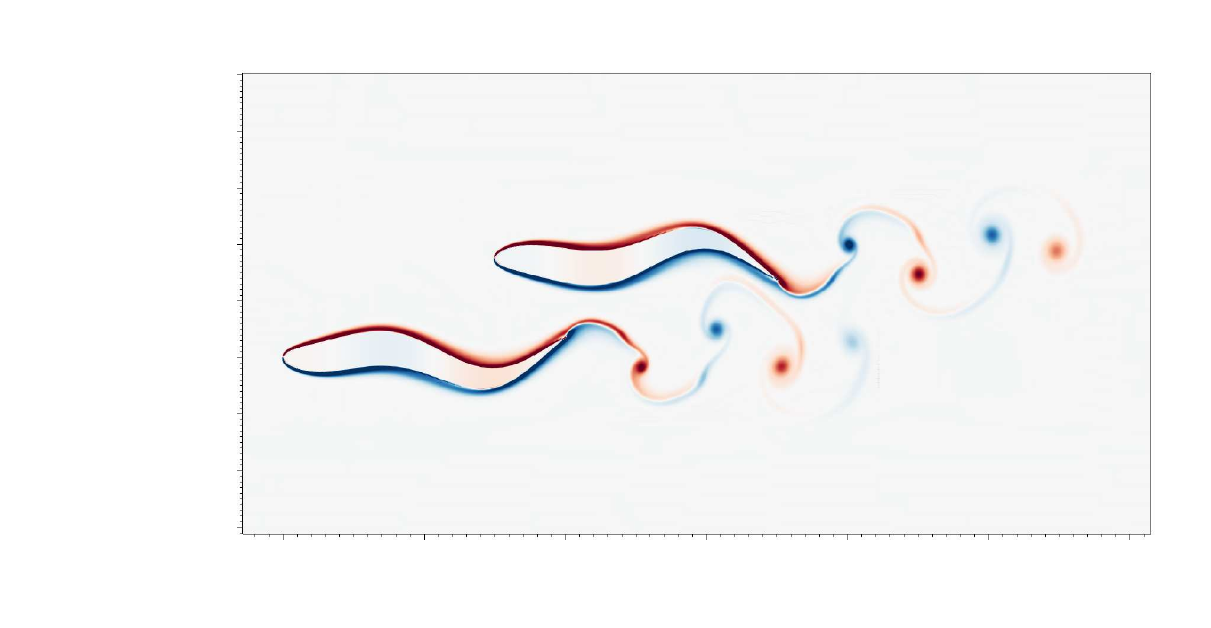}{(b) $ \lam = 0.8 $, $ t^*/T = 2.125 $}
		\addlabelvor{0.49}{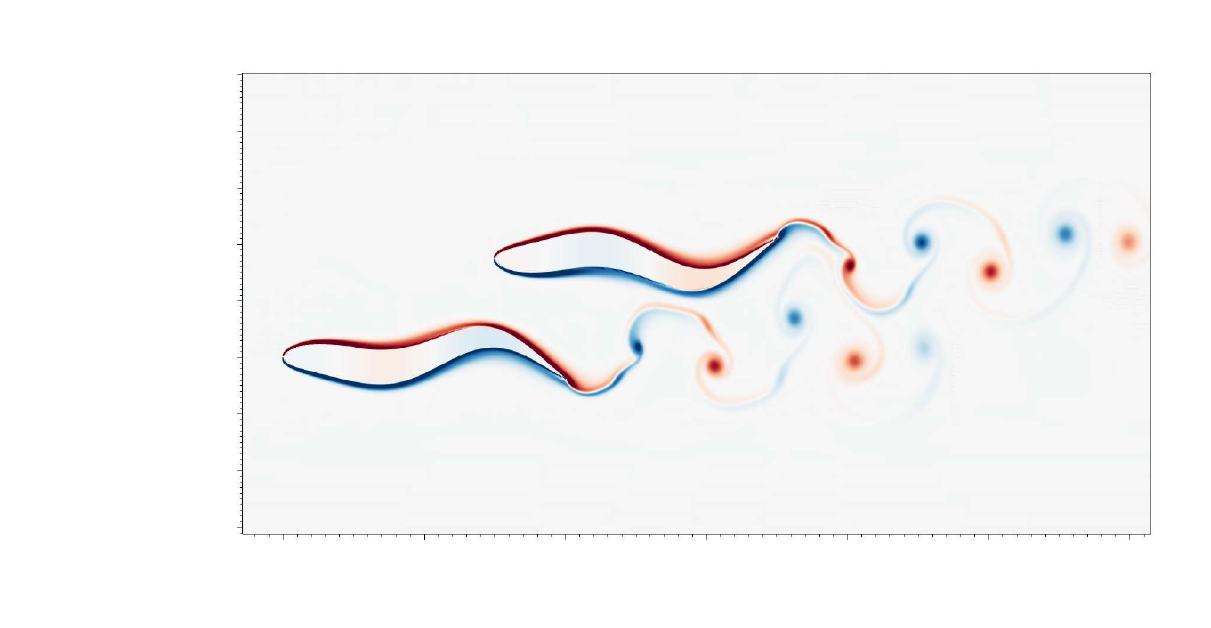}{(f) $ \lam = 0.8 $, $ t^*/T = 2.625 $}
		
		\addlabelvor{0.49}{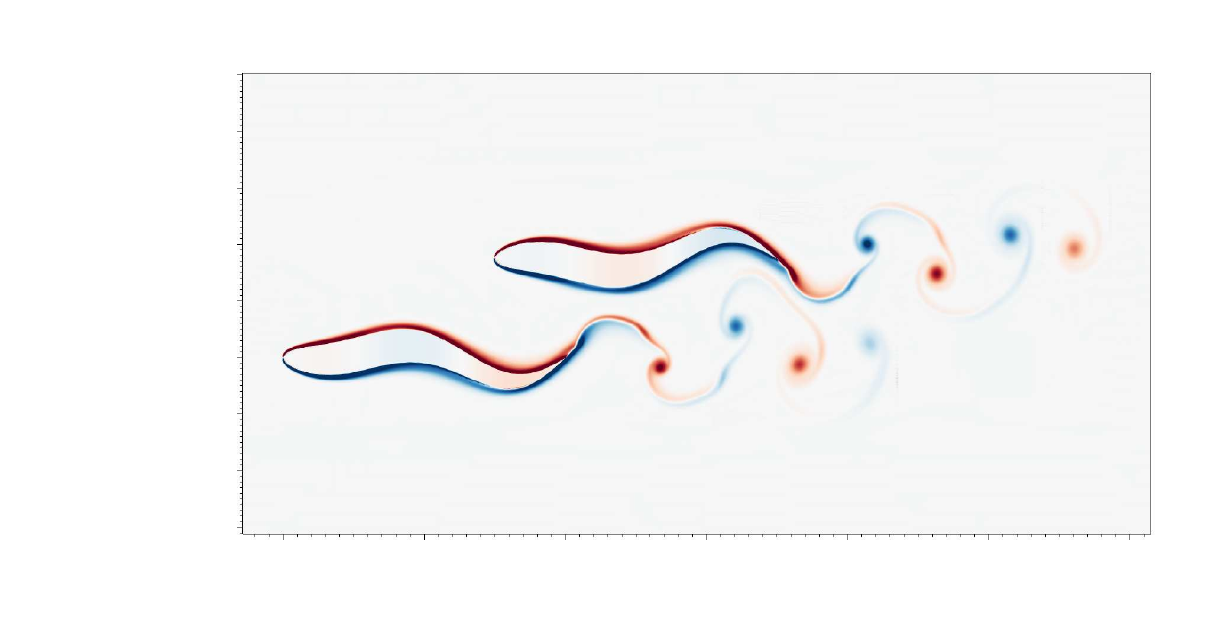}{(c) $ \lam = 0.8 $, $ t^*/T = 2.25 $}
		\addlabelvor{0.49}{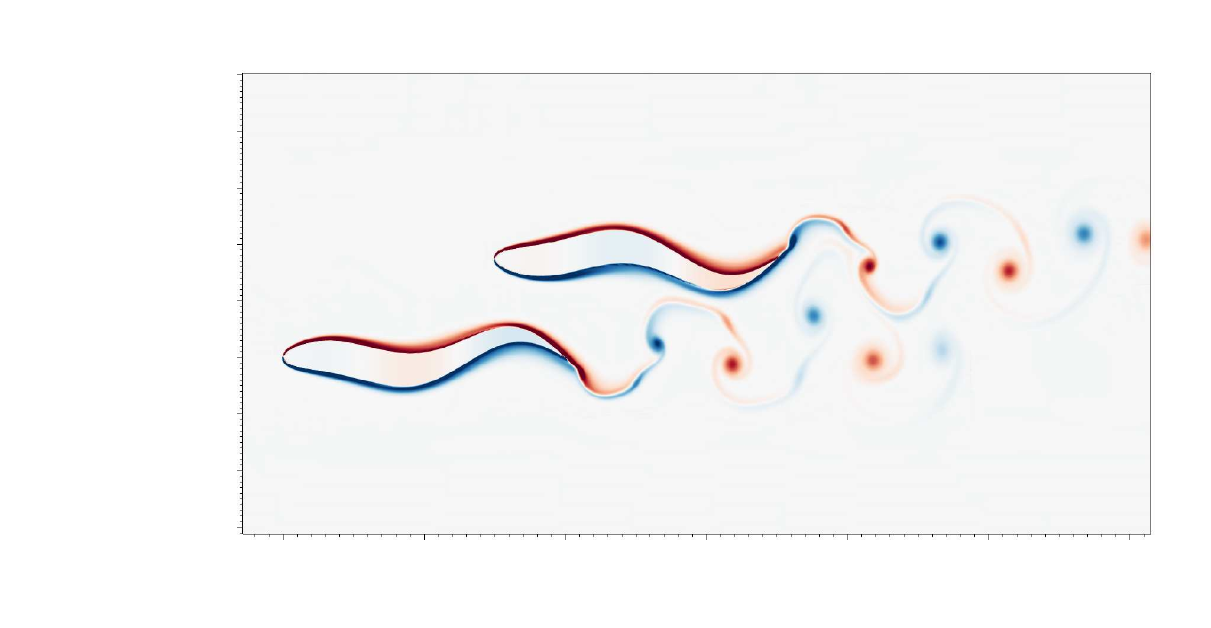}{(g) $ \lam = 0.8 $, $ t^*/T = 2.75 $}
		
		\addlabelvor{0.49}{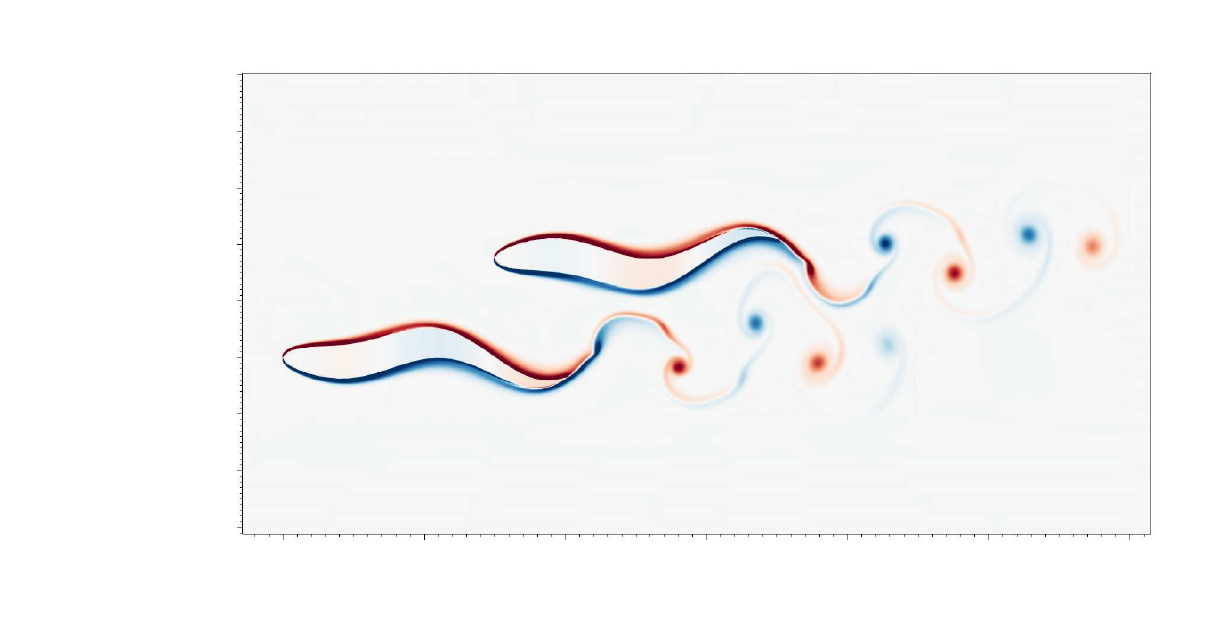}{(d) $ \lam = 0.8 $, $ t^*/T = 2.375 $}
		\addlabelvor{0.49}{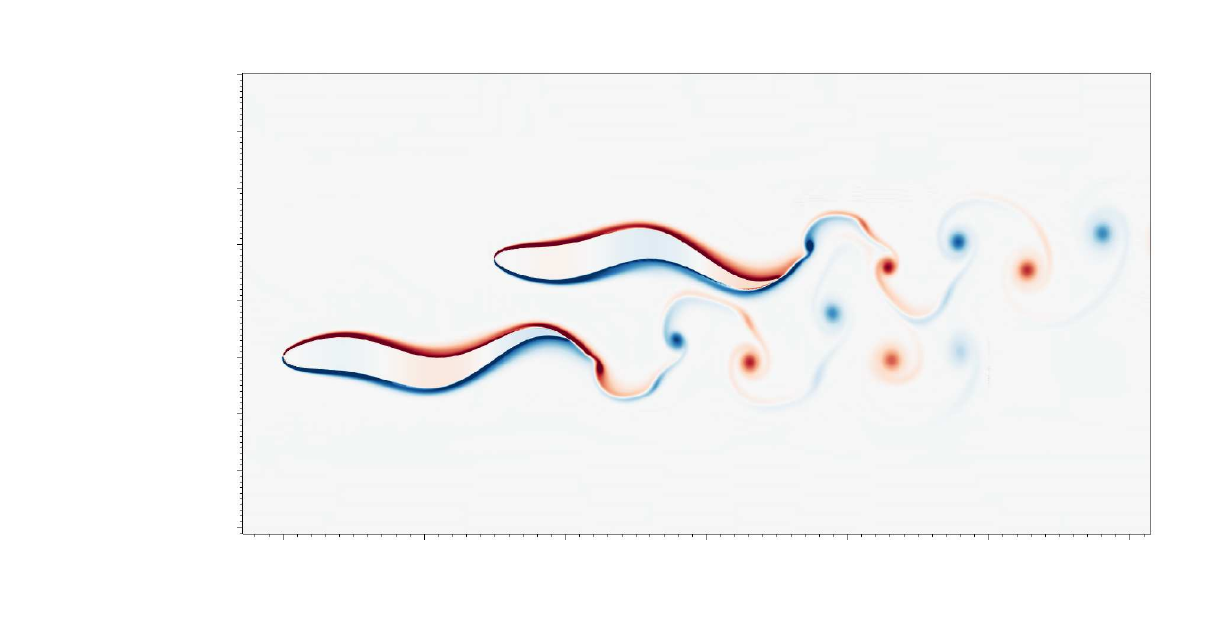}{(h) $ \lam = 0.8 $, $ t^*/T = 2.875 $}
		
		\addlabelnotrim{0.49}{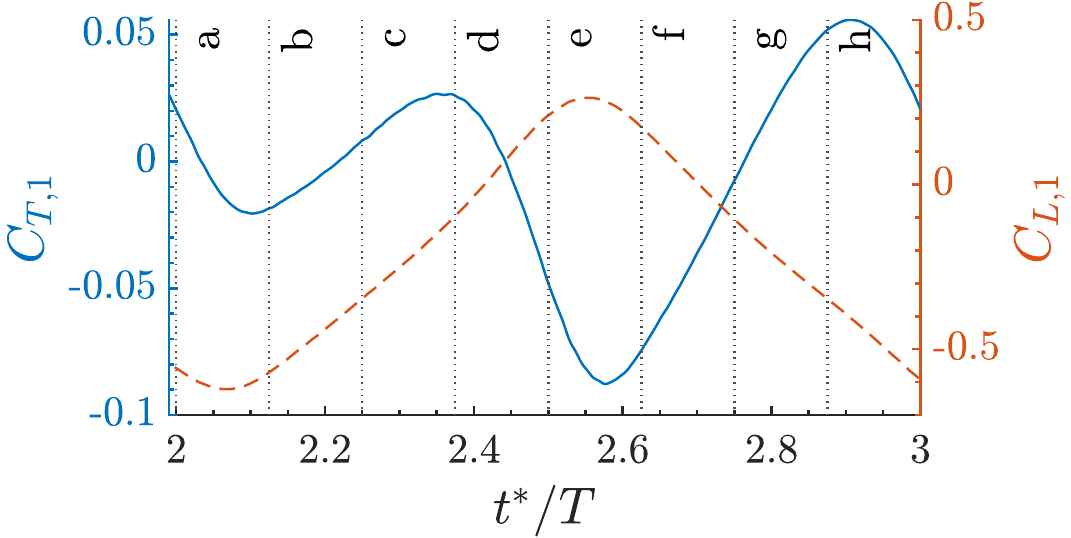}{(i) Leader}
		\addlabelnotrim{0.49}{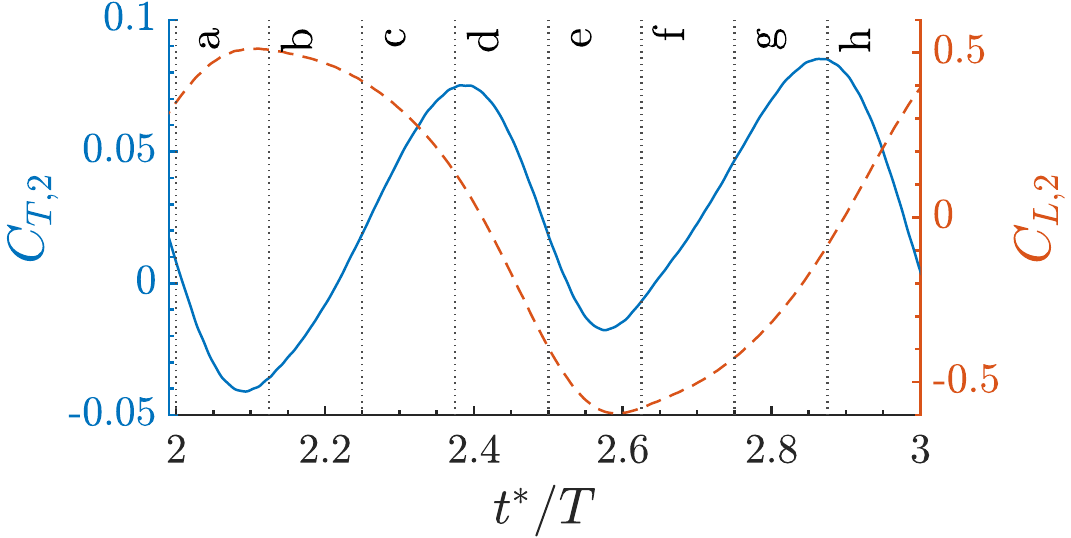}{(j) Follower}
		
		\caption{Vorticity contours and hydrofoil deformation with wavelength $ \lam = 0.8 $, side-by-side arrangement $ D = 0.75 $, lateral gap $ G = 0.35 $, anti-phase $ \ppi = 1.0 $ at instants of a typical period
			(a-h) $ t^*/T = 2.00-2.875 $. Time histories of thrust and lift coefficient for the (i) Leader and (j) Follower swimmers.}
		\label{fig:vor_1T_D0d75_lam0d8}
	\end{figure}
	
	At $ D =0.75,\ \lam = 0.8 $, the wake vorticity pattern looks as if each of the two foils swims as a single foil, \ie interference between the wake flows by two foils are visibly insignificant, as viewed in \Cref{fig:vor_1T_D0d75_lam0d8}a-\ref{fig:vor_1T_D0d75_lam0d8}h.
	However, large discrepancy of thrust force $ C_T $ is discovered between the two swimmers; the leader's thrust force is generally negative, whereas the follower's is on the whole positive, as seen in \Cref{fig:vor_1T_D0d75_lam0d8}i and \ref{fig:vor_1T_D0d75_lam0d8}j, indicating that the two foils are attracted towards each other due to the flow-mediated interaction.
	The lateral force $ C_L $ of the two foils is dissimilar from each other, which is different from the almost symmetrical lateral force time history at higher wavelength $ \lam = 2,\ 8 $. In other words, the lateral force is more sensitive to the flow-mediated interaction between the two swimmers. 
	This thrust and lateral force discrepancy may be further relevant to the pressure suction mechanism \citep{Blickhan1992} that is most typical in low wavelength swimmers.

	\renewcommand{\addlabelvor}[3]{%
		\begin{tikzpicture}
			\node[anchor=south west,inner sep=0] (image) at (0,0) 
			{\includegraphics[width=#1\linewidth, trim={4.05cm 1cm 1cm 1.25cm},clip]{#2}};%
			\begin{scope}[x={(image.south east)},y={(image.north west)}]
				\node[anchor=south west] at (0.00,0.75) {\footnotesize #3};	%
			\end{scope}
		\end{tikzpicture}%
	}
	\begin{figure}
		\centering
		\setcounter{testaa}{0}
		\addlabelvor{0.49}{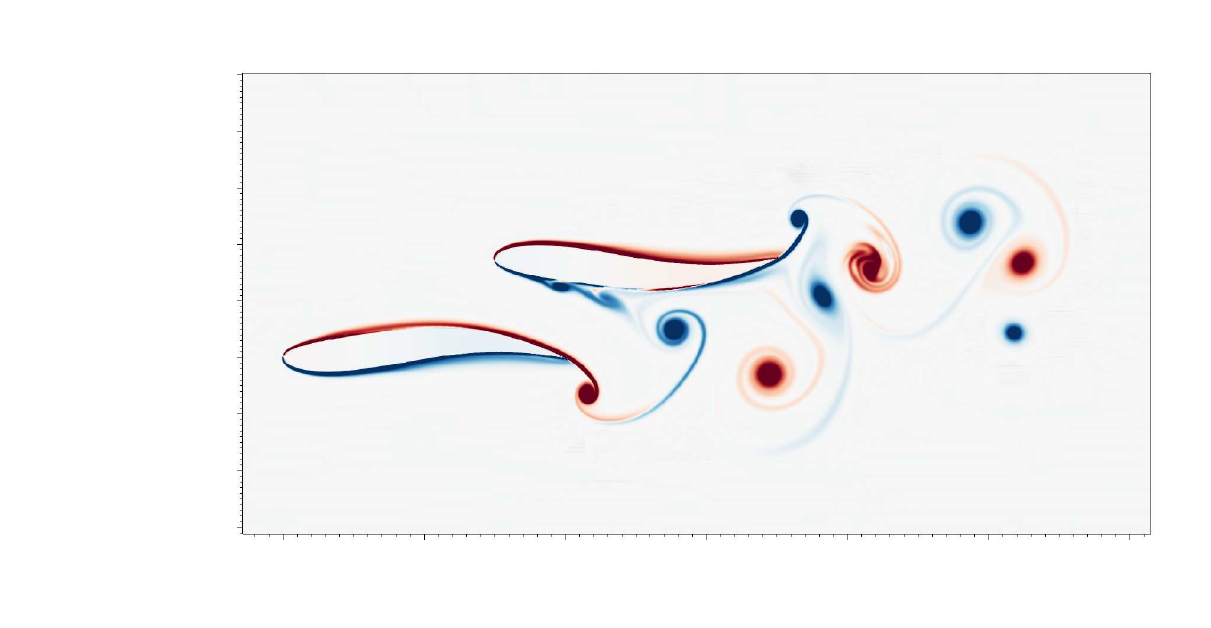}{(a) $ \lam = 2.0 $, $ t^*/T = 2.00 $}
		\addlabelvor{0.49}{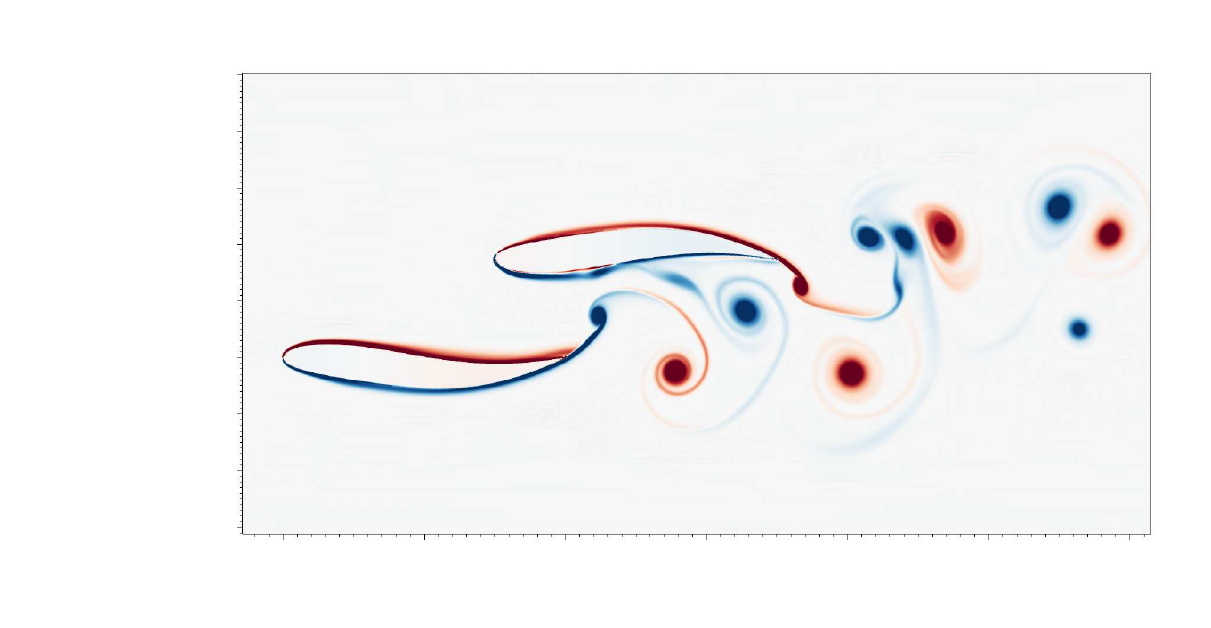}{(e) $ \lam = 2.0 $, $ t^*/T = 2.50 $}
		
		\addlabelvor{0.49}{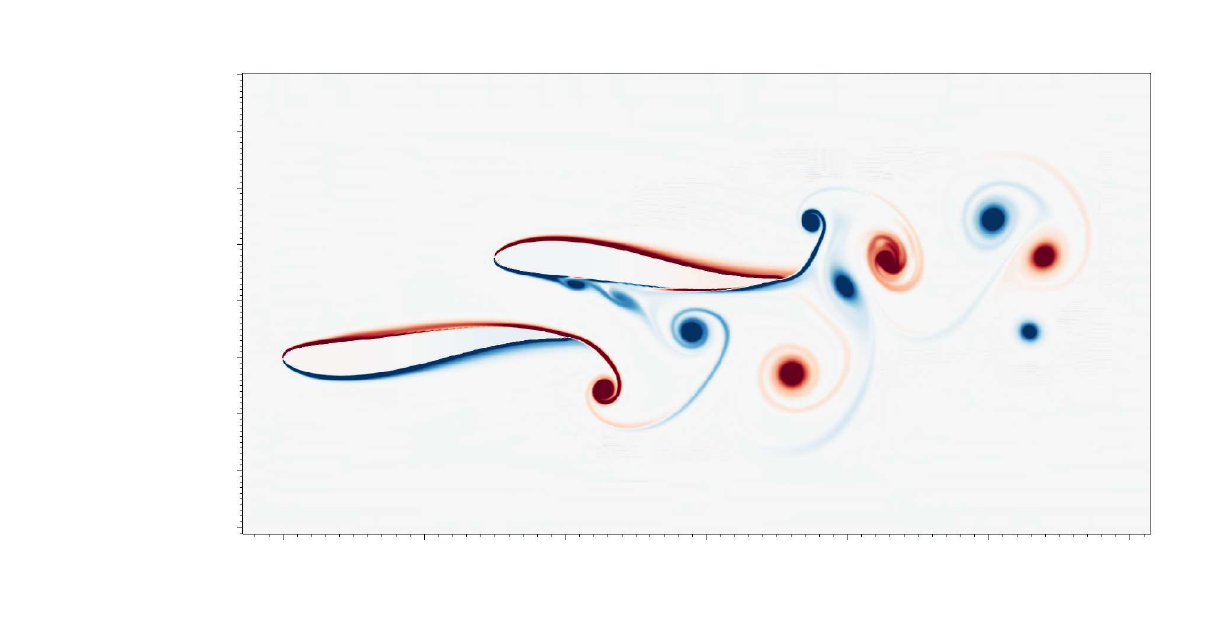}{(b) $ \lam = 2.0 $, $ t^*/T = 2.125 $}
		\addlabelvor{0.49}{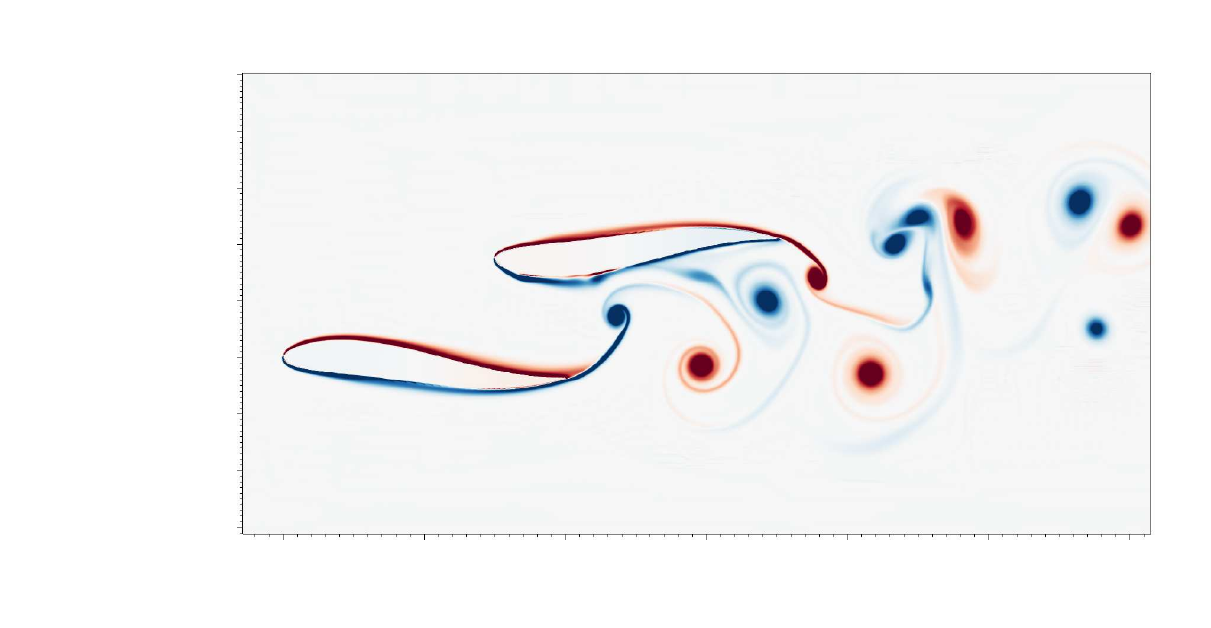}{(f) $ \lam = 2.0 $, $ t^*/T = 2.625 $}
		
		\addlabelvor{0.49}{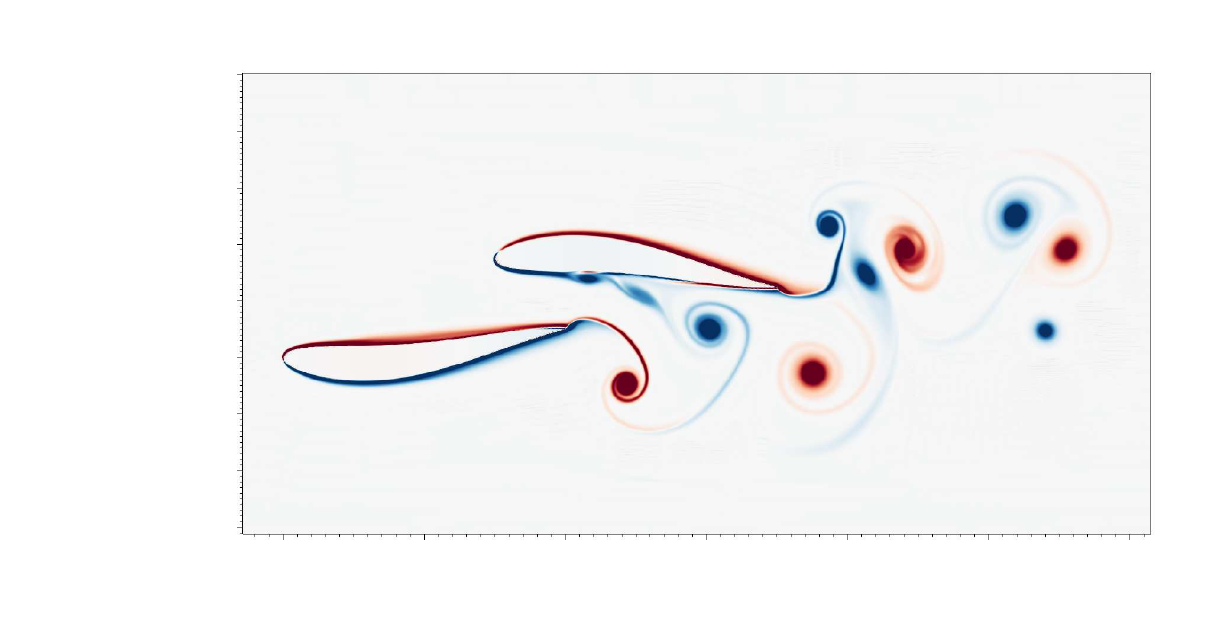}{(c) $ \lam = 2.0 $, $ t^*/T = 2.25 $}
		\addlabelvor{0.49}{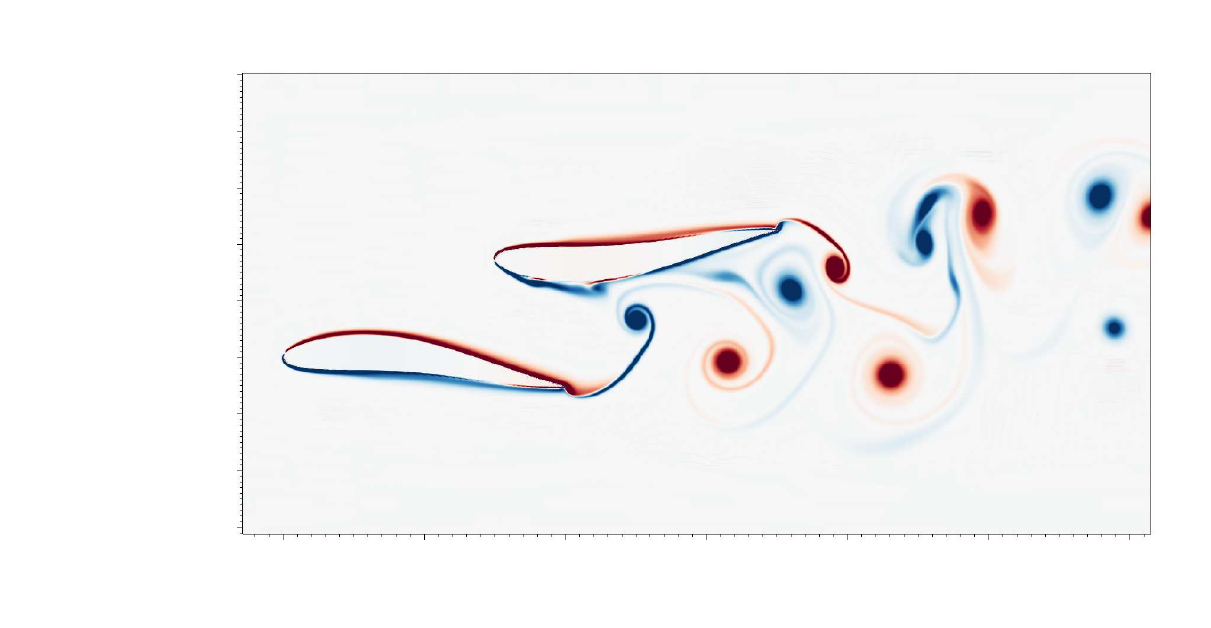}{(g) $ \lam = 2.0 $, $ t^*/T = 2.75 $}
		
		\addlabelvor{0.49}{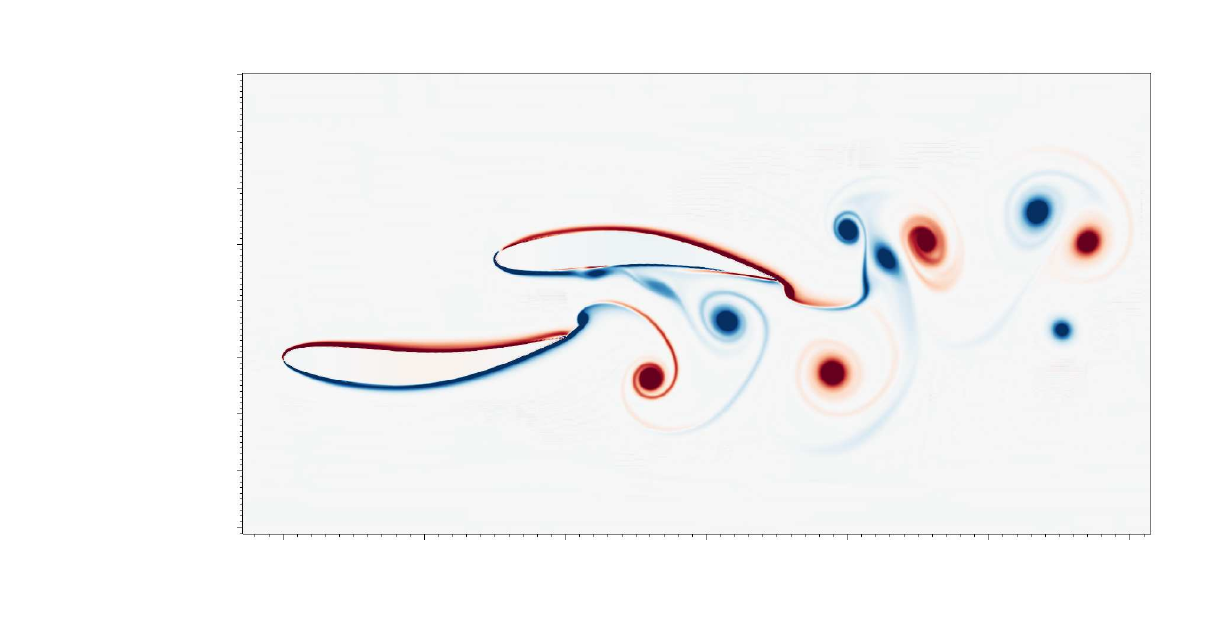}{(d) $ \lam = 2.0 $, $ t^*/T = 2.375 $}
		\addlabelvor{0.49}{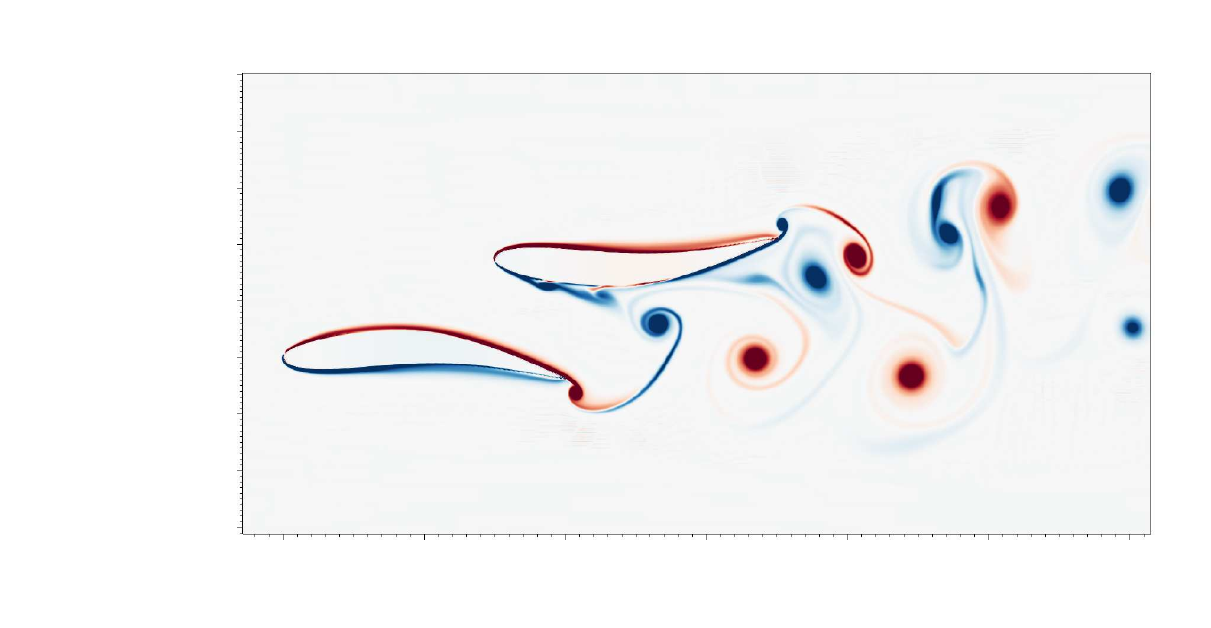}{(h) $ \lam = 2.0 $, $ t^*/T = 2.875 $}
		
		\addlabelnotrim{0.49}{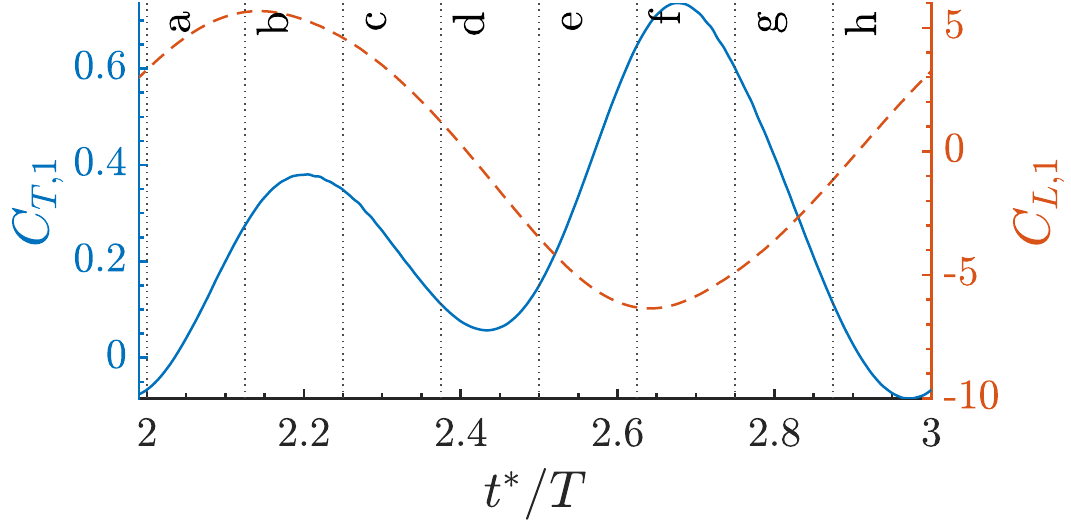}{(i) Leader}
		\addlabelnotrim{0.49}{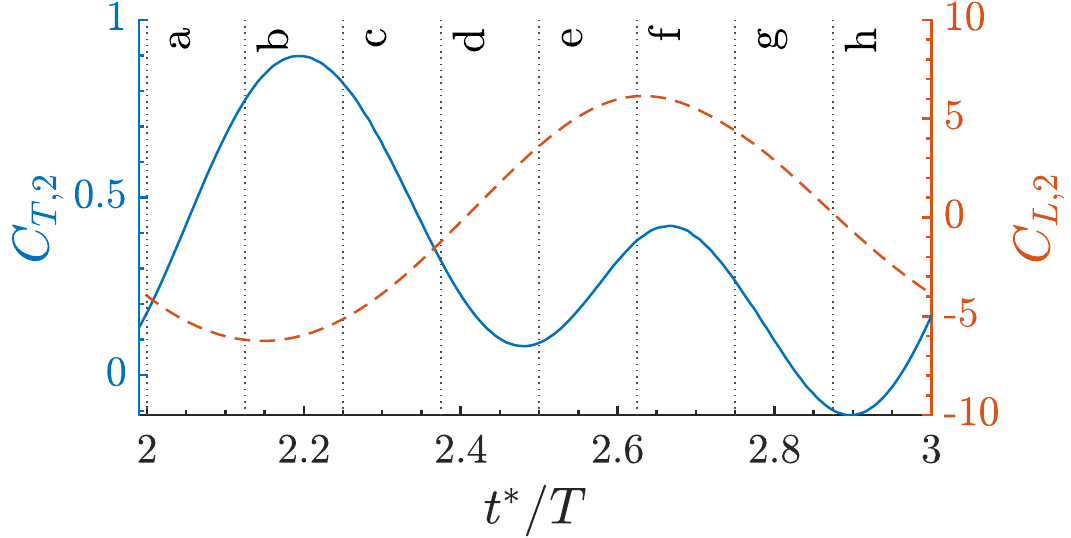}{(j) Follower}
		
		\caption{Vorticity contours and hydrofoil deformation with wavelength $ \lam = 2.0 $, side-by-side arrangement $ D = 0.75 $, lateral gap $ G = 0.35 $, anti-phase $ \ppi = 1.0 $ at instants of a typical period
			(a-h) $ t^*/T = 2.00-2.875 $. Time histories of thrust and lift coefficient for the (i) Leader and (j) Follower swimmers.}
		\label{fig:vor_1T_D0d75_lam2d0}
	\end{figure}

	At $ D =0.75,\ \lam = 2 $, the leader's upward vortices collides with the the follower's vortex street, causing great disturbance in the wake flow of the follower, as seen in \Cref{fig:vor_1T_D0d75_lam2d0}.
	"Vortex swapping" periodically takes place between the two foils, which is the swapping of positive vortices between the two swimmers: the leader's positive vortex from previous undulating cycle is entrained by the upward motion the follower's tail tip; eventually the leader's positive vortex pairs up with the follower's negative one, whereas the follower's positive vortex moves downward to pair with the leader's negative one.
	The thrust force $ C_{T} $ of the two swimmers demonstrates a phase difference of about $ T/2 $, although the lateral force amplitude upon the follower is about 40\% larger than the that upon the leader. Similar pattern is also observed in the case at $ \lam = 8 $, which will be discussed later.

	\renewcommand{\addlabelvor}[3]{%
		\begin{tikzpicture}
			\node[anchor=south west,inner sep=0] (image) at (0,0) 
			{\includegraphics[width=#1\linewidth, trim={4.05cm 1cm 1cm 1.25cm},clip]{#2}};%
			\begin{scope}[x={(image.south east)},y={(image.north west)}]
				\node[anchor=south west] at (0.00,0.75) {\footnotesize #3};	%
			\end{scope}
		\end{tikzpicture}%
	}
	\begin{figure}
		\centering
		\setcounter{testaa}{0}
		\addlabelvor{0.49}{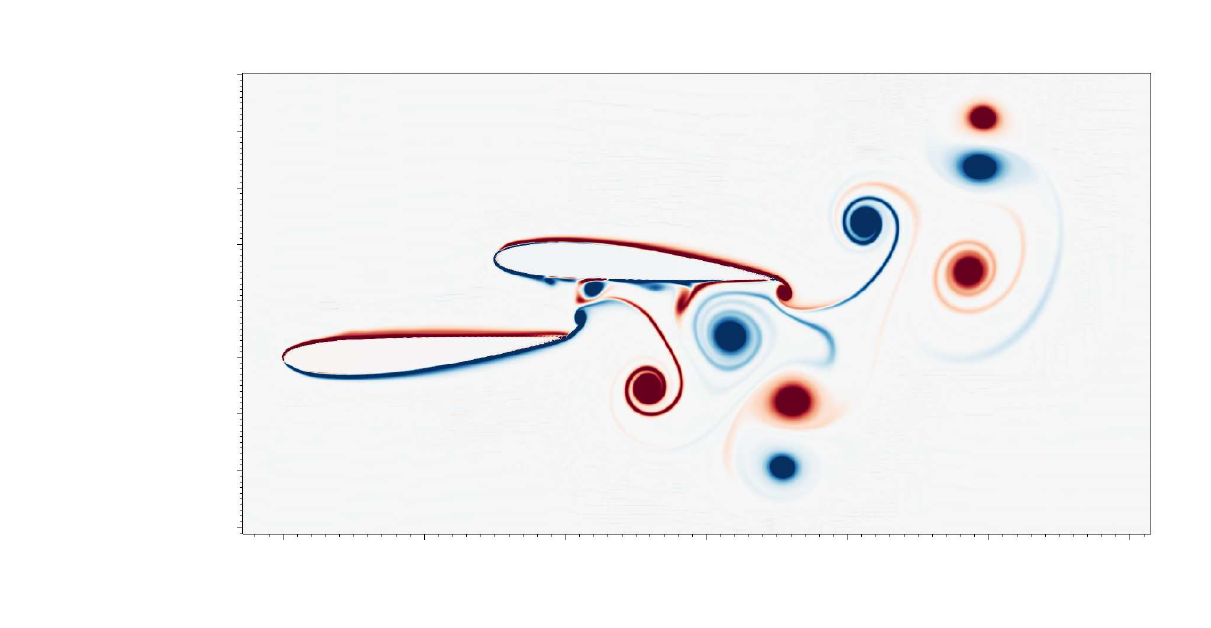}{(a) $ \lam = 8.0 $, $ t^*/T = 2.00 $}
		\addlabelvor{0.49}{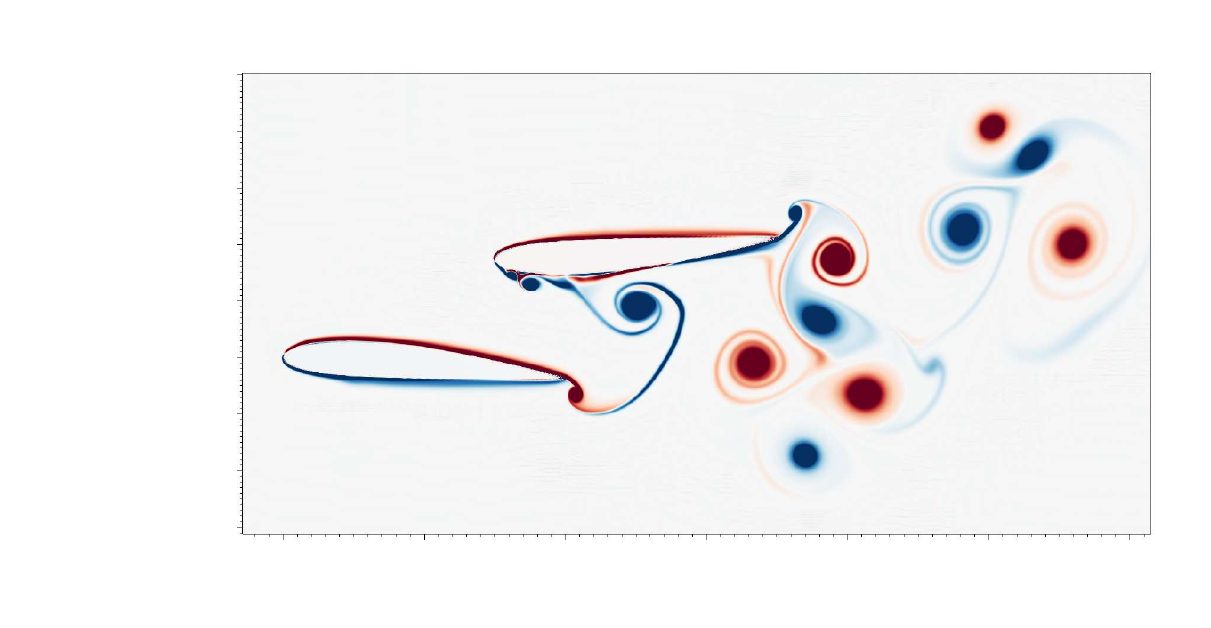}{(e) $ \lam = 8.0 $, $ t^*/T = 2.50 $}
		
		\addlabelvor{0.49}{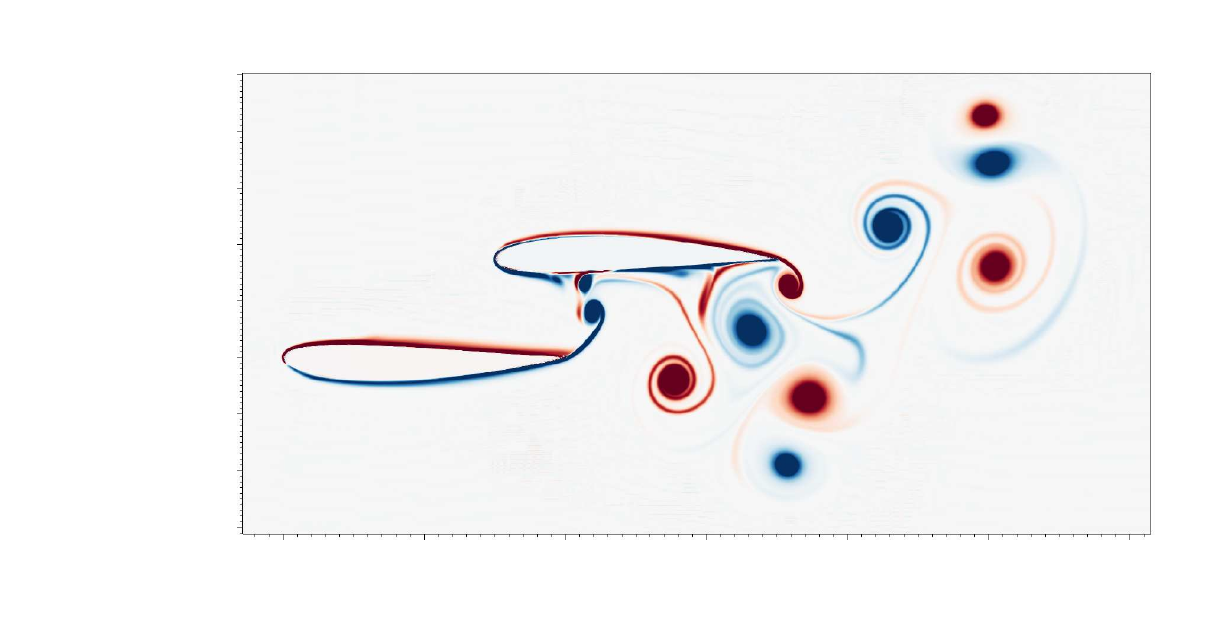}{(b) $ \lam = 8.0 $, $ t^*/T = 2.125 $}
		\addlabelvor{0.49}{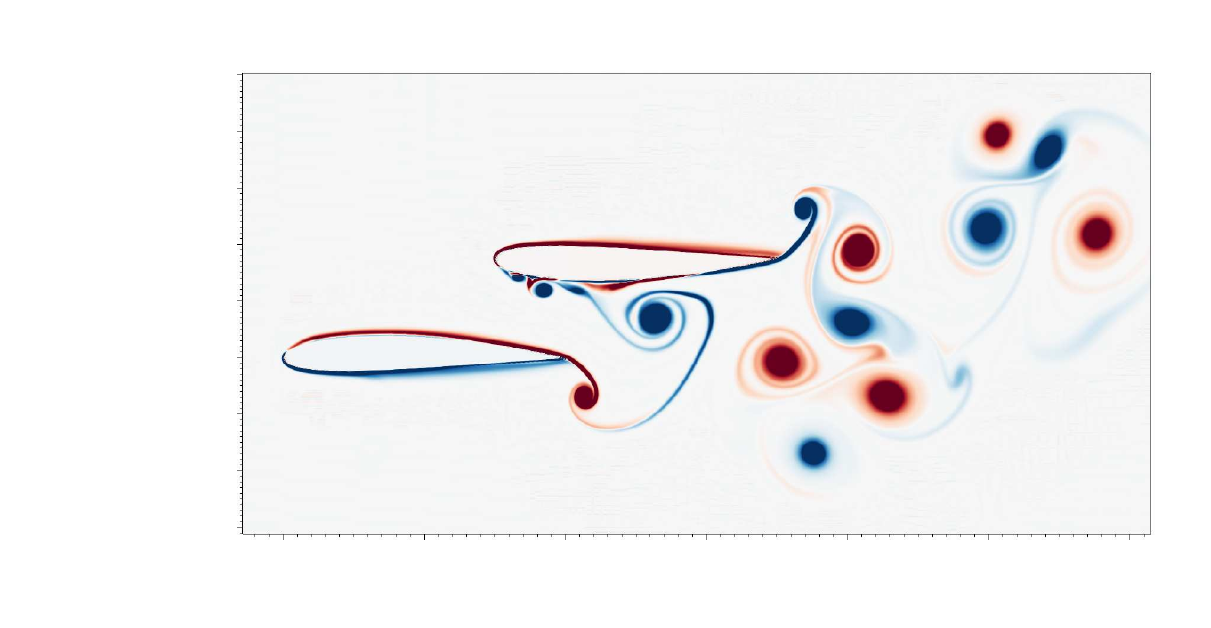}{(f) $ \lam = 8.0 $, $ t^*/T = 2.625 $}
		
		\addlabelvor{0.49}{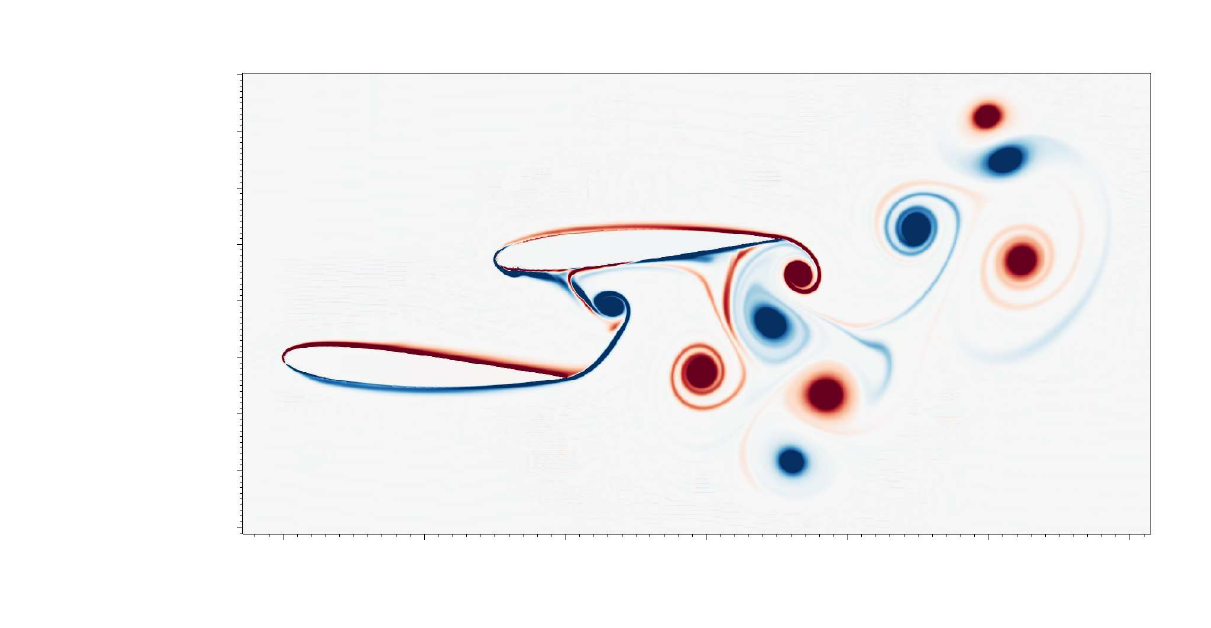}{(c) $ \lam = 8.0 $, $ t^*/T = 2.25 $}
		\addlabelvor{0.49}{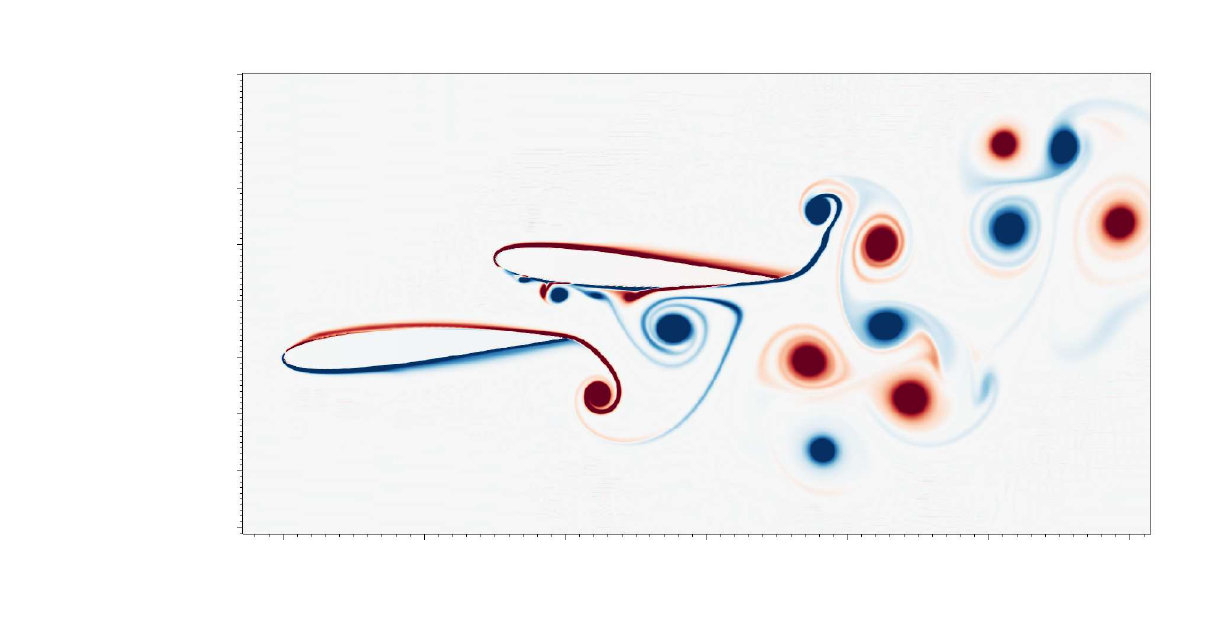}{(g) $ \lam = 8.0 $, $ t^*/T = 2.75 $}
		
		\addlabelvor{0.49}{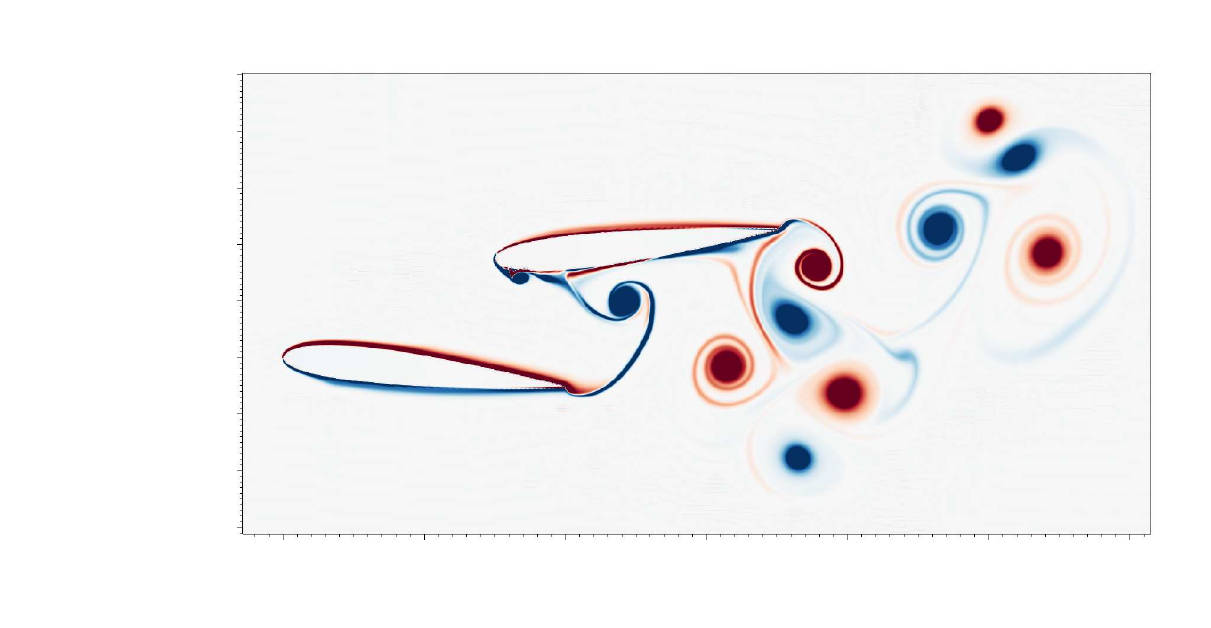}{(d) $ \lam = 8.0 $, $ t^*/T = 2.375 $}
		\addlabelvor{0.49}{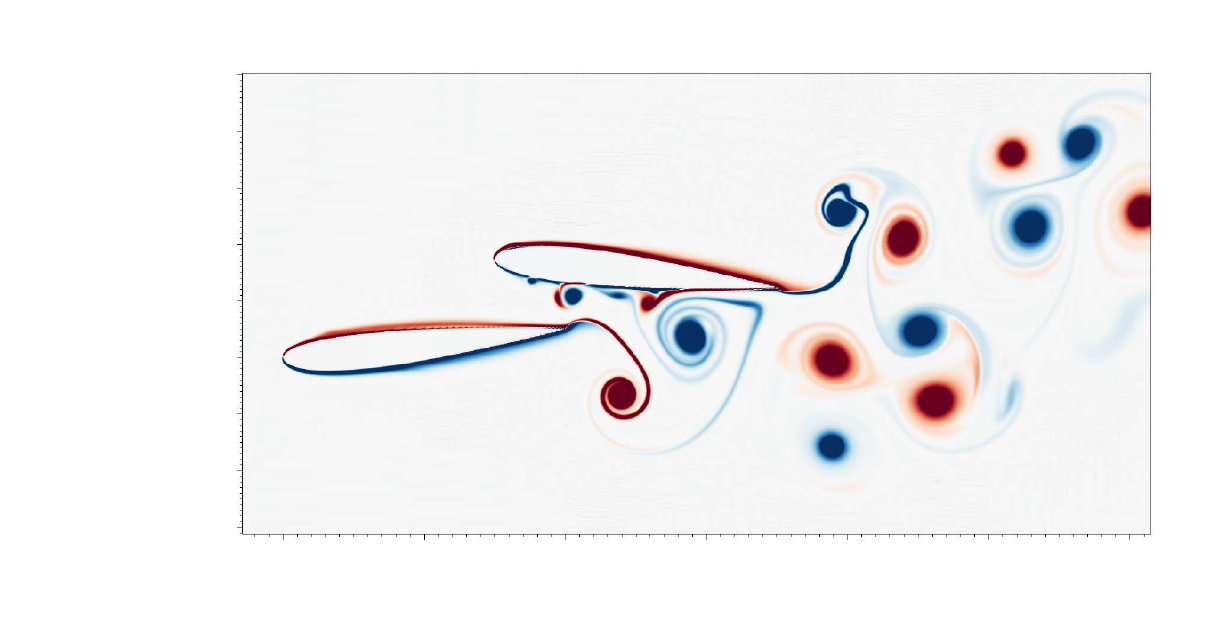}{(h) $ \lam = 8.0 $, $ t^*/T = 2.875 $}
		
		\addlabelnotrim{0.49}{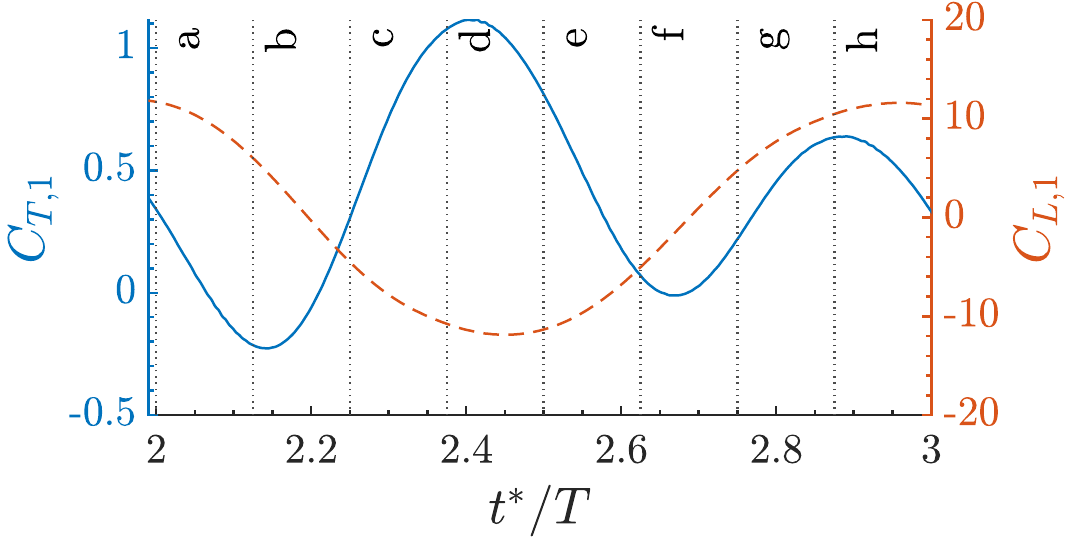}{(i) Leader}
		\addlabelnotrim{0.49}{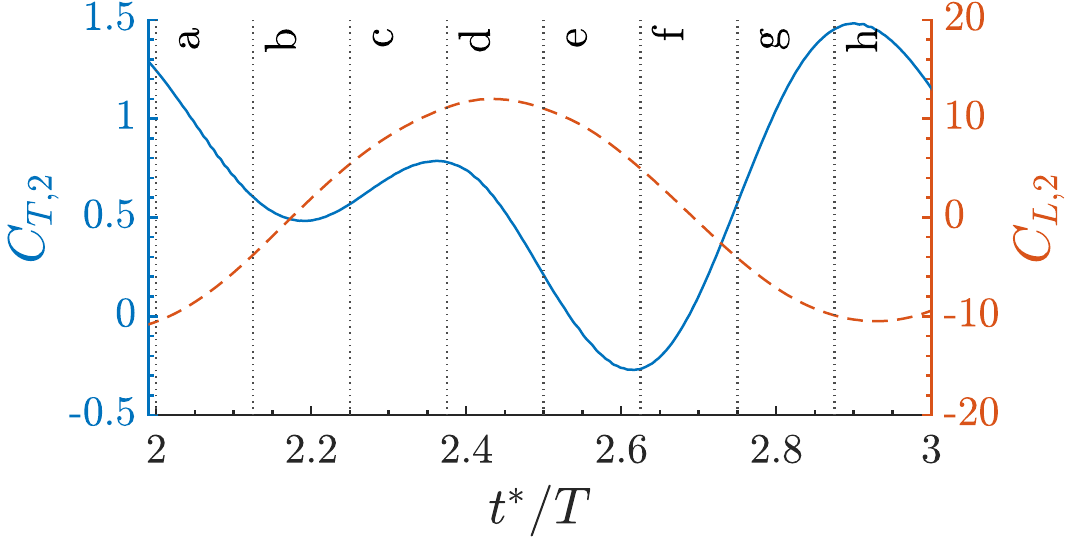}{(j) Follower}
		
		\caption{Vorticity contours and hydrofoil deformation with wavelength $ \lam = 8.0 $, side-by-side arrangement $ D = 0.75 $, lateral gap $ G = 0.35 $, anti-phase $ \ppi = 1.0 $ at instants of a typical period
			(a-h) $ t^*/T = 2.00-2.875 $. Time histories of thrust and lift coefficient for the (i) Leader and (j) Follower swimmers.}
		\label{fig:vor_1T_D0d75_lam8d0}
	\end{figure}

	At $ D =0.75,\ \lam = 8 $, the general flow structure is composed of several large vortices, as shown in \Cref{fig:vor_1T_D0d75_lam8d0}, rather than being broken into numerous small ones, as previously seen in \Cref{fig:vor_1T_D0_lam8d0}. Periodical interaction occurs in the region between the two foils, whereas in the wake flow, the vortex interaction is irregular and unpredictable.
	In the region between the two foils, the vortex pattern near the tail tip of the leader is very similar to that of a single pitching foil \citep{Thekkethil2018}, whereas a small vortex is generated from the left side of the follower's head, as seen in \Cref{fig:vor_1T_D0d75_lam8d0}a-\ref{fig:vor_1T_D0d75_lam8d0}h and then in the next cycle, this small vortex merges with the large vortex produced by the tail pitching of the leader. The merging of small and large vortices is stably repeated in every cycle of pitching in spite of the irregular wake flow.
	In the wake flow region, with the strong disturbance produced by $ \lam = 8 $, the front-back distance $ D =0.75 $ allows the interaction of vortices generated from different cycles and different swimmers. The vortex dipoles may swap their partners if collision between the dipoles occurs; the consequent new pair may draw a unique trajectory that further disturbs the wake flow.
	The thrust and lateral force upon the two foils is smooth in general, as seen in \Cref{fig:vor_1T_D0d75_lam8d0}i and \ref{fig:vor_1T_D0d75_lam8d0}j, corresponding to the relatively regular flow structure near the tail regions of the two foils. The lift force of the two foils are just opposite to each other $ C_{L,1} = -C_{L,2} $, \ie the amplitude of lift force is almost identical, yet the thrust force of the follower is generally larger than that of the leader. It is also interesting to note a half period phase difference between the leader and the follower's thrust force, which can only be caused by the interaction of vortex.

	\subsubsection{Summary for staggered $ D > 0 $ cases}
	
	This section provides a summary of the hydrodynamic characteristics for two foils swimming in staggered arrangement $ D > 0 $. The upstream swimmer is identified as the leader, while the downstream one as the follower.

	Thrust force $ \ctm $ for either swimmer generally increases with wavelengths $ \lam $.
	The follower's thrust force can be much larger than the leader's, \ie $ C_{Tm,follower} > C_{Tm,leader} $, especially at high $ \lam $ and $ \ppi = 1,\ 1.5 $.
	For the follower, the maximum achievable thrust force by tuning phase lag $ \phi $ decreases with both lateral $ G $ and front-back distances $ D $.
	The RMS of {lateral force} $ \clrms $ increases monotonically with wavelength $ \lam $, reaching as high as $ \clrms = 18 $ for both swimmers at short distances $ G = 0.25 $, $ D = 0.25 $ and anti-phase $ \ppi = 1 $.

	Propeller efficiency of either foil $ \eta_{1,2} $ or two foils as a group $ \etag $ reaches minimum and maximum at $ \lam = 0.8 $ and $ \lam = 2 $, respectively, and then gradually decreases at $ \lam > 2 $.
	The follower's efficiency $ \eta_{follower} $ is generally higher than the leader's $ \eta_{leader} $.
	Phase lag $ \phi $ is more influential to the follower's efficiency than the leader's, especially at high wavelength $ \lam > 2 $ and close front-back distance $ D \leq 0.5 $.
	The follower's propeller efficiency significantly decreases with a further front-back distance $ D $, while slightly increasing the leader's.

	Group efficiency $ \etag $ reaches a maximum of $ 33.3\% $ at $ D = 0.75 $ and $ \lam = 2.0 $.
	Group efficiency $ \etag $ generally increases with lateral gap $ G $ and front-back distance $ D $ across various wavelengths $ \lam $.
	At low wavelength $ \lam = 0.8 $, the group efficiency is especially sensitive to front-back distance $ D $, whereas effects of lateral gap $ G $ is less prominent.
	Phase lag $ \phi $ is more influential to group efficiency $ \etag $ at close distances across tested wavelengths $ \lam = 0.8 - 8 $.

	Vorticity distribution is examined to understand how various non-dimensional parameters affect the flow structure surrounding the two foils.
	The overall vorticity strength and the degree of irregularity both increase with the wavelengths $ \lam $ with various front-back distance $ D $, whereas the general wake flow pattern is only slightly affected by the phase difference $ \phi $.
	At $ \lam = 0.8 - 3.2 $, $ D = 0.25 $, the vortex dipoles shed by each hydrofoil do not mix in the wake flow, but bifurcating towards 2 distinct directions, forming a mirror symmetry pattern.
	At $ D \geq 0.50 $, the wake flow structure is more irregular due to unsteady interaction among dipoles that vortices can merge together to form a larger one.

	Flow-mediated interaction mechanism is examined in 8 instants within 1 period at $ D = 0.75 $ with low, intermediate, and high wavelengths $ \lam = 0.8,\ 2,\ 8 $.
	The staggered arrangement $ D > 0 $ allows vortices generated from different cycles of undulation/pitching to interact with each other, leading to a phase lag in the thrust force history. Distinct flow structure and force variation emerges with the change of wavelengths $ \lam $.
	At $ D =0.75,\ \lam = 0.8 $, the wake vorticity pattern looks as if each of the two foils swims as a single foil.
	At $ D =0.75,\ \lam = 2 $, the leader's upward vortices collides with the the follower's vortex street, causing great disturbance in the wake flow of the follower.
	\textit{Vortex swapping} periodically takes place between the two foils.
	The thrust force $ C_{T} $ of the two swimmers demonstrates a phase difference of about $ T/2 $.
	At $ D =0.75,\ \lam = 8 $, the general flow structure is composed of several large vortices. Periodical interaction occurs in the region between the two foils, whereas in the wake flow, the vortex interaction is irregular and unpredictable.
	In the region between the two foils, the vortex pattern near the tail tip of the leader is very similar to that of a single pitching foil, whereas a small vortex is generated from the left side of the follower's head that eventually merges with the large vortex at tail. 
	In the wake flow region, with the strong disturbance produced by $ \lam = 8 $, the front-back distance $ D =0.75 $ allows the interaction of vortices generated from different cycles and different swimmers. Vortex dipoles may swap their paired partners if collision occurs, further disturbing the wake flow.
	The thrust and lateral force upon the two foils is smooth in general, corresponding to the relatively regular flow structure near the tail regions of the two foils. The lift force of the two foils are just opposite to each other $ C_{L,1} = -C_{L,2} $. It is also interesting to note a half period phase difference between the leader and the follower's thrust force, which can only be caused by the interaction of vortex.

	\section{Conclusion}
	
	Fish swimming is a classic topic that involves rich physics \citep{Webb1984,Triantafyllou2000,Liao2007,Ashraf2017} with applications in biomimetics \citep{Duraisamy2019,Fish2020} and fish farming \citep{Webb2011}.
	Among the problems of fish swimming, abundant research exists for both fish schooling \citep{weihs1973hydromechanics,Weihs1975,Ashraf2016,Li2020} and swimming styles \citep{Sfakiotakis1999,Tytell2010,Cui2018,Thekkethil2017,Thekkethil2018,Thekkethil2020}.
	However, the combined effect of swimming style and schooling upon hydrodynamics of BCF swimmers has never been systematically examined in details.
	In the present paper, we investigate how swimming style affects fish schooling by a representative problem setup consisting of two NACA0012 hydrofoils undulating at various wavelengths $ \lam = 0.8 - 8 $, front-back distance $ D = 0,\ 0.25,\ 0.5,\ 0.75 $, phase difference $ \ppi = 0,\ 0.5,\ 1,\ 1.5 $, and lateral gap distance $ G = 0.25,\ 0.3,\ 0.35 $ with fixed Reynolds number $ Re = 5000 $, Strouhal number $ St = 0.4 $, and maximum amplitude $ A_{max} = 0.1 $. In total, 336 combinations were simulated by ConstraintIB module of IBAMR. Here, we classify the results by the relative front-back distance between the two foils as side-by-side $ D = 0 $ and staggered $ D > 0 $ conditions.

	The swimming style of BCF swimmers is represented by wavelengths $ \lam $. Low wavelength $ \lam < 1 $ corresponds to anguilliform swimmers, whereas high wavelength $ \lam \gg 1 $ for the thunniform ones.
	In the tested parametric space,
	the increase in wavelength $ \lam $ results in larger thrust $ C_{Tm} $ and monotonic increase of lateral $ C_{Lrms} $ force.
	Propeller efficiency of either foil $ \eta_{1,2} $ or two foils as a group $ \etag $ reaches minimum and maximum at $ \lam = 0.8 $ and $ \lam = 2 $, respectively, and then gradually decreases or remains constant at $ \lam > 2 $.
	It indicates high efficiency but low acceleration for swimmers of intermediate wavelength $ \lam = 2 $ and vice versa for high wavelength swimmers $ \lam > 2 $. Low wavelength $ \lam = 0.8 $ causes both low efficiency and low acceleration due to the fixed Strouhal number $ St = 0.4 $.
	The increase in wavelength also results in strong vortex and irregular flow structure.
	These effects of wavelength for schooling foils is consistent with that for the single foil cases by \cite{Thekkethil2018}.

	Phase difference $ \phi $ and front-back distance $ D $ can affect the interaction between vortex shed by the leader and undulating body of the follower \citep{Li2020}, thus greatly impact the consequent thrust force and propeller efficiency.
	The undulation phase difference $ \phi $ between the two schooling hydrofoils $ \phi $ can significantly affect the thrust/lateral force and the individual/group propeller efficiency, especially at close distances.
	The follower's force and efficiency is most sensitive to phase difference at intermediate or high wavelength $ \lam > 1 $,
	whereas group efficiency is most influenced by phase difference at low wavelength $ \lam < 1 $.
	In other words, by tuning the phase difference, low wavelength swimmers can school with significantly higher group efficiency. For the follower swimming at a high wavelength and close distances, phase tuning can effectively improve its acceleration and efficiency.
	The influence of phase difference $ \ppi $ can be more significant than wavelength $ \lam $ in some cases.

	Front-back distance $ D $ can greatly affect the follower's thrust force and propeller efficiency while exerting considerable impact on group efficiency at low wavelength $ \lam < 1 $ as well.
	A closer front-back distance generally results in advantageous acceleration and high efficiency for the follower, especially at high wavelength $ \lam \gg 1 $, \ie thunniform swimming style; however, a further distance is more beneficial to the group efficiency, especially at low wavelength $ \lam < 1 $, \ie anguilliform swimming style.
	At close front-back distances, phase lag $ \phi $ becomes more influential to group efficiency $ \etag $ across $ \lam = 0.8 - 8 $.
	It is an indication that, for schooling anguilliform swimmers, it can be critical to keep an appropriate front-back distance $ D $, which may even reverse the collective propulsive direction of the swimmer group; in contrast, the schooling performance for thunniform swimmers with high wavelength $ \lam > 6 $ is more stable; its group efficiency does not vary significantly with phase lag $ \phi $, especially at further front-back distance $ D \geq 0.75 $. This point is in support of the hypothesis that thunniform swimmers are more suitable for schooling in contrast with anguilliform swimmers.

	Lateral gap distance $ G $ can only slightly influence the leader-follower interaction in the tested range of $ G = 0.25 - 0.35 $.
	For the follower, the maximum achievable thrust force by tuning $ \phi $ and $ \lam $ slightly decreases with lateral gap $ G $.
	Group efficiency $ \etag $ generally increases with lateral gap $ G $ across various wavelengths $ \lam $.

	The leader and the follower are affected by these parameters in different ways.
	The follower's propulsive efficiency $ \eta_{follower} $ is generally higher than the leader's $ \eta_{leader} $.
	Phase difference $ \phi $ is more influential to the follower's efficiency than the leader's, especially at high wavelength $ \lam > 2 $ and close front-back distance $ D \leq 0.5 $.
	At high $ \lam \gg 1 $ and $ \ppi = 1,\ 1.5 $, the follower's thrust force can be much larger than the leader's.
	Vorticity distribution $ \omega^* $ is reviewed to understand how various non-dimensional parameters affect the flow structure surrounding two foils.
	The overall vorticity strength and the flow irregularity both increase with the wavelengths $ \lam $ regardless of front-back distance $ D $, whereas the general wake flow pattern is only slightly affected by the phase difference $ \phi $.
	Vortex dipoles are generated during foil undulation/pitching.
	At side-by-side $ D = 0 $ and anti-phase $ \ppi = 1 $, vorticity distribution can be symmetrical, since the two swimmers form a mirror symmetry geometry at any moment during the undulation process.
	However, symmetry breaking can occur after a number of initial periods, especially at high wavelength.
	At $ D = 0.25 $, $ \lam = 0.8 - 3.2 $, the shed vortex dipoles bifurcates towards 2 distinct directions, forming a mirror symmetry pattern.
	At $ D \geq 0.50 $, $ \lam = 0.8 - 3.2 $, the wake flow structure becomes irregular due to unsteady interaction among vortex dipoles.
	At low wavelength $ \lam < 1 $, the gap distance $ G $ and phase difference $ \phi $ can affect the skewness, symmetry and regularity of the wake pattern.
	At $ \ppi = 0.5 $ and $ \ppi = 1.5 $, the vortex shedding direction is slightly skewed towards the right and left sides of the swimming direction, respectively.
	The intensity of the vortices decreases with gap distance $ G $.
	Concentration of dynamic energy is discovered at low gap distance and when the two hydrofoils swim in anti-phase.
	Flow structure surrounding the two foils is examined at 8 consecutive instants within the 3rd cycle of undulation/pitching, \ie $ \tdT = 2-3 $, together with time histories of thrust $ C_T $ and lateral $ C_L $ force within that cycle. Side-by-side $ D = 0 $ and staggered $ D = 0.75 $ cases are studied at low $ \lam = 0.8 $, intermediate $ \lam = 2 $ and high $ \lam = 8 $ wavelengths, whereas the foil-to-foil phase difference remains constant at anti-phase $ \ppi = 1 $.
	
	At side-by-side $ D = 0 $ arrangement and anti-phase $ \ppi = 1 $, the most distinct flow characteristics is the \textit{symmetrical} flow pattern, which breaks easily at high wavelength. At {low wavelength} $ \lam = 0.8 $, the flow structure is highly symmetrical.
	At {intermediate wavelength} $ \lam = 2 $, where the highest energy efficiency is obtained, vortex dipoles form a streaming direction that directly points downstream.
	At {large wavelength} $ \lam = 8.0 $, the flow symmetry is broken, causing a complicated flow structure, whereas two additional small vortices emerge from the outer sides of two foils.
	
	At staggered arrangement $ D = 0.75 $ and anti-phase $ \ppi = 1 $, vortices shed from the leader can interact with those from the follower by from different cycles due to the front-back distance, leading to a delayed impact upon the foil body, thus a phase lag in the eventual thrust force history.
	At low wavelength $ \lam = 0.8 $, the vorticity pattern looks as if the each foil swims as a single foil.
	At intermediate wavelength $ \lam = 2 $, \textit{Vortex swapping} periodically takes place between the two foils.
	At high wavelength $ \lam = 8 $, the general flow structure consists of several large vortices. Periodical interaction occurs near the two foils despite the irregular wake flow.

	The hydrodynamic thrust and lateral force upon the two foils is smooth in general, except for the cases with side-by-side $ D = 0 $ arrangement and high wavelength $ \lam = 8 $, where the flow structure near the foils are highly fractured.
	At either side-by-side $ D = 0 $ or staggered $ D = 0.75 $ arrangement, the lift force of the two foils are generally opposite to each other $ C_{L,1} = -C_{L,2} $, especially at $ \lam > 2 $.
	At side-by-side $ D = 0 $ arrangement, thrust force upon the two foils is generally identical through the time histories as $ C_{T,1} = C_{T,2} $. In contrast, at staggered arrangement $ D = 0.75 $, a half period phase difference is discovered between the leader and the follower's thrust force, caused by the delayed vortex interaction due to front-back distance; the thrust force of the follower is generally higher than that of the leader.

	The low and high wavelength in this paper corresponds to anguilliform and thunniform swimmers, respectively.
	One attempt of the present paper is to test the hypothesis that thunniform swimmers are more adapted for schooling locomotion than anguilliform swimmers.
	Several biological implications can be derived from the above analysis.
	The group energy efficiency of anguilliform swimmers is more sensitive to relative distance and undulation phase difference, implying a less stable collective efficiency than the thunniform swimmers.
	While schooling together, the anguilliform swimmers tend to produce a more stable wake flow than the thunniform swimmers, disturbing less area of the fluid. This implies a better stealth performance for the anguilliform swimmers.
	At side-by-side arrangement, the thrust force produced by various swimming styles is roughly equivalent, given the in-phase or anti-phase coordination. This implies the side-by-side arrangement can be a stable formation regardless of swimming styles, given the maintenance of appropriate lateral distance. However, at such stable formation, anguilliform swimmers do not produce a beneficial propeller efficiency. This means that the anguilliform swimmers can be locked in a stable formation with undesirable efficiency, making it less suitable for schooling locomotion.
	Thunniform swimmers produce better thrust force, although not obtaining the best locomotion efficiency, which is coherent with single swimming foil as studied in \cite{Thekkethil2018}.
	For the above implications, the current results are in support of the hypothesis that thunniform swimmers are more hydrodynamically adapted to schooling locomotion than the anguilliform swimmers. These conclusions can also be useful for the schooling locomotion of fish-like robots.
	In the future, we intend to continue the present study by three-dimensional and self-propelled simulations.

	\section{Acknowledgement}
	We acknowledge the help from Yu Jiang for data processing. This work was funded by
	China Postdoctoral Science Foundation under Grant Number 2021M691865
	and
	Science and Technology Major Project of Fujian Province under Grant Number 2021NZ033016.
	A.P.S.B acknowledges support from US National Science Foundation award OAC 1931368.

	\FloatBarrier
	\clearpage
	
	\bibliographystyle{elsarticle-harv} 
	\bibliography{Fish_paper_1_submit_for_arXiv}

\end{document}